%% file: mcdermott.tex
\documentclass{JHEP}

\usepackage{epsfig}

\def\lsim{\mathrel{\rlap{\lower4pt\hbox{\hskip1pt$\sim$}}
    \raise1pt\hbox{$<$}}}                
\def\gsim{\mathrel{\rlap{\lower4pt\hbox{\hskip1pt$\sim$}}
    \raise1pt\hbox{$>$}}}                
\def\pmodd{\mathrel{\rlap{\lower4pt\hbox{\small{\hskip1pt$-0.39$}}}
    \raise4pt\hbox{\small{$+0.30$}}}}           
\newcommand{\alp}{\ensuremath{\alpha_{I\!\!P}}}
\newcommand{\pomeron}{I\!\!P}            

\title{A fresh look at diffractive $J/\psi$ photoproduction at HERA, with 
predictions for THERA.}

\author{L.~Frankfurt\\
        Nuclear Physics Dept., School of Physics and Astronomy,\\
        Tel Aviv University, 69978 Tel Aviv, Israel\\
        E-mail: \email{frankfur@lev.tau.ac.il}}
\author{M.~McDermott\\
        Division of Theoretical Physics,\\
        Dept. Mathematical Sciences, Liverpool University\\
        Liverpool L69 3BX, England\\
        E-mail: \email{martinmc@amtp.liv.ac.uk}}
\author{M.~Strikman\\
        Department of Physics, Penn State University,\\
        University Park, PA, 16802-6300, USA.\\
        E-mail: \email{strikman@phys.psu.edu}}
        
\preprint{\hepph{0009086}}

\abstract{
We quantify perturbative and non-perturbative QCD effects in the exclusive 
$J/\psi$-photoproduction cross section, and in the shrinkage of the 
differential cross section with respect to momentum transfer, $t$. 
We predict that in the high energy THERA region there will always be a significant contribution to this process that rises quickly with energy.
This implies that the taming of the rise of the cross section with energy, due to both 
the expansion of spatially-small fluctuations in the photon
and to higher twist effects, is rather gradual.}


\begin{document}

\section{Introduction}

One of the major challenges facing particle physics is to understand the interplay of 
perturbative and non-perturbative QCD effects in the high-energy production of heavy quark bound states   
and in the interaction of such states with hadrons. 
Understanding the details of this interplay is interesting in its own right because it reveals the 
practical boundaries of applicability of perturbation theory. It is also  
necessary to achieve an unambiguous interpretation of many related phenomena,  
including the suppression of the yield of $J/\psi$ mesons produced 
in nucleus-nucleus collisions. The investigation of hard exclusive processes,  
such as the exclusive photoproduction of the $J/\psi$ meson in photon-proton collisions 
considered here, gives a unique opportunity to quantify this physics.

For several years, high energy, hard exclusive and semi-inclusive 
processes, which include photo- and electroproduction of heavy vector 
mesons and deep inelastic production of light vector mesons and real 
photons (DVCS), have been modelled reasonably successfully using the 
exchange of two gluons in a colour singlet configuration 
\cite{fs89,ryskin,brod,fks1,fks2,fms,suzuki,hoyer}.

For vector meson production initiated by longitudinally polarised photons 
\cite{brod,cfs} and for DVCS \cite{cf} such calculations have now been placed 
on a firmer theoretical footing by the proof of QCD factorization theorems 
which show that perturbative two gluon exchange is the dominant process 
in the asymptotic limit ($Q^2 \to \infty$).

To make a process ``hard'' it is necessary to provide a large momentum scale 
(either a heavy quark mass or $Q^2$, or both) which squeezes the hadronic 
fluctuation of the photon so that small perturbative $q {\bar q}$ 
configurations are responsible for the dominant contribution. 
For the diffractive photoproduction of $J/\psi$ this hard scale is thought 
to be provided by the charm mass. However, since the charm quark is 
rather light ($m_c \approx 1.5$~GeV), and 
transverse polarisations of the quasi-real photon dominate, 
there are likely to be significant contributions 
from non-perturbative regions in which the hadronic fluctuation of the photon 
has a large transverse size  (these become progressively 
less important as the photon virtuality increases).
This interplay between soft and hard contributions has been seen 
in a phenomenological way by the success of the two-Pomeron fit of 
Donnachie and Landshoff \cite{dljpsi} in which the soft Pomeron term, 
associated with non-perturbative effects, appears make a significant 
contribution to $J/\psi$ photoproduction at HERA energies.

At very high energies, or 
small $x_{{\tiny \mbox{Bj}}}$, the time taken for a given fluctuation in the 
photon to interact with the proton target is considerably smaller 
than its formation time and the time required to produce the hadronic 
final state. This implies universality for the interaction cross section
over a wide range of inclusive and exclusive processes. 
In \cite{mfgs} we introduced a model for this universal cross section, 
${\hat \sigma}$, for all transverse size fluctuations\footnote{Modelling is currently unavoidable in QCD due to the necessity to take non-perturbative confinement effects into account. We defer a more detailed modelling of 
the strong interaction in the region of small $\alpha_s$,  
which seems to be required for the very small-$x$ regime in which Structure Functions may achieve the unitarity limit, for future studies.}.
We produced a satisfactory 
description of the inclusive cross section data, indicating that 
our model is reasonable. In this paper, having made some suitable 
minor adjustments,  we apply the model to $J/\psi$ photoproduction. Our 
aim is twofold, to provide a good description of this process 
(and hence to elucidate the role of non-perturbative physics 
in $J/\psi$ photoproduction) and
to further constrain the model in order to be able to make 
better predictions for other exclusive processes in the future, using 
the same framework.

Our model for the interaction cross-section is based on the well-known 
leading-log perturbative QCD result for 
the interaction of  a small transverse-size 
$q {\bar q}$ dipole which proceeds via two-gluon exchange \cite{bbfs}:

\begin{equation}
{\hat \sigma}_{pQCD} (b^2,x) = \frac{\pi^2}{3} \, b^2 \, \alpha_s 
({\bar Q^2}) \, x g(x^{\prime},{\bar Q^2}) \, , 
\label{sighat}
\end{equation}
where scales $x^{\prime}$ and ${\bar Q}^2$, which depend on 
transverse size $b$, are described 
in Subsection \ref{sect3.1}. This form is applicable for transverse sizes 
$b < b_{Q0} \approx 0.4$~fm. For larger transverse sizes we 
introduced an ansatz based on the known behaviour of soft 
hadronic interactions 
(we matched on to the measured $\pi p$ cross section at 
$b = b_{\pi} = 0.65 $~fm) and introduced a suitable interpolation for 
$b_{Q0} < b < b_{\pi}$ and an extrapolation for $b > b_{\pi}$. In 
this paper we slightly 
simplify the model for ${\hat \sigma}$ by connecting the points 
$b_{Q0}$ and $b_{\pi}$ with 
a straight line (instead of extrapolating a fit to the known $b$-shape just 
below $b_{Q0}$ as in \cite{mfgs}). At very high energies, we tame the steep 
increase due to the rapid rise of the small-$x$ gluon density
by imposing a unitarity restriction, 
$\sigma(b_1,W^2) = \sigma(b_{\pi},W^2)/2 $, and connect the point 
$b_1 < b_{Q0}$ to $b_{\pi}$ using a straight line 
(see figs.(\ref{sigspdfc},\ref{fig:sighat}) and Subsection \ref{sectskew} for more details).

We use this model, based on eq.(\ref{sighat}) with CTEQ4L \cite{cteq4l} 
gluon input 
density, evolved using skewed evolution, to investigate 
the interplay of perturbative
 and non-perturbative effects in $J/\psi$ 
photoproduction\footnote{The value of ${\hat \sigma}_{pQCD} (b^2,x) $ 
should not change significantly within the next-to-leading order 
approximation, as compared to the leading order, because it is determined from 
fitting the same value of $F_{2}$. However, both the relative size of  
${\hat \sigma}_{pQCD} (b^2,x) $ and $xg(x,Q^2)$ and the numerical value of 
$xg(x,Q^2)$ itself may change rather significantly.}.
 At the higher photon-proton energies of the 
HERA range ($W \approx 250$~GeV) this involves 
probing the gluon distribution at very small $x \approx 10^{-4}$, i.e. 
outside of the range in which it has been tested directly. In this region, 
it is mainly constrained indirectly 
by its effect on the observed scaling violations of the inclusive 
structure function $F_2$, predicted by DGLAP \cite{dglap} evolution. It 
should be remembered that the  
predictions we make would change if the input gluon density is changed.
Indeed, in our final plot for the cross section we also show the results using 
the latest MRST 
leading order partons \cite{dick}\footnote{This leading order 
input distribution 
is more recent than, but similar to, that found in \cite{mrstl}.}. 
Taken together, the predictions using both CTEQ4L and MRST produce a 
spread which spans the currently available data. 
Other models for the interaction 
cross section, ${\hat \sigma}$,  may be found in the literature 
(see e.g. \cite{dipoles, amirim} and references therein).

The amplitude for exclusive processes involves a convolution 
of ${\hat \sigma}$ with 
light-cone wavefunctions for the initial-state 
photon (known from QED \cite{photqed} 
in the $q {\bar q}$ case) and for the final-state 
diffractively-produced object.
For the light-cone wavefunction of the $J/\psi$ we use 
a hybrid wavefunction introduced in \cite{fks1}. 
This is derived by solving the Schr\"{o}dinger 
equation using the Buchmueller-Tye potential 
model \cite{btye} (constituent quark mass $m_c = 1.48$~GeV),
boosting the resulting Schr\"{o}dinger wavefunction to a fast moving frame.
The wave function at small $b < 0.3 $~fm is fixed by imposing QCD behaviour 
($\phi_V(z,b=0) \propto z(1-z)$) and normalized using the known leptonic 
decay rate of $J/\psi$.  In this way we account for the QCD radiative
corrections to the $q\bar q$ component of the charmonium 
wavefunction for small transverse sizes $b < 0.3 $~fm.
This effectively takes into account the strong modification of
non-relativistic charmonium-model predictions for the leptonic width due to
QCD radiative corrections (for a recent review see \cite{ynd}). 
The exclusive formation of the heavy bound state, 
the details of which are embodied in the light-cone wavefunction,  
suppresses the contribution of higher order Fock states 
(such as $|c {\bar c} g\!>$) in the virtual photon, making this 
a particularly good process in which to test the universality of the 
dipole picture. 

For small enough $x$, the rapid rise of ${\hat \sigma}$ due 
to the small-$x$ rise of the  gluon density should be tamed
to avoid violating the unitarity restriction for the interaction of spatially-small 
partonic fluctuations of the photon. Two types of effects act to tame the increase of high energy processes. As the energy increases, spatially-small partonic fluctuations in the photon 
typically expand rapidly in transverse size due to the increased phase-space for radiation. 
This expansion in size, known as Gribov diffusion, is a consequence of the randomness of radiation.
For large size configurations perturbative QCD is inapplicable so the amplitudes no longer 
have this fast increase with energy. 
We account for Gribov diffusion to some extent by implementing a behaviour typical of 
soft hadronic cross sections (i.e. with only mild increase with energy) at large transverse sizes.
Even if ${\hat \sigma}$ is completely independent of $b^2$ at sufficiently 
small-$x$ the Structure Function of a hadronic target should retain a residual 
$\ln(1/x)$ increase with energy as a result of the infinite normalization of 
the wavefunction of  a virtual photon (the infinite renormalization of electric 
charge is due to hadronic vacuum polarisation). Another effect is taming, or shadowing, 
of the rapid rise with energy due to higher twist effects. So, for small enough $x$, 
unitarity corrections were introduced in \cite{mfgs} to tame the rapid rise of 
${\hat \sigma}$ due to the small-$x$ rise of the gluon distribution. 
These affect larger transverse sizes in the perturbative domain first.
If perturbative QCD effects are tamed by the unitarity of S-matrix only, 
this mechanism would lead to an increase of the typical impact parameters, $\rho$, 
involved in the scattering of $q {\bar q}$ dipole from a nucleon as
$\rho^2 \propto \ln^2 1/x$ and leads to a related increase of Structure Functions
as $F_2 \propto (\ln 1/x)^3$ (and $\alpha'~\propto~\ln^2 1/x$). 

In this paper, we illustrate that the photoproduction of the relatively 
light $J/\psi$ can act as a precursor for this taming since it is sensitive 
to relatively large transverse sizes, not just to the wavefunction 
at the origin \cite{fks1,fks2}.
Thus, we consider the behaviour of the energy dependence of the cross section of
$J/\psi$-photoproduction (especially that of the slope of the momentum transfer, $t$, which is 
parameterised by $\alpha_{\pomeron}^{\prime}$) to be crucial in establishing the existence of 
a new regime in which the standard DGLAP approximation is violated. 
Remember that the slowing down of the increase of the Structure 
Function alone does not necessarily imply violation of DGLAP, because it may be 
due to Gribov diffusion, which is a leading twist phenomenon.
We note in passing that an important role of large transverse distance 
effects (soft QCD) reveals itself in the energy dependence of the slope for 
the photoproduction of $J/\psi$ mesons which, within our model, 
should be intermediate between the soft regime, 
$\alp^{\prime} (\mbox{soft}) \sim 0.25$~GeV$^{-2}$, 
and the perturbative regime, $\alp^{\prime} \ll  \alp^{\prime}(\mbox{soft})$ 
(see Subsection \ref{sectbw} and Section 6).

We predict a reduction in the steepness of the energy dependence of
amplitudes of hard exclusive processes which will begin to take effect at
the higher HERA energies and in the THERA region ($250 \le W \ge 1000$~GeV).
The precise details are of course specific features of our model,which
incorporates a simple taming ansatz in the small $x$ region.
More generally, taming corrections are eventually expected to reduce the
rise with energy of all hard small-$x$ cross sections at given impact parameter
\footnote{However, it is unclear whether the unitarity of $S$-matrix for the interaction of 
small size fluctuations should tame the increase with energy of Structure 
Functions, because of the related increase of the important impact parameters with energy.
This question will be investigated in a separate publication.}.

This paper extends the work of \cite{fks2,fms} in two original directions. 
Firstly, it 
explicitly includes the non-perturbative component coming 
from large transverse sizes and provides a 
reasonable unified description of the process from the low 
energies measured at fixed target experiments  
\cite{fixedtarget} as well as the HERA data \cite{h100,bruni,osaka}. 
Within the framework of our analysis 
the relative contribution of non-perturbative effects may be 
quantized, albeit in an inevitably model-dependent way.
Secondly, we make predictions for energies beyond the HERA range 
$W \gsim 300$~GeV
that may eventually be tested at a higher energy $ep$ collider such 
as the proposed 
Tesla-HERA, or THERA, project (see e.g. \cite{mklein}). The logic 
of \cite{mfgs} dictates that 
taming corrections are required in this high energy region and 
we qualify and quantify 
the expected effect of these corrections on $J/\psi$ photoproduction.

The paper is organized as follows. Section 2 contains the basic 
formula for the cross section. 
Section 3 investigates various issues surrounding the 
implementation of this formula including 
setting scales in the gluon distribution, running quark mass 
effects, $t-$dependence, skewedness and 
the calculation of the real part of the amplitude. In Section 4 
we illustrate the 
effect of changing the rescaling parameter $\lambda$, in the 
model for the dipole cross section.
Section 4 also contains our predictions for the THERA energy range. 
Following a discussion and evaluation of $\alpha_{\pomeron}^{\prime}$
in Section 5, we include a general discussion in 
Section 6 and conclude in Section 7.

\section{Basic Formulae in the photoproduction limit}

From eq.(50) of \cite{fks2} in the limit $Q^2 \to 0$, neglecting the 
real part of the amplitude, the differential cross section 
for the photoproduction of $J/\psi$ reads:
\begin{equation}
\frac{d \sigma}{dt} \left|_{t=0} \right. = \frac{12 \pi^3 \Gamma_V M_V^3}
{\alpha_{e.m} (4m_{c}^2)^4} 
~\left[\frac{\pi^2}{3} \alpha_s (Q^2_{\mbox{eff}}) xg(x,Q^2_{\mbox{eff}}) \right]^2  
~(\frac{3}{\pi^2})^2   ~C(Q^2=0) \, , 
\label{dst1}
\end{equation}

\noindent where $M_V, \Gamma_V$ are the mass and leptonic decay width of the vector meson and $m_c = 1.5$~GeV is the charm quark mass. Using the notation of \cite{fks2} (in particular eqs.(25,26,51)) we have for the overall 
dimensionless suppression factor
\begin{eqnarray}
C(Q^2=0) & = & \left( \frac{\eta_V m_c^4}{3} \right)^2  T(0) R(0) \label{cq0} \, , \\
T(0) R(0) & = & \frac{1}{M_V^4} \frac{\left[\int \frac{dz}{z(1-z)}
 ~\int db ~b^3 ~m_{c,r}^2 ~\phi_V ~\phi^{T}_{\gamma} \right]^2}{\left[\int \frac{dz}{z(1-z)} 
~\phi_{V}(z,b=0) \right]^2} \label{tr0} \, , \\
\left(\frac{\eta_V}{3} \right)^2 & = & \left( \frac{ \int \frac{dz}{z(1-z)} 
~\phi_V (z,b=0) }{6 \int dz ~\phi_V (z, b=0)} \right)^2 \label{eta3} \, , 
\end{eqnarray}
\noindent where $m_{c,r} (b)$ is the running charm quark mass and $\phi^{T}_{\gamma}, \phi_{V}$ are the light-cone wavefunctions for the transversely-polarised photon and vector meson, respectively. The latter depend on transverse size 
$b$, and on $z$, 
the momentum fraction of the photon energy carried by the quark. 
In eq.(\ref{dst1}) the gluon density and $\alpha_{s}$ have been 
extracted at the average point, $<\! b \!>$, of the integration over $b$ in the amplitude, using the relationship $Q^2_{\mbox{eff}} = \lambda / \left<b^2\right>$, with  $\lambda \approx 10$.
In \cite{fms} we attempted to go further than this average approximation 
by sampling these functions at the  ${\bar Q^2} = \lambda/b^2$, underneath the integral in transverse size $b$.
The factor $m_{c,r}^4$ implicitly depends on $b$, so rightfully also belongs  
underneath the integral in $b$, sampled at an appropriate scale.

Reinstating ${\hat \sigma}$ underneath the integral according to this 
procedure leads to a modified version of eq.(\ref{dst1}):
\begin{eqnarray}
\frac{d \sigma}{dt} |_{t=0} & = & \frac{12 \pi^3 \Gamma_V } {\alpha_{e.m} 6^2 4^4 M_V} 
\frac{9}{\pi^4} \frac{\left[\int \frac{dz}{z(1-z)}
 \int b ~db   ~m_{c,r}^2 ~\phi^{T}_{\gamma} ~{\hat \sigma} ~\phi_V \right]^2}
{\left[\int dz \phi_{V}(z,b=0) \right]^2}  \nonumber \, , \\
                            & = & \frac{1}{16 \pi} \frac{3}{16} 
\left[ \frac{\Gamma_V}{\alpha_{e.m} M_V [\int dz \phi_V]^2 } \right] 
|\Im m~{\cal A}|^2 \nonumber \, , \\
                            & = & \frac{1}{16 \pi} \frac{3}{16} 
\left[ \frac{32 \pi \alpha_{e.m} e_c^2}{M_V^2} \right] |\Im m ~{\cal A}|^2 \nonumber \, , \\
                            & = & \frac{N^2}{16 \pi} |\Im m ~{\cal A}|^2 \, ,
\label{dst2}
\end{eqnarray}
\noindent where the penultimate line makes use of eq.(40) of \cite{fks2}. The 
imaginary part of the amplitude, $\Im m ~{\cal A}$, and its normalization, $N$, 
are given by
\begin{eqnarray}
N^2 & = & \frac{6 \alpha_{e.m} \pi e_c^2 m_c^4}{M_V^2} \label{nsq} \, , \\
\Im m ~{\cal A}   & = &  \int b ~db ~I_{z} (b) ~{\hat \sigma} \, , \label{ima} \\ 
I_{z} (b) & = & \int_{0}^{1} \frac{dz}{z(1-z)} ~(\frac{m_{c,r} (b)}{m_c})^2 ~\phi^{T}_{\gamma} ~\phi_V \, , 
\label{eqintz} 
\end{eqnarray}
where the light-cone wavefunction for the photon, 
$\phi_{\gamma}^T = K_0 (b ~m_{c,r})$, is purely transverse for 
photoproduction. 
In this equation we have chosen to separate out a piece of the $b$-integral, 
$I_{z} (b)$,  which only depends on the light-cone wavefunctions of the vector meson and photon and 
is independent of energy.

For small dipole sizes $b < 0.3$~fm QCD behaviour is imposed \cite{fks2} on 
$\phi_{J/\psi}$, so it is appropriate to run the charm quark mass underneath 
the b-integral too, in both the overall $m_{c,r}^2$ factor and in the 
argument of the Bessel function, using the appropriate renormalization 
group equation for masses (see Subsection \ref{sectmr}).

Finally, assuming the usual exponential fall-off in $t$ we have
\begin{eqnarray}
\sigma( \gamma P \to J/\psi ~P)  = \frac{N^2 (1 + \beta^2) |\Im m ~{\cal A}|^2}{16 \pi B} \, ,
\label{photonaive}
\end{eqnarray}
\noindent where the real part of the amplitude has been reinstated via 
$\beta = \Re e ~{\cal A} / \Im m ~{\cal A} $ (see Subsection \ref{sectre}). 
The H1 collaboration recently reported \cite{h100} a value of 
$B = 4.73 \pm 0.25 \pmodd $~GeV$^{-2}$.  Recently ZEUS 
reported \cite{bruni} an improved measurement of $B$ 
in $J/\psi$ photoproduction and found it to depend on energy. 
We will incorporate this shrinkage using a simple form 
(see Subsection \ref{sectbw}).

\section{Improvements to the basic formula: uncertainties in the cross section}

In this section we expand and explain various features and uncertainties in the 
basic formulae of Section 2. Many crucial issues involve the $b$-integral 
in eq.(\ref{ima}).  Firstly we explain the choice of scales, $x^{\prime}, {\bar Q}^2$ used in 
the dipole cross section and make a first comparison to the available data, using 
very basic assumptions.  
Then we consider the changes induced by considering a more careful 
treatment of running mass effects, shrinkage, non-zero real part and skewedness. 
The latter requires that we replace the ordinary gluon with the skewed gluon 
in eqs.(\ref{sighat},\ref{ima}). We follow the usual choice of conventional input 
distributions evolved using skewed evolution. 

Lastly, in the next Section, 
we decrease the parameter $\lambda$ which relates transverse sizes to four-momentum scales, via $Q^2 = \lambda/b^2$, from $10$ to $4$.
This has the effect of increasing the non-perturbative contribution to the 
cross section and leads to a much better description of the current data. 
We also give predictions for THERA for both values of $\lambda$.

\subsection{Scale setting in the gluon density}
\label{sect3.1}

In \cite{mfgs} we examined inclusive structure functions, $F_L$ and $F_2$, using our model for ${\hat \sigma}$. 
We employed the following prescription for the b-dependence of 
the scales $x^{\prime}$ and ${\bar Q}^2$:
\begin{eqnarray}
x^{\prime} & = & x_{min}^{\prime} (1 + 0.75 \frac{<\!b\!>^2}{b^2}) \, , \label{eqxprim} \\
{\bar Q}^2 &=& \frac{\lambda}{b^2} \, , \label{qbar}
\end{eqnarray}
\noindent with, 
\begin{eqnarray}
x^{\prime}_{min} &=& x_{Bj} ~(1 + \frac{4 m_q^2}{Q^2}) \, , \nonumber \\
<\!b\!>^2   &=& \frac{\lambda}{Q^2 + 4 m_{q}^{2}} \, .
\label{eqdet}
\end{eqnarray}

By examining the standard perturbative QCD formula for $F_L$ for light quarks we found $<\!x\!> \approx  1.75 ~x$ (we designed the ansatz to reproduce this for large 
$Q^2$ when $b = \left<b\right>$). We also found that a value of  $\lambda = 10$ 
reproduced the perturbative QCD results fairly well. 
However the results turned out to be rather insensitive to the precise 
value of $\lambda$. In this paper we  
explicitly examine the sensitivity of $J/\psi$ photoproduction on $\lambda$ 
and find that it is much more sensitive to it than $F_{L}$ and $F_{2}$ (see Section 4). 
This is interesting because the value of $\lambda$ determines the 
dividing line between perturbative and non-perturbative physics.

In the photoproduction limit, $Q^2 \to 0$, for charm 
eqs.(\ref{eqxprim}, \ref{eqdet}) lead to
\begin{equation}
x^{\prime} \to \frac{4 m_{c}^{2}}{W^2} (1 + 0.75 \frac{\lambda}{b^2 4m_{c}^{2}}) \, ,
\label{eqxprimlim}
\end{equation}
\noindent for the $b-$dependent momentum fraction of the incoming gluon. This will be 
used for the  sampling of ${\hat \sigma}$ underneath the $b$-integral in eq.(\ref{ima}).

\begin{figure}[htbp]
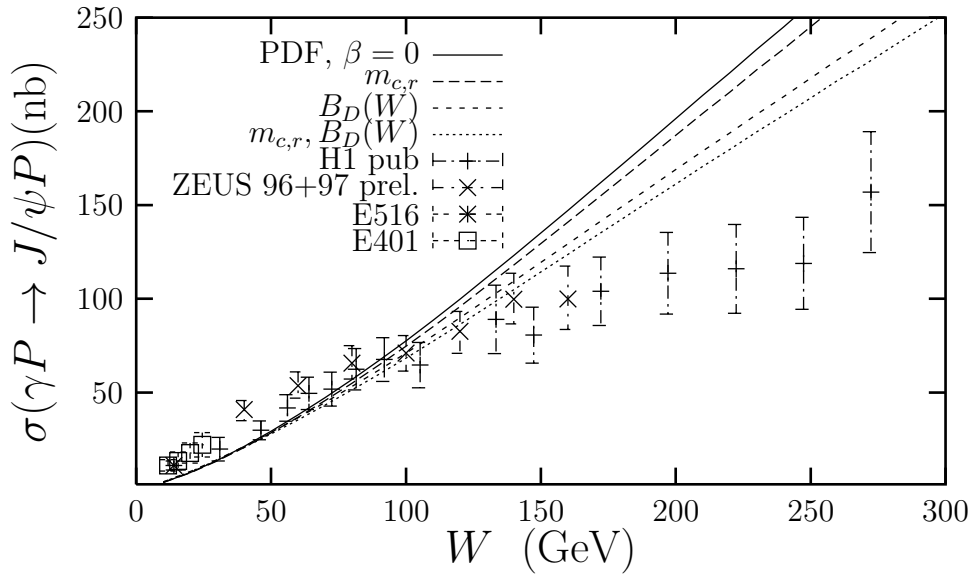

\begin{center}
\include{crosspdf}
\caption{A comparison of the $J/\psi$ photoproduction cross section, using conventional CTEQ4L parton distribution function  (PDF) for the gluon and $\beta = 0$, with data \cite{fixedtarget,h100,bruni,osaka}. The solid curve has fixed mass and slope $B = 4.0 $~GeV$^{-2}$, the long and short dashed curves include running charm quark mass and $W$-dependent slope, respectively. 
The dotted curve includes both effects.}%
 \label{sigpdf} 
\end{center}
\end{figure}

The solid line in fig.(\ref{sigpdf}) shows the resulting $J/\psi$ photoproduction cross section 
using eq.(\ref{photonaive}), with CTEQ4L gluon density, $\beta = 0$, a fixed slope parameter of 
$B = 4.0 $ ~GeV$^{-2}$ and fixed quark mass of $1.5$~GeV. 
The available data \cite{fixedtarget,h100,bruni,osaka} is also shown. 
This curve undershoots the data at low energies and overshoots at high energies, so it appears to rise too steeply with energy to provide a good description of all available data. We now consider several features which improve 
the shape of the energy dependence.

\subsection{Running quark mass}
\label{sectmr}

In \cite{fks2} hard QCD corrections were introduced in the vector meson wavefunction: 
for $b<b_0$ ($b_0 = 0.3 $ fm was chosen for $J/\psi$) QCD-behaviour 
was imposed ($\phi_V \propto z(1-z)$) and normalized to the measured leptonic decay width. 
For consistency it was then necessary to replace the constituent quark mass with 
the running quark mass for small $b$. 
This was implemented in the following way: (cf. eq.(38) of \cite{fks2}) 
\begin{eqnarray}
m^2  & \to m_{c,r}^2 (Q^{2}_{\mbox{eff}}) & = m_{c}^2 
~(1 - \frac{8 \alpha_s (Q^2_{\mbox{eff}})}{3 \pi}) \, . \nonumber
\label{msqq}
\end{eqnarray}

\begin{figure}[htbp]
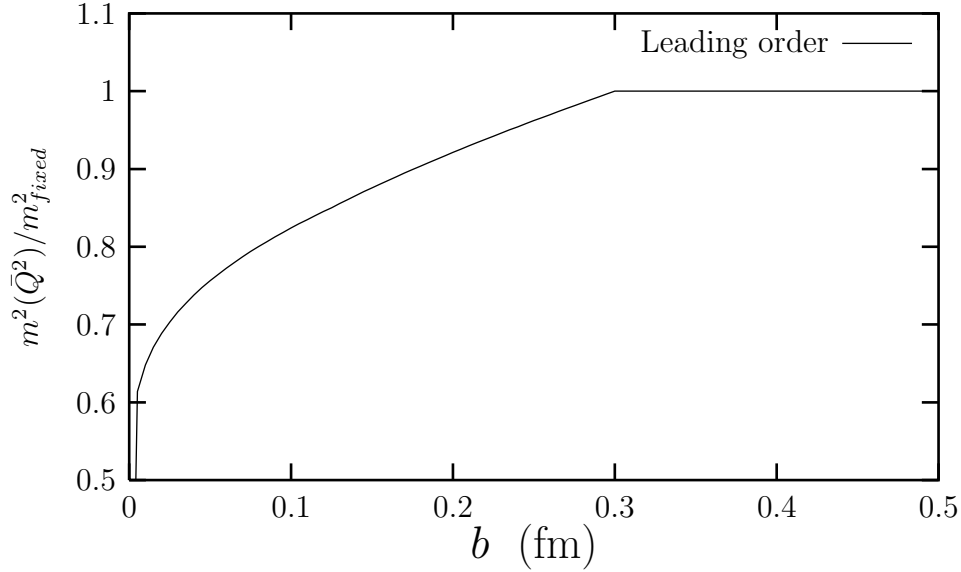

  \begin{center}
        \include{mrun}
\caption{The ratio of running charm quark mass squared to fixed mass squared as a function of transverse size using eq.(\ref{mqrun}) and incorporating four light flavours, $n_{f} = 4$.}
    \label{mrun}
  \end{center}
\end{figure}

\begin{figure}[htbp]
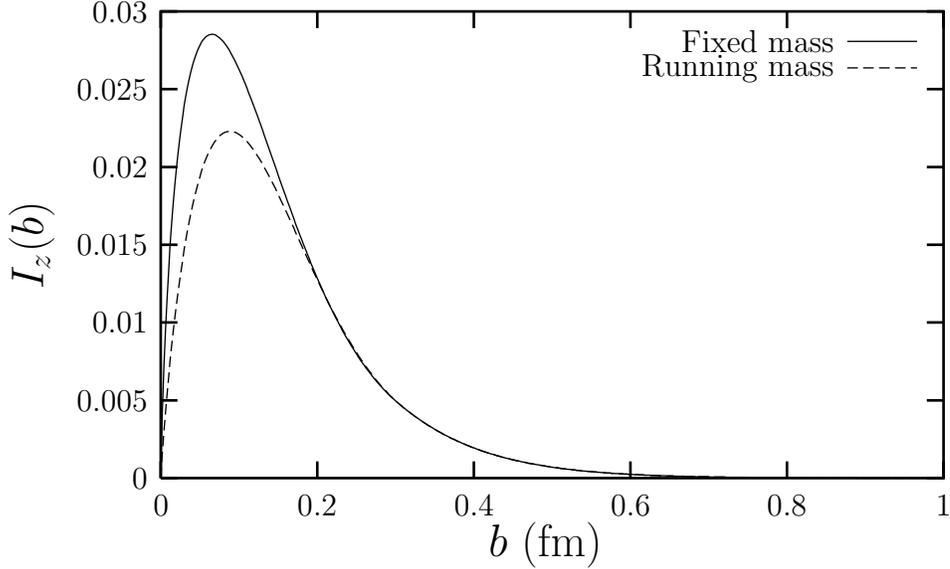

  \begin{center}
        \include{intz}
\caption{The integrand multiplying ${\hat \sigma}$ for fixed and 
running charm quark mass. 
The running mass effect is implemented for $b < 0.3 $~fm and 
causes a suppression of the small $b$ region.}
    \label{intz}
  \end{center}
\end{figure}

In this paper, we demand that the mass satisfies the renormalization group 
formula for quark masses (see e.g \cite{peskin}): 
\begin{equation}
m_{c,r} ({\bar Q}^2) = m_{c} \left( \frac{\alpha_s({\bar Q}^2)}{\alpha_s({\bar Q}_{f}^2)}
\right)^{\frac{12}{33 - 2 n_{f}}} \, ,
\label{mqrun}
\end{equation}
\noindent where $Q_f \approx 2.0$~GeV corresponds to the matching scale, 
$b_0 = 0.3$ ~fm, at which hard corrections are applied. We set the number of flavours $n_f = 4$ both in the exponent and in the one-loop beta-function of $\alpha_s$ (choosing $n_f=3$ instead would make very little difference).  
This effect suppresses the small $b$ region relative to the case in 
which the mass is taken as fixed, imposing the desired QCD behaviour 
$m_c \to 0 $ for ${\bar Q}^2 \to \infty $. 
Fig.(\ref{mrun}) shows $m_{c,r}^2(b)/m_{c}^2$ versus $b$. 
We use $m_{c} = 1.5$ ~GeV, which is approximately the constituent quark mass of 
Buchmueller-Tye potential model \cite{btye},  
which we use to construct $\phi_V$. Fig.(\ref{intz}) illustrates the size of the effect at the 
amplitude level. Running the quark mass influences both the argument of the $K_0$ Bessel function and 
the overall multiplicative factor in eq.(\ref{eqintz}). 
It turns out that there is only significant suppression for $b < 0.2 $~fm 
(rather than $b < b_{0} =0.3 $ fm). A comparison of the solid and long-dashed lines of 
fig.(\ref{sigpdf}) illustrates the rather small decrease that this change  
induces in the cross section 
as a result of the reduced contribution from small $b$.

\subsection{Energy dependent slope}
\label{sectbw}
We shall evaluate this effect in detail in the section 5. Here we give
a brief outline of the significance of the energy dependence of the slope 
of the $t$-dependence. The recent presentation of ZEUS data, 
from the 1996-97 running period, 
at DIS2000 \cite{bruni, osaka} indicated evidence for shrinkage. 
Following a fit to $d \sigma /dt$ of the form 
$\exp(B t)$ in each of the seven bins in $W$, a value 
of $\alp^{\prime} = 0.098 \pm 0.035 \pm 0.05 $ ~GeV$^{-2}$ is 
obtained from examining the energy dependence of $B$. 
In order to take this into account we use the following form 
for the energy-dependent slope parameter
\begin{equation}
B(W) = B(W_{0}) + 2 ~\alp^{\prime} \, \ln (W/W_{0})^2 \, ,
\label{bslopew}
\end{equation}
\noindent with $B(W_{0}) = 4.0 $ ~GeV$^{-2}$ at an input scale 
of $W_{0} = 40 $~GeV and following ZEUS 
we take $\alp^{\prime} = 0.1$~GeV$^{-2}$. This clearly implies a 
reduction in the overall normalization 
of the cross section at large $W$, relative to the case in which 
a constant $B$ is employed 
(for $W = 200$ GeV the reduction is about 14 \% ).

A comparison of the short-dashed and solid lines of fig.(\ref{sigpdf}) show 
the significant effect of including the $W$-dependent slope according to  
eq.(\ref{bslopew}). The dotted curve shows the combined effect of 
running quark mass and $B(W)$ on the solid curve (cf. also eq.(\ref{mqrun})).  
Both effects reduce the steepness of the energy dependence bringing 
the theory curves closer to the data. 

\subsection{Skewedness}
\label{sectskew}

Strictly speaking for exclusive processes we need to replace the ordinary 
gluon in eq.(\ref{sighat}) with the skewed gluon $G(x_1, \delta, {\bar Q}^2)$ 
(see \cite{ji}, and references therein, for a review of skewed parton distributions). 
For $J/\psi$ this should have a rather small effect, for $\Upsilon$, with its much larger 
mass, it is certainly a large effect \cite{fms}. 
It is not obvious precisely how to sample the skewed gluon. 
Assuming collinearity of the incoming and returning gluons (carrying momentum fractions $x_1$ and $-x_2$ of the proton, respectively) we have $ (x_1 P + q)^2 = M^{2}_{c {\bar c}} \, \,  , \, \, ((x_1 - x_2) ~P + q)^2 = M_V^2$. In the photoproduction limit for the skewedness parameter, $\delta$, this gives  
\begin{eqnarray}
\delta & =  x_1 - x_2 &= \frac{M_V^2}{W^2} \, ,\\
x_1 & = \left. \frac{M_{c {\bar c}}^2}{W^2} \right. &= \delta \, 
\frac{(m_c^2 + k_t^2)/~z(1-z)}{M_V^2} \, .
\end{eqnarray}
\noindent At first sight it appears there may be a danger of  
entering the ERBL \cite{erbl} region ($x_1 < \delta, x_2 < 0$) 
in certain points in the phase space (in particular for symmetric, $z \approx 1/2$, configurations  with small $k_t^2$). 
However, our ansatz protects us from this, in the case of $J/\psi$, since we only use eq.(\ref{sighat}) 
for  $b < b_{Q0}$, which corresponds roughly to $k_t^2 > Q_0^2$.  
Even for an input scale of $1.0$ GeV$^2$ this implies 
$M_{c {\bar c}}^2 > 4 ~( m_c^2 + Q_0^2 ) > 13.0$ ~GeV$^2$. This is 
much bigger than the 
square of the $J/\psi$ mass:  $M^2_{\psi} (1S) = 9.59$ ~GeV$^2$. The 
2S-state, $\psi^{\prime}$, 
with a mass of $M^2_{\psi^{\prime}} (2S) = 13.59$~GeV$^2$ is a 
more marginal case, which 
merits further investigation.

For $J/\psi$ photoproduction we will use 
\begin{equation}
x^{\prime} = x_1 = \delta ~(1 + 0.75 \frac{\lambda}{b^2 ~4m_{c}^2}) \, \label{eqprimc}
\end{equation}
\noindent (cf. eq.(\ref{eqxprim}) with $ x^{\prime}_{min} = \delta = M_{V}^2/W^2$). 
This choice obviously guarantees that $x^{\prime} > \delta$ for all $b$, so restricts us just 
to the DGLAP region (with this assumption we can never enter the ERBL 
region).  With $\lambda = 10, m_{c} = 1.5$ ~GeV, $ \left<b\right>^2 = 10 * (hc)^2 /4m_{c}^2 \approx $ (0.21 fm)$^2$ and this reduces to
\begin{equation}
x^{\prime} (b) = \delta \, ( 1 + \frac{0.032}{(b~(\mbox{fm}))^2}) \, .
\end{equation}

\begin{figure}[htbp]
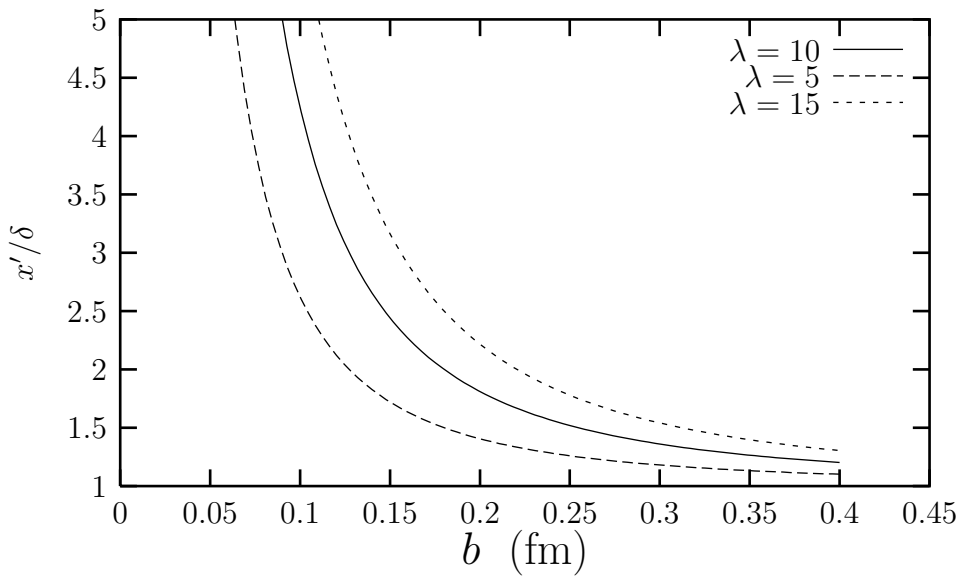

  \begin{center}
        \include{yeff}
\caption{The effective momentum fraction at which the gluon is sampled, divided by $\delta$, for several values of the scaling parameter $\lambda$.}
    \label{yeff}
  \end{center}
\end{figure}

\begin{figure}[htbp]
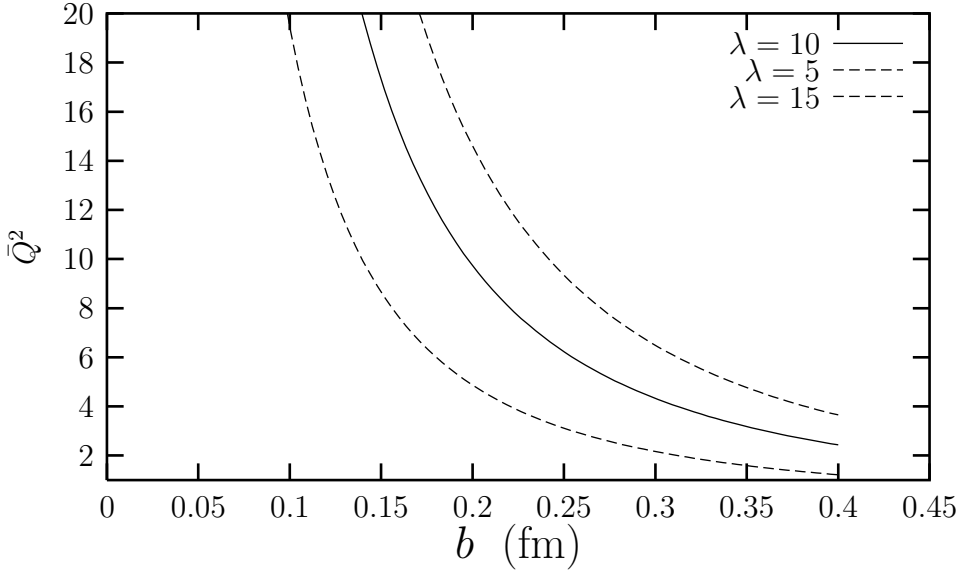

  \begin{center}
        \include{qeff}
\caption{The effective four-momentum scale at which the gluon is 
sampled for several values of the scaling parameter $\lambda$.}
    \label{qeff}
  \end{center}
\end{figure}

In our computer codes, the divergence at $b \to 0$, in the 
numerically unimportant very small $b$-region, is regulated by hand by adding a very small number to $b^2$ in the denominator. The skewed gluon density, 
$G (x^{\prime}, \delta, {\bar Q}^2)$, is sampled at four-momentum scale 
${\bar Q}^2 = \lambda/b^2  \approx 0.39/$($b$ (fm))$^2 $ GeV$^2$, 
hence some regulation of this scale is also implemented at very small $b^2$ (in practice we don't let the scale get larger than ${\bar Q}^2 = 100$~GeV$^2$). 
Figs.(\ref{yeff},\ref{qeff}) illustrate the $b$-dependence of $x^{\prime}/\delta$ 
and ${\bar Q}^2$ for different values of parameter $\lambda$.

\begin{figure}[htbp]
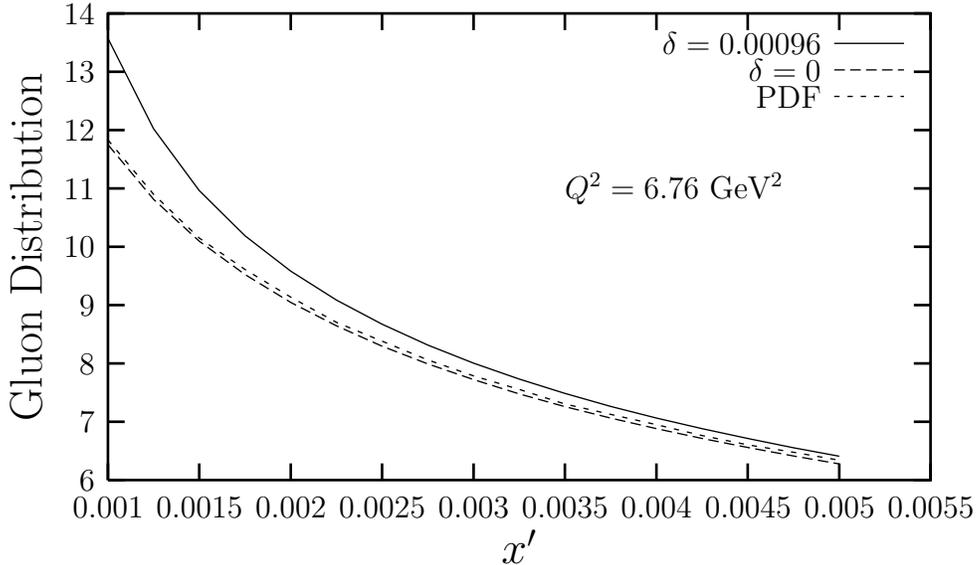

  \begin{center}
        \include{partons}
\caption{Skewed and conventional (PDF) gluon distributions at a scale typical of $J/\psi$ photoproduction,  $Q^2 = 6.76 $~GeV$^2$. The value of $\delta = 0.00096$ corresponds to $W=100$~GeV. This figure explicitly illustrates that the skewed distribution reduces to the standard one in the limit of zero skewedness.}
    \label{partons}
  \end{center}
\end{figure}

In order to implement skewedness we use the evolution package 
developed by Freund and Guzey \cite{fgcode}, which is based 
on the CTEQ code \cite{cteqsite}.
The skewed parton distributions (SPDs) are implemented in a 
systematic way underneath 
the integral in $b$. For a particular value of $W$, $\delta$ is fixed. 
For this fixed $\delta$, the skewed gluon is sampled on a grid, 
$G(x^{\prime},\delta,Q^{2 \prime})$ and the values saved to an array. 
The integration over 
$b$ is performed numerically folding in the $W^2$-independent 
piece from the wavefunctions with ${\hat \sigma}$ calculated by interpolating 
the skewed gluon array appropriately using a 
splines-based interpolation routine. 
Fig.(\ref{partons}) shows the skewed gluon versus the conventional 
one at a scale typical 
for $J/\psi$ photoproduction for $x^{\prime}$ close to, but larger 
than, $\delta = 9.6 \times 10^{-4}$
(the value which corresponds to $W= 100$~GeV). Also shown is 
the SPD with very small $\delta = 10^{-7}$ which coincides with 
the conventional PDF, illustrating the limit $G \to xg $ as $\delta \to 0$.

\begin{figure}[htbp]
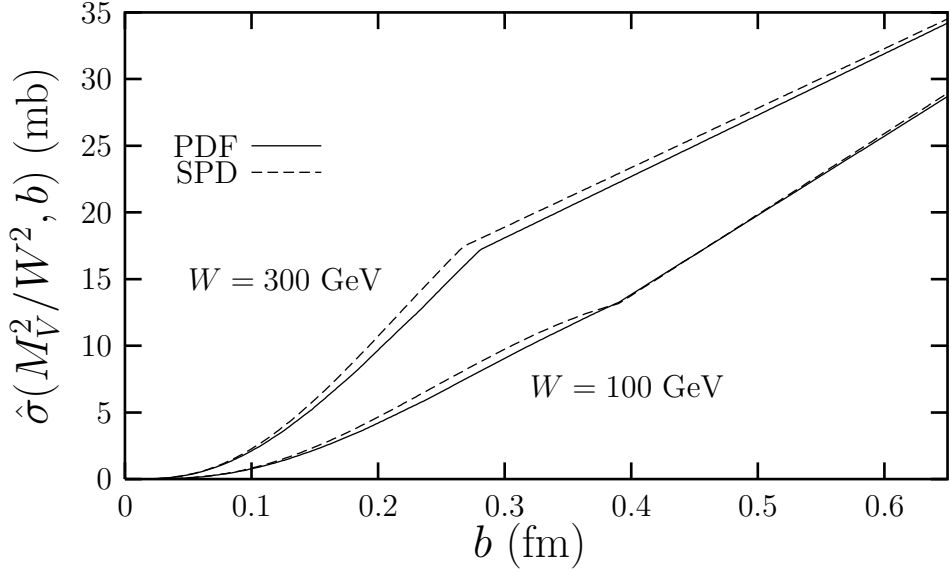

  \begin{center}
        \include{sigspdf}
\caption{The dipole cross section ${\hat \sigma}$ including skewedness (dashed curves) and without (solid curves) for two photon-proton energies.}
    \label{sigspdfc}
  \end{center}
\end{figure}

Following closely the prescription for 
${\hat \sigma}$ given in \cite{mfgs} we impose the 
following behaviour at $b=b_{\pi}$:
\begin{equation}
\sigma(b_{\pi},W^2) = \sigma (b_{\pi}, W_0) ~\left(\frac{W^2}{W_{0}^2} \right)^{\epsilon}
\end{equation} 
\noindent with $\sigma (b_{\pi}, W_0) = 24 $~mb, $\epsilon = 0.08$ and $W_0 = 31$~GeV. 
The latter value is chosen to coincide with the choice $x_0 = 0.01$ made in \cite{mfgs}.
A simple linear ansatz is used for the skewed ${\hat \sigma}$ in the region 
$b_{crit} < b < b_{\pi}$ (we connect $b_{crit}$ and $b_{\pi}$ with a straight line,    
$b_{crit} $ is the point in $b$ at which $\sigma(b,W^2) = \sigma(b_{\pi},W^2)/2 $ ). 
For small $W^2$ taming isn't required in the perturbative region, $b < b_{Q0} $, 
so the straight line starts at $b_{Q0}$. 
The dashed lines of fig.(\ref{sigspdfc}) show the effect of including skewedness
in the dipole cross section relative to using the standard PDF (solid lines) at 
$W = 100, 300$~GeV.

\begin{figure}[htbp]
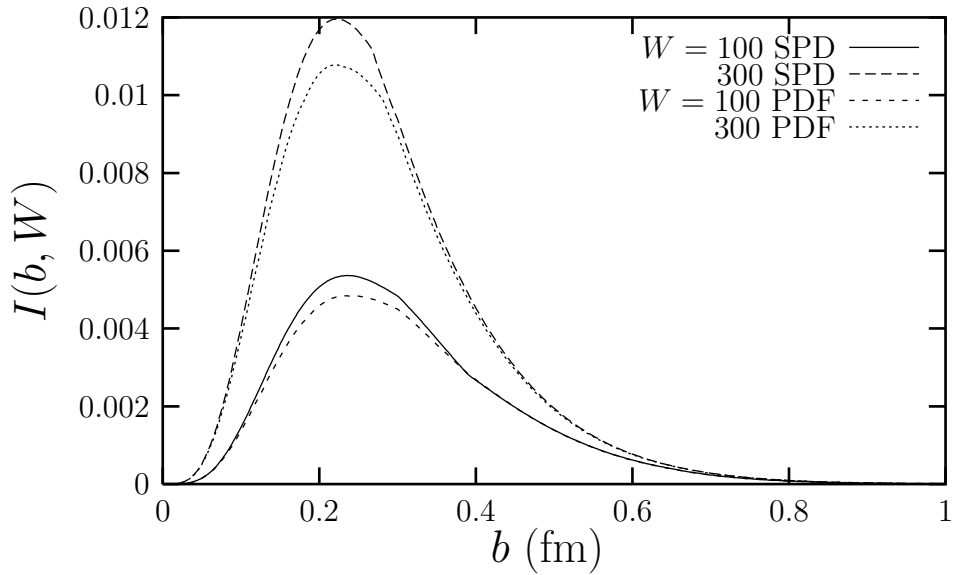

  \begin{center}
        \include{skewint}
\caption{A comparison of the $b$-integrand of eq.(\ref{ima}) 
with (SPD) and without (PDF) skewedness 
at two different photon-proton energies.}
    \label{skewint}
  \end{center}
\end{figure}

Close to the boundary at 
$b = b_{Q0} = 0.39$~fm, $x^{\prime} \approx \delta$ and 
${\bar Q}^2 \approx Q_0^2$. If one assumes the standard input at the boundary 
for skewed evolution as we do, the difference between 
the use of skewed and standard distributions 
is minimal close to $b_{Q0}$. For very small $b$, although 
there is a large evolution scale,  
${\bar Q}^2 \gg Q_{0}^2$, the gluon is sampled at 
$x^{\prime} \gg \delta$, so the overall effect 
of skewedness is expected to be fairly small everywhere. This 
is illustrated explicitly in 
fig.(\ref{skewint}) which shows the integrand of 
eq.(\ref{ima}) using skewed and standard gluons 
at two different energies ($W = 100, 300$~GeV). The maximum 
effect, of about $10\%$, is seen close 
to the peak. 

\begin{figure}[htbp]
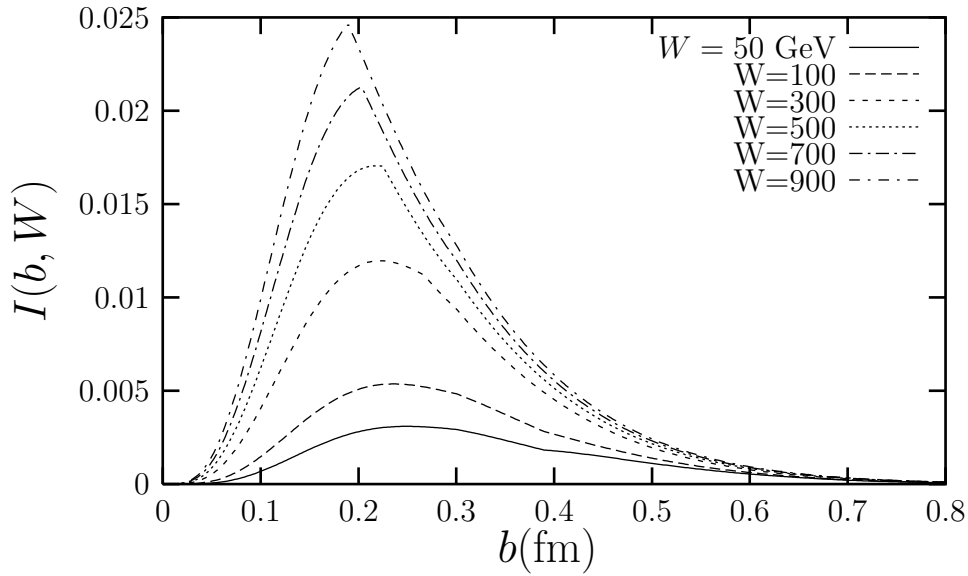

  \begin{center}
        \include{bndnew}
\caption{The evolution of the $b$-integrand with energy using 
skewed evolution and running quark mass in ${\hat \sigma}$. At high 
energies the effects of unitarity corrections are clearly seen in the shape}
    \label{bndnew}
  \end{center}
\end{figure}

\begin{figure}[htbp]
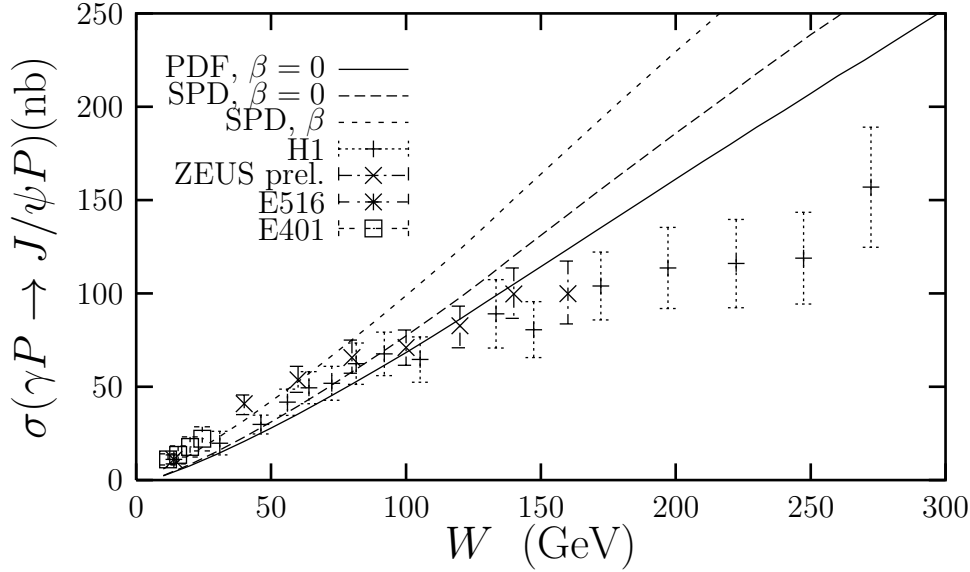

  \begin{center}
        \include{crossspd5}
\caption{The effect of including both skewedness and 
$\beta$ at the cross section level. The solid curve uses the ordinary 
gluon PDF with running mass and W-dependent slope included. The 
dashed curves include skewedness. The short-dashed curve also 
includes the real part of the amplitude.}
    \label{spd}
  \end{center}
\end{figure}

\begin{figure}[htbp]
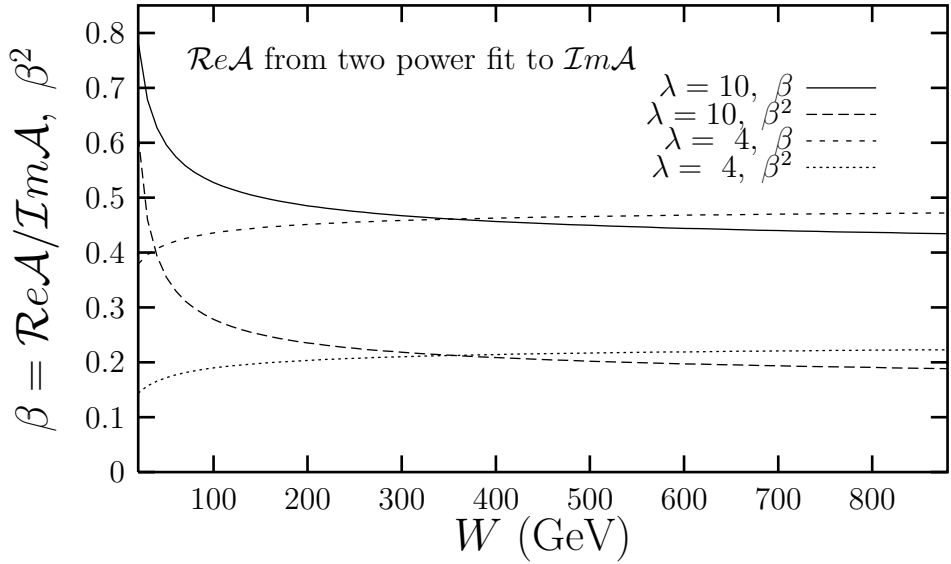

  \begin{center}
        \include{betanew}
\caption{A plot to illustrate the relative size of the real part of the amplitude as a function of energy (for CTEQ4L partons). In the case $\lambda = 4$ a two-power fit to MINUIT fit produces two positive terms with positive powers, hence $\beta$ calculated from the fit increases slightly as a function of energy (the fit parameters in this case are: $a_1~=~0.000386, ~p_1~=~0.291, ~a_2~=~0.000186, ~p_2~=~0.0272$). }
    \label{beta}
  \end{center}
\end{figure}

\begin{figure}[htbp]
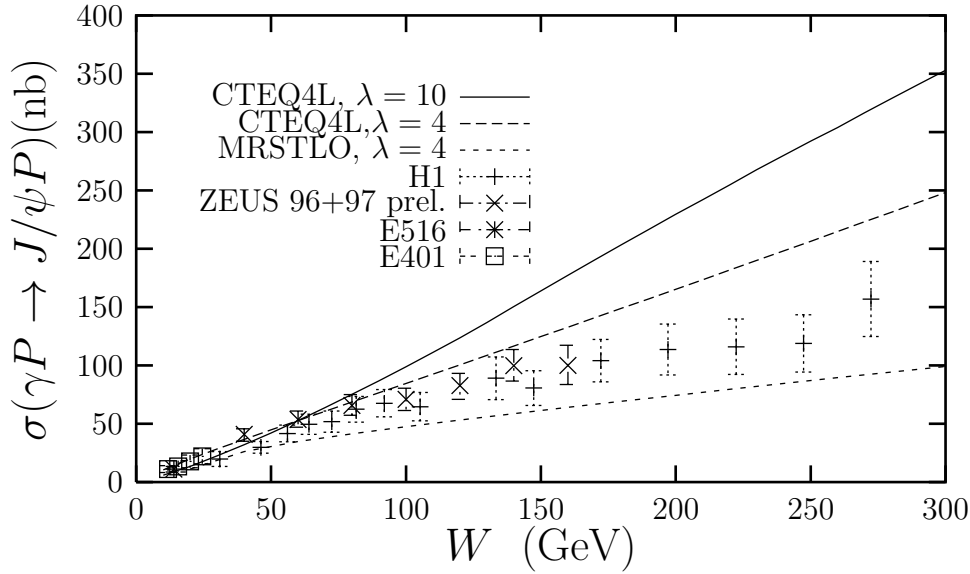

  \begin{center}
        \include{crossspd4}
\caption{Decreasing the value of $\lambda$ from 10 to 4 improves 
the agreement with data dramatically for CTEQ4L partons. The 
short-dashed implements MRST input partons, 
with $\lambda=4$.}
    \label{lam4}
  \end{center}
\end{figure}

For the skewed case it is interesting to show how the $b$-distribution changes with 
energy,  since this reflects the interplay of short and long distance contributions 
at different energies. One observes in fig.(\ref{bndnew}) that the peak shifts to the left 
as the energy increases and becomes more narrow, indicating an increase in the 
relative importance of short distance effects in this region. 
Examining such plots at very high energies reveals how the unitarity corrections begin to set in. 
One can start to see this in the shape of the curves at $ W=300,500 $ ~GeV, where the taming 
restriction begins to remove part of the distribution to the right of the peak. 
Although the taming corrections start to bite around $300$ GeV they take a long 
time to tame the majority of the (fairly broad) peak in $b$. 
In practice this implies that we expect the energy dependence to be tamed very 
gradually to the soft Pomeron one.

Finally, in fig.(\ref{spd}) we show the effect of including skewedness at 
the cross section level. The solid line corresponds to standard gluon PDF in 
${\hat \sigma}$ (with running mass and shrinkage implemented).
The long-dashed line corresponds to replacing the ordinary gluon distribution with the skewed one.
The overall effect, an enhancement of approximately $10\%$, is seen strongest at 
high energies where the small dipoles play an increasingly important role (see fig.(\ref{bndnew})).  
The shape in $W$ still appears to be too steep. It is possible that if we allow unitarity 
corrections to play a role earlier we may be able to fit the high energy data better. 

\subsection{Including the real part of the amplitude}
\label{sectre}

Having implemented the skewed gluon in the imaginary part of the amplitude we now 
reconstruct the real part using the analytic properties of the amplitude \cite{gribov}. 
Numerically we achieve this by performing a two-power fit to ${\cal I} m ~A (W)$ over 
a very wide range in $W$ ( $10 < W < 900 $ ~GeV) using the form
\begin{equation}
{\cal I} m ~A (W) = a_1 ~(\frac{W^2}{W^2_0})^{p_1} + a_2 ~(\frac{W^2}{W^2_0})^{p_2} \, ,
\end{equation}
\noindent the real part is then given by 
\begin{equation}
{\cal R} e ~A (W) = a_1 ~(\frac{W^2}{W^2_0})^{p_1} ~\tan (\frac{\pi p_{1}}{2}) + a_2 ~(\frac{W^2}{W^2_0})^{p_2}  \tan(\frac{\pi p_{2}}{2}) \, .
\end{equation}
The ratio of real to imaginary parts, $\beta$ is then obviously given by
\begin{equation}
\beta = \frac{\tan(\pi p_{1}/2) + a_2/a_1 ~(W^2/W^2_0)^{p_2-p_1}  \tan(\pi p_{2}/2)}{1 + a_2/a_1 ~(W^2/W^2_0)^{p_2-p_1} } \, .
\end{equation}
\noindent This increases the normalization of the cross section by 
$\beta^2 $ \% (cf. eq.(\ref{photonaive})).

\noindent We achieve an excellent two-power fit using MINUIT \cite{minuit} and get the following values for the fit parameters:
\begin{eqnarray}
a_1  = 0.00104,  & \,  a_2 = -0.000794 \nonumber \\
p_1 = 0.226,     & \,  p_2 =-0.0775 \, \, .
\end{eqnarray}

Fig.(\ref{beta}) illustrates that $\beta$ decreases as a function of $W$ as the 
smaller power becomes less significant (see curves labelled $\lambda = 10$).  
The short dashed line of fig.(\ref{spd}) shows the overall energy-dependent enhancement 
(about $20\%$ at high energies) when this implemented at the cross section level. 
This improves the description of the low energy data and in terms of shape, but leads to 
a worse overshoot at high energies.

\section{Decreasing $\lambda$ and predictions for THERA}

\begin{figure}[htbp]
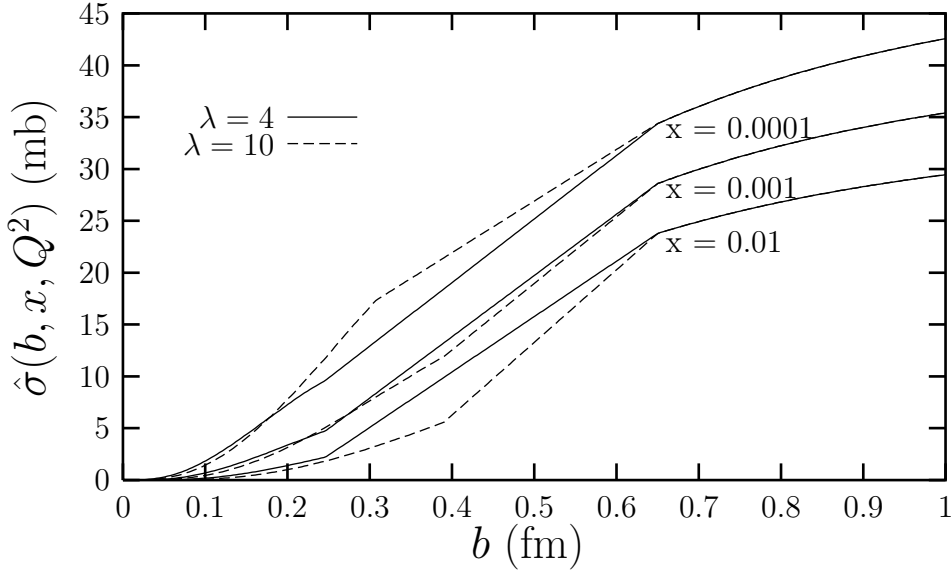

\begin{center}%
\include{sighat}
\caption{The interaction cross-section, ${\hat \sigma}$ for CTEQ4L, linear ansatz, for $x=0.01, 0.001, 0.0001$, $\lambda = 4, 10$. A value of $Q^2 = Q^2_{0} = 2.56$~GeV$^2$ is used in the ansatz for $b$-dependent scales in ${\hat \sigma}$.}
\label{fig:sighat}
\end{center}
\end{figure}

At large $Q^2$, within the leading and next-to-leading logarithmic 
approximations of perturbative QCD, there exists a rather straightforward relationship between 
transverse size and the relevant four-momentum scale for the process 
concerned, i.e. that they are inversely proportional \cite{fks1,fks2}.  
The constant of proportionality, $\lambda$,  
was determined in \cite{fks1} by an iterative averaging procedure 
involving the integral in $b$ for the structure function 
$F_L$: $\lambda = \left<b_{F_L}\right>^2 Q^2$. 
A value of $\lambda \approx 9$ was obtained 
from this method and this was found not to vary too much 
with $x$ and $Q^2$ at small $x \approx 10^{-3}$, 
provided $Q^2 \gsim 10$ GeV$^2$. 
At larger $b^2$, where applicability of perturbative QCD can not be justified, this relationship may break down, i.e. $\lambda $ may be found to depend 
on $b^2$ or equivalently on $Q^2$.  Indeed, in \cite{fms} we observed  
that the value coming from this procedure was seen to deviate when a 
wider kinematic range was considered. During the work carried out 
for \cite{mfgs} it was found that although $\lambda$ was changing, if kept 
within reasonable limits ($4 \lsim \lambda \lsim 15$) 
the result for $F_L$ only changed by a few percent 
(see fig.(5) of \cite{mfgs} for an explanation of this 
approximate scaling). Having made this observation, a 
value of $\lambda = 10$ was chosen for convenience (and because it most closely 
corresponded with average value of $b$ in $F_L$) in the analysis of 
inclusive Structure Functions. 

At smaller $Q^2$, we need to account for non-perturbative QCD effects, where
the relationship between $b^2$ and $Q^2$ is not so straightforward (in particular, 
$b^2$ does not tend to infinity as $Q^2\rightarrow 0$). 
To account for this slower dependence on $Q^2$, for smaller $Q^2$, and 
to test the sensitivity to non-perturbative QCD effects we diminish
$\lambda$ at all $b$. It is logical that decreasing $\lambda$ will have 
a much bigger effect on processes such as $J/\psi$ photoproduction which 
contain greater contamination from large distances and 
are sensitive to scales at which the gluon density is changing rapidly 
at small $x$  ($ Q_0^2 < {\bar Q}^2 < 10 $~GeV$^2$). 

Changing the value of $\lambda$ has several effects. Firstly, the 
position of the input scale, $b_{Q0}$, in 
$b$ shifts which affects ${\hat \sigma}$ 
directly (recall that $b_{\pi}$ which also specifies ${\hat \sigma}$ is fixed).
It also directly influences the scales at which the gluon 
distribution is sampled, ${\bar Q}^2$ and $x^{\prime}$, 
(cf. eqs.(\ref{qbar}, \ref{eqprimc}) and figs.(\ref{yeff},\ref{qeff})). 
Since the light-cone wavefunctions do not depend on $\lambda$ 
(except implicitly though the 
${\bar Q}^2$-dependence of $m_{c,r}$), 
this change has the effect of squeezing or dilating 
the perturbative region in $b$. Decreasing $\lambda$ 
decreases $b_{Q0} = \sqrt{\lambda}/Q_{0}$ 
and so diminishes the perturbative region almost 
without modifying ${\hat \sigma}$ in the perturbative regime.
The effect on ${\hat \sigma}$ is explicitly illustrated in fig.(\ref{fig:sighat}) which shows 
the dipole cross section for several values of $x$ and $\lambda= 4,10$ 
(for the purposes of this figure we neglect the $10\%$ skewedness effect).
This should make the cross section rise less steeply with energy as it 
enhances the non-perturbative piece. To test this we reran our code 
using the lowest value which left the results for $F_L$ unaffected,  
i.e. $\lambda = 4$ \footnote{Incidentally and accidentally, this is the value 
advocated in the papers of Gotsman, Levin and Maor and collaborators
for all $Q^2$. See \cite{gotsman} for the latest version of their model 
which includes shadowing corrections. It is applied to $J/\psi$ 
production in \cite{glmjpsi}.}. 

The effect on the cross section was rather dramatic and 
is shown in fig.({\ref{lam4}) (the real part of the amplitude, 
calculated as described above, is included, cf. fig.(\ref{beta})). The 
cross section is increased in the fixed target region and 
suppressed in the high energy region. Both effects move 
it in the direction of the data. 
From fig.(\ref{fig:sighat}) we can see that big differences between $\lambda=4$ and
$\lambda=10$ only emerge in the region around $b=0.3$~fm. Hence, for $200 < W < 300 $~GeV, 
where the main contribution to the cross section originates from this region in $b$, 
the largest relative difference between the final cross sections is observed 
(see fig.(\ref{lam4})). The difference does not increase further with energy since 
the contribution from smaller $b$ gradually becomes dominant (see fig.(\ref{bndnew})).

As an additional cross check we re-examined our 
description of $F_2 (x,Q^2)$, using $\lambda$ = 4. 
The results for selected values of $Q^2$ are shown by the dashed 
lines in fig.(\ref{f2lam4}). A comparison with the solid lines ($\lambda =10$) 
illustrates that $F_2$ is also fairly insensitive to this change in 
$\lambda$ and we still provide a reasonable description of the HERA data.
Also shown (dotted curves) is the evaluation of $F_2$ using the latest MRST 
leading-order gluon distribution \cite{mrstl}, with $\lambda=4$. This 
appears to get closer to the data than the earlier CTEQ4L distributions  
(at least in the deep inelastic region). Taken together both the MRST and 
CTEQ4L curves illustrate the spread of predictions available 
from modern leading-log fits and hence give some indication of the 
theoretical uncertainty of the description.

\begin{figure}[htbp]
  \begin{center}
     \includegraphics[height=15cm,width=15cm]{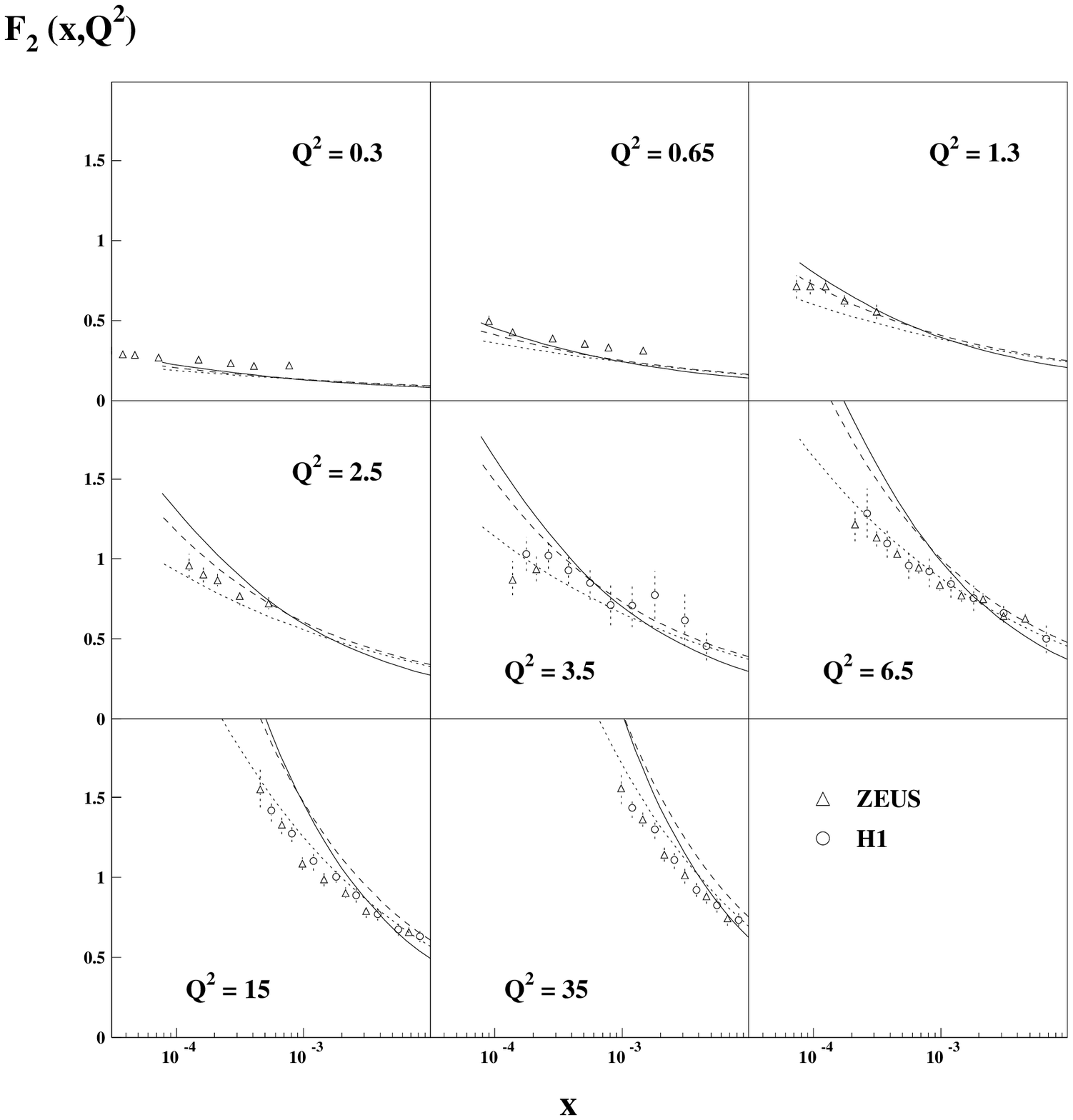}
    \caption{A comparison of the inclusive cross section $F_2$ using 
the dipole cross section with $\lambda = 4$ (dashed line) and  
$\lambda = 10$ (solid line), with a selection of the 
ZEUS \cite{z9497} and H1 \cite{H194} data. Also shown (dotted curves) are the values obtained using leading-order MRST partons with $\lambda = 4$.} 
    \label{f2lam4}
  \end{center}
\end{figure}

Extrapolating to higher energies we observe a rather broad spread of 
predictions for $J/\psi$-photoproduction in the THERA range in 
Fig.(\ref{highw}). Since, for CTEQ4L partons, the case 
$\lambda = 4$ does a better job on the lower energy data, 
we favour the dashed curve as our prediction for THERA. 
Despite the effect of the unitarity correction in the integrands, 
evident in fig.(\ref{bndnew}), the overall taming effect on the 
energy dependence at the cross section level appears to be rather mild 
within the considered model. This is because the contribution of very 
small $b$ for which taming is still not important becomes more and 
more significant as the energy increases. In order to illustrate the 
sensitivity of the predictions to the choice of input parton density set 
we also show in Figs.(\ref{lam4},\ref{highw}) (short-dashed) 
curves for the latest MRST leading-order partons \cite{dick}. 
The latter assume an analytic form for the small-$x$ behaviour of the input gluon 
which decreases as a function of $1/x$, which explains the milder energy dependence.

\begin{figure}[htbp]
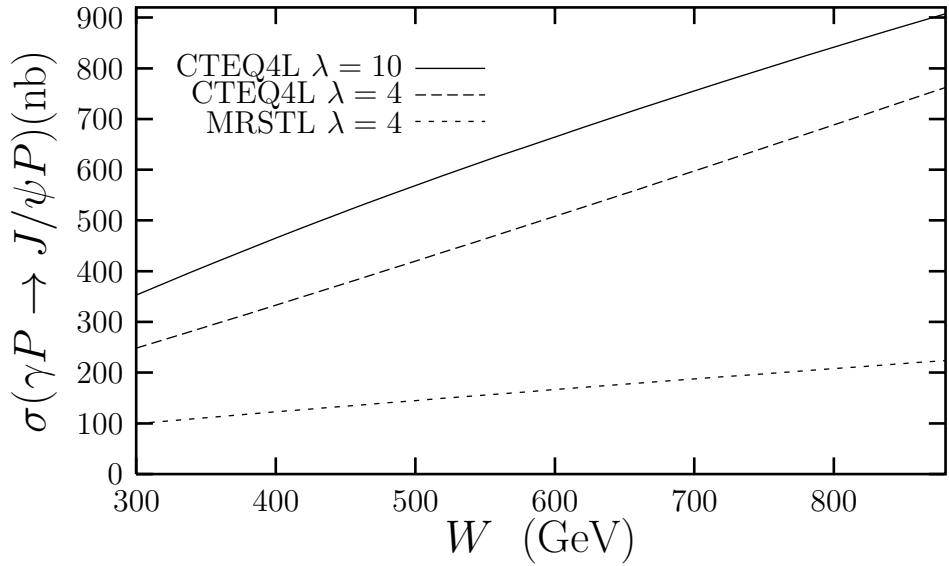

  \begin{center}
        \include{highw4}
\caption{The photoproduction cross section for $J/\psi$ in the THERA 
region using two different values for the scaling parameter $\lambda$ 
and CTEQ4L partons. The short-dashed curve shows the prediction using 
MRST leading order partons and $\lambda = 4$.}
    \label{highw}
  \end{center}
\end{figure}

Small $\lambda$ (e.g 4-5) appear from the $J/\Psi$ analysis to be required
where large $b$ is significant. This is strongly related to the relative 
influence of perturbative and non-perturbative contributions.
It seems that the current uncertainty associated with the small-$x$ gluon
distribution prevent any strong statement being made about $\lambda$
at small $b$. One may choose $\lambda$ in the range 5-15 and the results for
physical quantities don't change in quality 
(at present this applies to $F_2$ and hard vector meson production, since 
$F_L$ is not yet measured). 
Making this change is perfectly allowed within the leading-log 
accuracy of the model (it corresponds to an attempt to mimic some NLO 
corrections). This observation shows that an estimate of the region 
where unitarity corrections become significant \cite{fks2,mfgs,afs} will contain large uncertainties before a NLO
calculation of cross section is made.

\section{Evaluation of $\alpha_{\pomeron}^{\prime}$}

Since $\alp^{\prime}$ is rather sensitive to the physics relevant for the
taming of parton distributions, in this section we develop a more sophisticated
approach to calculating it.

We would like to point out that the measurement of 
$\alpha_{\pomeron}^{\prime}$ recently reported by ZEUS \cite{bruni} is 
entirely consistent with the results of our model. 
For large transverse sizes one expects a contribution of 
$\alpha_{\pomeron}^{\prime} (\mbox{soft}) = 0.25$~GeV$^{-2}$, whereas for very small sizes 
a negligible contribution is expected. 
We can see this by introducing a simple, but reasonable, model based on 
dimension analysis: 
\begin{equation}
\alp^{\prime} (b) = 0.5 ~(\frac{b^2}{b^2 + b_{\pi}^2}) ~~\mbox{GeV$^{-2}$}. 
\label{alpr}
\end{equation} 
\noindent 
This model is designed to give $\alpha_{\pomeron}^{\prime} (b=b_{\pi}) = \alp^{\prime} (\mbox{soft}) = 0.25$~GeV$^{-2}$, and to tend to zero quadratically at small $b$. There is a gradual decrease of $\alp^{\prime}$ with increase of energy, as small $b$ configurations become more important. We quantify this in fig.(\ref{alprime}) by plotting the 
energy dependence of the average of $\alp^{\prime}$ defined as 
(cf. eq.(\ref{ima}))
\begin{equation}
<\! \alpha_{\pomeron}^{\prime} \!> = \frac{\int b~db~\alpha_{\pomeron}^{\prime} (b) I_{z} (b) {\hat \sigma }}
{\int b~db~I_{z} (b) {\hat \sigma }} \, ,
\label{aval}
\end{equation}
\noindent for $\lambda = 4, 10$. We see that this model is 
in broad agreement with the ZEUS value \cite{bruni,osaka}: $\alp^{\prime} = 
0.098 \pm 0.035 \pm 0.05 $ ~GeV$^{-2}$. 

Strictly speaking, the energy dependence apparent in fig.(\ref{alprime}) 
may be slightly too strong when considering its effect on the $t$-slope 
parameter $B$. The average $b$, at a particular energy, is 
built in to the value of $B (W_0)$ at the normalization 
point (cf. eq.(\ref{bslopew})), but is expected to decrease with energy. 
We illustrate the point using the following 
phenomenological parameterisation for the slope of $t$-dependence: 
\begin{equation}
B(W^2,Q^2)=\frac{(<\!b^2\!>/4+r_N^2)}{3} + 2 \alpha_{\pomeron}^{\prime} \ln W^2/W_{0}^2
\end{equation}
\noindent Here $<\!b^2\!>$ is the average distance between $c$ and ${\bar c}$, defined in an analogous way to the average $\alpha_{\pomeron}^{\prime}$ 
(cf. eq.(\ref{aval})). In this phenomenological model a possible presence of $\ln b^2/b_{0}^2 $ terms is ignored. This formulae may be thought of as 
defining a convention for a new parameter, $\alpha^{\prime}$, related directly to the logarithmic energy dependence of the slope parameter, $B$:
\begin{eqnarray}
\alpha' & \equiv & \frac{\partial B} {\partial ~(4 ~\ln (W/W_0))} \, , \nonumber \\ 
\alpha' & = & \frac{<\!b^2 (W)\!> -  <\!b^2 (W_0)\!>}{48 \ln W/W_0} + \alpha_{\pomeron}^{\prime} \, .
\end{eqnarray}
\noindent However, it turns out that in HERA kinematics the difference between 
$\alpha^{\prime}$  and $\alpha_{\pomeron}^{\prime}$ from this effect is 
only about $-0.01$~GeV$^{-2}$.

Thus, the observed energy dependence of the average value of $b$
in hard amplitudes (which are influenced by DGLAP evolution)
leads to a more complicated and process-dependent energy dependence of 
the slope parameter, $B$, than that which is predicted by universal 
soft Pomeron exchange. At THERA energies and beyond, where the taming region 
starts to dominate, one expects that $\alp^{\prime}$ should start to 
increase again. 

\begin{figure}[htbp]
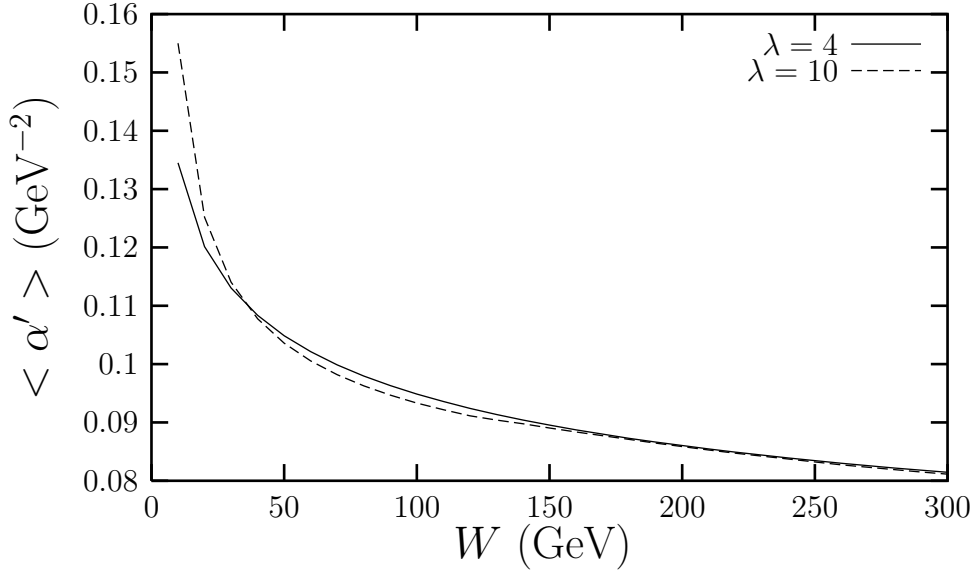

  \begin{center}%
        \include{alpri3}
\caption{The average shrinkage parameter, $\alpha^{\prime}$ 
(using the simple model of eq.(\ref{alpr})) as a function of 
$W$ (GeV), for two values of $\lambda$.}
    \label{alprime}
  \end{center}
\end{figure}

In our picture we expect both $B$ and $\alp^{\prime}$ to change 
with energy and photon virtuality as 
the balance between short and long distance contributions 
shifts. A dedicated forward detector 
for measuring scattered protons, such as the one recently 
proposed by the H1 collaboration \cite{favart},
would allow this issue to be investigated in detail.

\section{Discussion and open questions}

Using $\lambda = 4$ and CTEQ4L partons our model still seems to 
overshoot the available data somewhat 
(cf. long-dashed line in fig.(\ref{lam4})), whereas MRST 
undershoot ((cf. short-dashed line). We would like to reiterate 
the point that the gluon distribution 
at small $x \approx 10^{-4}$ is not very well constrained by the 
current data (which mainly provides 
an indirect constraint via DGLAP-driven scaling 
violations of $F_2$). This fact is reflected in the 
wide spread of numerical values for the gluon 
distribution at small $x$ in the currently available 
partons distributions
\footnote{See http://durpdg.dur.ac.uk/HEPDATA/PDF3.html}, of 
which we have shown only two here. Hence we are not too concerned 
by the fact that our curve appears to overshoot 
particularly the H1 data \cite{h100} at the highest 
HERA energies using CTEQ4L. The framework we have 
described is general and can clearly use any 
leading-log parton density set that is available.
However, we are encouraged by the improvement in 
agreement in the overall shape, which allows us to get 
closer to the data over a wide range in energy.

It is natural to ask if higher order 
Fock states ($|c {\bar c} g\!\!>$ etc), which 
are only formally suppressed by $\alpha_s$, are 
important in the photoproduction of $J/\psi$. 
The rationale employed in \cite{fks1} was that using the solution 
to the Schr\"{o}dinger equation (boosted to light cone) should 
take into account the most 
important corrections for large transverse distances. 
Recent studies of the effect of radiative corrections 
on the leptonic decay width 
\cite{ynd} indicate that perturbative corrections 
for small distances may well be large. 
At present it is unclear what effect this will have on 
the photoproduction of $J/\psi$.
From the theoretical side what is required is a complete 
next-to-leading-log calculation. 
This question could be addressed phenomenologically 
by examining the relative characteristics 
of various diffractive charm measurements (exclusive $J/\psi$, $J/\psi$ + 
1 jet, and open charm).  
The question is even more urgent in the case 
of $\psi^{\prime}$ which is obviously 
a larger bound state and contains a node.

\section{Conclusions}

We have investigated the photoproduction of $J/\psi$ in 
the context of the QCD-improved dipole model 
introduced in \cite{mfgs}. This model directly 
incorporates a contribution from long distances 
which is responsible for the low energy production. 
As the energy increases, the short distance mechanism of 
perturbative two-gluon exchange becomes 
increasingly important.  Overall our description of the data over the whole 
measured range is improved relative to analyses 
which only take the perturbative contribution 
into account. At very high energies we incorporate 
taming or unitarity corrections and present 
predictions for the cross section in the THERA range ($250 < W < 900$~GeV). 
In our model, unitarity corrections affect smaller and smaller transverse 
sizes at progressively higher energies. 
For $J/\psi$ photoproduction, which is sensitive to a broad range in transverse sizes,
it turns out that in the THERA region there is always a significant contribution to the 
cross section that rise quickly with energy. This implies that the taming 
of the growth of the cross section with energy is rather gradual 
(the precise details depend on the choice of input density). 

We show that the skewedness of the amplitude induces a relatively 
small enhancement in the cross section (approximately $10\%$). 
This effect is likely to be swamped by other uncertainties associated 
with the $t$-dependence and the light-cone wavefunction of the vector meson.

Of much greater numerical importance in describing 
the available data is to modify the balance 
between short and long distances contributions.
In our model, this may be controlled by changing 
the scaling parameter $\lambda$ which sets 
the relationship between four-momentum scales and 
transverse sizes in the interaction cross section. 
For CTEQ4L partons, the shape in energy of the $J/\psi$ photoproduction data appears to favour a change 
to a lower value ($\lambda = 4$) for large $b$ than that which was derived in 
\cite{fks1,fks2,mfgs} for small $b$ ($\lambda = 10$). The quality of the 
description of the structure function $F_2$ is relatively unchanged by this 
modification. The largest uncertainty comes from poor knowledge of the 
numerical size of the leading-log small-$x$ gluon distribution. 
This leads to a rather broad band of predictions for $J/\psi$ photoproduction in THERA region. 
Studies of the photo- and electroproduction of other vector mesons, 
within the same framework, is underway.

\section*{Acknowledgements}
We thank Sandy Donnachie, Vadim Guzey and Andreas Freund for 
useful contributions. We would also like to thank the referee for his helpful 
comments which led to several improvements to this text.

\end{document}

%% file: crosspdf.tex
\begingroup%
  \makeatletter%
  \newcommand{\GNUPLOTspecial}{%
    \@sanitize\catcode`\%=14\relax\special}%
  \setlength{\unitlength}{0.1bp}%
{\GNUPLOTspecial{!
/gnudict 256 dict def
gnudict begin
/Color false def
/Solid false def
/gnulinewidth 5.000 def
/userlinewidth gnulinewidth def
/vshift -33 def
/dl {10 mul} def
/hpt_ 31.5 def
/vpt_ 31.5 def
/hpt hpt_ def
/vpt vpt_ def
/M {moveto} bind def
/L {lineto} bind def
/R {rmoveto} bind def
/V {rlineto} bind def
/vpt2 vpt 2 mul def
/hpt2 hpt 2 mul def
/Lshow { currentpoint stroke M
  0 vshift R show } def
/Rshow { currentpoint stroke M
  dup stringwidth pop neg vshift R show } def
/Cshow { currentpoint stroke M
  dup stringwidth pop -2 div vshift R show } def
/UP { dup vpt_ mul /vpt exch def hpt_ mul /hpt exch def
  /hpt2 hpt 2 mul def /vpt2 vpt 2 mul def } def
/DL { Color {setrgbcolor Solid {pop []} if 0 setdash }
 {pop pop pop Solid {pop []} if 0 setdash} ifelse } def
/BL { stroke userlinewidth 2 mul setlinewidth } def
/AL { stroke userlinewidth 2 div setlinewidth } def
/UL { dup gnulinewidth mul /userlinewidth exch def
      10 mul /udl exch def } def
/PL { stroke userlinewidth setlinewidth } def
/LTb { BL [] 0 0 0 DL } def
/LTa { AL [1 udl mul 2 udl mul] 0 setdash 0 0 0 setrgbcolor } def
/LT0 { PL [] 1 0 0 DL } def
/LT1 { PL [4 dl 2 dl] 0 1 0 DL } def
/LT2 { PL [2 dl 3 dl] 0 0 1 DL } def
/LT3 { PL [1 dl 1.5 dl] 1 0 1 DL } def
/LT4 { PL [5 dl 2 dl 1 dl 2 dl] 0 1 1 DL } def
/LT5 { PL [4 dl 3 dl 1 dl 3 dl] 1 1 0 DL } def
/LT6 { PL [2 dl 2 dl 2 dl 4 dl] 0 0 0 DL } def
/LT7 { PL [2 dl 2 dl 2 dl 2 dl 2 dl 4 dl] 1 0.3 0 DL } def
/LT8 { PL [2 dl 2 dl 2 dl 2 dl 2 dl 2 dl 2 dl 4 dl] 0.5 0.5 0.5 DL } def
/Pnt { stroke [] 0 setdash
   gsave 1 setlinecap M 0 0 V stroke grestore } def
/Dia { stroke [] 0 setdash 2 copy vpt add M
  hpt neg vpt neg V hpt vpt neg V
  hpt vpt V hpt neg vpt V closepath stroke
  Pnt } def
/Pls { stroke [] 0 setdash vpt sub M 0 vpt2 V
  currentpoint stroke M
  hpt neg vpt neg R hpt2 0 V stroke
  } def
/Box { stroke [] 0 setdash 2 copy exch hpt sub exch vpt add M
  0 vpt2 neg V hpt2 0 V 0 vpt2 V
  hpt2 neg 0 V closepath stroke
  Pnt } def
/Crs { stroke [] 0 setdash exch hpt sub exch vpt add M
  hpt2 vpt2 neg V currentpoint stroke M
  hpt2 neg 0 R hpt2 vpt2 V stroke } def
/TriU { stroke [] 0 setdash 2 copy vpt 1.12 mul add M
  hpt neg vpt -1.62 mul V
  hpt 2 mul 0 V
  hpt neg vpt 1.62 mul V closepath stroke
  Pnt  } def
/Star { 2 copy Pls Crs } def
/BoxF { stroke [] 0 setdash exch hpt sub exch vpt add M
  0 vpt2 neg V  hpt2 0 V  0 vpt2 V
  hpt2 neg 0 V  closepath fill } def
/TriUF { stroke [] 0 setdash vpt 1.12 mul add M
  hpt neg vpt -1.62 mul V
  hpt 2 mul 0 V
  hpt neg vpt 1.62 mul V closepath fill } def
/TriD { stroke [] 0 setdash 2 copy vpt 1.12 mul sub M
  hpt neg vpt 1.62 mul V
  hpt 2 mul 0 V
  hpt neg vpt -1.62 mul V closepath stroke
  Pnt  } def
/TriDF { stroke [] 0 setdash vpt 1.12 mul sub M
  hpt neg vpt 1.62 mul V
  hpt 2 mul 0 V
  hpt neg vpt -1.62 mul V closepath fill} def
/DiaF { stroke [] 0 setdash vpt add M
  hpt neg vpt neg V hpt vpt neg V
  hpt vpt V hpt neg vpt V closepath fill } def
/Pent { stroke [] 0 setdash 2 copy gsave
  translate 0 hpt M 4 {72 rotate 0 hpt L} repeat
  closepath stroke grestore Pnt } def
/PentF { stroke [] 0 setdash gsave
  translate 0 hpt M 4 {72 rotate 0 hpt L} repeat
  closepath fill grestore } def
/Circle { stroke [] 0 setdash 2 copy
  hpt 0 360 arc stroke Pnt } def
/CircleF { stroke [] 0 setdash hpt 0 360 arc fill } def
/C0 { BL [] 0 setdash 2 copy moveto vpt 90 450  arc } bind def
/C1 { BL [] 0 setdash 2 copy        moveto
       2 copy  vpt 0 90 arc closepath fill
               vpt 0 360 arc closepath } bind def
/C2 { BL [] 0 setdash 2 copy moveto
       2 copy  vpt 90 180 arc closepath fill
               vpt 0 360 arc closepath } bind def
/C3 { BL [] 0 setdash 2 copy moveto
       2 copy  vpt 0 180 arc closepath fill
               vpt 0 360 arc closepath } bind def
/C4 { BL [] 0 setdash 2 copy moveto
       2 copy  vpt 180 270 arc closepath fill
               vpt 0 360 arc closepath } bind def
/C5 { BL [] 0 setdash 2 copy moveto
       2 copy  vpt 0 90 arc
       2 copy moveto
       2 copy  vpt 180 270 arc closepath fill
               vpt 0 360 arc } bind def
/C6 { BL [] 0 setdash 2 copy moveto
      2 copy  vpt 90 270 arc closepath fill
              vpt 0 360 arc closepath } bind def
/C7 { BL [] 0 setdash 2 copy moveto
      2 copy  vpt 0 270 arc closepath fill
              vpt 0 360 arc closepath } bind def
/C8 { BL [] 0 setdash 2 copy moveto
      2 copy vpt 270 360 arc closepath fill
              vpt 0 360 arc closepath } bind def
/C9 { BL [] 0 setdash 2 copy moveto
      2 copy  vpt 270 450 arc closepath fill
              vpt 0 360 arc closepath } bind def
/C10 { BL [] 0 setdash 2 copy 2 copy moveto vpt 270 360 arc closepath fill
       2 copy moveto
       2 copy vpt 90 180 arc closepath fill
               vpt 0 360 arc closepath } bind def
/C11 { BL [] 0 setdash 2 copy moveto
       2 copy  vpt 0 180 arc closepath fill
       2 copy moveto
       2 copy  vpt 270 360 arc closepath fill
               vpt 0 360 arc closepath } bind def
/C12 { BL [] 0 setdash 2 copy moveto
       2 copy  vpt 180 360 arc closepath fill
               vpt 0 360 arc closepath } bind def
/C13 { BL [] 0 setdash  2 copy moveto
       2 copy  vpt 0 90 arc closepath fill
       2 copy moveto
       2 copy  vpt 180 360 arc closepath fill
               vpt 0 360 arc closepath } bind def
/C14 { BL [] 0 setdash 2 copy moveto
       2 copy  vpt 90 360 arc closepath fill
               vpt 0 360 arc } bind def
/C15 { BL [] 0 setdash 2 copy vpt 0 360 arc closepath fill
               vpt 0 360 arc closepath } bind def
/Rec   { newpath 4 2 roll moveto 1 index 0 rlineto 0 exch rlineto
       neg 0 rlineto closepath } bind def
/Square { dup Rec } bind def
/Bsquare { vpt sub exch vpt sub exch vpt2 Square } bind def
/S0 { BL [] 0 setdash 2 copy moveto 0 vpt rlineto BL Bsquare } bind def
/S1 { BL [] 0 setdash 2 copy vpt Square fill Bsquare } bind def
/S2 { BL [] 0 setdash 2 copy exch vpt sub exch vpt Square fill Bsquare } bind def
/S3 { BL [] 0 setdash 2 copy exch vpt sub exch vpt2 vpt Rec fill Bsquare } bind def
/S4 { BL [] 0 setdash 2 copy exch vpt sub exch vpt sub vpt Square fill Bsquare } bind def
/S5 { BL [] 0 setdash 2 copy 2 copy vpt Square fill
       exch vpt sub exch vpt sub vpt Square fill Bsquare } bind def
/S6 { BL [] 0 setdash 2 copy exch vpt sub exch vpt sub vpt vpt2 Rec fill Bsquare } bind def
/S7 { BL [] 0 setdash 2 copy exch vpt sub exch vpt sub vpt vpt2 Rec fill
       2 copy vpt Square fill
       Bsquare } bind def
/S8 { BL [] 0 setdash 2 copy vpt sub vpt Square fill Bsquare } bind def
/S9 { BL [] 0 setdash 2 copy vpt sub vpt vpt2 Rec fill Bsquare } bind def
/S10 { BL [] 0 setdash 2 copy vpt sub vpt Square fill 2 copy exch vpt sub exch vpt Square fill
       Bsquare } bind def
/S11 { BL [] 0 setdash 2 copy vpt sub vpt Square fill 2 copy exch vpt sub exch vpt2 vpt Rec fill
       Bsquare } bind def
/S12 { BL [] 0 setdash 2 copy exch vpt sub exch vpt sub vpt2 vpt Rec fill Bsquare } bind def
/S13 { BL [] 0 setdash 2 copy exch vpt sub exch vpt sub vpt2 vpt Rec fill
       2 copy vpt Square fill Bsquare } bind def
/S14 { BL [] 0 setdash 2 copy exch vpt sub exch vpt sub vpt2 vpt Rec fill
       2 copy exch vpt sub exch vpt Square fill Bsquare } bind def
/S15 { BL [] 0 setdash 2 copy Bsquare fill Bsquare } bind def
/D0 { gsave translate 45 rotate 0 0 S0 stroke grestore } bind def
/D1 { gsave translate 45 rotate 0 0 S1 stroke grestore } bind def
/D2 { gsave translate 45 rotate 0 0 S2 stroke grestore } bind def
/D3 { gsave translate 45 rotate 0 0 S3 stroke grestore } bind def
/D4 { gsave translate 45 rotate 0 0 S4 stroke grestore } bind def
/D5 { gsave translate 45 rotate 0 0 S5 stroke grestore } bind def
/D6 { gsave translate 45 rotate 0 0 S6 stroke grestore } bind def
/D7 { gsave translate 45 rotate 0 0 S7 stroke grestore } bind def
/D8 { gsave translate 45 rotate 0 0 S8 stroke grestore } bind def
/D9 { gsave translate 45 rotate 0 0 S9 stroke grestore } bind def
/D10 { gsave translate 45 rotate 0 0 S10 stroke grestore } bind def
/D11 { gsave translate 45 rotate 0 0 S11 stroke grestore } bind def
/D12 { gsave translate 45 rotate 0 0 S12 stroke grestore } bind def
/D13 { gsave translate 45 rotate 0 0 S13 stroke grestore } bind def
/D14 { gsave translate 45 rotate 0 0 S14 stroke grestore } bind def
/D15 { gsave translate 45 rotate 0 0 S15 stroke grestore } bind def
/DiaE { stroke [] 0 setdash vpt add M
  hpt neg vpt neg V hpt vpt neg V
  hpt vpt V hpt neg vpt V closepath stroke } def
/BoxE { stroke [] 0 setdash exch hpt sub exch vpt add M
  0 vpt2 neg V hpt2 0 V 0 vpt2 V
  hpt2 neg 0 V closepath stroke } def
/TriUE { stroke [] 0 setdash vpt 1.12 mul add M
  hpt neg vpt -1.62 mul V
  hpt 2 mul 0 V
  hpt neg vpt 1.62 mul V closepath stroke } def
/TriDE { stroke [] 0 setdash vpt 1.12 mul sub M
  hpt neg vpt 1.62 mul V
  hpt 2 mul 0 V
  hpt neg vpt -1.62 mul V closepath stroke } def
/PentE { stroke [] 0 setdash gsave
  translate 0 hpt M 4 {72 rotate 0 hpt L} repeat
  closepath stroke grestore } def
/CircE { stroke [] 0 setdash 
  hpt 0 360 arc stroke } def
/Opaque { gsave closepath 1 setgray fill grestore 0 setgray closepath } def
/DiaW { stroke [] 0 setdash vpt add M
  hpt neg vpt neg V hpt vpt neg V
  hpt vpt V hpt neg vpt V Opaque stroke } def
/BoxW { stroke [] 0 setdash exch hpt sub exch vpt add M
  0 vpt2 neg V hpt2 0 V 0 vpt2 V
  hpt2 neg 0 V Opaque stroke } def
/TriUW { stroke [] 0 setdash vpt 1.12 mul add M
  hpt neg vpt -1.62 mul V
  hpt 2 mul 0 V
  hpt neg vpt 1.62 mul V Opaque stroke } def
/TriDW { stroke [] 0 setdash vpt 1.12 mul sub M
  hpt neg vpt 1.62 mul V
  hpt 2 mul 0 V
  hpt neg vpt -1.62 mul V Opaque stroke } def
/PentW { stroke [] 0 setdash gsave
  translate 0 hpt M 4 {72 rotate 0 hpt L} repeat
  Opaque stroke grestore } def
/CircW { stroke [] 0 setdash 
  hpt 0 360 arc Opaque stroke } def
/BoxFill { gsave Rec 1 setgray fill grestore } def
end
}}%
\begin{picture}(3600,2160)(0,0)%
{\GNUPLOTspecial{"
gnudict begin
gsave
0 0 translate
0.100 0.100 scale
0 setgray
newpath
1.000 UL
LTb
400 646 M
63 0 V
2987 0 R
-63 0 V
400 1000 M
63 0 V
2987 0 R
-63 0 V
400 1353 M
63 0 V
2987 0 R
-63 0 V
400 1707 M
63 0 V
2987 0 R
-63 0 V
400 2060 M
63 0 V
2987 0 R
-63 0 V
400 300 M
0 63 V
0 1697 R
0 -63 V
908 300 M
0 63 V
0 1697 R
0 -63 V
1417 300 M
0 63 V
0 1697 R
0 -63 V
1925 300 M
0 63 V
0 1697 R
0 -63 V
2433 300 M
0 63 V
0 1697 R
0 -63 V
2942 300 M
0 63 V
0 1697 R
0 -63 V
3450 300 M
0 63 V
0 1697 R
0 -63 V
1.000 UL
LTb
400 300 M
3050 0 V
0 1760 V
-3050 0 V
400 300 L
1.000 UL
LT0
1518 1919 M
263 0 V
502 307 M
101 37 V
102 46 V
102 53 V
101 57 V
102 62 V
102 65 V
101 68 V
102 72 V
102 74 V
101 77 V
102 80 V
102 83 V
101 82 V
102 85 V
102 85 V
101 86 V
102 86 V
102 87 V
101 87 V
102 88 V
102 85 V
101 88 V
102 85 V
40 35 V
1.000 UL
LT1
1518 1819 M
263 0 V
502 307 M
101 36 V
102 45 V
102 51 V
101 55 V
102 59 V
102 63 V
101 65 V
102 69 V
102 71 V
101 73 V
102 76 V
102 79 V
101 78 V
102 81 V
102 81 V
101 81 V
102 82 V
102 82 V
101 82 V
102 83 V
102 80 V
101 84 V
102 79 V
102 84 V
42 34 V
1.000 UL
LT2
1518 1719 M
263 0 V
502 309 M
101 39 V
102 45 V
102 50 V
101 53 V
102 55 V
102 58 V
101 60 V
102 62 V
102 64 V
101 65 V
102 68 V
102 70 V
101 68 V
102 71 V
102 70 V
101 70 V
102 70 V
102 70 V
101 70 V
102 70 V
102 68 V
101 70 V
102 67 V
102 70 V
101 69 V
102 65 V
102 69 V
36 25 V
1.000 UL
LT3
1518 1619 M
263 0 V
502 309 M
101 38 V
102 44 V
102 48 V
101 51 V
102 53 V
102 56 V
101 57 V
102 59 V
102 61 V
101 63 V
102 64 V
102 66 V
101 65 V
102 67 V
102 66 V
101 67 V
102 66 V
102 67 V
101 65 V
102 67 V
102 64 V
101 66 V
102 63 V
102 65 V
101 66 V
102 60 V
102 66 V
101 64 V
75 47 V
1.000 UP
1.000 UL
LT4
1518 1519 M
263 0 V
-263 31 R
0 -62 V
263 62 R
0 -62 V
715 477 M
0 -89 V
-31 89 R
62 0 V
684 388 M
62 0 V
870 539 M
0 -71 V
-31 71 R
62 0 V
839 468 M
62 0 V
69 170 R
0 -100 V
939 638 M
62 0 V
939 538 M
62 0 V
50 166 R
0 -122 V
-31 122 R
62 0 V
1020 582 M
62 0 V
55 141 R
0 -128 V
-31 128 R
62 0 V
1106 595 M
62 0 V
61 217 R
0 -156 V
-31 156 R
62 0 V
1198 656 M
62 0 V
75 197 R
0 -165 V
-31 165 R
62 0 V
1304 688 M
62 0 V
104 147 R
0 -171 V
-31 171 R
62 0 V
1439 664 M
62 0 V
255 387 R
0 -258 V
-31 258 R
62 0 V
1725 793 M
62 0 V
111 175 R
0 -211 V
-31 211 R
62 0 V
1867 757 M
62 0 V
222 400 R
0 -258 V
-31 258 R
62 0 V
2120 899 M
62 0 V
222 351 R
0 -308 V
-31 308 R
62 0 V
2373 942 M
62 0 V
226 338 R
0 -335 V
-31 335 R
62 0 V
2630 945 M
62 0 V
223 362 R
0 -347 V
-31 347 R
62 0 V
2884 960 M
62 0 V
223 670 R
0 -456 V
-31 456 R
62 0 V
-62 -456 R
62 0 V
715 433 Pls
870 504 Pls
970 588 Pls
1051 643 Pls
1137 659 Pls
1229 734 Pls
1335 771 Pls
1470 750 Pls
1756 922 Pls
1898 863 Pls
2151 1028 Pls
2404 1096 Pls
2661 1113 Pls
2915 1133 Pls
3169 1402 Pls
1649 1519 Pls
1.000 UP
1.000 UL
LT5
1518 1419 M
263 0 V
-263 31 R
0 -62 V
263 62 R
0 -62 V
807 616 M
0 -76 V
-31 76 R
62 0 V
776 540 M
62 0 V
172 184 R
0 -99 V
-31 99 R
62 0 V
979 625 M
62 0 V
172 198 R
0 -126 V
-31 126 R
62 0 V
1182 697 M
62 0 V
173 164 R
0 -134 V
-31 134 R
62 0 V
1386 727 M
62 0 V
172 225 R
0 -158 V
-31 158 R
62 0 V
1589 794 M
62 0 V
172 302 R
0 -191 V
-31 191 R
62 0 V
1792 905 M
62 0 V
173 218 R
0 -239 V
-31 239 R
62 0 V
1996 884 M
62 0 V
807 582 Crs
1010 672 Crs
1213 757 Crs
1417 795 Crs
1620 877 Crs
1823 998 Crs
2027 999 Crs
1649 1419 Crs
1.000 UP
1.000 UL
LT6
1518 1319 M
263 0 V
-263 31 R
0 -62 V
263 62 R
0 -62 V
542 388 M
0 -34 V
-31 34 R
62 0 V
511 354 M
62 0 V
542 371 Star
1649 1319 Star
1.000 UP
1.000 UL
LT7
1518 1219 M
263 0 V
-263 31 R
0 -62 V
263 62 R
0 -62 V
520 393 M
0 -43 V
-31 43 R
62 0 V
489 350 M
62 0 V
7 72 R
0 -68 V
-31 68 R
62 0 V
527 354 M
62 0 V
14 102 R
0 -76 V
-31 76 R
62 0 V
572 380 M
62 0 V
15 115 R
0 -92 V
-31 92 R
62 0 V
618 403 M
62 0 V
520 371 Box
558 388 Box
603 418 Box
649 449 Box
1649 1219 Box
stroke
grestore
end
showpage
}}%
\put(1468,1219){\makebox(0,0)[r]{E401}}%
\put(1468,1319){\makebox(0,0)[r]{E516}}%
\put(1468,1419){\makebox(0,0)[r]{ZEUS 96+97 prel.}}%
\put(1468,1519){\makebox(0,0)[r]{H1 pub}}%
\put(1468,1619){\makebox(0,0)[r]{$m_{c,r}, B_D (W)$}}%
\put(1468,1719){\makebox(0,0)[r]{$B_D(W)$}}%
\put(1468,1819){\makebox(0,0)[r]{$m_{c,r}$}}%
\put(1468,1919){\makebox(0,0)[r]{PDF, $\beta =0$}}%
\put(1925,50){\makebox(0,0){\Large $W$ ~(GeV)}}%
\put(100,1180){%
\special{ps: gsave currentpoint currentpoint translate
270 rotate neg exch neg exch translate}%
\makebox(0,0)[b]{\shortstack{\Large$ \sigma (\gamma P \rightarrow J/\psi P) $(nb)}}%
\special{ps: currentpoint grestore moveto}%
}%
\put(3450,200){\makebox(0,0){300}}%
\put(2942,200){\makebox(0,0){250}}%
\put(2433,200){\makebox(0,0){200}}%
\put(1925,200){\makebox(0,0){150}}%
\put(1417,200){\makebox(0,0){100}}%
\put(908,200){\makebox(0,0){50}}%
\put(400,200){\makebox(0,0){0}}%
\put(350,2060){\makebox(0,0)[r]{250}}%
\put(350,1707){\makebox(0,0)[r]{200}}%
\put(350,1353){\makebox(0,0)[r]{150}}%
\put(350,1000){\makebox(0,0)[r]{100}}%
\put(350,646){\makebox(0,0)[r]{50}}%
\end{picture}%
\endgroup
 

%% file: mrun.tex
\begingroup%
  \makeatletter%
  \newcommand{\GNUPLOTspecial}{%
    \@sanitize\catcode`\%=14\relax\special}%
  \setlength{\unitlength}{0.1bp}%
{\GNUPLOTspecial{!
/gnudict 256 dict def
gnudict begin
/Color false def
/Solid false def
/gnulinewidth 5.000 def
/userlinewidth gnulinewidth def
/vshift -33 def
/dl {10 mul} def
/hpt_ 31.5 def
/vpt_ 31.5 def
/hpt hpt_ def
/vpt vpt_ def
/M {moveto} bind def
/L {lineto} bind def
/R {rmoveto} bind def
/V {rlineto} bind def
/vpt2 vpt 2 mul def
/hpt2 hpt 2 mul def
/Lshow { currentpoint stroke M
  0 vshift R show } def
/Rshow { currentpoint stroke M
  dup stringwidth pop neg vshift R show } def
/Cshow { currentpoint stroke M
  dup stringwidth pop -2 div vshift R show } def
/UP { dup vpt_ mul /vpt exch def hpt_ mul /hpt exch def
  /hpt2 hpt 2 mul def /vpt2 vpt 2 mul def } def
/DL { Color {setrgbcolor Solid {pop []} if 0 setdash }
 {pop pop pop Solid {pop []} if 0 setdash} ifelse } def
/BL { stroke gnulinewidth 2 mul setlinewidth } def
/AL { stroke gnulinewidth 2 div setlinewidth } def
/UL { gnulinewidth mul /userlinewidth exch def } def
/PL { stroke userlinewidth setlinewidth } def
/LTb { BL [] 0 0 0 DL } def
/LTa { AL [1 dl 2 dl] 0 setdash 0 0 0 setrgbcolor } def
/LT0 { PL [] 1 0 0 DL } def
/LT1 { PL [4 dl 2 dl] 0 1 0 DL } def
/LT2 { PL [2 dl 3 dl] 0 0 1 DL } def
/LT3 { PL [1 dl 1.5 dl] 1 0 1 DL } def
/LT4 { PL [5 dl 2 dl 1 dl 2 dl] 0 1 1 DL } def
/LT5 { PL [4 dl 3 dl 1 dl 3 dl] 1 1 0 DL } def
/LT6 { PL [2 dl 2 dl 2 dl 4 dl] 0 0 0 DL } def
/LT7 { PL [2 dl 2 dl 2 dl 2 dl 2 dl 4 dl] 1 0.3 0 DL } def
/LT8 { PL [2 dl 2 dl 2 dl 2 dl 2 dl 2 dl 2 dl 4 dl] 0.5 0.5 0.5 DL } def
/Pnt { stroke [] 0 setdash
   gsave 1 setlinecap M 0 0 V stroke grestore } def
/Dia { stroke [] 0 setdash 2 copy vpt add M
  hpt neg vpt neg V hpt vpt neg V
  hpt vpt V hpt neg vpt V closepath stroke
  Pnt } def
/Pls { stroke [] 0 setdash vpt sub M 0 vpt2 V
  currentpoint stroke M
  hpt neg vpt neg R hpt2 0 V stroke
  } def
/Box { stroke [] 0 setdash 2 copy exch hpt sub exch vpt add M
  0 vpt2 neg V hpt2 0 V 0 vpt2 V
  hpt2 neg 0 V closepath stroke
  Pnt } def
/Crs { stroke [] 0 setdash exch hpt sub exch vpt add M
  hpt2 vpt2 neg V currentpoint stroke M
  hpt2 neg 0 R hpt2 vpt2 V stroke } def
/TriU { stroke [] 0 setdash 2 copy vpt 1.12 mul add M
  hpt neg vpt -1.62 mul V
  hpt 2 mul 0 V
  hpt neg vpt 1.62 mul V closepath stroke
  Pnt  } def
/Star { 2 copy Pls Crs } def
/BoxF { stroke [] 0 setdash exch hpt sub exch vpt add M
  0 vpt2 neg V  hpt2 0 V  0 vpt2 V
  hpt2 neg 0 V  closepath fill } def
/TriUF { stroke [] 0 setdash vpt 1.12 mul add M
  hpt neg vpt -1.62 mul V
  hpt 2 mul 0 V
  hpt neg vpt 1.62 mul V closepath fill } def
/TriD { stroke [] 0 setdash 2 copy vpt 1.12 mul sub M
  hpt neg vpt 1.62 mul V
  hpt 2 mul 0 V
  hpt neg vpt -1.62 mul V closepath stroke
  Pnt  } def
/TriDF { stroke [] 0 setdash vpt 1.12 mul sub M
  hpt neg vpt 1.62 mul V
  hpt 2 mul 0 V
  hpt neg vpt -1.62 mul V closepath fill} def
/DiaF { stroke [] 0 setdash vpt add M
  hpt neg vpt neg V hpt vpt neg V
  hpt vpt V hpt neg vpt V closepath fill } def
/Pent { stroke [] 0 setdash 2 copy gsave
  translate 0 hpt M 4 {72 rotate 0 hpt L} repeat
  closepath stroke grestore Pnt } def
/PentF { stroke [] 0 setdash gsave
  translate 0 hpt M 4 {72 rotate 0 hpt L} repeat
  closepath fill grestore } def
/Circle { stroke [] 0 setdash 2 copy
  hpt 0 360 arc stroke Pnt } def
/CircleF { stroke [] 0 setdash hpt 0 360 arc fill } def
/C0 { BL [] 0 setdash 2 copy moveto vpt 90 450  arc } bind def
/C1 { BL [] 0 setdash 2 copy        moveto
       2 copy  vpt 0 90 arc closepath fill
               vpt 0 360 arc closepath } bind def
/C2 { BL [] 0 setdash 2 copy moveto
       2 copy  vpt 90 180 arc closepath fill
               vpt 0 360 arc closepath } bind def
/C3 { BL [] 0 setdash 2 copy moveto
       2 copy  vpt 0 180 arc closepath fill
               vpt 0 360 arc closepath } bind def
/C4 { BL [] 0 setdash 2 copy moveto
       2 copy  vpt 180 270 arc closepath fill
               vpt 0 360 arc closepath } bind def
/C5 { BL [] 0 setdash 2 copy moveto
       2 copy  vpt 0 90 arc
       2 copy moveto
       2 copy  vpt 180 270 arc closepath fill
               vpt 0 360 arc } bind def
/C6 { BL [] 0 setdash 2 copy moveto
      2 copy  vpt 90 270 arc closepath fill
              vpt 0 360 arc closepath } bind def
/C7 { BL [] 0 setdash 2 copy moveto
      2 copy  vpt 0 270 arc closepath fill
              vpt 0 360 arc closepath } bind def
/C8 { BL [] 0 setdash 2 copy moveto
      2 copy vpt 270 360 arc closepath fill
              vpt 0 360 arc closepath } bind def
/C9 { BL [] 0 setdash 2 copy moveto
      2 copy  vpt 270 450 arc closepath fill
              vpt 0 360 arc closepath } bind def
/C10 { BL [] 0 setdash 2 copy 2 copy moveto vpt 270 360 arc closepath fill
       2 copy moveto
       2 copy vpt 90 180 arc closepath fill
               vpt 0 360 arc closepath } bind def
/C11 { BL [] 0 setdash 2 copy moveto
       2 copy  vpt 0 180 arc closepath fill
       2 copy moveto
       2 copy  vpt 270 360 arc closepath fill
               vpt 0 360 arc closepath } bind def
/C12 { BL [] 0 setdash 2 copy moveto
       2 copy  vpt 180 360 arc closepath fill
               vpt 0 360 arc closepath } bind def
/C13 { BL [] 0 setdash  2 copy moveto
       2 copy  vpt 0 90 arc closepath fill
       2 copy moveto
       2 copy  vpt 180 360 arc closepath fill
               vpt 0 360 arc closepath } bind def
/C14 { BL [] 0 setdash 2 copy moveto
       2 copy  vpt 90 360 arc closepath fill
               vpt 0 360 arc } bind def
/C15 { BL [] 0 setdash 2 copy vpt 0 360 arc closepath fill
               vpt 0 360 arc closepath } bind def
/Rec   { newpath 4 2 roll moveto 1 index 0 rlineto 0 exch rlineto
       neg 0 rlineto closepath } bind def
/Square { dup Rec } bind def
/Bsquare { vpt sub exch vpt sub exch vpt2 Square } bind def
/S0 { BL [] 0 setdash 2 copy moveto 0 vpt rlineto BL Bsquare } bind def
/S1 { BL [] 0 setdash 2 copy vpt Square fill Bsquare } bind def
/S2 { BL [] 0 setdash 2 copy exch vpt sub exch vpt Square fill Bsquare } bind def
/S3 { BL [] 0 setdash 2 copy exch vpt sub exch vpt2 vpt Rec fill Bsquare } bind def
/S4 { BL [] 0 setdash 2 copy exch vpt sub exch vpt sub vpt Square fill Bsquare } bind def
/S5 { BL [] 0 setdash 2 copy 2 copy vpt Square fill
       exch vpt sub exch vpt sub vpt Square fill Bsquare } bind def
/S6 { BL [] 0 setdash 2 copy exch vpt sub exch vpt sub vpt vpt2 Rec fill Bsquare } bind def
/S7 { BL [] 0 setdash 2 copy exch vpt sub exch vpt sub vpt vpt2 Rec fill
       2 copy vpt Square fill
       Bsquare } bind def
/S8 { BL [] 0 setdash 2 copy vpt sub vpt Square fill Bsquare } bind def
/S9 { BL [] 0 setdash 2 copy vpt sub vpt vpt2 Rec fill Bsquare } bind def
/S10 { BL [] 0 setdash 2 copy vpt sub vpt Square fill 2 copy exch vpt sub exch vpt Square fill
       Bsquare } bind def
/S11 { BL [] 0 setdash 2 copy vpt sub vpt Square fill 2 copy exch vpt sub exch vpt2 vpt Rec fill
       Bsquare } bind def
/S12 { BL [] 0 setdash 2 copy exch vpt sub exch vpt sub vpt2 vpt Rec fill Bsquare } bind def
/S13 { BL [] 0 setdash 2 copy exch vpt sub exch vpt sub vpt2 vpt Rec fill
       2 copy vpt Square fill Bsquare } bind def
/S14 { BL [] 0 setdash 2 copy exch vpt sub exch vpt sub vpt2 vpt Rec fill
       2 copy exch vpt sub exch vpt Square fill Bsquare } bind def
/S15 { BL [] 0 setdash 2 copy Bsquare fill Bsquare } bind def
/D0 { gsave translate 45 rotate 0 0 S0 stroke grestore } bind def
/D1 { gsave translate 45 rotate 0 0 S1 stroke grestore } bind def
/D2 { gsave translate 45 rotate 0 0 S2 stroke grestore } bind def
/D3 { gsave translate 45 rotate 0 0 S3 stroke grestore } bind def
/D4 { gsave translate 45 rotate 0 0 S4 stroke grestore } bind def
/D5 { gsave translate 45 rotate 0 0 S5 stroke grestore } bind def
/D6 { gsave translate 45 rotate 0 0 S6 stroke grestore } bind def
/D7 { gsave translate 45 rotate 0 0 S7 stroke grestore } bind def
/D8 { gsave translate 45 rotate 0 0 S8 stroke grestore } bind def
/D9 { gsave translate 45 rotate 0 0 S9 stroke grestore } bind def
/D10 { gsave translate 45 rotate 0 0 S10 stroke grestore } bind def
/D11 { gsave translate 45 rotate 0 0 S11 stroke grestore } bind def
/D12 { gsave translate 45 rotate 0 0 S12 stroke grestore } bind def
/D13 { gsave translate 45 rotate 0 0 S13 stroke grestore } bind def
/D14 { gsave translate 45 rotate 0 0 S14 stroke grestore } bind def
/D15 { gsave translate 45 rotate 0 0 S15 stroke grestore } bind def
/DiaE { stroke [] 0 setdash vpt add M
  hpt neg vpt neg V hpt vpt neg V
  hpt vpt V hpt neg vpt V closepath stroke } def
/BoxE { stroke [] 0 setdash exch hpt sub exch vpt add M
  0 vpt2 neg V hpt2 0 V 0 vpt2 V
  hpt2 neg 0 V closepath stroke } def
/TriUE { stroke [] 0 setdash vpt 1.12 mul add M
  hpt neg vpt -1.62 mul V
  hpt 2 mul 0 V
  hpt neg vpt 1.62 mul V closepath stroke } def
/TriDE { stroke [] 0 setdash vpt 1.12 mul sub M
  hpt neg vpt 1.62 mul V
  hpt 2 mul 0 V
  hpt neg vpt -1.62 mul V closepath stroke } def
/PentE { stroke [] 0 setdash gsave
  translate 0 hpt M 4 {72 rotate 0 hpt L} repeat
  closepath stroke grestore } def
/CircE { stroke [] 0 setdash 
  hpt 0 360 arc stroke } def
/Opaque { gsave closepath 1 setgray fill grestore 0 setgray closepath } def
/DiaW { stroke [] 0 setdash vpt add M
  hpt neg vpt neg V hpt vpt neg V
  hpt vpt V hpt neg vpt V Opaque stroke } def
/BoxW { stroke [] 0 setdash exch hpt sub exch vpt add M
  0 vpt2 neg V hpt2 0 V 0 vpt2 V
  hpt2 neg 0 V Opaque stroke } def
/TriUW { stroke [] 0 setdash vpt 1.12 mul add M
  hpt neg vpt -1.62 mul V
  hpt 2 mul 0 V
  hpt neg vpt 1.62 mul V Opaque stroke } def
/TriDW { stroke [] 0 setdash vpt 1.12 mul sub M
  hpt neg vpt 1.62 mul V
  hpt 2 mul 0 V
  hpt neg vpt -1.62 mul V Opaque stroke } def
/PentW { stroke [] 0 setdash gsave
  translate 0 hpt M 4 {72 rotate 0 hpt L} repeat
  Opaque stroke grestore } def
/CircW { stroke [] 0 setdash 
  hpt 0 360 arc Opaque stroke } def
/BoxFill { gsave Rec 1 setgray fill grestore } def
end
}}%
\begin{picture}(3600,2160)(0,0)%
{\GNUPLOTspecial{"
gnudict begin
gsave
0 0 translate
0.100 0.100 scale
0 setgray
newpath
1.000 UL
LTb
400 300 M
63 0 V
2987 0 R
-63 0 V
400 593 M
63 0 V
2987 0 R
-63 0 V
400 887 M
63 0 V
2987 0 R
-63 0 V
400 1180 M
63 0 V
2987 0 R
-63 0 V
400 1473 M
63 0 V
2987 0 R
-63 0 V
400 1767 M
63 0 V
2987 0 R
-63 0 V
400 2060 M
63 0 V
2987 0 R
-63 0 V
400 300 M
0 63 V
0 1697 R
0 -63 V
1010 300 M
0 63 V
0 1697 R
0 -63 V
1620 300 M
0 63 V
0 1697 R
0 -63 V
2230 300 M
0 63 V
0 1697 R
0 -63 V
2840 300 M
0 63 V
0 1697 R
0 -63 V
3450 300 M
0 63 V
0 1697 R
0 -63 V
1.000 UL
LTb
400 300 M
3050 0 V
0 1760 V
-3050 0 V
400 300 L
1.000 UL
LT0
3087 1947 M
263 0 V
425 300 M
6 334 V
30 100 V
30 67 V
31 52 V
31 43 V
30 38 V
31 33 V
30 31 V
30 28 V
31 26 V
31 24 V
30 23 V
31 22 V
30 21 V
30 20 V
31 19 V
31 18 V
30 18 V
31 17 V
30 17 V
30 16 V
31 15 V
31 16 V
30 15 V
31 14 V
30 16 V
31 15 V
30 15 V
30 15 V
31 14 V
30 14 V
31 14 V
31 14 V
30 14 V
30 13 V
31 13 V
30 13 V
31 13 V
31 13 V
30 13 V
30 12 V
31 13 V
30 12 V
31 12 V
31 12 V
30 12 V
30 12 V
31 11 V
30 12 V
31 12 V
31 11 V
30 11 V
31 12 V
30 11 V
31 11 V
30 11 V
30 11 V
31 11 V
30 11 V
31 11 V
30 0 V
31 0 V
31 0 V
30 0 V
31 0 V
30 0 V
31 0 V
30 0 V
30 0 V
31 0 V
30 0 V
31 0 V
30 0 V
31 0 V
31 0 V
30 0 V
31 0 V
30 0 V
31 0 V
30 0 V
31 0 V
30 0 V
30 0 V
31 0 V
30 0 V
31 0 V
30 0 V
31 0 V
31 0 V
30 0 V
31 0 V
30 0 V
31 0 V
30 0 V
30 0 V
31 0 V
30 0 V
31 0 V
30 0 V
31 0 V
stroke
grestore
end
showpage
}}%
\put(3037,1947){\makebox(0,0)[r]{Leading order}}%
\put(1925,50){\makebox(0,0){\Large $b$ ~(fm)}}%
\put(100,1180){%
\special{ps: gsave currentpoint currentpoint translate
270 rotate neg exch neg exch translate}%
\makebox(0,0)[b]{\shortstack{$m^2({\bar Q}^2)/m_{fixed}^2$}}%
\special{ps: currentpoint grestore moveto}%
}%
\put(3450,200){\makebox(0,0){0.5}}%
\put(2840,200){\makebox(0,0){0.4}}%
\put(2230,200){\makebox(0,0){0.3}}%
\put(1620,200){\makebox(0,0){0.2}}%
\put(1010,200){\makebox(0,0){0.1}}%
\put(400,200){\makebox(0,0){0}}%
\put(350,2060){\makebox(0,0)[r]{1.1}}%
\put(350,1767){\makebox(0,0)[r]{1}}%
\put(350,1473){\makebox(0,0)[r]{0.9}}%
\put(350,1180){\makebox(0,0)[r]{0.8}}%
\put(350,887){\makebox(0,0)[r]{0.7}}%
\put(350,593){\makebox(0,0)[r]{0.6}}%
\put(350,300){\makebox(0,0)[r]{0.5}}%
\end{picture}%
\endgroup
 

%% file: intz.tex
\begingroup%
  \makeatletter%
  \newcommand{\GNUPLOTspecial}{%
    \@sanitize\catcode`\%=14\relax\special}%
  \setlength{\unitlength}{0.1bp}%
{\GNUPLOTspecial{!
/gnudict 256 dict def
gnudict begin
/Color false def
/Solid false def
/gnulinewidth 5.000 def
/userlinewidth gnulinewidth def
/vshift -33 def
/dl {10 mul} def
/hpt_ 31.5 def
/vpt_ 31.5 def
/hpt hpt_ def
/vpt vpt_ def
/M {moveto} bind def
/L {lineto} bind def
/R {rmoveto} bind def
/V {rlineto} bind def
/vpt2 vpt 2 mul def
/hpt2 hpt 2 mul def
/Lshow { currentpoint stroke M
  0 vshift R show } def
/Rshow { currentpoint stroke M
  dup stringwidth pop neg vshift R show } def
/Cshow { currentpoint stroke M
  dup stringwidth pop -2 div vshift R show } def
/UP { dup vpt_ mul /vpt exch def hpt_ mul /hpt exch def
  /hpt2 hpt 2 mul def /vpt2 vpt 2 mul def } def
/DL { Color {setrgbcolor Solid {pop []} if 0 setdash }
 {pop pop pop Solid {pop []} if 0 setdash} ifelse } def
/BL { stroke userlinewidth 2 mul setlinewidth } def
/AL { stroke userlinewidth 2 div setlinewidth } def
/UL { dup gnulinewidth mul /userlinewidth exch def
      10 mul /udl exch def } def
/PL { stroke userlinewidth setlinewidth } def
/LTb { BL [] 0 0 0 DL } def
/LTa { AL [1 udl mul 2 udl mul] 0 setdash 0 0 0 setrgbcolor } def
/LT0 { PL [] 1 0 0 DL } def
/LT1 { PL [4 dl 2 dl] 0 1 0 DL } def
/LT2 { PL [2 dl 3 dl] 0 0 1 DL } def
/LT3 { PL [1 dl 1.5 dl] 1 0 1 DL } def
/LT4 { PL [5 dl 2 dl 1 dl 2 dl] 0 1 1 DL } def
/LT5 { PL [4 dl 3 dl 1 dl 3 dl] 1 1 0 DL } def
/LT6 { PL [2 dl 2 dl 2 dl 4 dl] 0 0 0 DL } def
/LT7 { PL [2 dl 2 dl 2 dl 2 dl 2 dl 4 dl] 1 0.3 0 DL } def
/LT8 { PL [2 dl 2 dl 2 dl 2 dl 2 dl 2 dl 2 dl 4 dl] 0.5 0.5 0.5 DL } def
/Pnt { stroke [] 0 setdash
   gsave 1 setlinecap M 0 0 V stroke grestore } def
/Dia { stroke [] 0 setdash 2 copy vpt add M
  hpt neg vpt neg V hpt vpt neg V
  hpt vpt V hpt neg vpt V closepath stroke
  Pnt } def
/Pls { stroke [] 0 setdash vpt sub M 0 vpt2 V
  currentpoint stroke M
  hpt neg vpt neg R hpt2 0 V stroke
  } def
/Box { stroke [] 0 setdash 2 copy exch hpt sub exch vpt add M
  0 vpt2 neg V hpt2 0 V 0 vpt2 V
  hpt2 neg 0 V closepath stroke
  Pnt } def
/Crs { stroke [] 0 setdash exch hpt sub exch vpt add M
  hpt2 vpt2 neg V currentpoint stroke M
  hpt2 neg 0 R hpt2 vpt2 V stroke } def
/TriU { stroke [] 0 setdash 2 copy vpt 1.12 mul add M
  hpt neg vpt -1.62 mul V
  hpt 2 mul 0 V
  hpt neg vpt 1.62 mul V closepath stroke
  Pnt  } def
/Star { 2 copy Pls Crs } def
/BoxF { stroke [] 0 setdash exch hpt sub exch vpt add M
  0 vpt2 neg V  hpt2 0 V  0 vpt2 V
  hpt2 neg 0 V  closepath fill } def
/TriUF { stroke [] 0 setdash vpt 1.12 mul add M
  hpt neg vpt -1.62 mul V
  hpt 2 mul 0 V
  hpt neg vpt 1.62 mul V closepath fill } def
/TriD { stroke [] 0 setdash 2 copy vpt 1.12 mul sub M
  hpt neg vpt 1.62 mul V
  hpt 2 mul 0 V
  hpt neg vpt -1.62 mul V closepath stroke
  Pnt  } def
/TriDF { stroke [] 0 setdash vpt 1.12 mul sub M
  hpt neg vpt 1.62 mul V
  hpt 2 mul 0 V
  hpt neg vpt -1.62 mul V closepath fill} def
/DiaF { stroke [] 0 setdash vpt add M
  hpt neg vpt neg V hpt vpt neg V
  hpt vpt V hpt neg vpt V closepath fill } def
/Pent { stroke [] 0 setdash 2 copy gsave
  translate 0 hpt M 4 {72 rotate 0 hpt L} repeat
  closepath stroke grestore Pnt } def
/PentF { stroke [] 0 setdash gsave
  translate 0 hpt M 4 {72 rotate 0 hpt L} repeat
  closepath fill grestore } def
/Circle { stroke [] 0 setdash 2 copy
  hpt 0 360 arc stroke Pnt } def
/CircleF { stroke [] 0 setdash hpt 0 360 arc fill } def
/C0 { BL [] 0 setdash 2 copy moveto vpt 90 450  arc } bind def
/C1 { BL [] 0 setdash 2 copy        moveto
       2 copy  vpt 0 90 arc closepath fill
               vpt 0 360 arc closepath } bind def
/C2 { BL [] 0 setdash 2 copy moveto
       2 copy  vpt 90 180 arc closepath fill
               vpt 0 360 arc closepath } bind def
/C3 { BL [] 0 setdash 2 copy moveto
       2 copy  vpt 0 180 arc closepath fill
               vpt 0 360 arc closepath } bind def
/C4 { BL [] 0 setdash 2 copy moveto
       2 copy  vpt 180 270 arc closepath fill
               vpt 0 360 arc closepath } bind def
/C5 { BL [] 0 setdash 2 copy moveto
       2 copy  vpt 0 90 arc
       2 copy moveto
       2 copy  vpt 180 270 arc closepath fill
               vpt 0 360 arc } bind def
/C6 { BL [] 0 setdash 2 copy moveto
      2 copy  vpt 90 270 arc closepath fill
              vpt 0 360 arc closepath } bind def
/C7 { BL [] 0 setdash 2 copy moveto
      2 copy  vpt 0 270 arc closepath fill
              vpt 0 360 arc closepath } bind def
/C8 { BL [] 0 setdash 2 copy moveto
      2 copy vpt 270 360 arc closepath fill
              vpt 0 360 arc closepath } bind def
/C9 { BL [] 0 setdash 2 copy moveto
      2 copy  vpt 270 450 arc closepath fill
              vpt 0 360 arc closepath } bind def
/C10 { BL [] 0 setdash 2 copy 2 copy moveto vpt 270 360 arc closepath fill
       2 copy moveto
       2 copy vpt 90 180 arc closepath fill
               vpt 0 360 arc closepath } bind def
/C11 { BL [] 0 setdash 2 copy moveto
       2 copy  vpt 0 180 arc closepath fill
       2 copy moveto
       2 copy  vpt 270 360 arc closepath fill
               vpt 0 360 arc closepath } bind def
/C12 { BL [] 0 setdash 2 copy moveto
       2 copy  vpt 180 360 arc closepath fill
               vpt 0 360 arc closepath } bind def
/C13 { BL [] 0 setdash  2 copy moveto
       2 copy  vpt 0 90 arc closepath fill
       2 copy moveto
       2 copy  vpt 180 360 arc closepath fill
               vpt 0 360 arc closepath } bind def
/C14 { BL [] 0 setdash 2 copy moveto
       2 copy  vpt 90 360 arc closepath fill
               vpt 0 360 arc } bind def
/C15 { BL [] 0 setdash 2 copy vpt 0 360 arc closepath fill
               vpt 0 360 arc closepath } bind def
/Rec   { newpath 4 2 roll moveto 1 index 0 rlineto 0 exch rlineto
       neg 0 rlineto closepath } bind def
/Square { dup Rec } bind def
/Bsquare { vpt sub exch vpt sub exch vpt2 Square } bind def
/S0 { BL [] 0 setdash 2 copy moveto 0 vpt rlineto BL Bsquare } bind def
/S1 { BL [] 0 setdash 2 copy vpt Square fill Bsquare } bind def
/S2 { BL [] 0 setdash 2 copy exch vpt sub exch vpt Square fill Bsquare } bind def
/S3 { BL [] 0 setdash 2 copy exch vpt sub exch vpt2 vpt Rec fill Bsquare } bind def
/S4 { BL [] 0 setdash 2 copy exch vpt sub exch vpt sub vpt Square fill Bsquare } bind def
/S5 { BL [] 0 setdash 2 copy 2 copy vpt Square fill
       exch vpt sub exch vpt sub vpt Square fill Bsquare } bind def
/S6 { BL [] 0 setdash 2 copy exch vpt sub exch vpt sub vpt vpt2 Rec fill Bsquare } bind def
/S7 { BL [] 0 setdash 2 copy exch vpt sub exch vpt sub vpt vpt2 Rec fill
       2 copy vpt Square fill
       Bsquare } bind def
/S8 { BL [] 0 setdash 2 copy vpt sub vpt Square fill Bsquare } bind def
/S9 { BL [] 0 setdash 2 copy vpt sub vpt vpt2 Rec fill Bsquare } bind def
/S10 { BL [] 0 setdash 2 copy vpt sub vpt Square fill 2 copy exch vpt sub exch vpt Square fill
       Bsquare } bind def
/S11 { BL [] 0 setdash 2 copy vpt sub vpt Square fill 2 copy exch vpt sub exch vpt2 vpt Rec fill
       Bsquare } bind def
/S12 { BL [] 0 setdash 2 copy exch vpt sub exch vpt sub vpt2 vpt Rec fill Bsquare } bind def
/S13 { BL [] 0 setdash 2 copy exch vpt sub exch vpt sub vpt2 vpt Rec fill
       2 copy vpt Square fill Bsquare } bind def
/S14 { BL [] 0 setdash 2 copy exch vpt sub exch vpt sub vpt2 vpt Rec fill
       2 copy exch vpt sub exch vpt Square fill Bsquare } bind def
/S15 { BL [] 0 setdash 2 copy Bsquare fill Bsquare } bind def
/D0 { gsave translate 45 rotate 0 0 S0 stroke grestore } bind def
/D1 { gsave translate 45 rotate 0 0 S1 stroke grestore } bind def
/D2 { gsave translate 45 rotate 0 0 S2 stroke grestore } bind def
/D3 { gsave translate 45 rotate 0 0 S3 stroke grestore } bind def
/D4 { gsave translate 45 rotate 0 0 S4 stroke grestore } bind def
/D5 { gsave translate 45 rotate 0 0 S5 stroke grestore } bind def
/D6 { gsave translate 45 rotate 0 0 S6 stroke grestore } bind def
/D7 { gsave translate 45 rotate 0 0 S7 stroke grestore } bind def
/D8 { gsave translate 45 rotate 0 0 S8 stroke grestore } bind def
/D9 { gsave translate 45 rotate 0 0 S9 stroke grestore } bind def
/D10 { gsave translate 45 rotate 0 0 S10 stroke grestore } bind def
/D11 { gsave translate 45 rotate 0 0 S11 stroke grestore } bind def
/D12 { gsave translate 45 rotate 0 0 S12 stroke grestore } bind def
/D13 { gsave translate 45 rotate 0 0 S13 stroke grestore } bind def
/D14 { gsave translate 45 rotate 0 0 S14 stroke grestore } bind def
/D15 { gsave translate 45 rotate 0 0 S15 stroke grestore } bind def
/DiaE { stroke [] 0 setdash vpt add M
  hpt neg vpt neg V hpt vpt neg V
  hpt vpt V hpt neg vpt V closepath stroke } def
/BoxE { stroke [] 0 setdash exch hpt sub exch vpt add M
  0 vpt2 neg V hpt2 0 V 0 vpt2 V
  hpt2 neg 0 V closepath stroke } def
/TriUE { stroke [] 0 setdash vpt 1.12 mul add M
  hpt neg vpt -1.62 mul V
  hpt 2 mul 0 V
  hpt neg vpt 1.62 mul V closepath stroke } def
/TriDE { stroke [] 0 setdash vpt 1.12 mul sub M
  hpt neg vpt 1.62 mul V
  hpt 2 mul 0 V
  hpt neg vpt -1.62 mul V closepath stroke } def
/PentE { stroke [] 0 setdash gsave
  translate 0 hpt M 4 {72 rotate 0 hpt L} repeat
  closepath stroke grestore } def
/CircE { stroke [] 0 setdash 
  hpt 0 360 arc stroke } def
/Opaque { gsave closepath 1 setgray fill grestore 0 setgray closepath } def
/DiaW { stroke [] 0 setdash vpt add M
  hpt neg vpt neg V hpt vpt neg V
  hpt vpt V hpt neg vpt V Opaque stroke } def
/BoxW { stroke [] 0 setdash exch hpt sub exch vpt add M
  0 vpt2 neg V hpt2 0 V 0 vpt2 V
  hpt2 neg 0 V Opaque stroke } def
/TriUW { stroke [] 0 setdash vpt 1.12 mul add M
  hpt neg vpt -1.62 mul V
  hpt 2 mul 0 V
  hpt neg vpt 1.62 mul V Opaque stroke } def
/TriDW { stroke [] 0 setdash vpt 1.12 mul sub M
  hpt neg vpt 1.62 mul V
  hpt 2 mul 0 V
  hpt neg vpt -1.62 mul V Opaque stroke } def
/PentW { stroke [] 0 setdash gsave
  translate 0 hpt M 4 {72 rotate 0 hpt L} repeat
  Opaque stroke grestore } def
/CircW { stroke [] 0 setdash 
  hpt 0 360 arc Opaque stroke } def
/BoxFill { gsave Rec 1 setgray fill grestore } def
end
}}%
\begin{picture}(3600,2160)(0,0)%
{\GNUPLOTspecial{"
gnudict begin
gsave
0 0 translate
0.100 0.100 scale
0 setgray
newpath
1.000 UL
LTb
500 300 M
63 0 V
2887 0 R
-63 0 V
500 593 M
63 0 V
2887 0 R
-63 0 V
500 887 M
63 0 V
2887 0 R
-63 0 V
500 1180 M
63 0 V
2887 0 R
-63 0 V
500 1473 M
63 0 V
2887 0 R
-63 0 V
500 1767 M
63 0 V
2887 0 R
-63 0 V
500 2060 M
63 0 V
2887 0 R
-63 0 V
500 300 M
0 63 V
0 1697 R
0 -63 V
1090 300 M
0 63 V
0 1697 R
0 -63 V
1680 300 M
0 63 V
0 1697 R
0 -63 V
2270 300 M
0 63 V
0 1697 R
0 -63 V
2860 300 M
0 63 V
0 1697 R
0 -63 V
3450 300 M
0 63 V
0 1697 R
0 -63 V
1.000 UL
LTb
500 300 M
2950 0 V
0 1760 V
-2950 0 V
500 300 L
1.000 UL
LT0
3087 1947 M
263 0 V
500 300 M
6 250 V
6 170 V
6 139 V
6 119 V
6 105 V
5 94 V
6 83 V
6 76 V
6 68 V
6 63 V
6 56 V
6 52 V
6 47 V
6 43 V
5 39 V
6 35 V
6 32 V
6 29 V
6 26 V
6 24 V
6 21 V
6 18 V
6 17 V
6 14 V
6 13 V
5 10 V
6 9 V
6 7 V
6 6 V
6 4 V
6 3 V
6 1 V
6 1 V
6 -1 V
6 -2 V
5 -4 V
6 -4 V
6 -5 V
6 -6 V
6 -6 V
6 -8 V
6 -8 V
6 -9 V
6 -10 V
5 -10 V
6 -11 V
6 -11 V
6 -12 V
6 -13 V
6 -13 V
6 -13 V
6 -13 V
6 -15 V
6 -14 V
6 -15 V
5 -15 V
6 -15 V
6 -15 V
6 -16 V
6 -16 V
6 -16 V
6 -16 V
6 -17 V
6 -16 V
6 -17 V
5 -16 V
6 -17 V
6 -17 V
6 -17 V
6 -17 V
6 -17 V
6 -17 V
6 -17 V
6 -17 V
5 -16 V
6 -17 V
6 -17 V
6 -17 V
6 -16 V
6 -17 V
6 -17 V
6 -16 V
6 -16 V
6 -17 V
6 -16 V
5 -16 V
6 -16 V
6 -15 V
6 -16 V
6 -15 V
6 -16 V
6 -15 V
6 -15 V
6 -15 V
6 -14 V
5 -15 V
6 -14 V
6 -14 V
6 -14 V
6 -14 V
6 -14 V
6 -13 V
6 -14 V
6 -13 V
5 -13 V
6 -12 V
6 -13 V
6 -12 V
6 -12 V
6 -12 V
6 -12 V
6 -12 V
6 -11 V
6 -12 V
6 -11 V
5 -11 V
6 -10 V
6 -11 V
6 -10 V
6 -10 V
6 -10 V
6 -10 V
6 -9 V
6 -10 V
6 -9 V
5 -9 V
6 -9 V
6 -9 V
6 -8 V
6 -8 V
6 -9 V
6 -8 V
6 -7 V
6 -8 V
6 -8 V
5 -7 V
6 -7 V
6 -7 V
6 -7 V
6 -7 V
6 -6 V
6 -7 V
6 -6 V
6 -6 V
5 -6 V
6 -6 V
6 -5 V
6 -6 V
6 -5 V
6 -6 V
6 -5 V
6 -5 V
6 -5 V
6 -5 V
5 -5 V
6 -4 V
6 -5 V
6 -5 V
6 -4 V
6 -5 V
6 -4 V
6 -5 V
6 -4 V
6 -4 V
6 -4 V
5 -4 V
6 -4 V
6 -4 V
6 -4 V
6 -4 V
6 -4 V
6 -3 V
6 -4 V
6 -3 V
5 -4 V
6 -3 V
6 -4 V
6 -3 V
6 -3 V
6 -4 V
6 -3 V
6 -3 V
6 -3 V
6 -3 V
5 -3 V
6 -3 V
6 -2 V
6 -3 V
6 -3 V
6 -3 V
6 -2 V
6 -3 V
6 -3 V
6 -2 V
6 -3 V
5 -2 V
6 -2 V
6 -3 V
6 -2 V
6 -2 V
6 -2 V
6 -3 V
6 -2 V
6 -2 V
5 -2 V
6 -2 V
6 -2 V
6 -2 V
6 -2 V
6 -2 V
6 -2 V
6 -1 V
6 -2 V
6 -2 V
5 -2 V
6 -1 V
6 -2 V
6 -2 V
6 -1 V
6 -2 V
6 -1 V
6 -2 V
6 -1 V
6 -2 V
6 -1 V
5 -1 V
6 -2 V
6 -1 V
6 -1 V
6 -2 V
6 -1 V
6 -1 V
6 -1 V
6 -2 V
5 -1 V
6 -1 V
6 -1 V
6 -1 V
6 -1 V
6 -1 V
6 -1 V
6 -1 V
6 -1 V
6 -1 V
5 -1 V
6 -1 V
6 -1 V
6 -1 V
6 -1 V
6 -1 V
6 -1 V
6 -1 V
6 0 V
6 -1 V
6 -1 V
5 -1 V
6 -1 V
6 0 V
6 -1 V
6 -1 V
6 0 V
6 -1 V
6 -1 V
6 0 V
6 -1 V
5 -1 V
6 0 V
6 -1 V
6 -1 V
6 0 V
6 -1 V
6 0 V
6 -1 V
6 0 V
6 -1 V
5 0 V
6 -1 V
6 0 V
6 -1 V
6 0 V
6 -1 V
6 0 V
6 -1 V
6 0 V
5 -1 V
6 0 V
6 0 V
6 -1 V
6 0 V
6 -1 V
6 0 V
6 0 V
6 -1 V
6 0 V
5 0 V
6 -1 V
6 0 V
6 0 V
6 -1 V
6 0 V
6 0 V
6 -1 V
6 0 V
6 0 V
5 0 V
6 -1 V
6 0 V
6 0 V
6 -1 V
6 0 V
6 0 V
6 0 V
6 -1 V
6 0 V
6 0 V
5 0 V
6 0 V
6 -1 V
6 0 V
6 0 V
6 0 V
6 0 V
6 -1 V
6 0 V
6 0 V
5 0 V
6 0 V
6 -1 V
6 0 V
6 0 V
6 0 V
6 0 V
6 0 V
6 0 V
6 -1 V
5 0 V
6 0 V
6 0 V
6 0 V
6 0 V
6 0 V
6 -1 V
6 0 V
6 0 V
5 0 V
6 0 V
6 0 V
6 0 V
6 0 V
6 0 V
6 -1 V
6 0 V
6 0 V
6 0 V
5 0 V
6 0 V
6 0 V
6 0 V
6 0 V
6 0 V
6 0 V
6 -1 V
6 0 V
6 0 V
5 0 V
6 0 V
6 0 V
6 0 V
6 0 V
6 0 V
6 0 V
6 0 V
6 0 V
6 0 V
6 0 V
5 -1 V
6 0 V
6 0 V
6 0 V
6 0 V
6 0 V
6 0 V
6 0 V
6 0 V
6 0 V
5 0 V
6 0 V
6 0 V
6 0 V
6 0 V
6 0 V
6 0 V
6 0 V
6 0 V
6 0 V
5 0 V
6 0 V
currentpoint stroke M
6 -1 V
6 0 V
6 0 V
6 0 V
6 0 V
6 0 V
6 0 V
6 0 V
5 0 V
6 0 V
6 0 V
6 0 V
6 0 V
6 0 V
6 0 V
6 0 V
6 0 V
5 0 V
6 0 V
6 0 V
6 0 V
6 0 V
6 0 V
6 0 V
6 0 V
6 0 V
6 0 V
5 0 V
6 0 V
6 0 V
6 0 V
6 0 V
6 0 V
6 0 V
6 0 V
6 0 V
6 0 V
5 0 V
6 0 V
6 0 V
6 0 V
6 0 V
6 0 V
6 0 V
6 0 V
6 0 V
6 -1 V
6 0 V
5 0 V
6 0 V
6 0 V
6 0 V
6 0 V
6 0 V
6 0 V
6 0 V
6 0 V
6 0 V
5 0 V
6 0 V
6 0 V
6 0 V
6 0 V
6 0 V
6 0 V
6 0 V
6 0 V
6 0 V
5 0 V
6 0 V
6 0 V
6 0 V
6 0 V
6 0 V
6 0 V
6 0 V
6 0 V
5 0 V
6 0 V
6 0 V
6 0 V
6 0 V
6 0 V
6 0 V
6 0 V
6 0 V
6 0 V
5 0 V
6 0 V
6 0 V
6 0 V
6 0 V
6 0 V
6 0 V
6 0 V
6 0 V
6 0 V
5 0 V
6 0 V
6 0 V
6 0 V
6 0 V
6 0 V
1.000 UL
LT1
3087 1847 M
263 0 V
500 300 M
6 94 V
6 81 V
6 73 V
6 69 V
6 64 V
5 61 V
6 57 V
6 54 V
6 51 V
6 49 V
6 46 V
6 44 V
6 41 V
6 40 V
5 37 V
6 35 V
6 34 V
6 31 V
6 30 V
6 28 V
6 27 V
6 25 V
6 23 V
6 22 V
6 21 V
5 19 V
6 18 V
6 17 V
6 15 V
6 14 V
6 13 V
6 11 V
6 11 V
6 10 V
6 8 V
5 8 V
6 6 V
6 6 V
6 4 V
6 4 V
6 3 V
6 2 V
6 1 V
6 1 V
5 -1 V
6 -1 V
6 -1 V
6 -2 V
6 -3 V
6 -4 V
6 -4 V
6 -5 V
6 -5 V
6 -6 V
6 -6 V
5 -7 V
6 -7 V
6 -8 V
6 -8 V
6 -9 V
6 -9 V
6 -9 V
6 -9 V
6 -10 V
6 -10 V
5 -10 V
6 -11 V
6 -11 V
6 -11 V
6 -11 V
6 -12 V
6 -12 V
6 -11 V
6 -13 V
5 -12 V
6 -12 V
6 -13 V
6 -12 V
6 -13 V
6 -13 V
6 -13 V
6 -13 V
6 -13 V
6 -13 V
6 -13 V
5 -13 V
6 -13 V
6 -13 V
6 -13 V
6 -13 V
6 -14 V
6 -13 V
6 -13 V
6 -13 V
6 -12 V
5 -13 V
6 -13 V
6 -13 V
6 -12 V
6 -13 V
6 -12 V
6 -13 V
6 -12 V
6 -12 V
5 -12 V
6 -12 V
6 -12 V
6 -12 V
6 -11 V
6 -12 V
6 -11 V
6 -11 V
6 -11 V
6 -11 V
6 -11 V
5 -10 V
6 -11 V
6 -10 V
6 -10 V
6 -10 V
6 -10 V
6 -10 V
6 -10 V
6 -9 V
6 -9 V
5 -9 V
6 -9 V
6 -9 V
6 -9 V
6 -8 V
6 -9 V
6 -8 V
6 -8 V
6 -8 V
6 -7 V
5 -8 V
6 -7 V
6 -8 V
6 -7 V
6 -7 V
6 -7 V
6 -6 V
6 -7 V
6 -6 V
5 -6 V
6 -7 V
6 -6 V
6 -5 V
6 -6 V
6 -6 V
6 -5 V
6 -5 V
6 -5 V
6 -5 V
5 -5 V
6 -4 V
6 -5 V
6 -5 V
6 -4 V
6 -5 V
6 -4 V
6 -5 V
6 -4 V
6 -4 V
6 -4 V
5 -4 V
6 -4 V
6 -4 V
6 -4 V
6 -4 V
6 -4 V
6 -3 V
6 -4 V
6 -3 V
5 -4 V
6 -3 V
6 -4 V
6 -3 V
6 -3 V
6 -4 V
6 -3 V
6 -3 V
6 -3 V
6 -3 V
5 -3 V
6 -3 V
6 -2 V
6 -3 V
6 -3 V
6 -3 V
6 -2 V
6 -3 V
6 -3 V
6 -2 V
6 -3 V
5 -2 V
6 -2 V
6 -3 V
6 -2 V
6 -2 V
6 -2 V
6 -3 V
6 -2 V
6 -2 V
5 -2 V
6 -2 V
6 -2 V
6 -2 V
6 -2 V
6 -2 V
6 -2 V
6 -1 V
6 -2 V
6 -2 V
5 -2 V
6 -1 V
6 -2 V
6 -2 V
6 -1 V
6 -2 V
6 -1 V
6 -2 V
6 -1 V
6 -2 V
6 -1 V
5 -1 V
6 -2 V
6 -1 V
6 -1 V
6 -2 V
6 -1 V
6 -1 V
6 -1 V
6 -2 V
5 -1 V
6 -1 V
6 -1 V
6 -1 V
6 -1 V
6 -1 V
6 -1 V
6 -1 V
6 -1 V
6 -1 V
5 -1 V
6 -1 V
6 -1 V
6 -1 V
6 -1 V
6 -1 V
6 -1 V
6 -1 V
6 0 V
6 -1 V
6 -1 V
5 -1 V
6 -1 V
6 0 V
6 -1 V
6 -1 V
6 0 V
6 -1 V
6 -1 V
6 0 V
6 -1 V
5 -1 V
6 0 V
6 -1 V
6 -1 V
6 0 V
6 -1 V
6 0 V
6 -1 V
6 0 V
6 -1 V
5 0 V
6 -1 V
6 0 V
6 -1 V
6 0 V
6 -1 V
6 0 V
6 -1 V
6 0 V
5 -1 V
6 0 V
6 0 V
6 -1 V
6 0 V
6 -1 V
6 0 V
6 0 V
6 -1 V
6 0 V
5 0 V
6 -1 V
6 0 V
6 0 V
6 -1 V
6 0 V
6 0 V
6 -1 V
6 0 V
6 0 V
5 0 V
6 -1 V
6 0 V
6 0 V
6 -1 V
6 0 V
6 0 V
6 0 V
6 -1 V
6 0 V
6 0 V
5 0 V
6 0 V
6 -1 V
6 0 V
6 0 V
6 0 V
6 0 V
6 -1 V
6 0 V
6 0 V
5 0 V
6 0 V
6 -1 V
6 0 V
6 0 V
6 0 V
6 0 V
6 0 V
6 0 V
6 -1 V
5 0 V
6 0 V
6 0 V
6 0 V
6 0 V
6 0 V
6 -1 V
6 0 V
6 0 V
5 0 V
6 0 V
6 0 V
6 0 V
6 0 V
6 0 V
6 -1 V
6 0 V
6 0 V
6 0 V
5 0 V
6 0 V
6 0 V
6 0 V
6 0 V
6 0 V
6 0 V
6 -1 V
6 0 V
6 0 V
5 0 V
6 0 V
6 0 V
6 0 V
6 0 V
6 0 V
6 0 V
6 0 V
6 0 V
6 0 V
6 0 V
5 -1 V
6 0 V
6 0 V
6 0 V
6 0 V
6 0 V
6 0 V
6 0 V
6 0 V
6 0 V
5 0 V
6 0 V
6 0 V
6 0 V
6 0 V
6 0 V
6 0 V
6 0 V
6 0 V
6 0 V
5 0 V
6 0 V
currentpoint stroke M
6 -1 V
6 0 V
6 0 V
6 0 V
6 0 V
6 0 V
6 0 V
6 0 V
5 0 V
6 0 V
6 0 V
6 0 V
6 0 V
6 0 V
6 0 V
6 0 V
6 0 V
5 0 V
6 0 V
6 0 V
6 0 V
6 0 V
6 0 V
6 0 V
6 0 V
6 0 V
6 0 V
5 0 V
6 0 V
6 0 V
6 0 V
6 0 V
6 0 V
6 0 V
6 0 V
6 0 V
6 0 V
5 0 V
6 0 V
6 0 V
6 0 V
6 0 V
6 0 V
6 0 V
6 0 V
6 0 V
6 -1 V
6 0 V
5 0 V
6 0 V
6 0 V
6 0 V
6 0 V
6 0 V
6 0 V
6 0 V
6 0 V
6 0 V
5 0 V
6 0 V
6 0 V
6 0 V
6 0 V
6 0 V
6 0 V
6 0 V
6 0 V
6 0 V
5 0 V
6 0 V
6 0 V
6 0 V
6 0 V
6 0 V
6 0 V
6 0 V
6 0 V
5 0 V
6 0 V
6 0 V
6 0 V
6 0 V
6 0 V
6 0 V
6 0 V
6 0 V
6 0 V
5 0 V
6 0 V
6 0 V
6 0 V
6 0 V
6 0 V
6 0 V
6 0 V
6 0 V
6 0 V
5 0 V
6 0 V
6 0 V
6 0 V
6 0 V
6 0 V
stroke
grestore
end
showpage
}}%
\put(3037,1847){\makebox(0,0)[r]{Running mass}}%
\put(3037,1947){\makebox(0,0)[r]{Fixed mass}}%
\put(1975,50){\makebox(0,0){\Large $b$ (fm) }}%
\put(100,1180){%
\special{ps: gsave currentpoint currentpoint translate
270 rotate neg exch neg exch translate}%
\makebox(0,0)[b]{\shortstack{\Large $ I_z (b) $}}%
\special{ps: currentpoint grestore moveto}%
}%
\put(3450,200){\makebox(0,0){1}}%
\put(2860,200){\makebox(0,0){0.8}}%
\put(2270,200){\makebox(0,0){0.6}}%
\put(1680,200){\makebox(0,0){0.4}}%
\put(1090,200){\makebox(0,0){0.2}}%
\put(500,200){\makebox(0,0){0}}%
\put(450,2060){\makebox(0,0)[r]{0.03}}%
\put(450,1767){\makebox(0,0)[r]{0.025}}%
\put(450,1473){\makebox(0,0)[r]{0.02}}%
\put(450,1180){\makebox(0,0)[r]{0.015}}%
\put(450,887){\makebox(0,0)[r]{0.01}}%
\put(450,593){\makebox(0,0)[r]{0.005}}%
\put(450,300){\makebox(0,0)[r]{0}}%
\end{picture}%
\endgroup
 

%% file: yeff.tex
\begingroup%
  \makeatletter%
  \newcommand{\GNUPLOTspecial}{%
    \@sanitize\catcode`\%=14\relax\special}%
  \setlength{\unitlength}{0.1bp}%
{\GNUPLOTspecial{!
/gnudict 256 dict def
gnudict begin
/Color false def
/Solid false def
/gnulinewidth 5.000 def
/userlinewidth gnulinewidth def
/vshift -33 def
/dl {10 mul} def
/hpt_ 31.5 def
/vpt_ 31.5 def
/hpt hpt_ def
/vpt vpt_ def
/M {moveto} bind def
/L {lineto} bind def
/R {rmoveto} bind def
/V {rlineto} bind def
/vpt2 vpt 2 mul def
/hpt2 hpt 2 mul def
/Lshow { currentpoint stroke M
  0 vshift R show } def
/Rshow { currentpoint stroke M
  dup stringwidth pop neg vshift R show } def
/Cshow { currentpoint stroke M
  dup stringwidth pop -2 div vshift R show } def
/UP { dup vpt_ mul /vpt exch def hpt_ mul /hpt exch def
  /hpt2 hpt 2 mul def /vpt2 vpt 2 mul def } def
/DL { Color {setrgbcolor Solid {pop []} if 0 setdash }
 {pop pop pop Solid {pop []} if 0 setdash} ifelse } def
/BL { stroke gnulinewidth 2 mul setlinewidth } def
/AL { stroke gnulinewidth 2 div setlinewidth } def
/UL { gnulinewidth mul /userlinewidth exch def } def
/PL { stroke userlinewidth setlinewidth } def
/LTb { BL [] 0 0 0 DL } def
/LTa { AL [1 dl 2 dl] 0 setdash 0 0 0 setrgbcolor } def
/LT0 { PL [] 1 0 0 DL } def
/LT1 { PL [4 dl 2 dl] 0 1 0 DL } def
/LT2 { PL [2 dl 3 dl] 0 0 1 DL } def
/LT3 { PL [1 dl 1.5 dl] 1 0 1 DL } def
/LT4 { PL [5 dl 2 dl 1 dl 2 dl] 0 1 1 DL } def
/LT5 { PL [4 dl 3 dl 1 dl 3 dl] 1 1 0 DL } def
/LT6 { PL [2 dl 2 dl 2 dl 4 dl] 0 0 0 DL } def
/LT7 { PL [2 dl 2 dl 2 dl 2 dl 2 dl 4 dl] 1 0.3 0 DL } def
/LT8 { PL [2 dl 2 dl 2 dl 2 dl 2 dl 2 dl 2 dl 4 dl] 0.5 0.5 0.5 DL } def
/Pnt { stroke [] 0 setdash
   gsave 1 setlinecap M 0 0 V stroke grestore } def
/Dia { stroke [] 0 setdash 2 copy vpt add M
  hpt neg vpt neg V hpt vpt neg V
  hpt vpt V hpt neg vpt V closepath stroke
  Pnt } def
/Pls { stroke [] 0 setdash vpt sub M 0 vpt2 V
  currentpoint stroke M
  hpt neg vpt neg R hpt2 0 V stroke
  } def
/Box { stroke [] 0 setdash 2 copy exch hpt sub exch vpt add M
  0 vpt2 neg V hpt2 0 V 0 vpt2 V
  hpt2 neg 0 V closepath stroke
  Pnt } def
/Crs { stroke [] 0 setdash exch hpt sub exch vpt add M
  hpt2 vpt2 neg V currentpoint stroke M
  hpt2 neg 0 R hpt2 vpt2 V stroke } def
/TriU { stroke [] 0 setdash 2 copy vpt 1.12 mul add M
  hpt neg vpt -1.62 mul V
  hpt 2 mul 0 V
  hpt neg vpt 1.62 mul V closepath stroke
  Pnt  } def
/Star { 2 copy Pls Crs } def
/BoxF { stroke [] 0 setdash exch hpt sub exch vpt add M
  0 vpt2 neg V  hpt2 0 V  0 vpt2 V
  hpt2 neg 0 V  closepath fill } def
/TriUF { stroke [] 0 setdash vpt 1.12 mul add M
  hpt neg vpt -1.62 mul V
  hpt 2 mul 0 V
  hpt neg vpt 1.62 mul V closepath fill } def
/TriD { stroke [] 0 setdash 2 copy vpt 1.12 mul sub M
  hpt neg vpt 1.62 mul V
  hpt 2 mul 0 V
  hpt neg vpt -1.62 mul V closepath stroke
  Pnt  } def
/TriDF { stroke [] 0 setdash vpt 1.12 mul sub M
  hpt neg vpt 1.62 mul V
  hpt 2 mul 0 V
  hpt neg vpt -1.62 mul V closepath fill} def
/DiaF { stroke [] 0 setdash vpt add M
  hpt neg vpt neg V hpt vpt neg V
  hpt vpt V hpt neg vpt V closepath fill } def
/Pent { stroke [] 0 setdash 2 copy gsave
  translate 0 hpt M 4 {72 rotate 0 hpt L} repeat
  closepath stroke grestore Pnt } def
/PentF { stroke [] 0 setdash gsave
  translate 0 hpt M 4 {72 rotate 0 hpt L} repeat
  closepath fill grestore } def
/Circle { stroke [] 0 setdash 2 copy
  hpt 0 360 arc stroke Pnt } def
/CircleF { stroke [] 0 setdash hpt 0 360 arc fill } def
/C0 { BL [] 0 setdash 2 copy moveto vpt 90 450  arc } bind def
/C1 { BL [] 0 setdash 2 copy        moveto
       2 copy  vpt 0 90 arc closepath fill
               vpt 0 360 arc closepath } bind def
/C2 { BL [] 0 setdash 2 copy moveto
       2 copy  vpt 90 180 arc closepath fill
               vpt 0 360 arc closepath } bind def
/C3 { BL [] 0 setdash 2 copy moveto
       2 copy  vpt 0 180 arc closepath fill
               vpt 0 360 arc closepath } bind def
/C4 { BL [] 0 setdash 2 copy moveto
       2 copy  vpt 180 270 arc closepath fill
               vpt 0 360 arc closepath } bind def
/C5 { BL [] 0 setdash 2 copy moveto
       2 copy  vpt 0 90 arc
       2 copy moveto
       2 copy  vpt 180 270 arc closepath fill
               vpt 0 360 arc } bind def
/C6 { BL [] 0 setdash 2 copy moveto
      2 copy  vpt 90 270 arc closepath fill
              vpt 0 360 arc closepath } bind def
/C7 { BL [] 0 setdash 2 copy moveto
      2 copy  vpt 0 270 arc closepath fill
              vpt 0 360 arc closepath } bind def
/C8 { BL [] 0 setdash 2 copy moveto
      2 copy vpt 270 360 arc closepath fill
              vpt 0 360 arc closepath } bind def
/C9 { BL [] 0 setdash 2 copy moveto
      2 copy  vpt 270 450 arc closepath fill
              vpt 0 360 arc closepath } bind def
/C10 { BL [] 0 setdash 2 copy 2 copy moveto vpt 270 360 arc closepath fill
       2 copy moveto
       2 copy vpt 90 180 arc closepath fill
               vpt 0 360 arc closepath } bind def
/C11 { BL [] 0 setdash 2 copy moveto
       2 copy  vpt 0 180 arc closepath fill
       2 copy moveto
       2 copy  vpt 270 360 arc closepath fill
               vpt 0 360 arc closepath } bind def
/C12 { BL [] 0 setdash 2 copy moveto
       2 copy  vpt 180 360 arc closepath fill
               vpt 0 360 arc closepath } bind def
/C13 { BL [] 0 setdash  2 copy moveto
       2 copy  vpt 0 90 arc closepath fill
       2 copy moveto
       2 copy  vpt 180 360 arc closepath fill
               vpt 0 360 arc closepath } bind def
/C14 { BL [] 0 setdash 2 copy moveto
       2 copy  vpt 90 360 arc closepath fill
               vpt 0 360 arc } bind def
/C15 { BL [] 0 setdash 2 copy vpt 0 360 arc closepath fill
               vpt 0 360 arc closepath } bind def
/Rec   { newpath 4 2 roll moveto 1 index 0 rlineto 0 exch rlineto
       neg 0 rlineto closepath } bind def
/Square { dup Rec } bind def
/Bsquare { vpt sub exch vpt sub exch vpt2 Square } bind def
/S0 { BL [] 0 setdash 2 copy moveto 0 vpt rlineto BL Bsquare } bind def
/S1 { BL [] 0 setdash 2 copy vpt Square fill Bsquare } bind def
/S2 { BL [] 0 setdash 2 copy exch vpt sub exch vpt Square fill Bsquare } bind def
/S3 { BL [] 0 setdash 2 copy exch vpt sub exch vpt2 vpt Rec fill Bsquare } bind def
/S4 { BL [] 0 setdash 2 copy exch vpt sub exch vpt sub vpt Square fill Bsquare } bind def
/S5 { BL [] 0 setdash 2 copy 2 copy vpt Square fill
       exch vpt sub exch vpt sub vpt Square fill Bsquare } bind def
/S6 { BL [] 0 setdash 2 copy exch vpt sub exch vpt sub vpt vpt2 Rec fill Bsquare } bind def
/S7 { BL [] 0 setdash 2 copy exch vpt sub exch vpt sub vpt vpt2 Rec fill
       2 copy vpt Square fill
       Bsquare } bind def
/S8 { BL [] 0 setdash 2 copy vpt sub vpt Square fill Bsquare } bind def
/S9 { BL [] 0 setdash 2 copy vpt sub vpt vpt2 Rec fill Bsquare } bind def
/S10 { BL [] 0 setdash 2 copy vpt sub vpt Square fill 2 copy exch vpt sub exch vpt Square fill
       Bsquare } bind def
/S11 { BL [] 0 setdash 2 copy vpt sub vpt Square fill 2 copy exch vpt sub exch vpt2 vpt Rec fill
       Bsquare } bind def
/S12 { BL [] 0 setdash 2 copy exch vpt sub exch vpt sub vpt2 vpt Rec fill Bsquare } bind def
/S13 { BL [] 0 setdash 2 copy exch vpt sub exch vpt sub vpt2 vpt Rec fill
       2 copy vpt Square fill Bsquare } bind def
/S14 { BL [] 0 setdash 2 copy exch vpt sub exch vpt sub vpt2 vpt Rec fill
       2 copy exch vpt sub exch vpt Square fill Bsquare } bind def
/S15 { BL [] 0 setdash 2 copy Bsquare fill Bsquare } bind def
/D0 { gsave translate 45 rotate 0 0 S0 stroke grestore } bind def
/D1 { gsave translate 45 rotate 0 0 S1 stroke grestore } bind def
/D2 { gsave translate 45 rotate 0 0 S2 stroke grestore } bind def
/D3 { gsave translate 45 rotate 0 0 S3 stroke grestore } bind def
/D4 { gsave translate 45 rotate 0 0 S4 stroke grestore } bind def
/D5 { gsave translate 45 rotate 0 0 S5 stroke grestore } bind def
/D6 { gsave translate 45 rotate 0 0 S6 stroke grestore } bind def
/D7 { gsave translate 45 rotate 0 0 S7 stroke grestore } bind def
/D8 { gsave translate 45 rotate 0 0 S8 stroke grestore } bind def
/D9 { gsave translate 45 rotate 0 0 S9 stroke grestore } bind def
/D10 { gsave translate 45 rotate 0 0 S10 stroke grestore } bind def
/D11 { gsave translate 45 rotate 0 0 S11 stroke grestore } bind def
/D12 { gsave translate 45 rotate 0 0 S12 stroke grestore } bind def
/D13 { gsave translate 45 rotate 0 0 S13 stroke grestore } bind def
/D14 { gsave translate 45 rotate 0 0 S14 stroke grestore } bind def
/D15 { gsave translate 45 rotate 0 0 S15 stroke grestore } bind def
/DiaE { stroke [] 0 setdash vpt add M
  hpt neg vpt neg V hpt vpt neg V
  hpt vpt V hpt neg vpt V closepath stroke } def
/BoxE { stroke [] 0 setdash exch hpt sub exch vpt add M
  0 vpt2 neg V hpt2 0 V 0 vpt2 V
  hpt2 neg 0 V closepath stroke } def
/TriUE { stroke [] 0 setdash vpt 1.12 mul add M
  hpt neg vpt -1.62 mul V
  hpt 2 mul 0 V
  hpt neg vpt 1.62 mul V closepath stroke } def
/TriDE { stroke [] 0 setdash vpt 1.12 mul sub M
  hpt neg vpt 1.62 mul V
  hpt 2 mul 0 V
  hpt neg vpt -1.62 mul V closepath stroke } def
/PentE { stroke [] 0 setdash gsave
  translate 0 hpt M 4 {72 rotate 0 hpt L} repeat
  closepath stroke grestore } def
/CircE { stroke [] 0 setdash 
  hpt 0 360 arc stroke } def
/Opaque { gsave closepath 1 setgray fill grestore 0 setgray closepath } def
/DiaW { stroke [] 0 setdash vpt add M
  hpt neg vpt neg V hpt vpt neg V
  hpt vpt V hpt neg vpt V Opaque stroke } def
/BoxW { stroke [] 0 setdash exch hpt sub exch vpt add M
  0 vpt2 neg V hpt2 0 V 0 vpt2 V
  hpt2 neg 0 V Opaque stroke } def
/TriUW { stroke [] 0 setdash vpt 1.12 mul add M
  hpt neg vpt -1.62 mul V
  hpt 2 mul 0 V
  hpt neg vpt 1.62 mul V Opaque stroke } def
/TriDW { stroke [] 0 setdash vpt 1.12 mul sub M
  hpt neg vpt 1.62 mul V
  hpt 2 mul 0 V
  hpt neg vpt -1.62 mul V Opaque stroke } def
/PentW { stroke [] 0 setdash gsave
  translate 0 hpt M 4 {72 rotate 0 hpt L} repeat
  Opaque stroke grestore } def
/CircW { stroke [] 0 setdash 
  hpt 0 360 arc Opaque stroke } def
/BoxFill { gsave Rec 1 setgray fill grestore } def
end
}}%
\begin{picture}(3600,2160)(0,0)%
{\GNUPLOTspecial{"
gnudict begin
gsave
0 0 translate
0.100 0.100 scale
0 setgray
newpath
1.000 UL
LTb
400 300 M
63 0 V
2987 0 R
-63 0 V
400 520 M
63 0 V
2987 0 R
-63 0 V
400 740 M
63 0 V
2987 0 R
-63 0 V
400 960 M
63 0 V
2987 0 R
-63 0 V
400 1180 M
63 0 V
2987 0 R
-63 0 V
400 1400 M
63 0 V
2987 0 R
-63 0 V
400 1620 M
63 0 V
2987 0 R
-63 0 V
400 1840 M
63 0 V
2987 0 R
-63 0 V
400 2060 M
63 0 V
2987 0 R
-63 0 V
400 300 M
0 63 V
0 1697 R
0 -63 V
739 300 M
0 63 V
0 1697 R
0 -63 V
1078 300 M
0 63 V
0 1697 R
0 -63 V
1417 300 M
0 63 V
0 1697 R
0 -63 V
1756 300 M
0 63 V
0 1697 R
0 -63 V
2094 300 M
0 63 V
0 1697 R
0 -63 V
2433 300 M
0 63 V
0 1697 R
0 -63 V
2772 300 M
0 63 V
0 1697 R
0 -63 V
3111 300 M
0 63 V
0 1697 R
0 -63 V
3450 300 M
0 63 V
0 1697 R
0 -63 V
1.000 UL
LTb
400 300 M
3050 0 V
0 1760 V
-3050 0 V
400 300 L
1.000 UL
LT0
3087 1947 M
263 0 V
1011 2060 M
17 -100 V
27 -132 V
27 -117 V
27 -105 V
26 -93 V
27 -83 V
27 -76 V
26 -68 V
27 -61 V
27 -56 V
26 -51 V
27 -47 V
27 -43 V
27 -39 V
26 -36 V
27 -34 V
27 -31 V
26 -28 V
27 -27 V
27 -25 V
26 -23 V
27 -21 V
27 -20 V
27 -19 V
26 -18 V
27 -17 V
27 -15 V
26 -15 V
27 -14 V
27 -13 V
26 -12 V
27 -12 V
27 -11 V
27 -10 V
26 -10 V
27 -10 V
27 -8 V
26 -9 V
27 -8 V
27 -8 V
26 -7 V
27 -7 V
27 -7 V
27 -6 V
26 -6 V
27 -6 V
27 -6 V
26 -5 V
27 -5 V
27 -5 V
27 -5 V
26 -4 V
27 -5 V
27 -4 V
26 -4 V
27 -4 V
27 -4 V
26 -3 V
27 -4 V
27 -3 V
27 -3 V
26 -3 V
27 -3 V
27 -3 V
26 -3 V
27 -3 V
27 -2 V
26 -3 V
27 -2 V
27 -3 V
27 -2 V
26 -2 V
27 -2 V
27 -2 V
26 -2 V
27 -2 V
27 -2 V
26 -2 V
27 -2 V
1.000 UL
LT1
3087 1847 M
263 0 V
832 2060 M
10 -78 V
26 -187 V
27 -157 V
27 -133 V
26 -115 V
27 -98 V
27 -86 V
26 -76 V
27 -66 V
27 -59 V
27 -52 V
26 -46 V
27 -42 V
27 -38 V
26 -34 V
27 -31 V
27 -28 V
26 -25 V
27 -23 V
27 -22 V
27 -19 V
26 -19 V
27 -16 V
27 -16 V
26 -14 V
27 -13 V
27 -13 V
26 -11 V
27 -11 V
27 -10 V
27 -10 V
26 -8 V
27 -9 V
27 -8 V
26 -7 V
27 -7 V
27 -6 V
26 -7 V
27 -5 V
27 -6 V
27 -5 V
26 -5 V
27 -5 V
27 -4 V
26 -4 V
27 -5 V
27 -3 V
26 -4 V
27 -4 V
27 -3 V
27 -3 V
26 -3 V
27 -3 V
27 -3 V
26 -3 V
27 -2 V
27 -3 V
27 -2 V
26 -2 V
27 -2 V
27 -2 V
26 -2 V
27 -2 V
27 -2 V
26 -2 V
27 -2 V
27 -1 V
27 -2 V
26 -2 V
27 -1 V
27 -2 V
26 -1 V
27 -1 V
27 -2 V
26 -1 V
27 -1 V
27 -1 V
27 -1 V
26 -2 V
27 -1 V
27 -1 V
26 -1 V
27 -1 V
27 -1 V
26 0 V
27 -1 V
1.000 UL
LT2
3087 1747 M
263 0 V
1148 2060 M
14 -66 V
27 -112 V
26 -102 V
27 -93 V
27 -84 V
26 -76 V
27 -70 V
27 -64 V
27 -59 V
26 -55 V
27 -50 V
27 -46 V
26 -43 V
27 -40 V
27 -37 V
26 -35 V
27 -32 V
27 -31 V
27 -28 V
26 -26 V
27 -25 V
27 -24 V
26 -22 V
27 -20 V
27 -20 V
26 -19 V
27 -17 V
27 -17 V
27 -15 V
26 -15 V
27 -14 V
27 -14 V
26 -12 V
27 -13 V
27 -11 V
26 -11 V
27 -11 V
27 -10 V
27 -9 V
26 -9 V
27 -9 V
27 -9 V
26 -8 V
27 -7 V
27 -8 V
27 -7 V
26 -7 V
27 -6 V
27 -6 V
26 -6 V
27 -6 V
27 -6 V
26 -5 V
27 -5 V
27 -5 V
27 -5 V
26 -5 V
27 -4 V
27 -5 V
26 -4 V
27 -4 V
27 -4 V
26 -4 V
27 -3 V
27 -4 V
27 -3 V
26 -4 V
27 -3 V
27 -3 V
26 -3 V
27 -3 V
27 -3 V
26 -2 V
27 -3 V
stroke
grestore
end
showpage
}}%
\put(3037,1747){\makebox(0,0)[r]{$\lambda =15$}}%
\put(3037,1847){\makebox(0,0)[r]{$\lambda =5$}}%
\put(3037,1947){\makebox(0,0)[r]{$\lambda =10$}}%
\put(1925,50){\makebox(0,0){\Large $b$ ~(fm)}}%
\put(100,1180){%
\special{ps: gsave currentpoint currentpoint translate
270 rotate neg exch neg exch translate}%
\makebox(0,0)[b]{\shortstack{$x'/\delta$}}%
\special{ps: currentpoint grestore moveto}%
}%
\put(3450,200){\makebox(0,0){0.45}}%
\put(3111,200){\makebox(0,0){0.4}}%
\put(2772,200){\makebox(0,0){0.35}}%
\put(2433,200){\makebox(0,0){0.3}}%
\put(2094,200){\makebox(0,0){0.25}}%
\put(1756,200){\makebox(0,0){0.2}}%
\put(1417,200){\makebox(0,0){0.15}}%
\put(1078,200){\makebox(0,0){0.1}}%
\put(739,200){\makebox(0,0){0.05}}%
\put(400,200){\makebox(0,0){0}}%
\put(350,2060){\makebox(0,0)[r]{5}}%
\put(350,1840){\makebox(0,0)[r]{4.5}}%
\put(350,1620){\makebox(0,0)[r]{4}}%
\put(350,1400){\makebox(0,0)[r]{3.5}}%
\put(350,1180){\makebox(0,0)[r]{3}}%
\put(350,960){\makebox(0,0)[r]{2.5}}%
\put(350,740){\makebox(0,0)[r]{2}}%
\put(350,520){\makebox(0,0)[r]{1.5}}%
\put(350,300){\makebox(0,0)[r]{1}}%
\end{picture}%
\endgroup
 

%% file: qeff.tex
\begingroup%
  \makeatletter%
  \newcommand{\GNUPLOTspecial}{%
    \@sanitize\catcode`\%=14\relax\special}%
  \setlength{\unitlength}{0.1bp}%
{\GNUPLOTspecial{!
/gnudict 256 dict def
gnudict begin
/Color false def
/Solid false def
/gnulinewidth 5.000 def
/userlinewidth gnulinewidth def
/vshift -33 def
/dl {10 mul} def
/hpt_ 31.5 def
/vpt_ 31.5 def
/hpt hpt_ def
/vpt vpt_ def
/M {moveto} bind def
/L {lineto} bind def
/R {rmoveto} bind def
/V {rlineto} bind def
/vpt2 vpt 2 mul def
/hpt2 hpt 2 mul def
/Lshow { currentpoint stroke M
  0 vshift R show } def
/Rshow { currentpoint stroke M
  dup stringwidth pop neg vshift R show } def
/Cshow { currentpoint stroke M
  dup stringwidth pop -2 div vshift R show } def
/UP { dup vpt_ mul /vpt exch def hpt_ mul /hpt exch def
  /hpt2 hpt 2 mul def /vpt2 vpt 2 mul def } def
/DL { Color {setrgbcolor Solid {pop []} if 0 setdash }
 {pop pop pop Solid {pop []} if 0 setdash} ifelse } def
/BL { stroke gnulinewidth 2 mul setlinewidth } def
/AL { stroke gnulinewidth 2 div setlinewidth } def
/UL { gnulinewidth mul /userlinewidth exch def } def
/PL { stroke userlinewidth setlinewidth } def
/LTb { BL [] 0 0 0 DL } def
/LTa { AL [1 dl 2 dl] 0 setdash 0 0 0 setrgbcolor } def
/LT0 { PL [] 1 0 0 DL } def
/LT1 { PL [4 dl 2 dl] 0 1 0 DL } def
/LT2 { PL [2 dl 3 dl] 0 0 1 DL } def
/LT3 { PL [1 dl 1.5 dl] 1 0 1 DL } def
/LT4 { PL [5 dl 2 dl 1 dl 2 dl] 0 1 1 DL } def
/LT5 { PL [4 dl 3 dl 1 dl 3 dl] 1 1 0 DL } def
/LT6 { PL [2 dl 2 dl 2 dl 4 dl] 0 0 0 DL } def
/LT7 { PL [2 dl 2 dl 2 dl 2 dl 2 dl 4 dl] 1 0.3 0 DL } def
/LT8 { PL [2 dl 2 dl 2 dl 2 dl 2 dl 2 dl 2 dl 4 dl] 0.5 0.5 0.5 DL } def
/Pnt { stroke [] 0 setdash
   gsave 1 setlinecap M 0 0 V stroke grestore } def
/Dia { stroke [] 0 setdash 2 copy vpt add M
  hpt neg vpt neg V hpt vpt neg V
  hpt vpt V hpt neg vpt V closepath stroke
  Pnt } def
/Pls { stroke [] 0 setdash vpt sub M 0 vpt2 V
  currentpoint stroke M
  hpt neg vpt neg R hpt2 0 V stroke
  } def
/Box { stroke [] 0 setdash 2 copy exch hpt sub exch vpt add M
  0 vpt2 neg V hpt2 0 V 0 vpt2 V
  hpt2 neg 0 V closepath stroke
  Pnt } def
/Crs { stroke [] 0 setdash exch hpt sub exch vpt add M
  hpt2 vpt2 neg V currentpoint stroke M
  hpt2 neg 0 R hpt2 vpt2 V stroke } def
/TriU { stroke [] 0 setdash 2 copy vpt 1.12 mul add M
  hpt neg vpt -1.62 mul V
  hpt 2 mul 0 V
  hpt neg vpt 1.62 mul V closepath stroke
  Pnt  } def
/Star { 2 copy Pls Crs } def
/BoxF { stroke [] 0 setdash exch hpt sub exch vpt add M
  0 vpt2 neg V  hpt2 0 V  0 vpt2 V
  hpt2 neg 0 V  closepath fill } def
/TriUF { stroke [] 0 setdash vpt 1.12 mul add M
  hpt neg vpt -1.62 mul V
  hpt 2 mul 0 V
  hpt neg vpt 1.62 mul V closepath fill } def
/TriD { stroke [] 0 setdash 2 copy vpt 1.12 mul sub M
  hpt neg vpt 1.62 mul V
  hpt 2 mul 0 V
  hpt neg vpt -1.62 mul V closepath stroke
  Pnt  } def
/TriDF { stroke [] 0 setdash vpt 1.12 mul sub M
  hpt neg vpt 1.62 mul V
  hpt 2 mul 0 V
  hpt neg vpt -1.62 mul V closepath fill} def
/DiaF { stroke [] 0 setdash vpt add M
  hpt neg vpt neg V hpt vpt neg V
  hpt vpt V hpt neg vpt V closepath fill } def
/Pent { stroke [] 0 setdash 2 copy gsave
  translate 0 hpt M 4 {72 rotate 0 hpt L} repeat
  closepath stroke grestore Pnt } def
/PentF { stroke [] 0 setdash gsave
  translate 0 hpt M 4 {72 rotate 0 hpt L} repeat
  closepath fill grestore } def
/Circle { stroke [] 0 setdash 2 copy
  hpt 0 360 arc stroke Pnt } def
/CircleF { stroke [] 0 setdash hpt 0 360 arc fill } def
/C0 { BL [] 0 setdash 2 copy moveto vpt 90 450  arc } bind def
/C1 { BL [] 0 setdash 2 copy        moveto
       2 copy  vpt 0 90 arc closepath fill
               vpt 0 360 arc closepath } bind def
/C2 { BL [] 0 setdash 2 copy moveto
       2 copy  vpt 90 180 arc closepath fill
               vpt 0 360 arc closepath } bind def
/C3 { BL [] 0 setdash 2 copy moveto
       2 copy  vpt 0 180 arc closepath fill
               vpt 0 360 arc closepath } bind def
/C4 { BL [] 0 setdash 2 copy moveto
       2 copy  vpt 180 270 arc closepath fill
               vpt 0 360 arc closepath } bind def
/C5 { BL [] 0 setdash 2 copy moveto
       2 copy  vpt 0 90 arc
       2 copy moveto
       2 copy  vpt 180 270 arc closepath fill
               vpt 0 360 arc } bind def
/C6 { BL [] 0 setdash 2 copy moveto
      2 copy  vpt 90 270 arc closepath fill
              vpt 0 360 arc closepath } bind def
/C7 { BL [] 0 setdash 2 copy moveto
      2 copy  vpt 0 270 arc closepath fill
              vpt 0 360 arc closepath } bind def
/C8 { BL [] 0 setdash 2 copy moveto
      2 copy vpt 270 360 arc closepath fill
              vpt 0 360 arc closepath } bind def
/C9 { BL [] 0 setdash 2 copy moveto
      2 copy  vpt 270 450 arc closepath fill
              vpt 0 360 arc closepath } bind def
/C10 { BL [] 0 setdash 2 copy 2 copy moveto vpt 270 360 arc closepath fill
       2 copy moveto
       2 copy vpt 90 180 arc closepath fill
               vpt 0 360 arc closepath } bind def
/C11 { BL [] 0 setdash 2 copy moveto
       2 copy  vpt 0 180 arc closepath fill
       2 copy moveto
       2 copy  vpt 270 360 arc closepath fill
               vpt 0 360 arc closepath } bind def
/C12 { BL [] 0 setdash 2 copy moveto
       2 copy  vpt 180 360 arc closepath fill
               vpt 0 360 arc closepath } bind def
/C13 { BL [] 0 setdash  2 copy moveto
       2 copy  vpt 0 90 arc closepath fill
       2 copy moveto
       2 copy  vpt 180 360 arc closepath fill
               vpt 0 360 arc closepath } bind def
/C14 { BL [] 0 setdash 2 copy moveto
       2 copy  vpt 90 360 arc closepath fill
               vpt 0 360 arc } bind def
/C15 { BL [] 0 setdash 2 copy vpt 0 360 arc closepath fill
               vpt 0 360 arc closepath } bind def
/Rec   { newpath 4 2 roll moveto 1 index 0 rlineto 0 exch rlineto
       neg 0 rlineto closepath } bind def
/Square { dup Rec } bind def
/Bsquare { vpt sub exch vpt sub exch vpt2 Square } bind def
/S0 { BL [] 0 setdash 2 copy moveto 0 vpt rlineto BL Bsquare } bind def
/S1 { BL [] 0 setdash 2 copy vpt Square fill Bsquare } bind def
/S2 { BL [] 0 setdash 2 copy exch vpt sub exch vpt Square fill Bsquare } bind def
/S3 { BL [] 0 setdash 2 copy exch vpt sub exch vpt2 vpt Rec fill Bsquare } bind def
/S4 { BL [] 0 setdash 2 copy exch vpt sub exch vpt sub vpt Square fill Bsquare } bind def
/S5 { BL [] 0 setdash 2 copy 2 copy vpt Square fill
       exch vpt sub exch vpt sub vpt Square fill Bsquare } bind def
/S6 { BL [] 0 setdash 2 copy exch vpt sub exch vpt sub vpt vpt2 Rec fill Bsquare } bind def
/S7 { BL [] 0 setdash 2 copy exch vpt sub exch vpt sub vpt vpt2 Rec fill
       2 copy vpt Square fill
       Bsquare } bind def
/S8 { BL [] 0 setdash 2 copy vpt sub vpt Square fill Bsquare } bind def
/S9 { BL [] 0 setdash 2 copy vpt sub vpt vpt2 Rec fill Bsquare } bind def
/S10 { BL [] 0 setdash 2 copy vpt sub vpt Square fill 2 copy exch vpt sub exch vpt Square fill
       Bsquare } bind def
/S11 { BL [] 0 setdash 2 copy vpt sub vpt Square fill 2 copy exch vpt sub exch vpt2 vpt Rec fill
       Bsquare } bind def
/S12 { BL [] 0 setdash 2 copy exch vpt sub exch vpt sub vpt2 vpt Rec fill Bsquare } bind def
/S13 { BL [] 0 setdash 2 copy exch vpt sub exch vpt sub vpt2 vpt Rec fill
       2 copy vpt Square fill Bsquare } bind def
/S14 { BL [] 0 setdash 2 copy exch vpt sub exch vpt sub vpt2 vpt Rec fill
       2 copy exch vpt sub exch vpt Square fill Bsquare } bind def
/S15 { BL [] 0 setdash 2 copy Bsquare fill Bsquare } bind def
/D0 { gsave translate 45 rotate 0 0 S0 stroke grestore } bind def
/D1 { gsave translate 45 rotate 0 0 S1 stroke grestore } bind def
/D2 { gsave translate 45 rotate 0 0 S2 stroke grestore } bind def
/D3 { gsave translate 45 rotate 0 0 S3 stroke grestore } bind def
/D4 { gsave translate 45 rotate 0 0 S4 stroke grestore } bind def
/D5 { gsave translate 45 rotate 0 0 S5 stroke grestore } bind def
/D6 { gsave translate 45 rotate 0 0 S6 stroke grestore } bind def
/D7 { gsave translate 45 rotate 0 0 S7 stroke grestore } bind def
/D8 { gsave translate 45 rotate 0 0 S8 stroke grestore } bind def
/D9 { gsave translate 45 rotate 0 0 S9 stroke grestore } bind def
/D10 { gsave translate 45 rotate 0 0 S10 stroke grestore } bind def
/D11 { gsave translate 45 rotate 0 0 S11 stroke grestore } bind def
/D12 { gsave translate 45 rotate 0 0 S12 stroke grestore } bind def
/D13 { gsave translate 45 rotate 0 0 S13 stroke grestore } bind def
/D14 { gsave translate 45 rotate 0 0 S14 stroke grestore } bind def
/D15 { gsave translate 45 rotate 0 0 S15 stroke grestore } bind def
/DiaE { stroke [] 0 setdash vpt add M
  hpt neg vpt neg V hpt vpt neg V
  hpt vpt V hpt neg vpt V closepath stroke } def
/BoxE { stroke [] 0 setdash exch hpt sub exch vpt add M
  0 vpt2 neg V hpt2 0 V 0 vpt2 V
  hpt2 neg 0 V closepath stroke } def
/TriUE { stroke [] 0 setdash vpt 1.12 mul add M
  hpt neg vpt -1.62 mul V
  hpt 2 mul 0 V
  hpt neg vpt 1.62 mul V closepath stroke } def
/TriDE { stroke [] 0 setdash vpt 1.12 mul sub M
  hpt neg vpt 1.62 mul V
  hpt 2 mul 0 V
  hpt neg vpt -1.62 mul V closepath stroke } def
/PentE { stroke [] 0 setdash gsave
  translate 0 hpt M 4 {72 rotate 0 hpt L} repeat
  closepath stroke grestore } def
/CircE { stroke [] 0 setdash 
  hpt 0 360 arc stroke } def
/Opaque { gsave closepath 1 setgray fill grestore 0 setgray closepath } def
/DiaW { stroke [] 0 setdash vpt add M
  hpt neg vpt neg V hpt vpt neg V
  hpt vpt V hpt neg vpt V Opaque stroke } def
/BoxW { stroke [] 0 setdash exch hpt sub exch vpt add M
  0 vpt2 neg V hpt2 0 V 0 vpt2 V
  hpt2 neg 0 V Opaque stroke } def
/TriUW { stroke [] 0 setdash vpt 1.12 mul add M
  hpt neg vpt -1.62 mul V
  hpt 2 mul 0 V
  hpt neg vpt 1.62 mul V Opaque stroke } def
/TriDW { stroke [] 0 setdash vpt 1.12 mul sub M
  hpt neg vpt 1.62 mul V
  hpt 2 mul 0 V
  hpt neg vpt -1.62 mul V Opaque stroke } def
/PentW { stroke [] 0 setdash gsave
  translate 0 hpt M 4 {72 rotate 0 hpt L} repeat
  Opaque stroke grestore } def
/CircW { stroke [] 0 setdash 
  hpt 0 360 arc Opaque stroke } def
/BoxFill { gsave Rec 1 setgray fill grestore } def
end
}}%
\begin{picture}(3600,2160)(0,0)%
{\GNUPLOTspecial{"
gnudict begin
gsave
0 0 translate
0.100 0.100 scale
0 setgray
newpath
1.000 UL
LTb
350 393 M
63 0 V
3037 0 R
-63 0 V
350 578 M
63 0 V
3037 0 R
-63 0 V
350 763 M
63 0 V
3037 0 R
-63 0 V
350 948 M
63 0 V
3037 0 R
-63 0 V
350 1134 M
63 0 V
3037 0 R
-63 0 V
350 1319 M
63 0 V
3037 0 R
-63 0 V
350 1504 M
63 0 V
3037 0 R
-63 0 V
350 1689 M
63 0 V
3037 0 R
-63 0 V
350 1875 M
63 0 V
3037 0 R
-63 0 V
350 2060 M
63 0 V
3037 0 R
-63 0 V
350 300 M
0 63 V
0 1697 R
0 -63 V
694 300 M
0 63 V
0 1697 R
0 -63 V
1039 300 M
0 63 V
0 1697 R
0 -63 V
1383 300 M
0 63 V
0 1697 R
0 -63 V
1728 300 M
0 63 V
0 1697 R
0 -63 V
2072 300 M
0 63 V
0 1697 R
0 -63 V
2417 300 M
0 63 V
0 1697 R
0 -63 V
2761 300 M
0 63 V
0 1697 R
0 -63 V
3106 300 M
0 63 V
0 1697 R
0 -63 V
3450 300 M
0 63 V
0 1697 R
0 -63 V
1.000 UL
LTb
350 300 M
3100 0 V
0 1760 V
-3100 0 V
350 300 L
1.000 UL
LT0
3087 1947 M
263 0 V
1311 2060 M
3 -12 V
28 -100 V
27 -91 V
27 -85 V
27 -78 V
27 -72 V
27 -68 V
27 -62 V
28 -59 V
27 -54 V
27 -51 V
27 -48 V
27 -44 V
27 -42 V
27 -40 V
28 -37 V
27 -35 V
27 -33 V
27 -31 V
27 -29 V
27 -28 V
27 -27 V
27 -25 V
28 -23 V
27 -23 V
27 -21 V
27 -21 V
27 -19 V
27 -19 V
27 -18 V
28 -17 V
27 -16 V
27 -15 V
27 -15 V
27 -14 V
27 -14 V
27 -12 V
28 -13 V
27 -12 V
27 -11 V
27 -11 V
27 -11 V
27 -10 V
27 -10 V
28 -9 V
27 -9 V
27 -9 V
27 -8 V
27 -9 V
27 -7 V
27 -8 V
27 -7 V
28 -7 V
27 -7 V
27 -7 V
27 -6 V
27 -6 V
27 -6 V
27 -6 V
28 -6 V
27 -5 V
27 -5 V
27 -5 V
27 -5 V
27 -5 V
27 -5 V
28 -4 V
1.000 UL
LT1
3087 1847 M
263 0 V
1030 2060 M
13 -71 V
27 -132 V
27 -117 V
27 -106 V
28 -95 V
27 -86 V
27 -77 V
27 -71 V
27 -64 V
27 -59 V
27 -55 V
28 -49 V
27 -46 V
27 -42 V
27 -39 V
27 -36 V
27 -34 V
27 -31 V
28 -30 V
27 -27 V
27 -25 V
27 -24 V
27 -22 V
27 -21 V
27 -20 V
28 -19 V
27 -17 V
27 -17 V
27 -15 V
27 -15 V
27 -14 V
27 -13 V
27 -13 V
28 -12 V
27 -11 V
27 -11 V
27 -10 V
27 -10 V
27 -9 V
27 -9 V
28 -8 V
27 -8 V
27 -8 V
27 -7 V
27 -7 V
27 -7 V
27 -7 V
28 -6 V
27 -6 V
27 -5 V
27 -6 V
27 -5 V
27 -5 V
27 -5 V
28 -5 V
27 -4 V
27 -5 V
27 -4 V
27 -4 V
27 -4 V
27 -4 V
27 -3 V
28 -4 V
27 -3 V
27 -4 V
27 -3 V
27 -3 V
27 -3 V
27 -3 V
28 -3 V
27 -2 V
27 -3 V
27 -2 V
27 -3 V
27 -2 V
27 -3 V
28 -2 V
1.000 UL
LT1
3087 1747 M
263 0 V
1527 2060 M
5 -13 V
27 -82 V
27 -76 V
27 -72 V
27 -67 V
27 -63 V
27 -59 V
28 -56 V
27 -52 V
27 -50 V
27 -46 V
27 -45 V
27 -41 V
27 -40 V
27 -38 V
28 -35 V
27 -34 V
27 -32 V
27 -31 V
27 -29 V
27 -28 V
27 -27 V
28 -25 V
27 -24 V
27 -23 V
27 -22 V
27 -22 V
27 -20 V
27 -19 V
28 -19 V
27 -18 V
27 -17 V
27 -17 V
27 -16 V
27 -15 V
27 -14 V
28 -15 V
27 -13 V
27 -13 V
27 -13 V
27 -12 V
27 -12 V
27 -11 V
27 -11 V
28 -11 V
27 -10 V
27 -10 V
27 -9 V
27 -10 V
27 -9 V
27 -8 V
28 -9 V
27 -8 V
27 -8 V
27 -7 V
27 -8 V
27 -7 V
27 -7 V
28 -6 V
stroke
grestore
end
showpage
}}%
\put(3037,1747){\makebox(0,0)[r]{$\lambda =15$}}%
\put(3037,1847){\makebox(0,0)[r]{$\lambda =5$}}%
\put(3037,1947){\makebox(0,0)[r]{$\lambda =10$}}%
\put(1900,50){\makebox(0,0){\Large $b$ ~(fm)}}%
\put(100,1180){%
\special{ps: gsave currentpoint currentpoint translate
270 rotate neg exch neg exch translate}%
\makebox(0,0)[b]{\shortstack{${\bar Q}^{2}$}}%
\special{ps: currentpoint grestore moveto}%
}%
\put(3450,200){\makebox(0,0){0.45}}%
\put(3106,200){\makebox(0,0){0.4}}%
\put(2761,200){\makebox(0,0){0.35}}%
\put(2417,200){\makebox(0,0){0.3}}%
\put(2072,200){\makebox(0,0){0.25}}%
\put(1728,200){\makebox(0,0){0.2}}%
\put(1383,200){\makebox(0,0){0.15}}%
\put(1039,200){\makebox(0,0){0.1}}%
\put(694,200){\makebox(0,0){0.05}}%
\put(350,200){\makebox(0,0){0}}%
\put(300,2060){\makebox(0,0)[r]{20}}%
\put(300,1875){\makebox(0,0)[r]{18}}%
\put(300,1689){\makebox(0,0)[r]{16}}%
\put(300,1504){\makebox(0,0)[r]{14}}%
\put(300,1319){\makebox(0,0)[r]{12}}%
\put(300,1134){\makebox(0,0)[r]{10}}%
\put(300,948){\makebox(0,0)[r]{8}}%
\put(300,763){\makebox(0,0)[r]{6}}%
\put(300,578){\makebox(0,0)[r]{4}}%
\put(300,393){\makebox(0,0)[r]{2}}%
\end{picture}%
\endgroup
 

%% file: partons.tex
\begingroup%
  \makeatletter%
  \newcommand{\GNUPLOTspecial}{%
    \@sanitize\catcode`\%=14\relax\special}%
  \setlength{\unitlength}{0.1bp}%
{\GNUPLOTspecial{!
/gnudict 256 dict def
gnudict begin
/Color false def
/Solid false def
/gnulinewidth 5.000 def
/userlinewidth gnulinewidth def
/vshift -33 def
/dl {10 mul} def
/hpt_ 31.5 def
/vpt_ 31.5 def
/hpt hpt_ def
/vpt vpt_ def
/M {moveto} bind def
/L {lineto} bind def
/R {rmoveto} bind def
/V {rlineto} bind def
/vpt2 vpt 2 mul def
/hpt2 hpt 2 mul def
/Lshow { currentpoint stroke M
  0 vshift R show } def
/Rshow { currentpoint stroke M
  dup stringwidth pop neg vshift R show } def
/Cshow { currentpoint stroke M
  dup stringwidth pop -2 div vshift R show } def
/UP { dup vpt_ mul /vpt exch def hpt_ mul /hpt exch def
  /hpt2 hpt 2 mul def /vpt2 vpt 2 mul def } def
/DL { Color {setrgbcolor Solid {pop []} if 0 setdash }
 {pop pop pop Solid {pop []} if 0 setdash} ifelse } def
/BL { stroke userlinewidth 2 mul setlinewidth } def
/AL { stroke userlinewidth 2 div setlinewidth } def
/UL { dup gnulinewidth mul /userlinewidth exch def
      10 mul /udl exch def } def
/PL { stroke userlinewidth setlinewidth } def
/LTb { BL [] 0 0 0 DL } def
/LTa { AL [1 udl mul 2 udl mul] 0 setdash 0 0 0 setrgbcolor } def
/LT0 { PL [] 1 0 0 DL } def
/LT1 { PL [4 dl 2 dl] 0 1 0 DL } def
/LT2 { PL [2 dl 3 dl] 0 0 1 DL } def
/LT3 { PL [1 dl 1.5 dl] 1 0 1 DL } def
/LT4 { PL [5 dl 2 dl 1 dl 2 dl] 0 1 1 DL } def
/LT5 { PL [4 dl 3 dl 1 dl 3 dl] 1 1 0 DL } def
/LT6 { PL [2 dl 2 dl 2 dl 4 dl] 0 0 0 DL } def
/LT7 { PL [2 dl 2 dl 2 dl 2 dl 2 dl 4 dl] 1 0.3 0 DL } def
/LT8 { PL [2 dl 2 dl 2 dl 2 dl 2 dl 2 dl 2 dl 4 dl] 0.5 0.5 0.5 DL } def
/Pnt { stroke [] 0 setdash
   gsave 1 setlinecap M 0 0 V stroke grestore } def
/Dia { stroke [] 0 setdash 2 copy vpt add M
  hpt neg vpt neg V hpt vpt neg V
  hpt vpt V hpt neg vpt V closepath stroke
  Pnt } def
/Pls { stroke [] 0 setdash vpt sub M 0 vpt2 V
  currentpoint stroke M
  hpt neg vpt neg R hpt2 0 V stroke
  } def
/Box { stroke [] 0 setdash 2 copy exch hpt sub exch vpt add M
  0 vpt2 neg V hpt2 0 V 0 vpt2 V
  hpt2 neg 0 V closepath stroke
  Pnt } def
/Crs { stroke [] 0 setdash exch hpt sub exch vpt add M
  hpt2 vpt2 neg V currentpoint stroke M
  hpt2 neg 0 R hpt2 vpt2 V stroke } def
/TriU { stroke [] 0 setdash 2 copy vpt 1.12 mul add M
  hpt neg vpt -1.62 mul V
  hpt 2 mul 0 V
  hpt neg vpt 1.62 mul V closepath stroke
  Pnt  } def
/Star { 2 copy Pls Crs } def
/BoxF { stroke [] 0 setdash exch hpt sub exch vpt add M
  0 vpt2 neg V  hpt2 0 V  0 vpt2 V
  hpt2 neg 0 V  closepath fill } def
/TriUF { stroke [] 0 setdash vpt 1.12 mul add M
  hpt neg vpt -1.62 mul V
  hpt 2 mul 0 V
  hpt neg vpt 1.62 mul V closepath fill } def
/TriD { stroke [] 0 setdash 2 copy vpt 1.12 mul sub M
  hpt neg vpt 1.62 mul V
  hpt 2 mul 0 V
  hpt neg vpt -1.62 mul V closepath stroke
  Pnt  } def
/TriDF { stroke [] 0 setdash vpt 1.12 mul sub M
  hpt neg vpt 1.62 mul V
  hpt 2 mul 0 V
  hpt neg vpt -1.62 mul V closepath fill} def
/DiaF { stroke [] 0 setdash vpt add M
  hpt neg vpt neg V hpt vpt neg V
  hpt vpt V hpt neg vpt V closepath fill } def
/Pent { stroke [] 0 setdash 2 copy gsave
  translate 0 hpt M 4 {72 rotate 0 hpt L} repeat
  closepath stroke grestore Pnt } def
/PentF { stroke [] 0 setdash gsave
  translate 0 hpt M 4 {72 rotate 0 hpt L} repeat
  closepath fill grestore } def
/Circle { stroke [] 0 setdash 2 copy
  hpt 0 360 arc stroke Pnt } def
/CircleF { stroke [] 0 setdash hpt 0 360 arc fill } def
/C0 { BL [] 0 setdash 2 copy moveto vpt 90 450  arc } bind def
/C1 { BL [] 0 setdash 2 copy        moveto
       2 copy  vpt 0 90 arc closepath fill
               vpt 0 360 arc closepath } bind def
/C2 { BL [] 0 setdash 2 copy moveto
       2 copy  vpt 90 180 arc closepath fill
               vpt 0 360 arc closepath } bind def
/C3 { BL [] 0 setdash 2 copy moveto
       2 copy  vpt 0 180 arc closepath fill
               vpt 0 360 arc closepath } bind def
/C4 { BL [] 0 setdash 2 copy moveto
       2 copy  vpt 180 270 arc closepath fill
               vpt 0 360 arc closepath } bind def
/C5 { BL [] 0 setdash 2 copy moveto
       2 copy  vpt 0 90 arc
       2 copy moveto
       2 copy  vpt 180 270 arc closepath fill
               vpt 0 360 arc } bind def
/C6 { BL [] 0 setdash 2 copy moveto
      2 copy  vpt 90 270 arc closepath fill
              vpt 0 360 arc closepath } bind def
/C7 { BL [] 0 setdash 2 copy moveto
      2 copy  vpt 0 270 arc closepath fill
              vpt 0 360 arc closepath } bind def
/C8 { BL [] 0 setdash 2 copy moveto
      2 copy vpt 270 360 arc closepath fill
              vpt 0 360 arc closepath } bind def
/C9 { BL [] 0 setdash 2 copy moveto
      2 copy  vpt 270 450 arc closepath fill
              vpt 0 360 arc closepath } bind def
/C10 { BL [] 0 setdash 2 copy 2 copy moveto vpt 270 360 arc closepath fill
       2 copy moveto
       2 copy vpt 90 180 arc closepath fill
               vpt 0 360 arc closepath } bind def
/C11 { BL [] 0 setdash 2 copy moveto
       2 copy  vpt 0 180 arc closepath fill
       2 copy moveto
       2 copy  vpt 270 360 arc closepath fill
               vpt 0 360 arc closepath } bind def
/C12 { BL [] 0 setdash 2 copy moveto
       2 copy  vpt 180 360 arc closepath fill
               vpt 0 360 arc closepath } bind def
/C13 { BL [] 0 setdash  2 copy moveto
       2 copy  vpt 0 90 arc closepath fill
       2 copy moveto
       2 copy  vpt 180 360 arc closepath fill
               vpt 0 360 arc closepath } bind def
/C14 { BL [] 0 setdash 2 copy moveto
       2 copy  vpt 90 360 arc closepath fill
               vpt 0 360 arc } bind def
/C15 { BL [] 0 setdash 2 copy vpt 0 360 arc closepath fill
               vpt 0 360 arc closepath } bind def
/Rec   { newpath 4 2 roll moveto 1 index 0 rlineto 0 exch rlineto
       neg 0 rlineto closepath } bind def
/Square { dup Rec } bind def
/Bsquare { vpt sub exch vpt sub exch vpt2 Square } bind def
/S0 { BL [] 0 setdash 2 copy moveto 0 vpt rlineto BL Bsquare } bind def
/S1 { BL [] 0 setdash 2 copy vpt Square fill Bsquare } bind def
/S2 { BL [] 0 setdash 2 copy exch vpt sub exch vpt Square fill Bsquare } bind def
/S3 { BL [] 0 setdash 2 copy exch vpt sub exch vpt2 vpt Rec fill Bsquare } bind def
/S4 { BL [] 0 setdash 2 copy exch vpt sub exch vpt sub vpt Square fill Bsquare } bind def
/S5 { BL [] 0 setdash 2 copy 2 copy vpt Square fill
       exch vpt sub exch vpt sub vpt Square fill Bsquare } bind def
/S6 { BL [] 0 setdash 2 copy exch vpt sub exch vpt sub vpt vpt2 Rec fill Bsquare } bind def
/S7 { BL [] 0 setdash 2 copy exch vpt sub exch vpt sub vpt vpt2 Rec fill
       2 copy vpt Square fill
       Bsquare } bind def
/S8 { BL [] 0 setdash 2 copy vpt sub vpt Square fill Bsquare } bind def
/S9 { BL [] 0 setdash 2 copy vpt sub vpt vpt2 Rec fill Bsquare } bind def
/S10 { BL [] 0 setdash 2 copy vpt sub vpt Square fill 2 copy exch vpt sub exch vpt Square fill
       Bsquare } bind def
/S11 { BL [] 0 setdash 2 copy vpt sub vpt Square fill 2 copy exch vpt sub exch vpt2 vpt Rec fill
       Bsquare } bind def
/S12 { BL [] 0 setdash 2 copy exch vpt sub exch vpt sub vpt2 vpt Rec fill Bsquare } bind def
/S13 { BL [] 0 setdash 2 copy exch vpt sub exch vpt sub vpt2 vpt Rec fill
       2 copy vpt Square fill Bsquare } bind def
/S14 { BL [] 0 setdash 2 copy exch vpt sub exch vpt sub vpt2 vpt Rec fill
       2 copy exch vpt sub exch vpt Square fill Bsquare } bind def
/S15 { BL [] 0 setdash 2 copy Bsquare fill Bsquare } bind def
/D0 { gsave translate 45 rotate 0 0 S0 stroke grestore } bind def
/D1 { gsave translate 45 rotate 0 0 S1 stroke grestore } bind def
/D2 { gsave translate 45 rotate 0 0 S2 stroke grestore } bind def
/D3 { gsave translate 45 rotate 0 0 S3 stroke grestore } bind def
/D4 { gsave translate 45 rotate 0 0 S4 stroke grestore } bind def
/D5 { gsave translate 45 rotate 0 0 S5 stroke grestore } bind def
/D6 { gsave translate 45 rotate 0 0 S6 stroke grestore } bind def
/D7 { gsave translate 45 rotate 0 0 S7 stroke grestore } bind def
/D8 { gsave translate 45 rotate 0 0 S8 stroke grestore } bind def
/D9 { gsave translate 45 rotate 0 0 S9 stroke grestore } bind def
/D10 { gsave translate 45 rotate 0 0 S10 stroke grestore } bind def
/D11 { gsave translate 45 rotate 0 0 S11 stroke grestore } bind def
/D12 { gsave translate 45 rotate 0 0 S12 stroke grestore } bind def
/D13 { gsave translate 45 rotate 0 0 S13 stroke grestore } bind def
/D14 { gsave translate 45 rotate 0 0 S14 stroke grestore } bind def
/D15 { gsave translate 45 rotate 0 0 S15 stroke grestore } bind def
/DiaE { stroke [] 0 setdash vpt add M
  hpt neg vpt neg V hpt vpt neg V
  hpt vpt V hpt neg vpt V closepath stroke } def
/BoxE { stroke [] 0 setdash exch hpt sub exch vpt add M
  0 vpt2 neg V hpt2 0 V 0 vpt2 V
  hpt2 neg 0 V closepath stroke } def
/TriUE { stroke [] 0 setdash vpt 1.12 mul add M
  hpt neg vpt -1.62 mul V
  hpt 2 mul 0 V
  hpt neg vpt 1.62 mul V closepath stroke } def
/TriDE { stroke [] 0 setdash vpt 1.12 mul sub M
  hpt neg vpt 1.62 mul V
  hpt 2 mul 0 V
  hpt neg vpt -1.62 mul V closepath stroke } def
/PentE { stroke [] 0 setdash gsave
  translate 0 hpt M 4 {72 rotate 0 hpt L} repeat
  closepath stroke grestore } def
/CircE { stroke [] 0 setdash 
  hpt 0 360 arc stroke } def
/Opaque { gsave closepath 1 setgray fill grestore 0 setgray closepath } def
/DiaW { stroke [] 0 setdash vpt add M
  hpt neg vpt neg V hpt vpt neg V
  hpt vpt V hpt neg vpt V Opaque stroke } def
/BoxW { stroke [] 0 setdash exch hpt sub exch vpt add M
  0 vpt2 neg V hpt2 0 V 0 vpt2 V
  hpt2 neg 0 V Opaque stroke } def
/TriUW { stroke [] 0 setdash vpt 1.12 mul add M
  hpt neg vpt -1.62 mul V
  hpt 2 mul 0 V
  hpt neg vpt 1.62 mul V Opaque stroke } def
/TriDW { stroke [] 0 setdash vpt 1.12 mul sub M
  hpt neg vpt 1.62 mul V
  hpt 2 mul 0 V
  hpt neg vpt -1.62 mul V Opaque stroke } def
/PentW { stroke [] 0 setdash gsave
  translate 0 hpt M 4 {72 rotate 0 hpt L} repeat
  Opaque stroke grestore } def
/CircW { stroke [] 0 setdash 
  hpt 0 360 arc Opaque stroke } def
/BoxFill { gsave Rec 1 setgray fill grestore } def
end
}}%
\begin{picture}(3600,2160)(0,0)%
{\GNUPLOTspecial{"
gnudict begin
gsave
0 0 translate
0.100 0.100 scale
0 setgray
newpath
1.000 UL
LTb
350 300 M
63 0 V
3037 0 R
-63 0 V
350 520 M
63 0 V
3037 0 R
-63 0 V
350 740 M
63 0 V
3037 0 R
-63 0 V
350 960 M
63 0 V
3037 0 R
-63 0 V
350 1180 M
63 0 V
3037 0 R
-63 0 V
350 1400 M
63 0 V
3037 0 R
-63 0 V
350 1620 M
63 0 V
3037 0 R
-63 0 V
350 1840 M
63 0 V
3037 0 R
-63 0 V
350 2060 M
63 0 V
3037 0 R
-63 0 V
350 300 M
0 63 V
0 1697 R
0 -63 V
694 300 M
0 63 V
0 1697 R
0 -63 V
1039 300 M
0 63 V
0 1697 R
0 -63 V
1383 300 M
0 63 V
0 1697 R
0 -63 V
1728 300 M
0 63 V
0 1697 R
0 -63 V
2072 300 M
0 63 V
0 1697 R
0 -63 V
2417 300 M
0 63 V
0 1697 R
0 -63 V
2761 300 M
0 63 V
0 1697 R
0 -63 V
3106 300 M
0 63 V
0 1697 R
0 -63 V
3450 300 M
0 63 V
0 1697 R
0 -63 V
1.000 UL
LTb
350 300 M
3100 0 V
0 1760 V
-3100 0 V
350 300 L
1.000 UL
LT0
3087 1947 M
263 0 V
350 1966 M
522 1625 L
694 1392 L
867 1221 L
172 -133 V
1211 979 L
172 -91 V
173 -79 V
172 -68 V
172 -60 V
172 -54 V
172 -49 V
173 -44 V
172 -40 V
172 -37 V
172 -35 V
173 -32 V
1.000 UL
LT1
3087 1847 M
263 0 V
350 1565 M
522 1359 L
694 1201 L
867 1075 L
1039 970 L
172 -88 V
172 -77 V
173 -67 V
172 -59 V
172 -54 V
172 -48 V
172 -43 V
173 -40 V
172 -37 V
172 -34 V
172 -32 V
173 -30 V
1.000 UL
LT2
3087 1747 M
263 0 V
350 1583 M
522 1378 L
694 1213 L
867 1094 L
1039 990 L
172 -95 V
172 -71 V
173 -71 V
172 -60 V
172 -51 V
172 -55 V
172 -39 V
173 -39 V
172 -41 V
172 -35 V
172 -28 V
173 -30 V
stroke
grestore
end
showpage
}}%
\put(3037,1747){\makebox(0,0)[r]{PDF}}%
\put(3037,1847){\makebox(0,0)[r]{$\delta = 0$}}%
\put(3037,1947){\makebox(0,0)[r]{$\delta = 0.00096$}}%
\put(2072,1400){\makebox(0,0)[l]{$Q^2 = 6.76$~GeV$^2$}}%
\put(1900,50){\makebox(0,0){\Large $x'$}}%
\put(100,1180){%
\special{ps: gsave currentpoint currentpoint translate
270 rotate neg exch neg exch translate}%
\makebox(0,0)[b]{\shortstack{\Large Gluon Distribution}}%
\special{ps: currentpoint grestore moveto}%
}%
\put(3450,200){\makebox(0,0){0.0055}}%
\put(3106,200){\makebox(0,0){0.005}}%
\put(2761,200){\makebox(0,0){0.0045}}%
\put(2417,200){\makebox(0,0){0.004}}%
\put(2072,200){\makebox(0,0){0.0035}}%
\put(1728,200){\makebox(0,0){0.003}}%
\put(1383,200){\makebox(0,0){0.0025}}%
\put(1039,200){\makebox(0,0){0.002}}%
\put(694,200){\makebox(0,0){0.0015}}%
\put(350,200){\makebox(0,0){0.001}}%
\put(300,2060){\makebox(0,0)[r]{14}}%
\put(300,1840){\makebox(0,0)[r]{13}}%
\put(300,1620){\makebox(0,0)[r]{12}}%
\put(300,1400){\makebox(0,0)[r]{11}}%
\put(300,1180){\makebox(0,0)[r]{10}}%
\put(300,960){\makebox(0,0)[r]{9}}%
\put(300,740){\makebox(0,0)[r]{8}}%
\put(300,520){\makebox(0,0)[r]{7}}%
\put(300,300){\makebox(0,0)[r]{6}}%
\end{picture}%
\endgroup
 

%% file: sigspdf.tex
\begingroup%
  \makeatletter%
  \newcommand{\GNUPLOTspecial}{%
    \@sanitize\catcode`\%=14\relax\special}%
  \setlength{\unitlength}{0.1bp}%
{\GNUPLOTspecial{!
/gnudict 256 dict def
gnudict begin
/Color false def
/Solid false def
/gnulinewidth 5.000 def
/userlinewidth gnulinewidth def
/vshift -33 def
/dl {10 mul} def
/hpt_ 31.5 def
/vpt_ 31.5 def
/hpt hpt_ def
/vpt vpt_ def
/M {moveto} bind def
/L {lineto} bind def
/R {rmoveto} bind def
/V {rlineto} bind def
/vpt2 vpt 2 mul def
/hpt2 hpt 2 mul def
/Lshow { currentpoint stroke M
  0 vshift R show } def
/Rshow { currentpoint stroke M
  dup stringwidth pop neg vshift R show } def
/Cshow { currentpoint stroke M
  dup stringwidth pop -2 div vshift R show } def
/UP { dup vpt_ mul /vpt exch def hpt_ mul /hpt exch def
  /hpt2 hpt 2 mul def /vpt2 vpt 2 mul def } def
/DL { Color {setrgbcolor Solid {pop []} if 0 setdash }
 {pop pop pop Solid {pop []} if 0 setdash} ifelse } def
/BL { stroke userlinewidth 2 mul setlinewidth } def
/AL { stroke userlinewidth 2 div setlinewidth } def
/UL { dup gnulinewidth mul /userlinewidth exch def
      10 mul /udl exch def } def
/PL { stroke userlinewidth setlinewidth } def
/LTb { BL [] 0 0 0 DL } def
/LTa { AL [1 udl mul 2 udl mul] 0 setdash 0 0 0 setrgbcolor } def
/LT0 { PL [] 1 0 0 DL } def
/LT1 { PL [4 dl 2 dl] 0 1 0 DL } def
/LT2 { PL [2 dl 3 dl] 0 0 1 DL } def
/LT3 { PL [1 dl 1.5 dl] 1 0 1 DL } def
/LT4 { PL [5 dl 2 dl 1 dl 2 dl] 0 1 1 DL } def
/LT5 { PL [4 dl 3 dl 1 dl 3 dl] 1 1 0 DL } def
/LT6 { PL [2 dl 2 dl 2 dl 4 dl] 0 0 0 DL } def
/LT7 { PL [2 dl 2 dl 2 dl 2 dl 2 dl 4 dl] 1 0.3 0 DL } def
/LT8 { PL [2 dl 2 dl 2 dl 2 dl 2 dl 2 dl 2 dl 4 dl] 0.5 0.5 0.5 DL } def
/Pnt { stroke [] 0 setdash
   gsave 1 setlinecap M 0 0 V stroke grestore } def
/Dia { stroke [] 0 setdash 2 copy vpt add M
  hpt neg vpt neg V hpt vpt neg V
  hpt vpt V hpt neg vpt V closepath stroke
  Pnt } def
/Pls { stroke [] 0 setdash vpt sub M 0 vpt2 V
  currentpoint stroke M
  hpt neg vpt neg R hpt2 0 V stroke
  } def
/Box { stroke [] 0 setdash 2 copy exch hpt sub exch vpt add M
  0 vpt2 neg V hpt2 0 V 0 vpt2 V
  hpt2 neg 0 V closepath stroke
  Pnt } def
/Crs { stroke [] 0 setdash exch hpt sub exch vpt add M
  hpt2 vpt2 neg V currentpoint stroke M
  hpt2 neg 0 R hpt2 vpt2 V stroke } def
/TriU { stroke [] 0 setdash 2 copy vpt 1.12 mul add M
  hpt neg vpt -1.62 mul V
  hpt 2 mul 0 V
  hpt neg vpt 1.62 mul V closepath stroke
  Pnt  } def
/Star { 2 copy Pls Crs } def
/BoxF { stroke [] 0 setdash exch hpt sub exch vpt add M
  0 vpt2 neg V  hpt2 0 V  0 vpt2 V
  hpt2 neg 0 V  closepath fill } def
/TriUF { stroke [] 0 setdash vpt 1.12 mul add M
  hpt neg vpt -1.62 mul V
  hpt 2 mul 0 V
  hpt neg vpt 1.62 mul V closepath fill } def
/TriD { stroke [] 0 setdash 2 copy vpt 1.12 mul sub M
  hpt neg vpt 1.62 mul V
  hpt 2 mul 0 V
  hpt neg vpt -1.62 mul V closepath stroke
  Pnt  } def
/TriDF { stroke [] 0 setdash vpt 1.12 mul sub M
  hpt neg vpt 1.62 mul V
  hpt 2 mul 0 V
  hpt neg vpt -1.62 mul V closepath fill} def
/DiaF { stroke [] 0 setdash vpt add M
  hpt neg vpt neg V hpt vpt neg V
  hpt vpt V hpt neg vpt V closepath fill } def
/Pent { stroke [] 0 setdash 2 copy gsave
  translate 0 hpt M 4 {72 rotate 0 hpt L} repeat
  closepath stroke grestore Pnt } def
/PentF { stroke [] 0 setdash gsave
  translate 0 hpt M 4 {72 rotate 0 hpt L} repeat
  closepath fill grestore } def
/Circle { stroke [] 0 setdash 2 copy
  hpt 0 360 arc stroke Pnt } def
/CircleF { stroke [] 0 setdash hpt 0 360 arc fill } def
/C0 { BL [] 0 setdash 2 copy moveto vpt 90 450  arc } bind def
/C1 { BL [] 0 setdash 2 copy        moveto
       2 copy  vpt 0 90 arc closepath fill
               vpt 0 360 arc closepath } bind def
/C2 { BL [] 0 setdash 2 copy moveto
       2 copy  vpt 90 180 arc closepath fill
               vpt 0 360 arc closepath } bind def
/C3 { BL [] 0 setdash 2 copy moveto
       2 copy  vpt 0 180 arc closepath fill
               vpt 0 360 arc closepath } bind def
/C4 { BL [] 0 setdash 2 copy moveto
       2 copy  vpt 180 270 arc closepath fill
               vpt 0 360 arc closepath } bind def
/C5 { BL [] 0 setdash 2 copy moveto
       2 copy  vpt 0 90 arc
       2 copy moveto
       2 copy  vpt 180 270 arc closepath fill
               vpt 0 360 arc } bind def
/C6 { BL [] 0 setdash 2 copy moveto
      2 copy  vpt 90 270 arc closepath fill
              vpt 0 360 arc closepath } bind def
/C7 { BL [] 0 setdash 2 copy moveto
      2 copy  vpt 0 270 arc closepath fill
              vpt 0 360 arc closepath } bind def
/C8 { BL [] 0 setdash 2 copy moveto
      2 copy vpt 270 360 arc closepath fill
              vpt 0 360 arc closepath } bind def
/C9 { BL [] 0 setdash 2 copy moveto
      2 copy  vpt 270 450 arc closepath fill
              vpt 0 360 arc closepath } bind def
/C10 { BL [] 0 setdash 2 copy 2 copy moveto vpt 270 360 arc closepath fill
       2 copy moveto
       2 copy vpt 90 180 arc closepath fill
               vpt 0 360 arc closepath } bind def
/C11 { BL [] 0 setdash 2 copy moveto
       2 copy  vpt 0 180 arc closepath fill
       2 copy moveto
       2 copy  vpt 270 360 arc closepath fill
               vpt 0 360 arc closepath } bind def
/C12 { BL [] 0 setdash 2 copy moveto
       2 copy  vpt 180 360 arc closepath fill
               vpt 0 360 arc closepath } bind def
/C13 { BL [] 0 setdash  2 copy moveto
       2 copy  vpt 0 90 arc closepath fill
       2 copy moveto
       2 copy  vpt 180 360 arc closepath fill
               vpt 0 360 arc closepath } bind def
/C14 { BL [] 0 setdash 2 copy moveto
       2 copy  vpt 90 360 arc closepath fill
               vpt 0 360 arc } bind def
/C15 { BL [] 0 setdash 2 copy vpt 0 360 arc closepath fill
               vpt 0 360 arc closepath } bind def
/Rec   { newpath 4 2 roll moveto 1 index 0 rlineto 0 exch rlineto
       neg 0 rlineto closepath } bind def
/Square { dup Rec } bind def
/Bsquare { vpt sub exch vpt sub exch vpt2 Square } bind def
/S0 { BL [] 0 setdash 2 copy moveto 0 vpt rlineto BL Bsquare } bind def
/S1 { BL [] 0 setdash 2 copy vpt Square fill Bsquare } bind def
/S2 { BL [] 0 setdash 2 copy exch vpt sub exch vpt Square fill Bsquare } bind def
/S3 { BL [] 0 setdash 2 copy exch vpt sub exch vpt2 vpt Rec fill Bsquare } bind def
/S4 { BL [] 0 setdash 2 copy exch vpt sub exch vpt sub vpt Square fill Bsquare } bind def
/S5 { BL [] 0 setdash 2 copy 2 copy vpt Square fill
       exch vpt sub exch vpt sub vpt Square fill Bsquare } bind def
/S6 { BL [] 0 setdash 2 copy exch vpt sub exch vpt sub vpt vpt2 Rec fill Bsquare } bind def
/S7 { BL [] 0 setdash 2 copy exch vpt sub exch vpt sub vpt vpt2 Rec fill
       2 copy vpt Square fill
       Bsquare } bind def
/S8 { BL [] 0 setdash 2 copy vpt sub vpt Square fill Bsquare } bind def
/S9 { BL [] 0 setdash 2 copy vpt sub vpt vpt2 Rec fill Bsquare } bind def
/S10 { BL [] 0 setdash 2 copy vpt sub vpt Square fill 2 copy exch vpt sub exch vpt Square fill
       Bsquare } bind def
/S11 { BL [] 0 setdash 2 copy vpt sub vpt Square fill 2 copy exch vpt sub exch vpt2 vpt Rec fill
       Bsquare } bind def
/S12 { BL [] 0 setdash 2 copy exch vpt sub exch vpt sub vpt2 vpt Rec fill Bsquare } bind def
/S13 { BL [] 0 setdash 2 copy exch vpt sub exch vpt sub vpt2 vpt Rec fill
       2 copy vpt Square fill Bsquare } bind def
/S14 { BL [] 0 setdash 2 copy exch vpt sub exch vpt sub vpt2 vpt Rec fill
       2 copy exch vpt sub exch vpt Square fill Bsquare } bind def
/S15 { BL [] 0 setdash 2 copy Bsquare fill Bsquare } bind def
/D0 { gsave translate 45 rotate 0 0 S0 stroke grestore } bind def
/D1 { gsave translate 45 rotate 0 0 S1 stroke grestore } bind def
/D2 { gsave translate 45 rotate 0 0 S2 stroke grestore } bind def
/D3 { gsave translate 45 rotate 0 0 S3 stroke grestore } bind def
/D4 { gsave translate 45 rotate 0 0 S4 stroke grestore } bind def
/D5 { gsave translate 45 rotate 0 0 S5 stroke grestore } bind def
/D6 { gsave translate 45 rotate 0 0 S6 stroke grestore } bind def
/D7 { gsave translate 45 rotate 0 0 S7 stroke grestore } bind def
/D8 { gsave translate 45 rotate 0 0 S8 stroke grestore } bind def
/D9 { gsave translate 45 rotate 0 0 S9 stroke grestore } bind def
/D10 { gsave translate 45 rotate 0 0 S10 stroke grestore } bind def
/D11 { gsave translate 45 rotate 0 0 S11 stroke grestore } bind def
/D12 { gsave translate 45 rotate 0 0 S12 stroke grestore } bind def
/D13 { gsave translate 45 rotate 0 0 S13 stroke grestore } bind def
/D14 { gsave translate 45 rotate 0 0 S14 stroke grestore } bind def
/D15 { gsave translate 45 rotate 0 0 S15 stroke grestore } bind def
/DiaE { stroke [] 0 setdash vpt add M
  hpt neg vpt neg V hpt vpt neg V
  hpt vpt V hpt neg vpt V closepath stroke } def
/BoxE { stroke [] 0 setdash exch hpt sub exch vpt add M
  0 vpt2 neg V hpt2 0 V 0 vpt2 V
  hpt2 neg 0 V closepath stroke } def
/TriUE { stroke [] 0 setdash vpt 1.12 mul add M
  hpt neg vpt -1.62 mul V
  hpt 2 mul 0 V
  hpt neg vpt 1.62 mul V closepath stroke } def
/TriDE { stroke [] 0 setdash vpt 1.12 mul sub M
  hpt neg vpt 1.62 mul V
  hpt 2 mul 0 V
  hpt neg vpt -1.62 mul V closepath stroke } def
/PentE { stroke [] 0 setdash gsave
  translate 0 hpt M 4 {72 rotate 0 hpt L} repeat
  closepath stroke grestore } def
/CircE { stroke [] 0 setdash 
  hpt 0 360 arc stroke } def
/Opaque { gsave closepath 1 setgray fill grestore 0 setgray closepath } def
/DiaW { stroke [] 0 setdash vpt add M
  hpt neg vpt neg V hpt vpt neg V
  hpt vpt V hpt neg vpt V Opaque stroke } def
/BoxW { stroke [] 0 setdash exch hpt sub exch vpt add M
  0 vpt2 neg V hpt2 0 V 0 vpt2 V
  hpt2 neg 0 V Opaque stroke } def
/TriUW { stroke [] 0 setdash vpt 1.12 mul add M
  hpt neg vpt -1.62 mul V
  hpt 2 mul 0 V
  hpt neg vpt 1.62 mul V Opaque stroke } def
/TriDW { stroke [] 0 setdash vpt 1.12 mul sub M
  hpt neg vpt 1.62 mul V
  hpt 2 mul 0 V
  hpt neg vpt -1.62 mul V Opaque stroke } def
/PentW { stroke [] 0 setdash gsave
  translate 0 hpt M 4 {72 rotate 0 hpt L} repeat
  Opaque stroke grestore } def
/CircW { stroke [] 0 setdash 
  hpt 0 360 arc Opaque stroke } def
/BoxFill { gsave Rec 1 setgray fill grestore } def
end
}}%
\begin{picture}(3600,2160)(0,0)%
{\GNUPLOTspecial{"
gnudict begin
gsave
0 0 translate
0.100 0.100 scale
0 setgray
newpath
1.000 UL
LTb
350 300 M
63 0 V
3037 0 R
-63 0 V
350 551 M
63 0 V
3037 0 R
-63 0 V
350 803 M
63 0 V
3037 0 R
-63 0 V
350 1054 M
63 0 V
3037 0 R
-63 0 V
350 1306 M
63 0 V
3037 0 R
-63 0 V
350 1557 M
63 0 V
3037 0 R
-63 0 V
350 1809 M
63 0 V
3037 0 R
-63 0 V
350 2060 M
63 0 V
3037 0 R
-63 0 V
350 300 M
0 63 V
0 1697 R
0 -63 V
827 300 M
0 63 V
0 1697 R
0 -63 V
1304 300 M
0 63 V
0 1697 R
0 -63 V
1781 300 M
0 63 V
0 1697 R
0 -63 V
2258 300 M
0 63 V
0 1697 R
0 -63 V
2735 300 M
0 63 V
0 1697 R
0 -63 V
3212 300 M
0 63 V
0 1697 R
0 -63 V
1.000 UL
LTb
350 300 M
3100 0 V
0 1760 V
-3100 0 V
350 300 L
1.000 UL
LT0
827 1557 M
263 0 V
350 300 M
10 0 V
9 0 V
10 0 V
9 0 V
10 0 V
9 0 V
10 0 V
9 0 V
10 0 V
9 0 V
10 0 V
9 0 V
10 0 V
10 1 V
9 0 V
10 0 V
9 0 V
10 0 V
9 1 V
10 0 V
9 1 V
10 0 V
9 0 V
10 1 V
9 1 V
10 0 V
10 1 V
9 1 V
10 1 V
9 0 V
10 1 V
9 1 V
10 1 V
9 1 V
10 2 V
9 1 V
10 1 V
9 2 V
10 1 V
10 1 V
9 2 V
10 1 V
9 2 V
10 2 V
9 2 V
10 2 V
9 1 V
10 2 V
9 3 V
10 2 V
9 2 V
10 2 V
10 2 V
9 3 V
10 2 V
9 3 V
10 2 V
9 3 V
10 2 V
9 3 V
10 3 V
9 3 V
10 2 V
9 3 V
10 3 V
10 4 V
9 3 V
10 3 V
9 3 V
10 4 V
9 3 V
10 3 V
9 3 V
10 4 V
9 3 V
10 4 V
9 3 V
10 4 V
10 4 V
9 4 V
10 3 V
9 4 V
10 4 V
9 4 V
10 4 V
9 4 V
10 4 V
9 4 V
10 4 V
9 4 V
10 4 V
10 4 V
9 4 V
10 4 V
9 5 V
10 4 V
9 5 V
10 4 V
9 5 V
10 4 V
9 5 V
10 4 V
9 5 V
10 5 V
10 4 V
9 5 V
10 4 V
9 5 V
10 5 V
9 4 V
10 5 V
9 5 V
10 4 V
9 5 V
10 4 V
9 5 V
10 5 V
10 4 V
9 5 V
10 5 V
9 5 V
10 5 V
9 5 V
10 5 V
9 5 V
10 5 V
9 5 V
10 5 V
9 5 V
10 6 V
10 5 V
9 5 V
10 5 V
9 5 V
10 5 V
9 5 V
10 5 V
9 5 V
10 5 V
9 5 V
10 5 V
9 5 V
10 5 V
10 5 V
9 5 V
10 5 V
9 5 V
10 5 V
9 5 V
10 5 V
9 5 V
10 5 V
9 5 V
10 5 V
9 5 V
10 5 V
10 5 V
9 5 V
10 5 V
9 4 V
10 5 V
9 5 V
10 5 V
9 5 V
10 4 V
9 5 V
10 5 V
9 5 V
10 4 V
10 5 V
9 5 V
10 4 V
9 5 V
10 5 V
9 4 V
10 5 V
9 5 V
10 5 V
9 4 V
10 5 V
9 5 V
10 4 V
10 5 V
9 5 V
10 4 V
9 5 V
10 4 V
9 5 V
10 4 V
9 5 V
10 4 V
9 4 V
10 5 V
9 4 V
10 4 V
10 6 V
9 6 V
10 6 V
9 6 V
10 6 V
9 6 V
10 6 V
9 6 V
10 6 V
9 6 V
10 6 V
9 6 V
10 6 V
10 6 V
9 6 V
10 6 V
9 6 V
10 6 V
9 6 V
10 6 V
9 6 V
10 6 V
9 6 V
10 6 V
9 6 V
10 6 V
10 6 V
9 6 V
10 5 V
9 6 V
10 6 V
9 6 V
10 6 V
9 6 V
10 6 V
9 6 V
10 6 V
9 6 V
10 6 V
10 6 V
9 6 V
10 6 V
9 6 V
10 6 V
9 6 V
10 6 V
9 6 V
10 6 V
9 6 V
10 6 V
9 6 V
10 6 V
10 6 V
9 6 V
10 6 V
9 6 V
10 6 V
9 6 V
10 6 V
9 6 V
10 6 V
9 6 V
10 5 V
9 6 V
10 6 V
10 6 V
9 6 V
10 6 V
9 6 V
10 6 V
9 6 V
10 6 V
9 6 V
10 6 V
9 6 V
10 6 V
9 6 V
10 6 V
10 6 V
9 6 V
10 6 V
9 6 V
10 6 V
9 6 V
10 6 V
9 6 V
10 6 V
9 6 V
10 6 V
9 6 V
10 6 V
10 6 V
9 6 V
10 6 V
9 6 V
10 5 V
9 6 V
10 6 V
9 6 V
10 6 V
9 6 V
10 6 V
9 6 V
10 6 V
10 6 V
9 6 V
10 6 V
9 6 V
10 6 V
9 6 V
10 6 V
9 6 V
10 6 V
9 6 V
10 6 V
9 6 V
10 6 V
10 6 V
9 6 V
10 6 V
9 6 V
10 6 V
9 6 V
10 6 V
9 6 V
10 6 V
9 6 V
10 6 V
9 6 V
10 5 V
1.000 UL
LT0
350 300 M
10 0 V
9 0 V
10 0 V
9 0 V
10 0 V
9 0 V
10 0 V
9 0 V
10 1 V
9 0 V
10 0 V
9 1 V
10 0 V
10 1 V
9 0 V
10 1 V
9 1 V
10 1 V
9 1 V
10 1 V
9 2 V
10 1 V
9 2 V
10 1 V
9 2 V
10 2 V
10 2 V
9 2 V
10 3 V
9 2 V
10 3 V
9 3 V
10 3 V
9 3 V
10 3 V
9 3 V
10 4 V
9 3 V
10 4 V
10 4 V
9 4 V
10 4 V
9 5 V
10 4 V
9 5 V
10 4 V
9 5 V
10 5 V
9 5 V
10 5 V
9 6 V
10 5 V
10 5 V
9 6 V
10 6 V
9 6 V
10 6 V
9 7 V
10 6 V
9 6 V
10 6 V
9 6 V
10 7 V
9 7 V
10 7 V
10 8 V
9 7 V
10 7 V
9 7 V
10 8 V
9 7 V
10 7 V
9 7 V
10 7 V
9 9 V
10 8 V
9 8 V
10 8 V
10 8 V
9 9 V
10 8 V
9 8 V
10 8 V
9 9 V
10 8 V
9 8 V
10 8 V
9 8 V
10 9 V
9 8 V
10 9 V
10 9 V
9 9 V
10 9 V
9 9 V
10 10 V
9 9 V
10 9 V
9 9 V
10 9 V
9 10 V
10 9 V
9 9 V
10 9 V
10 9 V
9 9 V
10 10 V
9 9 V
10 9 V
9 9 V
10 9 V
9 9 V
10 9 V
9 9 V
10 8 V
9 10 V
10 9 V
10 10 V
9 10 V
10 10 V
9 9 V
10 10 V
9 10 V
10 9 V
9 10 V
10 10 V
9 9 V
10 10 V
9 9 V
10 10 V
10 10 V
9 9 V
10 10 V
9 9 V
10 9 V
9 10 V
10 9 V
9 9 V
10 10 V
9 9 V
10 7 V
9 4 V
10 5 V
10 5 V
9 4 V
10 5 V
9 5 V
10 4 V
9 5 V
10 5 V
9 4 V
10 5 V
9 4 V
10 5 V
9 5 V
10 4 V
10 5 V
9 5 V
10 4 V
9 5 V
10 4 V
9 5 V
10 5 V
9 4 V
10 5 V
9 5 V
10 4 V
9 5 V
10 4 V
10 5 V
9 5 V
10 4 V
9 5 V
10 5 V
9 4 V
10 5 V
9 5 V
10 4 V
9 5 V
10 4 V
9 5 V
10 5 V
10 4 V
9 5 V
10 5 V
9 4 V
10 5 V
9 4 V
10 5 V
9 5 V
10 4 V
9 5 V
10 5 V
9 4 V
10 5 V
10 4 V
9 5 V
10 5 V
9 4 V
10 5 V
9 5 V
10 4 V
9 5 V
10 5 V
9 4 V
10 5 V
9 4 V
10 5 V
10 5 V
9 4 V
10 5 V
9 5 V
10 4 V
9 5 V
10 4 V
9 5 V
10 5 V
9 4 V
10 5 V
9 5 V
10 4 V
10 5 V
9 4 V
10 5 V
9 5 V
10 4 V
9 5 V
10 5 V
9 4 V
10 5 V
9 5 V
10 4 V
9 5 V
10 4 V
10 5 V
9 5 V
10 4 V
9 5 V
10 5 V
9 4 V
10 5 V
9 4 V
10 5 V
9 5 V
10 4 V
9 5 V
10 5 V
10 4 V
9 5 V
10 4 V
9 5 V
10 5 V
9 4 V
10 5 V
9 5 V
10 4 V
9 5 V
10 5 V
9 4 V
10 5 V
10 4 V
9 5 V
10 5 V
9 4 V
10 5 V
9 5 V
10 4 V
9 5 V
10 4 V
9 5 V
10 5 V
9 4 V
10 5 V
10 5 V
9 4 V
10 5 V
9 4 V
10 5 V
9 5 V
10 4 V
9 5 V
10 5 V
9 4 V
10 5 V
9 5 V
10 4 V
10 5 V
9 4 V
10 5 V
9 5 V
10 4 V
9 5 V
10 5 V
9 4 V
10 5 V
9 4 V
10 5 V
9 5 V
10 4 V
10 5 V
9 5 V
10 4 V
9 5 V
10 4 V
9 5 V
10 5 V
9 4 V
10 5 V
9 5 V
10 4 V
9 5 V
10 4 V
10 5 V
9 5 V
10 4 V
9 5 V
10 5 V
9 4 V
10 5 V
9 5 V
10 4 V
9 5 V
10 4 V
9 5 V
10 5 V
1.000 UL
LT1
827 1457 M
263 0 V
350 300 M
10 0 V
9 0 V
10 0 V
9 0 V
10 0 V
9 0 V
10 0 V
9 0 V
10 0 V
9 0 V
10 0 V
9 0 V
10 1 V
10 0 V
9 0 V
10 0 V
9 0 V
10 1 V
9 0 V
10 0 V
9 1 V
10 0 V
9 1 V
10 0 V
9 1 V
10 1 V
10 0 V
9 1 V
10 1 V
9 1 V
10 1 V
9 1 V
10 1 V
9 1 V
10 1 V
9 1 V
10 2 V
9 1 V
10 2 V
10 1 V
9 2 V
10 2 V
9 2 V
10 1 V
9 2 V
10 2 V
9 2 V
10 3 V
9 2 V
10 2 V
9 3 V
10 2 V
10 3 V
9 2 V
10 3 V
9 2 V
10 3 V
9 3 V
10 3 V
9 3 V
10 3 V
9 3 V
10 4 V
9 3 V
10 3 V
10 4 V
9 3 V
10 4 V
9 3 V
10 4 V
9 4 V
10 3 V
9 4 V
10 4 V
9 4 V
10 4 V
9 4 V
10 4 V
10 4 V
9 4 V
10 4 V
9 5 V
10 4 V
9 5 V
10 4 V
9 4 V
10 5 V
9 4 V
10 5 V
9 5 V
10 4 V
10 5 V
9 5 V
10 5 V
9 4 V
10 5 V
9 5 V
10 5 V
9 5 V
10 5 V
9 5 V
10 5 V
9 5 V
10 5 V
10 5 V
9 5 V
10 5 V
9 5 V
10 5 V
9 6 V
10 5 V
9 5 V
10 5 V
9 5 V
10 5 V
9 6 V
10 5 V
10 5 V
9 5 V
10 6 V
9 5 V
10 5 V
9 5 V
10 6 V
9 5 V
10 5 V
9 5 V
10 6 V
9 5 V
10 5 V
10 5 V
9 5 V
10 5 V
9 6 V
10 5 V
9 5 V
10 5 V
9 5 V
10 5 V
9 5 V
10 5 V
9 5 V
10 5 V
10 5 V
9 5 V
10 5 V
9 5 V
10 5 V
9 5 V
10 5 V
9 5 V
10 4 V
9 5 V
10 5 V
9 4 V
10 5 V
10 4 V
9 5 V
10 5 V
9 4 V
10 5 V
9 4 V
10 5 V
9 4 V
10 4 V
9 4 V
10 4 V
9 5 V
10 4 V
10 4 V
9 4 V
10 4 V
9 3 V
10 4 V
9 4 V
10 4 V
9 3 V
10 4 V
9 4 V
10 4 V
9 3 V
10 4 V
10 3 V
9 3 V
10 3 V
9 3 V
10 3 V
9 3 V
10 3 V
9 3 V
10 3 V
9 3 V
10 2 V
9 3 V
10 3 V
10 6 V
9 6 V
10 6 V
9 6 V
10 6 V
9 6 V
10 7 V
9 6 V
10 6 V
9 6 V
10 6 V
9 6 V
10 6 V
10 6 V
9 6 V
10 6 V
9 6 V
10 7 V
9 6 V
10 6 V
9 6 V
10 6 V
9 6 V
10 6 V
9 7 V
10 6 V
10 6 V
9 6 V
10 6 V
9 6 V
10 6 V
9 6 V
10 6 V
9 6 V
10 6 V
9 7 V
10 6 V
9 6 V
10 6 V
10 6 V
9 6 V
10 6 V
9 6 V
10 6 V
9 6 V
10 7 V
9 6 V
10 6 V
9 6 V
10 6 V
9 6 V
10 6 V
10 7 V
9 6 V
10 6 V
9 6 V
10 6 V
9 6 V
10 6 V
9 6 V
10 6 V
9 6 V
10 6 V
9 6 V
10 7 V
10 6 V
9 6 V
10 6 V
9 6 V
10 6 V
9 6 V
10 6 V
9 6 V
10 7 V
9 6 V
10 6 V
9 6 V
10 6 V
10 6 V
9 6 V
10 7 V
9 6 V
10 6 V
9 6 V
10 6 V
9 6 V
10 6 V
9 6 V
10 6 V
9 6 V
10 6 V
10 6 V
9 6 V
10 7 V
9 6 V
10 6 V
9 6 V
10 6 V
9 6 V
10 6 V
9 6 V
10 7 V
9 6 V
10 6 V
10 6 V
9 6 V
10 6 V
9 6 V
10 7 V
9 6 V
10 6 V
9 6 V
10 6 V
9 6 V
10 6 V
9 6 V
10 6 V
10 6 V
9 6 V
10 6 V
9 6 V
10 6 V
9 7 V
10 6 V
9 6 V
10 6 V
9 6 V
10 6 V
9 6 V
10 7 V
1.000 UL
LT1
350 300 M
10 0 V
9 0 V
10 0 V
9 0 V
10 0 V
9 0 V
10 0 V
9 0 V
10 1 V
9 0 V
10 0 V
9 1 V
10 0 V
10 1 V
9 1 V
10 0 V
9 1 V
10 1 V
9 1 V
10 2 V
9 1 V
10 1 V
9 2 V
10 2 V
9 2 V
10 2 V
10 2 V
9 2 V
10 3 V
9 2 V
10 3 V
9 3 V
10 3 V
9 4 V
10 3 V
9 4 V
10 3 V
9 4 V
10 4 V
10 5 V
9 4 V
10 4 V
9 5 V
10 5 V
9 5 V
10 5 V
9 5 V
10 6 V
9 5 V
10 6 V
9 6 V
10 6 V
10 6 V
9 6 V
10 7 V
9 6 V
10 7 V
9 7 V
10 7 V
9 7 V
10 7 V
9 8 V
10 7 V
9 8 V
10 7 V
10 8 V
9 8 V
10 8 V
9 8 V
10 9 V
9 8 V
10 9 V
9 8 V
10 9 V
9 8 V
10 9 V
9 9 V
10 9 V
10 9 V
9 9 V
10 9 V
9 10 V
10 9 V
9 9 V
10 10 V
9 9 V
10 10 V
9 10 V
10 9 V
9 10 V
10 10 V
10 10 V
9 10 V
10 10 V
9 9 V
10 10 V
9 10 V
10 10 V
9 11 V
10 10 V
9 10 V
10 10 V
9 10 V
10 10 V
10 11 V
9 10 V
10 10 V
9 10 V
10 11 V
9 10 V
10 10 V
9 10 V
10 10 V
9 10 V
10 11 V
9 10 V
10 10 V
10 10 V
9 10 V
10 10 V
9 10 V
10 10 V
9 10 V
10 10 V
9 11 V
10 9 V
9 10 V
10 10 V
9 10 V
10 9 V
10 10 V
9 10 V
10 9 V
9 8 V
10 5 V
9 5 V
10 4 V
9 4 V
10 5 V
9 4 V
10 5 V
9 4 V
10 5 V
10 4 V
9 5 V
10 4 V
9 5 V
10 4 V
9 5 V
10 4 V
9 4 V
10 5 V
9 4 V
10 5 V
9 5 V
10 4 V
10 5 V
9 4 V
10 5 V
9 4 V
10 5 V
9 4 V
10 5 V
9 4 V
10 4 V
9 5 V
10 4 V
9 5 V
10 4 V
10 5 V
9 4 V
10 5 V
9 4 V
10 5 V
9 5 V
10 4 V
9 4 V
10 5 V
9 4 V
10 5 V
9 4 V
10 5 V
10 4 V
9 5 V
10 4 V
9 5 V
10 4 V
9 5 V
10 4 V
9 4 V
10 5 V
9 4 V
10 5 V
9 5 V
10 4 V
10 5 V
9 4 V
10 5 V
9 4 V
10 5 V
9 4 V
10 4 V
9 5 V
10 4 V
9 5 V
10 4 V
9 5 V
10 4 V
10 5 V
9 4 V
10 5 V
9 4 V
10 5 V
9 5 V
10 4 V
9 4 V
10 5 V
9 4 V
10 5 V
9 4 V
10 5 V
10 4 V
9 5 V
10 4 V
9 5 V
10 4 V
9 5 V
10 4 V
9 4 V
10 5 V
9 4 V
10 5 V
9 5 V
10 4 V
10 5 V
9 4 V
10 5 V
9 4 V
10 5 V
9 4 V
10 4 V
9 5 V
10 4 V
9 5 V
10 4 V
9 5 V
10 4 V
10 5 V
9 4 V
10 5 V
9 4 V
10 5 V
9 5 V
10 4 V
9 4 V
10 5 V
9 4 V
10 5 V
9 4 V
10 5 V
10 4 V
9 5 V
10 4 V
9 5 V
10 4 V
9 4 V
10 5 V
9 4 V
10 5 V
9 4 V
10 5 V
9 5 V
10 4 V
10 5 V
9 4 V
10 5 V
9 4 V
10 5 V
9 4 V
10 4 V
9 5 V
10 4 V
9 5 V
10 4 V
9 5 V
10 4 V
10 5 V
9 4 V
10 5 V
9 4 V
10 5 V
9 4 V
10 5 V
9 4 V
10 5 V
9 4 V
10 5 V
9 4 V
10 5 V
10 4 V
9 5 V
10 4 V
9 5 V
10 4 V
9 4 V
10 5 V
9 4 V
10 5 V
9 4 V
10 5 V
9 5 V
10 4 V
10 5 V
9 4 V
10 5 V
9 4 V
10 5 V
9 4 V
10 4 V
9 5 V
10 4 V
9 5 V
10 4 V
9 5 V
10 4 V
stroke
grestore
end
showpage
}}%
\put(777,1457){\makebox(0,0)[r]{SPD}}%
\put(777,1557){\makebox(0,0)[r]{PDF}}%
\put(588,1054){\makebox(0,0)[l]{$W = 300$~GeV}}%
\put(1876,652){\makebox(0,0)[l]{$W = 100$~GeV}}%
\put(1900,50){\makebox(0,0){\Large $b$ (fm) }}%
\put(100,1180){%
\special{ps: gsave currentpoint currentpoint translate
270 rotate neg exch neg exch translate}%
\makebox(0,0)[b]{\shortstack{\Large ${\hat \sigma} (M_V^2/W^2,b) $ (mb)}}%
\special{ps: currentpoint grestore moveto}%
}%
\put(3212,200){\makebox(0,0){0.6}}%
\put(2735,200){\makebox(0,0){0.5}}%
\put(2258,200){\makebox(0,0){0.4}}%
\put(1781,200){\makebox(0,0){0.3}}%
\put(1304,200){\makebox(0,0){0.2}}%
\put(827,200){\makebox(0,0){0.1}}%
\put(350,200){\makebox(0,0){0}}%
\put(300,2060){\makebox(0,0)[r]{35}}%
\put(300,1809){\makebox(0,0)[r]{30}}%
\put(300,1557){\makebox(0,0)[r]{25}}%
\put(300,1306){\makebox(0,0)[r]{20}}%
\put(300,1054){\makebox(0,0)[r]{15}}%
\put(300,803){\makebox(0,0)[r]{10}}%
\put(300,551){\makebox(0,0)[r]{5}}%
\put(300,300){\makebox(0,0)[r]{0}}%
\end{picture}%
\endgroup
 

%% file: skewint.tex
\begingroup%
  \makeatletter%
  \newcommand{\GNUPLOTspecial}{%
    \@sanitize\catcode`\%=14\relax\special}%
  \setlength{\unitlength}{0.1bp}%
{\GNUPLOTspecial{!
/gnudict 256 dict def
gnudict begin
/Color false def
/Solid false def
/gnulinewidth 5.000 def
/userlinewidth gnulinewidth def
/vshift -33 def
/dl {10 mul} def
/hpt_ 31.5 def
/vpt_ 31.5 def
/hpt hpt_ def
/vpt vpt_ def
/M {moveto} bind def
/L {lineto} bind def
/R {rmoveto} bind def
/V {rlineto} bind def
/vpt2 vpt 2 mul def
/hpt2 hpt 2 mul def
/Lshow { currentpoint stroke M
  0 vshift R show } def
/Rshow { currentpoint stroke M
  dup stringwidth pop neg vshift R show } def
/Cshow { currentpoint stroke M
  dup stringwidth pop -2 div vshift R show } def
/UP { dup vpt_ mul /vpt exch def hpt_ mul /hpt exch def
  /hpt2 hpt 2 mul def /vpt2 vpt 2 mul def } def
/DL { Color {setrgbcolor Solid {pop []} if 0 setdash }
 {pop pop pop Solid {pop []} if 0 setdash} ifelse } def
/BL { stroke userlinewidth 2 mul setlinewidth } def
/AL { stroke userlinewidth 2 div setlinewidth } def
/UL { dup gnulinewidth mul /userlinewidth exch def
      10 mul /udl exch def } def
/PL { stroke userlinewidth setlinewidth } def
/LTb { BL [] 0 0 0 DL } def
/LTa { AL [1 udl mul 2 udl mul] 0 setdash 0 0 0 setrgbcolor } def
/LT0 { PL [] 1 0 0 DL } def
/LT1 { PL [4 dl 2 dl] 0 1 0 DL } def
/LT2 { PL [2 dl 3 dl] 0 0 1 DL } def
/LT3 { PL [1 dl 1.5 dl] 1 0 1 DL } def
/LT4 { PL [5 dl 2 dl 1 dl 2 dl] 0 1 1 DL } def
/LT5 { PL [4 dl 3 dl 1 dl 3 dl] 1 1 0 DL } def
/LT6 { PL [2 dl 2 dl 2 dl 4 dl] 0 0 0 DL } def
/LT7 { PL [2 dl 2 dl 2 dl 2 dl 2 dl 4 dl] 1 0.3 0 DL } def
/LT8 { PL [2 dl 2 dl 2 dl 2 dl 2 dl 2 dl 2 dl 4 dl] 0.5 0.5 0.5 DL } def
/Pnt { stroke [] 0 setdash
   gsave 1 setlinecap M 0 0 V stroke grestore } def
/Dia { stroke [] 0 setdash 2 copy vpt add M
  hpt neg vpt neg V hpt vpt neg V
  hpt vpt V hpt neg vpt V closepath stroke
  Pnt } def
/Pls { stroke [] 0 setdash vpt sub M 0 vpt2 V
  currentpoint stroke M
  hpt neg vpt neg R hpt2 0 V stroke
  } def
/Box { stroke [] 0 setdash 2 copy exch hpt sub exch vpt add M
  0 vpt2 neg V hpt2 0 V 0 vpt2 V
  hpt2 neg 0 V closepath stroke
  Pnt } def
/Crs { stroke [] 0 setdash exch hpt sub exch vpt add M
  hpt2 vpt2 neg V currentpoint stroke M
  hpt2 neg 0 R hpt2 vpt2 V stroke } def
/TriU { stroke [] 0 setdash 2 copy vpt 1.12 mul add M
  hpt neg vpt -1.62 mul V
  hpt 2 mul 0 V
  hpt neg vpt 1.62 mul V closepath stroke
  Pnt  } def
/Star { 2 copy Pls Crs } def
/BoxF { stroke [] 0 setdash exch hpt sub exch vpt add M
  0 vpt2 neg V  hpt2 0 V  0 vpt2 V
  hpt2 neg 0 V  closepath fill } def
/TriUF { stroke [] 0 setdash vpt 1.12 mul add M
  hpt neg vpt -1.62 mul V
  hpt 2 mul 0 V
  hpt neg vpt 1.62 mul V closepath fill } def
/TriD { stroke [] 0 setdash 2 copy vpt 1.12 mul sub M
  hpt neg vpt 1.62 mul V
  hpt 2 mul 0 V
  hpt neg vpt -1.62 mul V closepath stroke
  Pnt  } def
/TriDF { stroke [] 0 setdash vpt 1.12 mul sub M
  hpt neg vpt 1.62 mul V
  hpt 2 mul 0 V
  hpt neg vpt -1.62 mul V closepath fill} def
/DiaF { stroke [] 0 setdash vpt add M
  hpt neg vpt neg V hpt vpt neg V
  hpt vpt V hpt neg vpt V closepath fill } def
/Pent { stroke [] 0 setdash 2 copy gsave
  translate 0 hpt M 4 {72 rotate 0 hpt L} repeat
  closepath stroke grestore Pnt } def
/PentF { stroke [] 0 setdash gsave
  translate 0 hpt M 4 {72 rotate 0 hpt L} repeat
  closepath fill grestore } def
/Circle { stroke [] 0 setdash 2 copy
  hpt 0 360 arc stroke Pnt } def
/CircleF { stroke [] 0 setdash hpt 0 360 arc fill } def
/C0 { BL [] 0 setdash 2 copy moveto vpt 90 450  arc } bind def
/C1 { BL [] 0 setdash 2 copy        moveto
       2 copy  vpt 0 90 arc closepath fill
               vpt 0 360 arc closepath } bind def
/C2 { BL [] 0 setdash 2 copy moveto
       2 copy  vpt 90 180 arc closepath fill
               vpt 0 360 arc closepath } bind def
/C3 { BL [] 0 setdash 2 copy moveto
       2 copy  vpt 0 180 arc closepath fill
               vpt 0 360 arc closepath } bind def
/C4 { BL [] 0 setdash 2 copy moveto
       2 copy  vpt 180 270 arc closepath fill
               vpt 0 360 arc closepath } bind def
/C5 { BL [] 0 setdash 2 copy moveto
       2 copy  vpt 0 90 arc
       2 copy moveto
       2 copy  vpt 180 270 arc closepath fill
               vpt 0 360 arc } bind def
/C6 { BL [] 0 setdash 2 copy moveto
      2 copy  vpt 90 270 arc closepath fill
              vpt 0 360 arc closepath } bind def
/C7 { BL [] 0 setdash 2 copy moveto
      2 copy  vpt 0 270 arc closepath fill
              vpt 0 360 arc closepath } bind def
/C8 { BL [] 0 setdash 2 copy moveto
      2 copy vpt 270 360 arc closepath fill
              vpt 0 360 arc closepath } bind def
/C9 { BL [] 0 setdash 2 copy moveto
      2 copy  vpt 270 450 arc closepath fill
              vpt 0 360 arc closepath } bind def
/C10 { BL [] 0 setdash 2 copy 2 copy moveto vpt 270 360 arc closepath fill
       2 copy moveto
       2 copy vpt 90 180 arc closepath fill
               vpt 0 360 arc closepath } bind def
/C11 { BL [] 0 setdash 2 copy moveto
       2 copy  vpt 0 180 arc closepath fill
       2 copy moveto
       2 copy  vpt 270 360 arc closepath fill
               vpt 0 360 arc closepath } bind def
/C12 { BL [] 0 setdash 2 copy moveto
       2 copy  vpt 180 360 arc closepath fill
               vpt 0 360 arc closepath } bind def
/C13 { BL [] 0 setdash  2 copy moveto
       2 copy  vpt 0 90 arc closepath fill
       2 copy moveto
       2 copy  vpt 180 360 arc closepath fill
               vpt 0 360 arc closepath } bind def
/C14 { BL [] 0 setdash 2 copy moveto
       2 copy  vpt 90 360 arc closepath fill
               vpt 0 360 arc } bind def
/C15 { BL [] 0 setdash 2 copy vpt 0 360 arc closepath fill
               vpt 0 360 arc closepath } bind def
/Rec   { newpath 4 2 roll moveto 1 index 0 rlineto 0 exch rlineto
       neg 0 rlineto closepath } bind def
/Square { dup Rec } bind def
/Bsquare { vpt sub exch vpt sub exch vpt2 Square } bind def
/S0 { BL [] 0 setdash 2 copy moveto 0 vpt rlineto BL Bsquare } bind def
/S1 { BL [] 0 setdash 2 copy vpt Square fill Bsquare } bind def
/S2 { BL [] 0 setdash 2 copy exch vpt sub exch vpt Square fill Bsquare } bind def
/S3 { BL [] 0 setdash 2 copy exch vpt sub exch vpt2 vpt Rec fill Bsquare } bind def
/S4 { BL [] 0 setdash 2 copy exch vpt sub exch vpt sub vpt Square fill Bsquare } bind def
/S5 { BL [] 0 setdash 2 copy 2 copy vpt Square fill
       exch vpt sub exch vpt sub vpt Square fill Bsquare } bind def
/S6 { BL [] 0 setdash 2 copy exch vpt sub exch vpt sub vpt vpt2 Rec fill Bsquare } bind def
/S7 { BL [] 0 setdash 2 copy exch vpt sub exch vpt sub vpt vpt2 Rec fill
       2 copy vpt Square fill
       Bsquare } bind def
/S8 { BL [] 0 setdash 2 copy vpt sub vpt Square fill Bsquare } bind def
/S9 { BL [] 0 setdash 2 copy vpt sub vpt vpt2 Rec fill Bsquare } bind def
/S10 { BL [] 0 setdash 2 copy vpt sub vpt Square fill 2 copy exch vpt sub exch vpt Square fill
       Bsquare } bind def
/S11 { BL [] 0 setdash 2 copy vpt sub vpt Square fill 2 copy exch vpt sub exch vpt2 vpt Rec fill
       Bsquare } bind def
/S12 { BL [] 0 setdash 2 copy exch vpt sub exch vpt sub vpt2 vpt Rec fill Bsquare } bind def
/S13 { BL [] 0 setdash 2 copy exch vpt sub exch vpt sub vpt2 vpt Rec fill
       2 copy vpt Square fill Bsquare } bind def
/S14 { BL [] 0 setdash 2 copy exch vpt sub exch vpt sub vpt2 vpt Rec fill
       2 copy exch vpt sub exch vpt Square fill Bsquare } bind def
/S15 { BL [] 0 setdash 2 copy Bsquare fill Bsquare } bind def
/D0 { gsave translate 45 rotate 0 0 S0 stroke grestore } bind def
/D1 { gsave translate 45 rotate 0 0 S1 stroke grestore } bind def
/D2 { gsave translate 45 rotate 0 0 S2 stroke grestore } bind def
/D3 { gsave translate 45 rotate 0 0 S3 stroke grestore } bind def
/D4 { gsave translate 45 rotate 0 0 S4 stroke grestore } bind def
/D5 { gsave translate 45 rotate 0 0 S5 stroke grestore } bind def
/D6 { gsave translate 45 rotate 0 0 S6 stroke grestore } bind def
/D7 { gsave translate 45 rotate 0 0 S7 stroke grestore } bind def
/D8 { gsave translate 45 rotate 0 0 S8 stroke grestore } bind def
/D9 { gsave translate 45 rotate 0 0 S9 stroke grestore } bind def
/D10 { gsave translate 45 rotate 0 0 S10 stroke grestore } bind def
/D11 { gsave translate 45 rotate 0 0 S11 stroke grestore } bind def
/D12 { gsave translate 45 rotate 0 0 S12 stroke grestore } bind def
/D13 { gsave translate 45 rotate 0 0 S13 stroke grestore } bind def
/D14 { gsave translate 45 rotate 0 0 S14 stroke grestore } bind def
/D15 { gsave translate 45 rotate 0 0 S15 stroke grestore } bind def
/DiaE { stroke [] 0 setdash vpt add M
  hpt neg vpt neg V hpt vpt neg V
  hpt vpt V hpt neg vpt V closepath stroke } def
/BoxE { stroke [] 0 setdash exch hpt sub exch vpt add M
  0 vpt2 neg V hpt2 0 V 0 vpt2 V
  hpt2 neg 0 V closepath stroke } def
/TriUE { stroke [] 0 setdash vpt 1.12 mul add M
  hpt neg vpt -1.62 mul V
  hpt 2 mul 0 V
  hpt neg vpt 1.62 mul V closepath stroke } def
/TriDE { stroke [] 0 setdash vpt 1.12 mul sub M
  hpt neg vpt 1.62 mul V
  hpt 2 mul 0 V
  hpt neg vpt -1.62 mul V closepath stroke } def
/PentE { stroke [] 0 setdash gsave
  translate 0 hpt M 4 {72 rotate 0 hpt L} repeat
  closepath stroke grestore } def
/CircE { stroke [] 0 setdash 
  hpt 0 360 arc stroke } def
/Opaque { gsave closepath 1 setgray fill grestore 0 setgray closepath } def
/DiaW { stroke [] 0 setdash vpt add M
  hpt neg vpt neg V hpt vpt neg V
  hpt vpt V hpt neg vpt V Opaque stroke } def
/BoxW { stroke [] 0 setdash exch hpt sub exch vpt add M
  0 vpt2 neg V hpt2 0 V 0 vpt2 V
  hpt2 neg 0 V Opaque stroke } def
/TriUW { stroke [] 0 setdash vpt 1.12 mul add M
  hpt neg vpt -1.62 mul V
  hpt 2 mul 0 V
  hpt neg vpt 1.62 mul V Opaque stroke } def
/TriDW { stroke [] 0 setdash vpt 1.12 mul sub M
  hpt neg vpt 1.62 mul V
  hpt 2 mul 0 V
  hpt neg vpt -1.62 mul V Opaque stroke } def
/PentW { stroke [] 0 setdash gsave
  translate 0 hpt M 4 {72 rotate 0 hpt L} repeat
  Opaque stroke grestore } def
/CircW { stroke [] 0 setdash 
  hpt 0 360 arc Opaque stroke } def
/BoxFill { gsave Rec 1 setgray fill grestore } def
end
}}%
\begin{picture}(3600,2160)(0,0)%
{\GNUPLOTspecial{"
gnudict begin
gsave
0 0 translate
0.100 0.100 scale
0 setgray
newpath
1.000 UL
LTb
500 300 M
63 0 V
2887 0 R
-63 0 V
500 593 M
63 0 V
2887 0 R
-63 0 V
500 887 M
63 0 V
2887 0 R
-63 0 V
500 1180 M
63 0 V
2887 0 R
-63 0 V
500 1473 M
63 0 V
2887 0 R
-63 0 V
500 1767 M
63 0 V
2887 0 R
-63 0 V
500 2060 M
63 0 V
2887 0 R
-63 0 V
500 300 M
0 63 V
0 1697 R
0 -63 V
1090 300 M
0 63 V
0 1697 R
0 -63 V
1680 300 M
0 63 V
0 1697 R
0 -63 V
2270 300 M
0 63 V
0 1697 R
0 -63 V
2860 300 M
0 63 V
0 1697 R
0 -63 V
3450 300 M
0 63 V
0 1697 R
0 -63 V
1.000 UL
LTb
500 300 M
2950 0 V
0 1760 V
-2950 0 V
500 300 L
1.000 UL
LT0
3087 1947 M
263 0 V
500 300 M
6 0 V
6 0 V
6 0 V
6 0 V
6 0 V
5 0 V
6 0 V
6 0 V
6 0 V
6 1 V
6 0 V
6 0 V
6 1 V
6 0 V
5 1 V
6 1 V
6 1 V
6 1 V
6 2 V
6 2 V
6 2 V
6 2 V
6 3 V
6 2 V
6 4 V
5 3 V
6 4 V
6 4 V
6 4 V
6 5 V
6 5 V
6 6 V
6 6 V
6 6 V
6 7 V
5 7 V
6 8 V
6 7 V
6 9 V
6 8 V
6 9 V
6 9 V
6 10 V
6 10 V
5 10 V
6 10 V
6 11 V
6 11 V
6 11 V
6 11 V
6 12 V
6 12 V
6 12 V
6 12 V
6 12 V
5 13 V
6 12 V
6 13 V
6 12 V
6 13 V
6 13 V
6 13 V
6 13 V
6 13 V
6 13 V
5 13 V
6 13 V
6 13 V
6 12 V
6 13 V
6 13 V
6 12 V
6 12 V
6 12 V
5 12 V
6 12 V
6 11 V
6 12 V
6 11 V
6 11 V
6 10 V
6 10 V
6 11 V
6 10 V
6 9 V
5 10 V
6 9 V
6 8 V
6 9 V
6 8 V
6 8 V
6 8 V
6 7 V
6 7 V
6 7 V
5 6 V
6 6 V
6 6 V
6 5 V
6 6 V
6 4 V
6 5 V
6 4 V
6 4 V
5 4 V
6 3 V
6 3 V
6 3 V
6 2 V
6 3 V
6 1 V
6 2 V
6 1 V
6 2 V
6 0 V
5 1 V
6 0 V
6 1 V
6 -1 V
6 0 V
6 0 V
6 -1 V
6 -1 V
6 -1 V
6 -2 V
5 -1 V
6 -2 V
6 -2 V
6 -2 V
6 -2 V
6 -2 V
6 -3 V
6 -2 V
6 -3 V
6 -3 V
5 -2 V
6 -3 V
6 -4 V
6 -3 V
6 -3 V
6 -3 V
6 -3 V
6 -4 V
6 -3 V
5 -4 V
6 -3 V
6 -4 V
6 -3 V
6 -4 V
6 -4 V
6 -5 V
6 -6 V
6 -6 V
6 -6 V
5 -6 V
6 -6 V
6 -7 V
6 -6 V
6 -6 V
6 -7 V
6 -6 V
6 -7 V
6 -6 V
6 -7 V
6 -6 V
5 -7 V
6 -7 V
6 -7 V
6 -6 V
6 -7 V
6 -7 V
6 -7 V
6 -7 V
6 -7 V
5 -6 V
6 -7 V
6 -7 V
6 -7 V
6 -7 V
6 -7 V
6 -7 V
6 -6 V
6 -7 V
6 -7 V
5 -7 V
6 -7 V
6 -6 V
6 -7 V
6 -7 V
6 -6 V
6 -7 V
6 -7 V
6 -6 V
6 -7 V
6 -6 V
5 -4 V
6 -5 V
6 -4 V
6 -4 V
6 -4 V
6 -5 V
6 -4 V
6 -4 V
6 -4 V
5 -5 V
6 -4 V
6 -4 V
6 -4 V
6 -4 V
6 -4 V
6 -5 V
6 -4 V
6 -4 V
6 -4 V
5 -4 V
6 -4 V
6 -4 V
6 -4 V
6 -4 V
6 -4 V
6 -4 V
6 -4 V
6 -4 V
6 -4 V
6 -3 V
5 -4 V
6 -4 V
6 -4 V
6 -3 V
6 -4 V
6 -4 V
6 -3 V
6 -4 V
6 -3 V
5 -4 V
6 -4 V
6 -3 V
6 -3 V
6 -4 V
6 -3 V
6 -4 V
6 -3 V
6 -3 V
6 -3 V
5 -4 V
6 -3 V
6 -3 V
6 -3 V
6 -3 V
6 -3 V
6 -3 V
6 -3 V
6 -3 V
6 -3 V
6 -3 V
5 -3 V
6 -3 V
6 -2 V
6 -3 V
6 -3 V
6 -3 V
6 -2 V
6 -3 V
6 -2 V
6 -3 V
5 -2 V
6 -3 V
6 -2 V
6 -3 V
6 -2 V
6 -3 V
6 -2 V
6 -2 V
6 -3 V
6 -2 V
5 -2 V
6 -2 V
6 -2 V
6 -3 V
6 -2 V
6 -2 V
6 -2 V
6 -2 V
6 -2 V
5 -2 V
6 -2 V
6 -2 V
6 -2 V
6 -2 V
6 -1 V
6 -2 V
6 -2 V
6 -2 V
6 -1 V
5 -2 V
6 -2 V
6 -2 V
6 -1 V
6 -2 V
6 -1 V
6 -2 V
6 -1 V
6 -2 V
6 -1 V
5 -2 V
6 -1 V
6 -2 V
6 -1 V
6 -2 V
6 -1 V
6 -1 V
6 -2 V
6 -1 V
6 -1 V
6 -1 V
5 -2 V
6 -1 V
6 -1 V
6 -1 V
6 -1 V
6 -2 V
6 -1 V
6 -1 V
6 -1 V
6 -1 V
5 -1 V
6 -1 V
6 -1 V
6 -2 V
6 -1 V
6 -1 V
6 -1 V
6 -1 V
6 -1 V
6 -1 V
5 -1 V
6 -1 V
6 -1 V
6 -1 V
6 0 V
6 -1 V
6 -1 V
6 -1 V
6 -1 V
5 -1 V
6 0 V
6 -1 V
6 -1 V
6 -1 V
6 -1 V
6 0 V
6 -1 V
6 -1 V
6 0 V
5 -1 V
6 -1 V
6 0 V
6 -1 V
6 -1 V
6 0 V
6 -1 V
6 0 V
6 -1 V
6 0 V
5 -1 V
6 -1 V
6 0 V
6 -1 V
6 0 V
6 -1 V
6 0 V
6 -1 V
6 0 V
6 0 V
6 -1 V
5 0 V
6 -1 V
6 0 V
6 -1 V
6 0 V
6 0 V
6 -1 V
6 0 V
6 -1 V
6 0 V
5 0 V
6 -1 V
6 0 V
6 0 V
6 -1 V
6 0 V
6 0 V
6 0 V
6 -1 V
6 0 V
5 0 V
6 -1 V
currentpoint stroke M
6 0 V
6 0 V
6 0 V
6 -1 V
6 0 V
6 0 V
6 0 V
6 -1 V
5 0 V
6 0 V
6 0 V
6 -1 V
6 0 V
6 0 V
6 0 V
6 0 V
6 -1 V
5 0 V
6 0 V
6 0 V
6 0 V
6 -1 V
6 0 V
6 0 V
6 0 V
6 0 V
6 0 V
5 0 V
6 -1 V
6 0 V
6 0 V
6 0 V
6 0 V
6 0 V
6 0 V
6 -1 V
6 0 V
5 0 V
6 0 V
6 0 V
6 0 V
6 0 V
6 0 V
6 0 V
6 -1 V
6 0 V
6 0 V
6 0 V
5 0 V
6 0 V
6 0 V
6 0 V
6 0 V
6 0 V
6 -1 V
6 0 V
6 0 V
6 0 V
5 0 V
6 0 V
6 0 V
6 0 V
6 0 V
6 0 V
6 0 V
6 0 V
6 0 V
6 0 V
5 0 V
6 -1 V
6 0 V
6 0 V
6 0 V
6 0 V
6 0 V
6 0 V
6 0 V
5 0 V
6 0 V
6 0 V
6 0 V
6 0 V
6 0 V
6 0 V
6 0 V
6 0 V
6 0 V
5 0 V
6 0 V
6 0 V
6 -1 V
6 0 V
6 0 V
6 0 V
6 0 V
6 0 V
6 0 V
5 0 V
6 0 V
6 0 V
6 0 V
6 0 V
6 0 V
1.000 UL
LT1
3087 1847 M
263 0 V
500 300 M
6 0 V
6 0 V
6 0 V
6 0 V
6 0 V
5 0 V
6 1 V
6 0 V
6 1 V
6 1 V
6 1 V
6 2 V
6 2 V
6 2 V
5 3 V
6 3 V
6 4 V
6 5 V
6 5 V
6 6 V
6 7 V
6 7 V
6 8 V
6 9 V
6 10 V
5 11 V
6 12 V
6 13 V
6 13 V
6 15 V
6 16 V
6 15 V
6 18 V
6 18 V
6 19 V
5 20 V
6 20 V
6 21 V
6 22 V
6 23 V
6 24 V
6 24 V
6 24 V
6 26 V
5 26 V
6 26 V
6 27 V
6 28 V
6 27 V
6 28 V
6 29 V
6 29 V
6 29 V
6 29 V
6 29 V
5 30 V
6 29 V
6 30 V
6 30 V
6 29 V
6 30 V
6 29 V
6 29 V
6 30 V
6 29 V
5 29 V
6 29 V
6 28 V
6 28 V
6 28 V
6 27 V
6 27 V
6 26 V
6 25 V
5 25 V
6 25 V
6 24 V
6 23 V
6 22 V
6 22 V
6 21 V
6 21 V
6 20 V
6 19 V
6 19 V
5 18 V
6 17 V
6 18 V
6 14 V
6 17 V
6 13 V
6 14 V
6 12 V
6 13 V
6 11 V
5 11 V
6 9 V
6 10 V
6 8 V
6 9 V
6 5 V
6 8 V
6 4 V
6 6 V
5 4 V
6 3 V
6 3 V
6 3 V
6 2 V
6 0 V
6 1 V
6 0 V
6 -1 V
6 -2 V
6 -1 V
5 -2 V
6 -3 V
6 -4 V
6 -3 V
6 -4 V
6 -5 V
6 -6 V
6 -4 V
6 -6 V
6 -7 V
5 -6 V
6 -7 V
6 -8 V
6 -7 V
6 -7 V
6 -8 V
6 -9 V
6 -8 V
6 -12 V
6 -18 V
5 -17 V
6 -18 V
6 -18 V
6 -16 V
6 -17 V
6 -16 V
6 -17 V
6 -16 V
6 -16 V
5 -15 V
6 -16 V
6 -15 V
6 -14 V
6 -15 V
6 -14 V
6 -18 V
6 -18 V
6 -17 V
6 -18 V
5 -17 V
6 -17 V
6 -17 V
6 -17 V
6 -17 V
6 -17 V
6 -16 V
6 -17 V
6 -16 V
6 -16 V
6 -16 V
5 -15 V
6 -16 V
6 -15 V
6 -15 V
6 -16 V
6 -15 V
6 -14 V
6 -15 V
6 -14 V
5 -15 V
6 -14 V
6 -14 V
6 -14 V
6 -13 V
6 -14 V
6 -13 V
6 -13 V
6 -13 V
6 -13 V
5 -13 V
6 -13 V
6 -12 V
6 -12 V
6 -13 V
6 -12 V
6 -11 V
6 -12 V
6 -12 V
6 -11 V
6 -11 V
5 -11 V
6 -11 V
6 -11 V
6 -11 V
6 -11 V
6 -10 V
6 -10 V
6 -11 V
6 -10 V
5 -9 V
6 -10 V
6 -10 V
6 -10 V
6 -9 V
6 -9 V
6 -9 V
6 -9 V
6 -9 V
6 -9 V
5 -9 V
6 -8 V
6 -9 V
6 -8 V
6 -8 V
6 -8 V
6 -8 V
6 -8 V
6 -8 V
6 -7 V
6 -8 V
5 -7 V
6 -8 V
6 -7 V
6 -7 V
6 -7 V
6 -7 V
6 -7 V
6 -6 V
6 -7 V
5 -6 V
6 -7 V
6 -6 V
6 -6 V
6 -6 V
6 -6 V
6 -6 V
6 -6 V
6 -6 V
6 -5 V
5 -6 V
6 -5 V
6 -6 V
6 -5 V
6 -5 V
6 -5 V
6 -5 V
6 -5 V
6 -5 V
6 -5 V
6 -5 V
5 -4 V
6 -5 V
6 -4 V
6 -5 V
6 -4 V
6 -4 V
6 -5 V
6 -4 V
6 -4 V
6 -4 V
5 -4 V
6 -4 V
6 -3 V
6 -4 V
6 -4 V
6 -4 V
6 -3 V
6 -4 V
6 -3 V
6 -4 V
5 -3 V
6 -3 V
6 -3 V
6 -4 V
6 -3 V
6 -3 V
6 -3 V
6 -3 V
6 -3 V
5 -3 V
6 -2 V
6 -3 V
6 -3 V
6 -3 V
6 -2 V
6 -3 V
6 -2 V
6 -3 V
6 -2 V
5 -3 V
6 -2 V
6 -2 V
6 -3 V
6 -2 V
6 -2 V
6 -2 V
6 -2 V
6 -2 V
6 -2 V
5 -2 V
6 -2 V
6 -2 V
6 -2 V
6 -2 V
6 -2 V
6 -2 V
6 -2 V
6 -1 V
6 -2 V
6 -2 V
5 -1 V
6 -2 V
6 -2 V
6 -1 V
6 -2 V
6 -1 V
6 -2 V
6 -1 V
6 -1 V
6 -2 V
5 -1 V
6 -2 V
6 -1 V
6 -1 V
6 -2 V
6 -1 V
6 -1 V
6 -1 V
6 -2 V
6 -1 V
5 -1 V
6 -1 V
6 -1 V
6 -1 V
6 -1 V
6 -1 V
6 -1 V
6 -1 V
6 -1 V
5 -1 V
6 -1 V
6 -1 V
6 -1 V
6 -1 V
6 -1 V
6 -1 V
6 -1 V
6 0 V
6 -1 V
5 -1 V
6 -1 V
6 -1 V
6 0 V
6 -1 V
6 -1 V
6 -1 V
6 0 V
6 -1 V
6 -1 V
5 0 V
6 -1 V
6 0 V
6 -1 V
6 -1 V
6 0 V
6 -1 V
6 0 V
6 -1 V
6 -1 V
6 0 V
5 -1 V
6 0 V
6 -1 V
6 0 V
6 -1 V
6 0 V
6 0 V
6 -1 V
6 0 V
6 -1 V
5 0 V
6 -1 V
6 0 V
6 0 V
6 -1 V
6 0 V
6 -1 V
6 0 V
6 0 V
6 -1 V
5 0 V
6 0 V
currentpoint stroke M
6 -1 V
6 0 V
6 0 V
6 -1 V
6 0 V
6 0 V
6 0 V
6 -1 V
5 0 V
6 0 V
6 -1 V
6 0 V
6 0 V
6 0 V
6 -1 V
6 0 V
6 0 V
5 0 V
6 0 V
6 -1 V
6 0 V
6 0 V
6 0 V
6 0 V
6 -1 V
6 0 V
6 0 V
5 0 V
6 0 V
6 -1 V
6 0 V
6 0 V
6 0 V
6 0 V
6 0 V
6 -1 V
6 0 V
5 0 V
6 0 V
6 0 V
6 0 V
6 0 V
6 0 V
6 -1 V
6 0 V
6 0 V
6 0 V
6 0 V
5 0 V
6 0 V
6 0 V
6 -1 V
6 0 V
6 0 V
6 0 V
6 0 V
6 0 V
6 0 V
5 0 V
6 0 V
6 0 V
6 0 V
6 -1 V
6 0 V
6 0 V
6 0 V
6 0 V
6 0 V
5 0 V
6 0 V
6 0 V
6 0 V
6 0 V
6 0 V
6 0 V
6 0 V
6 -1 V
5 0 V
6 0 V
6 0 V
6 0 V
6 0 V
6 0 V
6 0 V
6 0 V
6 0 V
6 0 V
5 0 V
6 0 V
6 0 V
6 0 V
6 0 V
6 0 V
6 0 V
6 0 V
6 0 V
6 0 V
5 -1 V
6 0 V
6 0 V
6 0 V
6 0 V
6 0 V
1.000 UL
LT2
3087 1747 M
263 0 V
500 300 M
6 0 V
6 0 V
6 0 V
6 0 V
6 0 V
5 0 V
6 0 V
6 0 V
6 0 V
6 0 V
6 1 V
6 0 V
6 0 V
6 1 V
5 1 V
6 0 V
6 2 V
6 1 V
6 1 V
6 2 V
6 2 V
6 2 V
6 3 V
6 2 V
6 4 V
5 3 V
6 4 V
6 4 V
6 4 V
6 5 V
6 5 V
6 5 V
6 6 V
6 6 V
6 6 V
5 7 V
6 7 V
6 7 V
6 8 V
6 8 V
6 8 V
6 8 V
6 10 V
6 9 V
5 9 V
6 10 V
6 9 V
6 10 V
6 11 V
6 10 V
6 11 V
6 11 V
6 10 V
6 11 V
6 11 V
5 11 V
6 12 V
6 12 V
6 12 V
6 11 V
6 11 V
6 11 V
6 11 V
6 11 V
6 13 V
5 12 V
6 12 V
6 11 V
6 12 V
6 11 V
6 11 V
6 11 V
6 10 V
6 10 V
5 10 V
6 10 V
6 11 V
6 10 V
6 10 V
6 10 V
6 10 V
6 9 V
6 9 V
6 9 V
6 8 V
5 8 V
6 8 V
6 7 V
6 7 V
6 7 V
6 6 V
6 6 V
6 7 V
6 6 V
6 7 V
5 6 V
6 6 V
6 5 V
6 5 V
6 5 V
6 5 V
6 4 V
6 4 V
6 3 V
5 4 V
6 3 V
6 2 V
6 3 V
6 2 V
6 2 V
6 2 V
6 1 V
6 1 V
6 1 V
6 0 V
5 1 V
6 0 V
6 0 V
6 -1 V
6 1 V
6 0 V
6 0 V
6 0 V
6 -1 V
6 0 V
5 -1 V
6 -1 V
6 -1 V
6 -1 V
6 -1 V
6 -2 V
6 -1 V
6 -2 V
6 -2 V
6 -1 V
5 -2 V
6 -2 V
6 -3 V
6 -2 V
6 -2 V
6 -2 V
6 -3 V
6 -2 V
6 -3 V
5 -2 V
6 -3 V
6 -3 V
6 -2 V
6 -3 V
6 -2 V
6 -5 V
6 -5 V
6 -5 V
6 -4 V
5 -5 V
6 -5 V
6 -5 V
6 -5 V
6 -6 V
6 -5 V
6 -5 V
6 -6 V
6 -5 V
6 -6 V
6 -5 V
5 -6 V
6 -5 V
6 -6 V
6 -5 V
6 -6 V
6 -6 V
6 -5 V
6 -6 V
6 -6 V
5 -6 V
6 -5 V
6 -6 V
6 -5 V
6 -6 V
6 -5 V
6 -6 V
6 -5 V
6 -6 V
6 -6 V
5 -5 V
6 -6 V
6 -5 V
6 -6 V
6 -5 V
6 -6 V
6 -5 V
6 -6 V
6 -5 V
6 -6 V
6 -5 V
5 -5 V
6 -4 V
6 -4 V
6 -5 V
6 -4 V
6 -5 V
6 -4 V
6 -4 V
6 -5 V
5 -4 V
6 -4 V
6 -4 V
6 -5 V
6 -4 V
6 -4 V
6 -4 V
6 -5 V
6 -4 V
6 -4 V
5 -4 V
6 -4 V
6 -4 V
6 -4 V
6 -4 V
6 -4 V
6 -4 V
6 -4 V
6 -4 V
6 -4 V
6 -4 V
5 -4 V
6 -4 V
6 -4 V
6 -3 V
6 -4 V
6 -4 V
6 -3 V
6 -4 V
6 -4 V
5 -3 V
6 -4 V
6 -3 V
6 -4 V
6 -3 V
6 -4 V
6 -3 V
6 -3 V
6 -4 V
6 -3 V
5 -3 V
6 -3 V
6 -3 V
6 -4 V
6 -3 V
6 -3 V
6 -3 V
6 -3 V
6 -3 V
6 -3 V
6 -3 V
5 -2 V
6 -3 V
6 -3 V
6 -3 V
6 -3 V
6 -2 V
6 -3 V
6 -3 V
6 -2 V
6 -3 V
5 -2 V
6 -3 V
6 -2 V
6 -3 V
6 -2 V
6 -2 V
6 -3 V
6 -2 V
6 -2 V
6 -3 V
5 -2 V
6 -2 V
6 -2 V
6 -2 V
6 -2 V
6 -3 V
6 -2 V
6 -2 V
6 -2 V
5 -2 V
6 -2 V
6 -1 V
6 -2 V
6 -2 V
6 -2 V
6 -2 V
6 -2 V
6 -1 V
6 -2 V
5 -2 V
6 -2 V
6 -1 V
6 -2 V
6 -1 V
6 -2 V
6 -2 V
6 -1 V
6 -2 V
6 -1 V
5 -2 V
6 -1 V
6 -1 V
6 -2 V
6 -1 V
6 -2 V
6 -1 V
6 -1 V
6 -1 V
6 -2 V
6 -1 V
5 -1 V
6 -1 V
6 -2 V
6 -1 V
6 -1 V
6 -1 V
6 -1 V
6 -1 V
6 -1 V
6 -1 V
5 -2 V
6 -1 V
6 -1 V
6 -1 V
6 -1 V
6 -1 V
6 -1 V
6 -1 V
6 -1 V
6 -1 V
5 -1 V
6 -1 V
6 -1 V
6 -1 V
6 -1 V
6 -1 V
6 -1 V
6 0 V
6 -1 V
5 -1 V
6 -1 V
6 -1 V
6 0 V
6 -1 V
6 -1 V
6 -1 V
6 0 V
6 -1 V
6 -1 V
5 0 V
6 -1 V
6 -1 V
6 0 V
6 -1 V
6 0 V
6 -1 V
6 -1 V
6 0 V
6 -1 V
5 0 V
6 -1 V
6 0 V
6 -1 V
6 0 V
6 -1 V
6 0 V
6 -1 V
6 0 V
6 -1 V
6 0 V
5 -1 V
6 0 V
6 0 V
6 -1 V
6 0 V
6 -1 V
6 0 V
6 0 V
6 -1 V
6 0 V
5 0 V
6 -1 V
6 0 V
6 0 V
6 -1 V
6 0 V
6 0 V
6 -1 V
6 0 V
6 0 V
5 0 V
6 -1 V
currentpoint stroke M
6 0 V
6 0 V
6 -1 V
6 0 V
6 0 V
6 0 V
6 -1 V
6 0 V
5 0 V
6 0 V
6 0 V
6 -1 V
6 0 V
6 0 V
6 0 V
6 0 V
6 -1 V
5 0 V
6 0 V
6 0 V
6 0 V
6 -1 V
6 0 V
6 0 V
6 0 V
6 0 V
6 0 V
5 -1 V
6 0 V
6 0 V
6 0 V
6 0 V
6 0 V
6 0 V
6 0 V
6 -1 V
6 0 V
5 0 V
6 0 V
6 0 V
6 0 V
6 0 V
6 0 V
6 -1 V
6 0 V
6 0 V
6 0 V
6 0 V
5 0 V
6 0 V
6 0 V
6 0 V
6 0 V
6 0 V
6 -1 V
6 0 V
6 0 V
6 0 V
5 0 V
6 0 V
6 0 V
6 0 V
6 0 V
6 0 V
6 0 V
6 0 V
6 0 V
6 0 V
5 -1 V
6 0 V
6 0 V
6 0 V
6 0 V
6 0 V
6 0 V
6 0 V
6 0 V
5 0 V
6 0 V
6 0 V
6 0 V
6 0 V
6 0 V
6 0 V
6 0 V
6 0 V
6 0 V
5 0 V
6 0 V
6 -1 V
6 0 V
6 0 V
6 0 V
6 0 V
6 0 V
6 0 V
6 0 V
5 0 V
6 0 V
6 0 V
6 0 V
6 0 V
6 0 V
1.000 UL
LT3
3087 1647 M
263 0 V
500 300 M
6 0 V
6 0 V
6 0 V
6 0 V
6 0 V
5 0 V
6 1 V
6 0 V
6 1 V
6 0 V
6 2 V
6 1 V
6 2 V
6 2 V
5 3 V
6 3 V
6 4 V
6 5 V
6 5 V
6 6 V
6 6 V
6 8 V
6 8 V
6 9 V
6 9 V
5 11 V
6 11 V
6 11 V
6 14 V
6 14 V
6 14 V
6 15 V
6 17 V
6 17 V
6 17 V
5 17 V
6 20 V
6 21 V
6 20 V
6 20 V
6 21 V
6 24 V
6 23 V
6 23 V
5 23 V
6 23 V
6 25 V
6 26 V
6 26 V
6 26 V
6 26 V
6 25 V
6 24 V
6 27 V
6 27 V
5 28 V
6 27 V
6 27 V
6 26 V
6 26 V
6 24 V
6 25 V
6 26 V
6 28 V
6 27 V
5 27 V
6 26 V
6 25 V
6 25 V
6 23 V
6 23 V
6 23 V
6 21 V
6 21 V
5 23 V
6 23 V
6 22 V
6 22 V
6 20 V
6 20 V
6 19 V
6 18 V
6 17 V
6 17 V
6 15 V
5 15 V
6 13 V
6 14 V
6 12 V
6 11 V
6 14 V
6 13 V
6 12 V
6 12 V
6 11 V
5 10 V
6 9 V
6 9 V
6 8 V
6 7 V
6 7 V
6 5 V
6 5 V
6 5 V
5 3 V
6 3 V
6 3 V
6 1 V
6 1 V
6 1 V
6 0 V
6 -1 V
6 -1 V
6 -2 V
6 -2 V
5 -2 V
6 -1 V
6 -2 V
6 -2 V
6 -2 V
6 -3 V
6 -3 V
6 -3 V
6 -4 V
6 -4 V
5 -5 V
6 -5 V
6 -5 V
6 -5 V
6 -6 V
6 -6 V
6 -6 V
6 -7 V
6 -6 V
6 -7 V
5 -7 V
6 -7 V
6 -7 V
6 -8 V
6 -7 V
6 -11 V
6 -15 V
6 -15 V
6 -15 V
5 -14 V
6 -14 V
6 -14 V
6 -13 V
6 -14 V
6 -13 V
6 -16 V
6 -17 V
6 -16 V
6 -17 V
5 -16 V
6 -16 V
6 -16 V
6 -16 V
6 -16 V
6 -15 V
6 -16 V
6 -15 V
6 -15 V
6 -15 V
6 -15 V
5 -15 V
6 -14 V
6 -15 V
6 -14 V
6 -15 V
6 -14 V
6 -14 V
6 -13 V
6 -14 V
5 -14 V
6 -13 V
6 -13 V
6 -13 V
6 -13 V
6 -13 V
6 -13 V
6 -12 V
6 -13 V
6 -12 V
5 -12 V
6 -12 V
6 -12 V
6 -11 V
6 -12 V
6 -11 V
6 -12 V
6 -11 V
6 -11 V
6 -11 V
6 -11 V
5 -10 V
6 -11 V
6 -10 V
6 -10 V
6 -11 V
6 -10 V
6 -9 V
6 -10 V
6 -10 V
5 -9 V
6 -10 V
6 -9 V
6 -9 V
6 -9 V
6 -9 V
6 -9 V
6 -9 V
6 -8 V
6 -9 V
5 -8 V
6 -8 V
6 -8 V
6 -8 V
6 -8 V
6 -8 V
6 -8 V
6 -7 V
6 -8 V
6 -7 V
6 -7 V
5 -7 V
6 -7 V
6 -7 V
6 -7 V
6 -7 V
6 -7 V
6 -6 V
6 -7 V
6 -6 V
5 -6 V
6 -6 V
6 -6 V
6 -6 V
6 -6 V
6 -6 V
6 -6 V
6 -5 V
6 -6 V
6 -5 V
5 -6 V
6 -5 V
6 -5 V
6 -6 V
6 -5 V
6 -5 V
6 -4 V
6 -5 V
6 -5 V
6 -5 V
6 -4 V
5 -5 V
6 -4 V
6 -5 V
6 -4 V
6 -4 V
6 -4 V
6 -5 V
6 -4 V
6 -4 V
6 -3 V
5 -4 V
6 -4 V
6 -4 V
6 -4 V
6 -3 V
6 -4 V
6 -3 V
6 -4 V
6 -3 V
6 -3 V
5 -4 V
6 -3 V
6 -3 V
6 -3 V
6 -3 V
6 -3 V
6 -3 V
6 -3 V
6 -3 V
5 -3 V
6 -2 V
6 -3 V
6 -3 V
6 -2 V
6 -3 V
6 -3 V
6 -2 V
6 -2 V
6 -3 V
5 -2 V
6 -3 V
6 -2 V
6 -2 V
6 -2 V
6 -2 V
6 -3 V
6 -2 V
6 -2 V
6 -2 V
5 -2 V
6 -2 V
6 -1 V
6 -2 V
6 -2 V
6 -2 V
6 -2 V
6 -1 V
6 -2 V
6 -2 V
6 -2 V
5 -1 V
6 -2 V
6 -1 V
6 -2 V
6 -1 V
6 -2 V
6 -1 V
6 -2 V
6 -1 V
6 -1 V
5 -2 V
6 -1 V
6 -2 V
6 -1 V
6 -1 V
6 -1 V
6 -2 V
6 -1 V
6 -1 V
6 -1 V
5 -1 V
6 -1 V
6 -2 V
6 -1 V
6 -1 V
6 -1 V
6 -1 V
6 -1 V
6 -1 V
5 -1 V
6 -1 V
6 -1 V
6 -1 V
6 0 V
6 -1 V
6 -1 V
6 -1 V
6 -1 V
6 -1 V
5 0 V
6 -1 V
6 -1 V
6 -1 V
6 0 V
6 -1 V
6 -1 V
6 -1 V
6 0 V
6 -1 V
5 0 V
6 -1 V
6 -1 V
6 0 V
6 -1 V
6 -1 V
6 0 V
6 -1 V
6 0 V
6 -1 V
6 0 V
5 -1 V
6 0 V
6 -1 V
6 0 V
6 -1 V
6 0 V
6 -1 V
6 0 V
6 -1 V
6 0 V
5 0 V
6 -1 V
6 0 V
6 -1 V
6 0 V
6 0 V
6 -1 V
6 0 V
6 0 V
6 -1 V
5 0 V
6 0 V
currentpoint stroke M
6 -1 V
6 0 V
6 0 V
6 -1 V
6 0 V
6 0 V
6 -1 V
6 0 V
5 0 V
6 0 V
6 -1 V
6 0 V
6 0 V
6 0 V
6 -1 V
6 0 V
6 0 V
5 0 V
6 -1 V
6 0 V
6 0 V
6 0 V
6 0 V
6 -1 V
6 0 V
6 0 V
6 0 V
5 0 V
6 0 V
6 -1 V
6 0 V
6 0 V
6 0 V
6 0 V
6 0 V
6 -1 V
6 0 V
5 0 V
6 0 V
6 0 V
6 0 V
6 0 V
6 -1 V
6 0 V
6 0 V
6 0 V
6 0 V
6 0 V
5 0 V
6 0 V
6 0 V
6 -1 V
6 0 V
6 0 V
6 0 V
6 0 V
6 0 V
6 0 V
5 0 V
6 0 V
6 0 V
6 -1 V
6 0 V
6 0 V
6 0 V
6 0 V
6 0 V
6 0 V
5 0 V
6 0 V
6 0 V
6 0 V
6 0 V
6 0 V
6 0 V
6 0 V
6 -1 V
5 0 V
6 0 V
6 0 V
6 0 V
6 0 V
6 0 V
6 0 V
6 0 V
6 0 V
6 0 V
5 0 V
6 0 V
6 0 V
6 0 V
6 0 V
6 0 V
6 0 V
6 0 V
6 0 V
6 0 V
5 -1 V
6 0 V
6 0 V
6 0 V
6 0 V
6 0 V
stroke
grestore
end
showpage
}}%
\put(3037,1647){\makebox(0,0)[r]{$300$ PDF}}%
\put(3037,1747){\makebox(0,0)[r]{$W = 100$ PDF}}%
\put(3037,1847){\makebox(0,0)[r]{$300$ SPD}}%
\put(3037,1947){\makebox(0,0)[r]{$W = 100$ SPD}}%
\put(1975,50){\makebox(0,0){\Large $b$ (fm) }}%
\put(100,1180){%
\special{ps: gsave currentpoint currentpoint translate
270 rotate neg exch neg exch translate}%
\makebox(0,0)[b]{\shortstack{\Large $ I (b,W) $}}%
\special{ps: currentpoint grestore moveto}%
}%
\put(3450,200){\makebox(0,0){1}}%
\put(2860,200){\makebox(0,0){0.8}}%
\put(2270,200){\makebox(0,0){0.6}}%
\put(1680,200){\makebox(0,0){0.4}}%
\put(1090,200){\makebox(0,0){0.2}}%
\put(500,200){\makebox(0,0){0}}%
\put(450,2060){\makebox(0,0)[r]{0.012}}%
\put(450,1767){\makebox(0,0)[r]{0.01}}%
\put(450,1473){\makebox(0,0)[r]{0.008}}%
\put(450,1180){\makebox(0,0)[r]{0.006}}%
\put(450,887){\makebox(0,0)[r]{0.004}}%
\put(450,593){\makebox(0,0)[r]{0.002}}%
\put(450,300){\makebox(0,0)[r]{0}}%
\end{picture}%
\endgroup
 

%% file: bndnew.tex
\begingroup%
  \makeatletter%
  \newcommand{\GNUPLOTspecial}{%
    \@sanitize\catcode`\%=14\relax\special}%
  \setlength{\unitlength}{0.1bp}%
{\GNUPLOTspecial{!
/gnudict 256 dict def
gnudict begin
/Color false def
/Solid false def
/gnulinewidth 5.000 def
/userlinewidth gnulinewidth def
/vshift -33 def
/dl {10 mul} def
/hpt_ 31.5 def
/vpt_ 31.5 def
/hpt hpt_ def
/vpt vpt_ def
/M {moveto} bind def
/L {lineto} bind def
/R {rmoveto} bind def
/V {rlineto} bind def
/vpt2 vpt 2 mul def
/hpt2 hpt 2 mul def
/Lshow { currentpoint stroke M
  0 vshift R show } def
/Rshow { currentpoint stroke M
  dup stringwidth pop neg vshift R show } def
/Cshow { currentpoint stroke M
  dup stringwidth pop -2 div vshift R show } def
/UP { dup vpt_ mul /vpt exch def hpt_ mul /hpt exch def
  /hpt2 hpt 2 mul def /vpt2 vpt 2 mul def } def
/DL { Color {setrgbcolor Solid {pop []} if 0 setdash }
 {pop pop pop Solid {pop []} if 0 setdash} ifelse } def
/BL { stroke userlinewidth 2 mul setlinewidth } def
/AL { stroke userlinewidth 2 div setlinewidth } def
/UL { dup gnulinewidth mul /userlinewidth exch def
      10 mul /udl exch def } def
/PL { stroke userlinewidth setlinewidth } def
/LTb { BL [] 0 0 0 DL } def
/LTa { AL [1 udl mul 2 udl mul] 0 setdash 0 0 0 setrgbcolor } def
/LT0 { PL [] 1 0 0 DL } def
/LT1 { PL [4 dl 2 dl] 0 1 0 DL } def
/LT2 { PL [2 dl 3 dl] 0 0 1 DL } def
/LT3 { PL [1 dl 1.5 dl] 1 0 1 DL } def
/LT4 { PL [5 dl 2 dl 1 dl 2 dl] 0 1 1 DL } def
/LT5 { PL [4 dl 3 dl 1 dl 3 dl] 1 1 0 DL } def
/LT6 { PL [2 dl 2 dl 2 dl 4 dl] 0 0 0 DL } def
/LT7 { PL [2 dl 2 dl 2 dl 2 dl 2 dl 4 dl] 1 0.3 0 DL } def
/LT8 { PL [2 dl 2 dl 2 dl 2 dl 2 dl 2 dl 2 dl 4 dl] 0.5 0.5 0.5 DL } def
/Pnt { stroke [] 0 setdash
   gsave 1 setlinecap M 0 0 V stroke grestore } def
/Dia { stroke [] 0 setdash 2 copy vpt add M
  hpt neg vpt neg V hpt vpt neg V
  hpt vpt V hpt neg vpt V closepath stroke
  Pnt } def
/Pls { stroke [] 0 setdash vpt sub M 0 vpt2 V
  currentpoint stroke M
  hpt neg vpt neg R hpt2 0 V stroke
  } def
/Box { stroke [] 0 setdash 2 copy exch hpt sub exch vpt add M
  0 vpt2 neg V hpt2 0 V 0 vpt2 V
  hpt2 neg 0 V closepath stroke
  Pnt } def
/Crs { stroke [] 0 setdash exch hpt sub exch vpt add M
  hpt2 vpt2 neg V currentpoint stroke M
  hpt2 neg 0 R hpt2 vpt2 V stroke } def
/TriU { stroke [] 0 setdash 2 copy vpt 1.12 mul add M
  hpt neg vpt -1.62 mul V
  hpt 2 mul 0 V
  hpt neg vpt 1.62 mul V closepath stroke
  Pnt  } def
/Star { 2 copy Pls Crs } def
/BoxF { stroke [] 0 setdash exch hpt sub exch vpt add M
  0 vpt2 neg V  hpt2 0 V  0 vpt2 V
  hpt2 neg 0 V  closepath fill } def
/TriUF { stroke [] 0 setdash vpt 1.12 mul add M
  hpt neg vpt -1.62 mul V
  hpt 2 mul 0 V
  hpt neg vpt 1.62 mul V closepath fill } def
/TriD { stroke [] 0 setdash 2 copy vpt 1.12 mul sub M
  hpt neg vpt 1.62 mul V
  hpt 2 mul 0 V
  hpt neg vpt -1.62 mul V closepath stroke
  Pnt  } def
/TriDF { stroke [] 0 setdash vpt 1.12 mul sub M
  hpt neg vpt 1.62 mul V
  hpt 2 mul 0 V
  hpt neg vpt -1.62 mul V closepath fill} def
/DiaF { stroke [] 0 setdash vpt add M
  hpt neg vpt neg V hpt vpt neg V
  hpt vpt V hpt neg vpt V closepath fill } def
/Pent { stroke [] 0 setdash 2 copy gsave
  translate 0 hpt M 4 {72 rotate 0 hpt L} repeat
  closepath stroke grestore Pnt } def
/PentF { stroke [] 0 setdash gsave
  translate 0 hpt M 4 {72 rotate 0 hpt L} repeat
  closepath fill grestore } def
/Circle { stroke [] 0 setdash 2 copy
  hpt 0 360 arc stroke Pnt } def
/CircleF { stroke [] 0 setdash hpt 0 360 arc fill } def
/C0 { BL [] 0 setdash 2 copy moveto vpt 90 450  arc } bind def
/C1 { BL [] 0 setdash 2 copy        moveto
       2 copy  vpt 0 90 arc closepath fill
               vpt 0 360 arc closepath } bind def
/C2 { BL [] 0 setdash 2 copy moveto
       2 copy  vpt 90 180 arc closepath fill
               vpt 0 360 arc closepath } bind def
/C3 { BL [] 0 setdash 2 copy moveto
       2 copy  vpt 0 180 arc closepath fill
               vpt 0 360 arc closepath } bind def
/C4 { BL [] 0 setdash 2 copy moveto
       2 copy  vpt 180 270 arc closepath fill
               vpt 0 360 arc closepath } bind def
/C5 { BL [] 0 setdash 2 copy moveto
       2 copy  vpt 0 90 arc
       2 copy moveto
       2 copy  vpt 180 270 arc closepath fill
               vpt 0 360 arc } bind def
/C6 { BL [] 0 setdash 2 copy moveto
      2 copy  vpt 90 270 arc closepath fill
              vpt 0 360 arc closepath } bind def
/C7 { BL [] 0 setdash 2 copy moveto
      2 copy  vpt 0 270 arc closepath fill
              vpt 0 360 arc closepath } bind def
/C8 { BL [] 0 setdash 2 copy moveto
      2 copy vpt 270 360 arc closepath fill
              vpt 0 360 arc closepath } bind def
/C9 { BL [] 0 setdash 2 copy moveto
      2 copy  vpt 270 450 arc closepath fill
              vpt 0 360 arc closepath } bind def
/C10 { BL [] 0 setdash 2 copy 2 copy moveto vpt 270 360 arc closepath fill
       2 copy moveto
       2 copy vpt 90 180 arc closepath fill
               vpt 0 360 arc closepath } bind def
/C11 { BL [] 0 setdash 2 copy moveto
       2 copy  vpt 0 180 arc closepath fill
       2 copy moveto
       2 copy  vpt 270 360 arc closepath fill
               vpt 0 360 arc closepath } bind def
/C12 { BL [] 0 setdash 2 copy moveto
       2 copy  vpt 180 360 arc closepath fill
               vpt 0 360 arc closepath } bind def
/C13 { BL [] 0 setdash  2 copy moveto
       2 copy  vpt 0 90 arc closepath fill
       2 copy moveto
       2 copy  vpt 180 360 arc closepath fill
               vpt 0 360 arc closepath } bind def
/C14 { BL [] 0 setdash 2 copy moveto
       2 copy  vpt 90 360 arc closepath fill
               vpt 0 360 arc } bind def
/C15 { BL [] 0 setdash 2 copy vpt 0 360 arc closepath fill
               vpt 0 360 arc closepath } bind def
/Rec   { newpath 4 2 roll moveto 1 index 0 rlineto 0 exch rlineto
       neg 0 rlineto closepath } bind def
/Square { dup Rec } bind def
/Bsquare { vpt sub exch vpt sub exch vpt2 Square } bind def
/S0 { BL [] 0 setdash 2 copy moveto 0 vpt rlineto BL Bsquare } bind def
/S1 { BL [] 0 setdash 2 copy vpt Square fill Bsquare } bind def
/S2 { BL [] 0 setdash 2 copy exch vpt sub exch vpt Square fill Bsquare } bind def
/S3 { BL [] 0 setdash 2 copy exch vpt sub exch vpt2 vpt Rec fill Bsquare } bind def
/S4 { BL [] 0 setdash 2 copy exch vpt sub exch vpt sub vpt Square fill Bsquare } bind def
/S5 { BL [] 0 setdash 2 copy 2 copy vpt Square fill
       exch vpt sub exch vpt sub vpt Square fill Bsquare } bind def
/S6 { BL [] 0 setdash 2 copy exch vpt sub exch vpt sub vpt vpt2 Rec fill Bsquare } bind def
/S7 { BL [] 0 setdash 2 copy exch vpt sub exch vpt sub vpt vpt2 Rec fill
       2 copy vpt Square fill
       Bsquare } bind def
/S8 { BL [] 0 setdash 2 copy vpt sub vpt Square fill Bsquare } bind def
/S9 { BL [] 0 setdash 2 copy vpt sub vpt vpt2 Rec fill Bsquare } bind def
/S10 { BL [] 0 setdash 2 copy vpt sub vpt Square fill 2 copy exch vpt sub exch vpt Square fill
       Bsquare } bind def
/S11 { BL [] 0 setdash 2 copy vpt sub vpt Square fill 2 copy exch vpt sub exch vpt2 vpt Rec fill
       Bsquare } bind def
/S12 { BL [] 0 setdash 2 copy exch vpt sub exch vpt sub vpt2 vpt Rec fill Bsquare } bind def
/S13 { BL [] 0 setdash 2 copy exch vpt sub exch vpt sub vpt2 vpt Rec fill
       2 copy vpt Square fill Bsquare } bind def
/S14 { BL [] 0 setdash 2 copy exch vpt sub exch vpt sub vpt2 vpt Rec fill
       2 copy exch vpt sub exch vpt Square fill Bsquare } bind def
/S15 { BL [] 0 setdash 2 copy Bsquare fill Bsquare } bind def
/D0 { gsave translate 45 rotate 0 0 S0 stroke grestore } bind def
/D1 { gsave translate 45 rotate 0 0 S1 stroke grestore } bind def
/D2 { gsave translate 45 rotate 0 0 S2 stroke grestore } bind def
/D3 { gsave translate 45 rotate 0 0 S3 stroke grestore } bind def
/D4 { gsave translate 45 rotate 0 0 S4 stroke grestore } bind def
/D5 { gsave translate 45 rotate 0 0 S5 stroke grestore } bind def
/D6 { gsave translate 45 rotate 0 0 S6 stroke grestore } bind def
/D7 { gsave translate 45 rotate 0 0 S7 stroke grestore } bind def
/D8 { gsave translate 45 rotate 0 0 S8 stroke grestore } bind def
/D9 { gsave translate 45 rotate 0 0 S9 stroke grestore } bind def
/D10 { gsave translate 45 rotate 0 0 S10 stroke grestore } bind def
/D11 { gsave translate 45 rotate 0 0 S11 stroke grestore } bind def
/D12 { gsave translate 45 rotate 0 0 S12 stroke grestore } bind def
/D13 { gsave translate 45 rotate 0 0 S13 stroke grestore } bind def
/D14 { gsave translate 45 rotate 0 0 S14 stroke grestore } bind def
/D15 { gsave translate 45 rotate 0 0 S15 stroke grestore } bind def
/DiaE { stroke [] 0 setdash vpt add M
  hpt neg vpt neg V hpt vpt neg V
  hpt vpt V hpt neg vpt V closepath stroke } def
/BoxE { stroke [] 0 setdash exch hpt sub exch vpt add M
  0 vpt2 neg V hpt2 0 V 0 vpt2 V
  hpt2 neg 0 V closepath stroke } def
/TriUE { stroke [] 0 setdash vpt 1.12 mul add M
  hpt neg vpt -1.62 mul V
  hpt 2 mul 0 V
  hpt neg vpt 1.62 mul V closepath stroke } def
/TriDE { stroke [] 0 setdash vpt 1.12 mul sub M
  hpt neg vpt 1.62 mul V
  hpt 2 mul 0 V
  hpt neg vpt -1.62 mul V closepath stroke } def
/PentE { stroke [] 0 setdash gsave
  translate 0 hpt M 4 {72 rotate 0 hpt L} repeat
  closepath stroke grestore } def
/CircE { stroke [] 0 setdash 
  hpt 0 360 arc stroke } def
/Opaque { gsave closepath 1 setgray fill grestore 0 setgray closepath } def
/DiaW { stroke [] 0 setdash vpt add M
  hpt neg vpt neg V hpt vpt neg V
  hpt vpt V hpt neg vpt V Opaque stroke } def
/BoxW { stroke [] 0 setdash exch hpt sub exch vpt add M
  0 vpt2 neg V hpt2 0 V 0 vpt2 V
  hpt2 neg 0 V Opaque stroke } def
/TriUW { stroke [] 0 setdash vpt 1.12 mul add M
  hpt neg vpt -1.62 mul V
  hpt 2 mul 0 V
  hpt neg vpt 1.62 mul V Opaque stroke } def
/TriDW { stroke [] 0 setdash vpt 1.12 mul sub M
  hpt neg vpt 1.62 mul V
  hpt 2 mul 0 V
  hpt neg vpt -1.62 mul V Opaque stroke } def
/PentW { stroke [] 0 setdash gsave
  translate 0 hpt M 4 {72 rotate 0 hpt L} repeat
  Opaque stroke grestore } def
/CircW { stroke [] 0 setdash 
  hpt 0 360 arc Opaque stroke } def
/BoxFill { gsave Rec 1 setgray fill grestore } def
end
}}%
\begin{picture}(3600,2160)(0,0)%
{\GNUPLOTspecial{"
gnudict begin
gsave
0 0 translate
0.100 0.100 scale
0 setgray
newpath
1.000 UL
LTb
500 300 M
63 0 V
2887 0 R
-63 0 V
500 652 M
63 0 V
2887 0 R
-63 0 V
500 1004 M
63 0 V
2887 0 R
-63 0 V
500 1356 M
63 0 V
2887 0 R
-63 0 V
500 1708 M
63 0 V
2887 0 R
-63 0 V
500 2060 M
63 0 V
2887 0 R
-63 0 V
500 300 M
0 63 V
0 1697 R
0 -63 V
869 300 M
0 63 V
0 1697 R
0 -63 V
1238 300 M
0 63 V
0 1697 R
0 -63 V
1606 300 M
0 63 V
0 1697 R
0 -63 V
1975 300 M
0 63 V
0 1697 R
0 -63 V
2344 300 M
0 63 V
0 1697 R
0 -63 V
2712 300 M
0 63 V
0 1697 R
0 -63 V
3081 300 M
0 63 V
0 1697 R
0 -63 V
3450 300 M
0 63 V
0 1697 R
0 -63 V
1.000 UL
LTb
500 300 M
2950 0 V
0 1760 V
-2950 0 V
500 300 L
1.000 UL
LT0
3087 1947 M
263 0 V
500 300 M
7 0 V
8 0 V
7 0 V
8 0 V
7 0 V
7 0 V
8 0 V
7 0 V
7 0 V
8 0 V
7 0 V
8 0 V
7 0 V
7 0 V
8 0 V
7 0 V
7 1 V
8 0 V
7 0 V
8 0 V
7 1 V
7 0 V
8 1 V
7 0 V
7 1 V
8 1 V
7 0 V
8 1 V
7 1 V
7 1 V
8 1 V
7 1 V
7 1 V
8 2 V
7 1 V
7 2 V
8 2 V
7 1 V
8 2 V
7 2 V
7 2 V
8 2 V
7 3 V
7 2 V
8 2 V
7 3 V
8 2 V
7 3 V
7 3 V
8 3 V
7 3 V
7 3 V
8 3 V
7 3 V
8 3 V
7 3 V
7 3 V
8 3 V
7 4 V
7 3 V
8 3 V
7 4 V
8 3 V
7 4 V
7 3 V
8 4 V
7 3 V
8 4 V
7 3 V
7 4 V
8 3 V
7 4 V
7 3 V
8 4 V
7 3 V
7 3 V
8 4 V
7 3 V
8 3 V
7 3 V
7 3 V
8 3 V
7 3 V
8 3 V
7 3 V
7 3 V
8 3 V
7 3 V
7 2 V
8 3 V
7 2 V
7 3 V
8 2 V
7 2 V
8 3 V
7 2 V
7 2 V
8 2 V
7 2 V
8 1 V
7 2 V
7 2 V
8 1 V
7 2 V
7 1 V
8 1 V
7 2 V
7 1 V
8 1 V
7 1 V
8 1 V
7 1 V
7 1 V
8 0 V
7 1 V
8 0 V
7 1 V
7 0 V
8 1 V
7 0 V
7 0 V
8 1 V
7 0 V
7 0 V
8 0 V
7 0 V
8 0 V
7 -1 V
7 0 V
8 0 V
7 0 V
8 -1 V
7 0 V
7 0 V
8 -1 V
7 0 V
7 -1 V
8 0 V
7 -1 V
8 0 V
7 -1 V
7 0 V
8 -1 V
7 -1 V
7 0 V
8 -1 V
7 -1 V
7 -1 V
8 0 V
7 -1 V
8 -1 V
7 -2 V
7 -1 V
8 -2 V
7 -1 V
7 -2 V
8 -1 V
7 -2 V
8 -1 V
7 -2 V
7 -1 V
8 -2 V
7 -2 V
8 -1 V
7 -2 V
7 -2 V
8 -2 V
7 -1 V
7 -2 V
8 -2 V
7 -2 V
7 -1 V
8 -2 V
7 -2 V
8 -2 V
7 -2 V
7 -2 V
8 -1 V
7 -2 V
7 -2 V
8 -2 V
7 -2 V
8 -2 V
7 -1 V
7 -2 V
8 -2 V
7 -2 V
8 -2 V
7 -2 V
7 -2 V
8 -2 V
7 -1 V
7 -2 V
8 -2 V
7 -2 V
8 0 V
7 -1 V
7 0 V
8 -1 V
7 -1 V
7 0 V
8 -1 V
7 -1 V
7 0 V
8 -1 V
7 -1 V
8 -1 V
7 0 V
7 -1 V
8 -1 V
7 -1 V
7 -1 V
8 -1 V
7 -1 V
8 -1 V
7 -1 V
7 0 V
8 -1 V
7 -1 V
8 -1 V
7 -1 V
7 -1 V
8 -1 V
7 -1 V
7 -1 V
8 -1 V
7 -1 V
8 -1 V
7 -1 V
7 -1 V
8 -1 V
7 -1 V
7 -1 V
8 -2 V
7 -1 V
7 -1 V
8 -1 V
7 -1 V
8 -1 V
7 -1 V
7 -1 V
8 -1 V
7 -1 V
7 -1 V
8 -1 V
7 -1 V
8 -1 V
7 -1 V
7 -1 V
8 -1 V
7 -1 V
8 -1 V
7 -1 V
7 -1 V
8 -1 V
7 -1 V
7 0 V
8 -1 V
7 -1 V
8 -1 V
7 -1 V
7 -1 V
8 -1 V
7 -1 V
7 -1 V
8 -1 V
7 -1 V
8 0 V
7 -1 V
7 -1 V
8 -1 V
7 -1 V
7 -1 V
8 0 V
7 -1 V
8 -1 V
7 -1 V
7 -1 V
8 0 V
7 -1 V
7 -1 V
8 -1 V
7 -1 V
7 0 V
8 -1 V
7 -1 V
8 0 V
7 -1 V
7 -1 V
8 -1 V
7 0 V
7 -1 V
8 -1 V
7 0 V
8 -1 V
7 -1 V
7 0 V
8 -1 V
7 0 V
7 -1 V
8 -1 V
7 0 V
8 -1 V
7 0 V
7 -1 V
8 -1 V
7 0 V
7 -1 V
8 0 V
7 -1 V
8 0 V
7 -1 V
7 0 V
8 -1 V
7 0 V
8 -1 V
7 0 V
7 -1 V
8 0 V
7 -1 V
7 0 V
8 -1 V
7 0 V
8 -1 V
7 0 V
7 -1 V
8 0 V
7 -1 V
7 0 V
8 0 V
7 -1 V
8 0 V
7 -1 V
7 0 V
8 -1 V
7 0 V
7 -1 V
8 0 V
7 0 V
8 -1 V
7 0 V
7 0 V
8 -1 V
7 0 V
7 -1 V
8 0 V
7 0 V
7 -1 V
8 0 V
7 0 V
8 -1 V
7 0 V
7 0 V
8 0 V
7 -1 V
7 0 V
8 0 V
7 -1 V
8 0 V
7 0 V
7 0 V
8 -1 V
7 0 V
7 0 V
8 0 V
7 -1 V
8 0 V
7 0 V
7 0 V
8 0 V
7 -1 V
7 0 V
8 0 V
7 0 V
8 0 V
7 -1 V
7 0 V
8 0 V
7 0 V
8 0 V
7 0 V
7 -1 V
8 0 V
7 0 V
7 0 V
8 0 V
7 0 V
8 -1 V
7 0 V
7 0 V
8 0 V
7 0 V
7 0 V
8 0 V
7 0 V
8 -1 V
7 0 V
currentpoint stroke M
7 0 V
8 0 V
7 0 V
1.000 UL
LT1
3087 1847 M
263 0 V
500 300 M
7 0 V
8 0 V
7 0 V
8 0 V
7 0 V
7 0 V
8 0 V
7 0 V
7 0 V
8 0 V
7 0 V
8 1 V
7 0 V
7 0 V
8 0 V
7 1 V
7 0 V
8 1 V
7 1 V
8 1 V
7 1 V
7 1 V
8 1 V
7 1 V
7 2 V
8 2 V
7 1 V
8 2 V
7 2 V
7 3 V
8 2 V
7 3 V
7 3 V
8 3 V
7 3 V
7 4 V
8 3 V
7 4 V
8 4 V
7 4 V
7 4 V
8 5 V
7 4 V
7 5 V
8 5 V
7 5 V
8 5 V
7 5 V
7 6 V
8 5 V
7 6 V
7 5 V
8 6 V
7 6 V
8 6 V
7 6 V
7 6 V
8 6 V
7 6 V
7 6 V
8 6 V
7 6 V
8 7 V
7 6 V
7 6 V
8 6 V
7 7 V
8 6 V
7 6 V
7 6 V
8 6 V
7 6 V
7 6 V
8 6 V
7 5 V
7 6 V
8 5 V
7 6 V
8 5 V
7 5 V
7 5 V
8 5 V
7 5 V
8 5 V
7 5 V
7 4 V
8 5 V
7 4 V
7 4 V
8 4 V
7 4 V
7 3 V
8 4 V
7 3 V
8 3 V
7 3 V
7 3 V
8 3 V
7 3 V
8 2 V
7 3 V
7 2 V
8 2 V
7 2 V
7 1 V
8 2 V
7 1 V
7 2 V
8 1 V
7 1 V
8 1 V
7 1 V
7 0 V
8 1 V
7 0 V
8 0 V
7 1 V
7 0 V
8 0 V
7 -1 V
7 0 V
8 0 V
7 -1 V
7 0 V
8 -1 V
7 -1 V
8 0 V
7 -1 V
7 -1 V
8 -1 V
7 -1 V
8 -2 V
7 -1 V
7 -1 V
8 -1 V
7 -2 V
7 -1 V
8 -2 V
7 -1 V
8 -2 V
7 -1 V
7 -2 V
8 -2 V
7 -1 V
7 -2 V
8 -2 V
7 -1 V
7 -2 V
8 -2 V
7 -1 V
8 -3 V
7 -3 V
7 -3 V
8 -3 V
7 -3 V
7 -3 V
8 -3 V
7 -3 V
8 -3 V
7 -3 V
7 -3 V
8 -3 V
7 -3 V
8 -3 V
7 -4 V
7 -3 V
8 -3 V
7 -3 V
7 -3 V
8 -4 V
7 -3 V
7 -3 V
8 -4 V
7 -3 V
8 -3 V
7 -3 V
7 -4 V
8 -3 V
7 -3 V
7 -4 V
8 -3 V
7 -3 V
8 -3 V
7 -4 V
7 -3 V
8 -3 V
7 -3 V
8 -4 V
7 -3 V
7 -3 V
8 -3 V
7 -3 V
7 -4 V
8 -3 V
7 -3 V
8 -2 V
7 -2 V
7 -2 V
8 -2 V
7 -2 V
7 -2 V
8 -2 V
7 -2 V
7 -2 V
8 -2 V
7 -2 V
8 -2 V
7 -2 V
7 -2 V
8 -2 V
7 -2 V
7 -2 V
8 -2 V
7 -2 V
8 -2 V
7 -2 V
7 -2 V
8 -2 V
7 -2 V
8 -2 V
7 -2 V
7 -1 V
8 -2 V
7 -2 V
7 -2 V
8 -2 V
7 -2 V
8 -1 V
7 -2 V
7 -2 V
8 -2 V
7 -1 V
7 -2 V
8 -2 V
7 -2 V
7 -1 V
8 -2 V
7 -2 V
8 -1 V
7 -2 V
7 -1 V
8 -2 V
7 -2 V
7 -1 V
8 -2 V
7 -1 V
8 -2 V
7 -1 V
7 -2 V
8 -1 V
7 -2 V
8 -1 V
7 -1 V
7 -2 V
8 -1 V
7 -2 V
7 -1 V
8 -1 V
7 -2 V
8 -1 V
7 -1 V
7 -1 V
8 -2 V
7 -1 V
7 -1 V
8 -1 V
7 -2 V
8 -1 V
7 -1 V
7 -1 V
8 -1 V
7 -1 V
7 -2 V
8 -1 V
7 -1 V
8 -1 V
7 -1 V
7 -1 V
8 -1 V
7 -1 V
7 -1 V
8 -1 V
7 -1 V
7 -1 V
8 -1 V
7 -1 V
8 -1 V
7 -1 V
7 0 V
8 -1 V
7 -1 V
7 -1 V
8 -1 V
7 -1 V
8 -1 V
7 0 V
7 -1 V
8 -1 V
7 -1 V
7 0 V
8 -1 V
7 -1 V
8 -1 V
7 0 V
7 -1 V
8 -1 V
7 -1 V
7 0 V
8 -1 V
7 0 V
8 -1 V
7 -1 V
7 0 V
8 -1 V
7 -1 V
8 0 V
7 -1 V
7 0 V
8 -1 V
7 0 V
7 -1 V
8 -1 V
7 0 V
8 -1 V
7 0 V
7 -1 V
8 0 V
7 -1 V
7 0 V
8 -1 V
7 0 V
8 -1 V
7 0 V
7 -1 V
8 0 V
7 -1 V
7 0 V
8 -1 V
7 0 V
8 -1 V
7 0 V
7 0 V
8 -1 V
7 0 V
7 -1 V
8 0 V
7 0 V
7 -1 V
8 0 V
7 0 V
8 -1 V
7 0 V
7 0 V
8 -1 V
7 0 V
7 0 V
8 -1 V
7 0 V
8 0 V
7 -1 V
7 0 V
8 0 V
7 0 V
7 -1 V
8 0 V
7 0 V
8 -1 V
7 0 V
7 0 V
8 0 V
7 0 V
7 -1 V
8 0 V
7 0 V
8 0 V
7 -1 V
7 0 V
8 0 V
7 0 V
8 0 V
7 -1 V
7 0 V
8 0 V
7 0 V
7 0 V
8 0 V
7 -1 V
8 0 V
7 0 V
7 0 V
8 0 V
7 0 V
7 -1 V
8 0 V
7 0 V
8 0 V
7 0 V
currentpoint stroke M
7 0 V
8 0 V
7 0 V
1.000 UL
LT2
3087 1747 M
263 0 V
500 300 M
7 0 V
8 0 V
7 0 V
8 0 V
7 0 V
7 0 V
8 0 V
7 1 V
7 0 V
8 0 V
7 1 V
8 1 V
7 1 V
7 1 V
8 1 V
7 2 V
7 2 V
8 2 V
7 2 V
8 3 V
7 4 V
7 3 V
8 4 V
7 4 V
7 5 V
8 5 V
7 6 V
8 6 V
7 7 V
7 7 V
8 7 V
7 8 V
7 8 V
8 9 V
7 9 V
7 10 V
8 9 V
7 11 V
8 10 V
7 11 V
7 11 V
8 12 V
7 12 V
7 12 V
8 13 V
7 12 V
8 13 V
7 13 V
7 14 V
8 13 V
7 14 V
7 14 V
8 14 V
7 14 V
8 14 V
7 14 V
7 14 V
8 15 V
7 14 V
7 14 V
8 14 V
7 14 V
8 14 V
7 14 V
7 15 V
8 14 V
7 13 V
8 14 V
7 13 V
7 14 V
8 13 V
7 12 V
7 13 V
8 12 V
7 12 V
7 12 V
8 11 V
7 12 V
8 10 V
7 11 V
7 10 V
8 10 V
7 10 V
8 9 V
7 9 V
7 9 V
8 8 V
7 8 V
7 7 V
8 8 V
7 6 V
7 7 V
8 6 V
7 6 V
8 5 V
7 5 V
7 5 V
8 4 V
7 4 V
8 4 V
7 4 V
7 3 V
8 2 V
7 3 V
7 2 V
8 1 V
7 2 V
7 1 V
8 0 V
7 1 V
8 0 V
7 0 V
7 0 V
8 -1 V
7 -1 V
8 -1 V
7 -1 V
7 -2 V
8 -2 V
7 -2 V
7 -2 V
8 -2 V
7 -3 V
7 -3 V
8 -3 V
7 -3 V
8 -3 V
7 -4 V
7 -3 V
8 -4 V
7 -4 V
8 -4 V
7 -4 V
7 -5 V
8 -9 V
7 -9 V
7 -8 V
8 -8 V
7 -9 V
8 -8 V
7 -8 V
7 -7 V
8 -8 V
7 -8 V
7 -7 V
8 -7 V
7 -8 V
7 -7 V
8 -7 V
7 -6 V
8 -9 V
7 -9 V
7 -8 V
8 -8 V
7 -9 V
7 -8 V
8 -8 V
7 -8 V
8 -8 V
7 -8 V
7 -8 V
8 -8 V
7 -8 V
8 -8 V
7 -7 V
7 -8 V
8 -7 V
7 -8 V
7 -7 V
8 -7 V
7 -7 V
7 -7 V
8 -7 V
7 -7 V
8 -7 V
7 -7 V
7 -7 V
8 -6 V
7 -7 V
7 -6 V
8 -7 V
7 -6 V
8 -6 V
7 -7 V
7 -6 V
8 -6 V
7 -6 V
8 -6 V
7 -6 V
7 -5 V
8 -6 V
7 -6 V
7 -5 V
8 -6 V
7 -5 V
8 -5 V
7 -6 V
7 -5 V
8 -5 V
7 -5 V
7 -5 V
8 -5 V
7 -5 V
7 -5 V
8 -5 V
7 -4 V
8 -5 V
7 -4 V
7 -5 V
8 -4 V
7 -5 V
7 -4 V
8 -4 V
7 -5 V
8 -4 V
7 -4 V
7 -4 V
8 -4 V
7 -4 V
8 -4 V
7 -4 V
7 -3 V
8 -4 V
7 -4 V
7 -3 V
8 -4 V
7 -3 V
8 -4 V
7 -3 V
7 -4 V
8 -3 V
7 -3 V
7 -3 V
8 -3 V
7 -4 V
7 -3 V
8 -3 V
7 -3 V
8 -3 V
7 -2 V
7 -3 V
8 -3 V
7 -3 V
7 -2 V
8 -3 V
7 -3 V
8 -2 V
7 -3 V
7 -2 V
8 -3 V
7 -2 V
8 -3 V
7 -2 V
7 -2 V
8 -3 V
7 -2 V
7 -2 V
8 -2 V
7 -2 V
8 -2 V
7 -2 V
7 -2 V
8 -2 V
7 -2 V
7 -2 V
8 -2 V
7 -2 V
8 -2 V
7 -2 V
7 -1 V
8 -2 V
7 -2 V
7 -2 V
8 -1 V
7 -2 V
8 -2 V
7 -1 V
7 -2 V
8 -1 V
7 -2 V
7 -1 V
8 -2 V
7 -1 V
7 -1 V
8 -2 V
7 -1 V
8 -1 V
7 -2 V
7 -1 V
8 -1 V
7 -2 V
7 -1 V
8 -1 V
7 -1 V
8 -1 V
7 -1 V
7 -2 V
8 -1 V
7 -1 V
7 -1 V
8 -1 V
7 -1 V
8 -1 V
7 -1 V
7 -1 V
8 -1 V
7 -1 V
7 -1 V
8 0 V
7 -1 V
8 -1 V
7 -1 V
7 -1 V
8 -1 V
7 -1 V
8 0 V
7 -1 V
7 -1 V
8 -1 V
7 0 V
7 -1 V
8 -1 V
7 0 V
8 -1 V
7 -1 V
7 0 V
8 -1 V
7 -1 V
7 0 V
8 -1 V
7 -1 V
8 0 V
7 -1 V
7 0 V
8 -1 V
7 -1 V
7 0 V
8 -1 V
7 0 V
8 -1 V
7 0 V
7 -1 V
8 0 V
7 -1 V
7 0 V
8 -1 V
7 0 V
7 0 V
8 -1 V
7 0 V
8 -1 V
7 0 V
7 -1 V
8 0 V
7 0 V
7 -1 V
8 0 V
7 0 V
8 -1 V
7 0 V
7 0 V
8 -1 V
7 0 V
7 0 V
8 -1 V
7 0 V
8 0 V
7 -1 V
7 0 V
8 0 V
7 0 V
7 -1 V
8 0 V
7 0 V
8 -1 V
7 0 V
7 0 V
8 0 V
7 0 V
8 -1 V
7 0 V
7 0 V
8 0 V
7 -1 V
7 0 V
8 0 V
7 0 V
8 0 V
7 -1 V
7 0 V
8 0 V
7 0 V
7 0 V
8 0 V
7 -1 V
8 0 V
7 0 V
currentpoint stroke M
7 0 V
8 0 V
7 0 V
1.000 UL
LT3
3087 1647 M
263 0 V
500 300 M
7 0 V
8 0 V
7 0 V
8 0 V
7 0 V
7 0 V
8 1 V
7 0 V
7 1 V
8 0 V
7 1 V
8 2 V
7 1 V
7 2 V
8 2 V
7 3 V
7 3 V
8 4 V
7 4 V
8 5 V
7 5 V
7 6 V
8 6 V
7 7 V
7 8 V
8 8 V
7 9 V
8 10 V
7 10 V
7 12 V
8 12 V
7 11 V
7 13 V
8 14 V
7 14 V
7 14 V
8 15 V
7 15 V
8 16 V
7 17 V
7 17 V
8 17 V
7 18 V
7 18 V
8 19 V
7 19 V
8 19 V
7 19 V
7 20 V
8 20 V
7 20 V
7 20 V
8 20 V
7 21 V
8 20 V
7 21 V
7 20 V
8 21 V
7 20 V
7 21 V
8 20 V
7 20 V
8 20 V
7 20 V
7 20 V
8 20 V
7 19 V
8 20 V
7 19 V
7 18 V
8 18 V
7 18 V
7 18 V
8 17 V
7 16 V
7 16 V
8 16 V
7 15 V
8 15 V
7 14 V
7 14 V
8 14 V
7 13 V
8 12 V
7 12 V
7 11 V
8 11 V
7 11 V
7 10 V
8 9 V
7 9 V
7 8 V
8 8 V
7 8 V
8 6 V
7 7 V
7 5 V
8 6 V
7 5 V
8 4 V
7 4 V
7 3 V
8 3 V
7 2 V
7 2 V
8 2 V
7 1 V
7 0 V
8 0 V
7 0 V
8 0 V
7 -1 V
7 -7 V
8 -14 V
7 -13 V
8 -13 V
7 -13 V
7 -13 V
8 -13 V
7 -13 V
7 -13 V
8 -12 V
7 -13 V
7 -12 V
8 -13 V
7 -12 V
8 -12 V
7 -12 V
7 -12 V
8 -12 V
7 -12 V
8 -11 V
7 -11 V
7 -12 V
8 -11 V
7 -11 V
7 -10 V
8 -11 V
7 -10 V
8 -11 V
7 -10 V
7 -10 V
8 -10 V
7 -9 V
7 -10 V
8 -9 V
7 -9 V
7 -9 V
8 -9 V
7 -8 V
8 -11 V
7 -10 V
7 -11 V
8 -10 V
7 -11 V
7 -10 V
8 -10 V
7 -10 V
8 -10 V
7 -9 V
7 -10 V
8 -10 V
7 -9 V
8 -9 V
7 -10 V
7 -9 V
8 -9 V
7 -9 V
7 -9 V
8 -8 V
7 -9 V
7 -9 V
8 -8 V
7 -8 V
8 -9 V
7 -8 V
7 -8 V
8 -8 V
7 -8 V
7 -7 V
8 -8 V
7 -8 V
8 -7 V
7 -7 V
7 -8 V
8 -7 V
7 -7 V
8 -7 V
7 -7 V
7 -7 V
8 -6 V
7 -7 V
7 -7 V
8 -6 V
7 -6 V
8 -7 V
7 -6 V
7 -6 V
8 -6 V
7 -6 V
7 -6 V
8 -6 V
7 -5 V
7 -6 V
8 -6 V
7 -5 V
8 -6 V
7 -5 V
7 -5 V
8 -5 V
7 -5 V
7 -5 V
8 -5 V
7 -5 V
8 -5 V
7 -5 V
7 -4 V
8 -5 V
7 -5 V
8 -4 V
7 -5 V
7 -4 V
8 -4 V
7 -4 V
7 -5 V
8 -4 V
7 -4 V
8 -4 V
7 -4 V
7 -3 V
8 -4 V
7 -4 V
7 -4 V
8 -3 V
7 -4 V
7 -3 V
8 -4 V
7 -3 V
8 -3 V
7 -4 V
7 -3 V
8 -3 V
7 -3 V
7 -3 V
8 -3 V
7 -3 V
8 -3 V
7 -3 V
7 -3 V
8 -3 V
7 -2 V
8 -3 V
7 -3 V
7 -2 V
8 -3 V
7 -2 V
7 -3 V
8 -2 V
7 -3 V
8 -2 V
7 -2 V
7 -3 V
8 -2 V
7 -2 V
7 -2 V
8 -2 V
7 -3 V
8 -2 V
7 -2 V
7 -2 V
8 -2 V
7 -2 V
7 -1 V
8 -2 V
7 -2 V
8 -2 V
7 -2 V
7 -1 V
8 -2 V
7 -2 V
7 -1 V
8 -2 V
7 -2 V
7 -1 V
8 -2 V
7 -1 V
8 -2 V
7 -1 V
7 -2 V
8 -1 V
7 -1 V
7 -2 V
8 -1 V
7 -1 V
8 -2 V
7 -1 V
7 -1 V
8 -1 V
7 -2 V
7 -1 V
8 -1 V
7 -1 V
8 -1 V
7 -1 V
7 -1 V
8 -1 V
7 -1 V
7 -1 V
8 -1 V
7 -1 V
8 -1 V
7 -1 V
7 -1 V
8 -1 V
7 -1 V
8 -1 V
7 -1 V
7 -1 V
8 0 V
7 -1 V
7 -1 V
8 -1 V
7 0 V
8 -1 V
7 -1 V
7 -1 V
8 0 V
7 -1 V
7 -1 V
8 0 V
7 -1 V
8 -1 V
7 0 V
7 -1 V
8 -1 V
7 0 V
7 -1 V
8 0 V
7 -1 V
8 -1 V
7 0 V
7 -1 V
8 0 V
7 -1 V
7 0 V
8 -1 V
7 0 V
7 -1 V
8 0 V
7 -1 V
8 0 V
7 0 V
7 -1 V
8 0 V
7 -1 V
7 0 V
8 -1 V
7 0 V
8 0 V
7 -1 V
7 0 V
8 0 V
7 -1 V
7 0 V
8 0 V
7 -1 V
8 0 V
7 0 V
7 -1 V
8 0 V
7 0 V
7 -1 V
8 0 V
7 0 V
8 -1 V
7 0 V
7 0 V
8 0 V
7 -1 V
8 0 V
7 0 V
7 0 V
8 -1 V
7 0 V
7 0 V
8 0 V
7 0 V
8 -1 V
7 0 V
7 0 V
8 0 V
7 0 V
7 -1 V
8 0 V
7 0 V
8 0 V
7 0 V
currentpoint stroke M
7 -1 V
8 0 V
7 0 V
1.000 UL
LT4
3087 1547 M
263 0 V
500 300 M
7 0 V
8 0 V
7 0 V
8 0 V
7 0 V
7 0 V
8 1 V
7 0 V
7 1 V
8 1 V
7 2 V
8 2 V
7 2 V
7 2 V
8 4 V
7 3 V
7 5 V
8 4 V
7 6 V
8 6 V
7 7 V
7 8 V
8 9 V
7 9 V
7 11 V
8 11 V
7 12 V
8 13 V
7 14 V
7 15 V
8 16 V
7 15 V
7 18 V
8 17 V
7 18 V
7 19 V
8 20 V
7 20 V
8 21 V
7 21 V
7 22 V
8 22 V
7 23 V
7 24 V
8 24 V
7 24 V
8 24 V
7 25 V
7 25 V
8 26 V
7 25 V
7 26 V
8 26 V
7 26 V
8 26 V
7 26 V
7 26 V
8 26 V
7 25 V
7 26 V
8 25 V
7 26 V
8 25 V
7 25 V
7 25 V
8 25 V
7 24 V
8 24 V
7 24 V
7 23 V
8 22 V
7 22 V
7 22 V
8 21 V
7 20 V
7 20 V
8 19 V
7 19 V
8 18 V
7 18 V
7 16 V
8 17 V
7 15 V
8 15 V
7 15 V
7 14 V
8 13 V
7 12 V
7 12 V
8 11 V
7 11 V
7 9 V
8 9 V
7 9 V
8 8 V
7 7 V
7 7 V
8 5 V
7 6 V
8 4 V
7 5 V
7 -5 V
8 -15 V
7 -16 V
7 -15 V
8 -16 V
7 -16 V
7 -15 V
8 -16 V
7 -15 V
8 -16 V
7 -15 V
7 -15 V
8 -16 V
7 -15 V
8 -15 V
7 -15 V
7 -15 V
8 -15 V
7 -15 V
7 -14 V
8 -15 V
7 -14 V
7 -14 V
8 -14 V
7 -14 V
8 -14 V
7 -14 V
7 -14 V
8 -13 V
7 -13 V
8 -13 V
7 -13 V
7 -13 V
8 -12 V
7 -13 V
7 -12 V
8 -12 V
7 -12 V
8 -11 V
7 -12 V
7 -11 V
8 -11 V
7 -11 V
7 -11 V
8 -10 V
7 -10 V
7 -10 V
8 -10 V
7 -10 V
8 -12 V
7 -12 V
7 -11 V
8 -12 V
7 -11 V
7 -12 V
8 -11 V
7 -11 V
8 -11 V
7 -11 V
7 -11 V
8 -10 V
7 -11 V
8 -10 V
7 -11 V
7 -10 V
8 -10 V
7 -10 V
7 -10 V
8 -9 V
7 -10 V
7 -9 V
8 -10 V
7 -9 V
8 -9 V
7 -9 V
7 -9 V
8 -9 V
7 -8 V
7 -9 V
8 -8 V
7 -9 V
8 -8 V
7 -8 V
7 -8 V
8 -8 V
7 -8 V
8 -7 V
7 -8 V
7 -7 V
8 -8 V
7 -7 V
7 -7 V
8 -7 V
7 -7 V
8 -7 V
7 -7 V
7 -7 V
8 -6 V
7 -7 V
7 -6 V
8 -7 V
7 -6 V
7 -6 V
8 -6 V
7 -6 V
8 -6 V
7 -6 V
7 -5 V
8 -6 V
7 -6 V
7 -5 V
8 -6 V
7 -5 V
8 -5 V
7 -5 V
7 -5 V
8 -5 V
7 -5 V
8 -5 V
7 -5 V
7 -5 V
8 -4 V
7 -5 V
7 -4 V
8 -5 V
7 -4 V
8 -5 V
7 -4 V
7 -4 V
8 -4 V
7 -4 V
7 -4 V
8 -4 V
7 -4 V
7 -4 V
8 -3 V
7 -4 V
8 -4 V
7 -3 V
7 -4 V
8 -3 V
7 -3 V
7 -4 V
8 -3 V
7 -3 V
8 -3 V
7 -3 V
7 -3 V
8 -3 V
7 -3 V
8 -3 V
7 -3 V
7 -3 V
8 -3 V
7 -2 V
7 -3 V
8 -3 V
7 -2 V
8 -3 V
7 -2 V
7 -3 V
8 -2 V
7 -3 V
7 -2 V
8 -2 V
7 -2 V
8 -3 V
7 -2 V
7 -2 V
8 -2 V
7 -2 V
7 -2 V
8 -2 V
7 -2 V
8 -2 V
7 -2 V
7 -2 V
8 -2 V
7 -1 V
7 -2 V
8 -2 V
7 -2 V
7 -1 V
8 -2 V
7 -2 V
8 -1 V
7 -2 V
7 -1 V
8 -2 V
7 -1 V
7 -2 V
8 -1 V
7 -1 V
8 -2 V
7 -1 V
7 -1 V
8 -2 V
7 -1 V
7 -1 V
8 -2 V
7 -1 V
8 -1 V
7 -1 V
7 -1 V
8 -1 V
7 -1 V
7 -1 V
8 -2 V
7 -1 V
8 -1 V
7 -1 V
7 -1 V
8 -1 V
7 0 V
8 -1 V
7 -1 V
7 -1 V
8 -1 V
7 -1 V
7 -1 V
8 -1 V
7 0 V
8 -1 V
7 -1 V
7 -1 V
8 0 V
7 -1 V
7 -1 V
8 -1 V
7 0 V
8 -1 V
7 -1 V
7 0 V
8 -1 V
7 -1 V
7 0 V
8 -1 V
7 -1 V
8 0 V
7 -1 V
7 0 V
8 -1 V
7 0 V
7 -1 V
8 0 V
7 -1 V
7 0 V
8 -1 V
7 0 V
8 -1 V
7 0 V
7 -1 V
8 0 V
7 -1 V
7 0 V
8 -1 V
7 0 V
8 0 V
7 -1 V
7 0 V
8 -1 V
7 0 V
7 0 V
8 -1 V
7 0 V
8 0 V
7 -1 V
7 0 V
8 0 V
7 -1 V
7 0 V
8 0 V
7 -1 V
8 0 V
7 0 V
7 0 V
8 -1 V
7 0 V
8 0 V
7 0 V
7 -1 V
8 0 V
7 0 V
7 0 V
8 -1 V
7 0 V
8 0 V
7 0 V
7 -1 V
8 0 V
7 0 V
7 0 V
8 0 V
7 -1 V
8 0 V
7 0 V
currentpoint stroke M
7 0 V
8 0 V
7 0 V
1.000 UL
LT5
3087 1447 M
263 0 V
500 300 M
7 0 V
8 0 V
7 0 V
8 0 V
7 0 V
7 1 V
8 0 V
7 1 V
7 1 V
8 1 V
7 2 V
8 2 V
7 3 V
7 3 V
8 4 V
7 5 V
7 5 V
8 6 V
7 7 V
8 8 V
7 9 V
7 10 V
8 11 V
7 11 V
7 13 V
8 14 V
7 15 V
8 16 V
7 17 V
7 19 V
8 19 V
7 19 V
7 21 V
8 22 V
7 22 V
7 23 V
8 23 V
7 25 V
8 25 V
7 26 V
7 26 V
8 28 V
7 27 V
7 28 V
8 29 V
7 29 V
8 30 V
7 29 V
7 30 V
8 31 V
7 30 V
7 31 V
8 31 V
7 31 V
8 31 V
7 30 V
7 31 V
8 31 V
7 31 V
7 30 V
8 30 V
7 30 V
8 30 V
7 29 V
7 30 V
8 29 V
7 28 V
8 29 V
7 27 V
7 27 V
8 27 V
7 25 V
7 26 V
8 24 V
7 24 V
7 23 V
8 23 V
7 21 V
8 21 V
7 20 V
7 20 V
8 19 V
7 18 V
8 17 V
7 16 V
7 16 V
8 15 V
7 14 V
7 14 V
8 12 V
7 12 V
7 11 V
8 10 V
7 10 V
8 1 V
7 -17 V
7 -17 V
8 -17 V
7 -17 V
8 -17 V
7 -18 V
7 -17 V
8 -17 V
7 -17 V
7 -18 V
8 -17 V
7 -17 V
7 -17 V
8 -18 V
7 -17 V
8 -17 V
7 -17 V
7 -17 V
8 -17 V
7 -16 V
8 -17 V
7 -16 V
7 -17 V
8 -16 V
7 -16 V
7 -17 V
8 -15 V
7 -16 V
7 -16 V
8 -15 V
7 -16 V
8 -15 V
7 -15 V
7 -15 V
8 -14 V
7 -15 V
8 -14 V
7 -14 V
7 -14 V
8 -14 V
7 -13 V
7 -14 V
8 -13 V
7 -13 V
8 -12 V
7 -13 V
7 -12 V
8 -12 V
7 -12 V
7 -11 V
8 -12 V
7 -11 V
7 -11 V
8 -10 V
7 -11 V
8 -13 V
7 -13 V
7 -12 V
8 -13 V
7 -12 V
7 -13 V
8 -12 V
7 -12 V
8 -12 V
7 -11 V
7 -12 V
8 -11 V
7 -12 V
8 -11 V
7 -11 V
7 -11 V
8 -11 V
7 -10 V
7 -11 V
8 -10 V
7 -11 V
7 -10 V
8 -10 V
7 -10 V
8 -10 V
7 -9 V
7 -10 V
8 -9 V
7 -9 V
7 -10 V
8 -9 V
7 -9 V
8 -8 V
7 -9 V
7 -9 V
8 -8 V
7 -8 V
8 -9 V
7 -8 V
7 -8 V
8 -8 V
7 -7 V
7 -8 V
8 -8 V
7 -7 V
8 -7 V
7 -8 V
7 -7 V
8 -7 V
7 -7 V
7 -7 V
8 -6 V
7 -7 V
7 -7 V
8 -6 V
7 -7 V
8 -6 V
7 -6 V
7 -6 V
8 -6 V
7 -6 V
7 -6 V
8 -6 V
7 -5 V
8 -6 V
7 -5 V
7 -6 V
8 -5 V
7 -5 V
8 -5 V
7 -6 V
7 -5 V
8 -4 V
7 -5 V
7 -5 V
8 -5 V
7 -4 V
8 -5 V
7 -4 V
7 -5 V
8 -4 V
7 -4 V
7 -5 V
8 -4 V
7 -4 V
7 -4 V
8 -4 V
7 -4 V
8 -4 V
7 -3 V
7 -4 V
8 -4 V
7 -3 V
7 -4 V
8 -3 V
7 -3 V
8 -4 V
7 -3 V
7 -3 V
8 -4 V
7 -3 V
8 -3 V
7 -3 V
7 -3 V
8 -3 V
7 -2 V
7 -3 V
8 -3 V
7 -3 V
8 -2 V
7 -3 V
7 -3 V
8 -2 V
7 -3 V
7 -2 V
8 -3 V
7 -2 V
8 -2 V
7 -3 V
7 -2 V
8 -2 V
7 -2 V
7 -2 V
8 -2 V
7 -2 V
8 -2 V
7 -2 V
7 -2 V
8 -2 V
7 -2 V
7 -2 V
8 -2 V
7 -2 V
7 -1 V
8 -2 V
7 -2 V
8 -2 V
7 -1 V
7 -2 V
8 -1 V
7 -2 V
7 -1 V
8 -2 V
7 -1 V
8 -2 V
7 -1 V
7 -2 V
8 -1 V
7 -1 V
7 -2 V
8 -1 V
7 -1 V
8 -1 V
7 -2 V
7 -1 V
8 -1 V
7 -1 V
7 -1 V
8 -1 V
7 -1 V
8 -1 V
7 -2 V
7 -1 V
8 -1 V
7 -1 V
8 0 V
7 -1 V
7 -1 V
8 -1 V
7 -1 V
7 -1 V
8 -1 V
7 -1 V
8 -1 V
7 0 V
7 -1 V
8 -1 V
7 -1 V
7 -1 V
8 0 V
7 -1 V
8 -1 V
7 0 V
7 -1 V
8 -1 V
7 0 V
7 -1 V
8 -1 V
7 0 V
8 -1 V
7 -1 V
7 0 V
8 -1 V
7 0 V
7 -1 V
8 0 V
7 -1 V
7 0 V
8 -1 V
7 0 V
8 -1 V
7 0 V
7 -1 V
8 0 V
7 -1 V
7 0 V
8 -1 V
7 0 V
8 -1 V
7 0 V
7 0 V
8 -1 V
7 0 V
7 -1 V
8 0 V
7 0 V
8 -1 V
7 0 V
7 0 V
8 -1 V
7 0 V
7 0 V
8 -1 V
7 0 V
8 0 V
7 -1 V
7 0 V
8 0 V
7 0 V
8 -1 V
7 0 V
7 0 V
8 -1 V
7 0 V
7 0 V
8 0 V
7 0 V
8 -1 V
7 0 V
7 0 V
8 0 V
7 -1 V
7 0 V
8 0 V
7 0 V
8 0 V
7 -1 V
currentpoint stroke M
7 0 V
8 0 V
7 0 V
stroke
grestore
end
showpage
}}%
\put(3037,1447){\makebox(0,0)[r]{W=900}}%
\put(3037,1547){\makebox(0,0)[r]{W=700}}%
\put(3037,1647){\makebox(0,0)[r]{W=500}}%
\put(3037,1747){\makebox(0,0)[r]{W=300}}%
\put(3037,1847){\makebox(0,0)[r]{W=100}}%
\put(3037,1947){\makebox(0,0)[r]{$W=50$~GeV}}%
\put(1975,50){\makebox(0,0){\Large $b$(fm) }}%
\put(100,1180){%
\special{ps: gsave currentpoint currentpoint translate
270 rotate neg exch neg exch translate}%
\makebox(0,0)[b]{\shortstack{\Large $ I(b, W)$ }}%
\special{ps: currentpoint grestore moveto}%
}%
\put(3450,200){\makebox(0,0){0.8}}%
\put(3081,200){\makebox(0,0){0.7}}%
\put(2712,200){\makebox(0,0){0.6}}%
\put(2344,200){\makebox(0,0){0.5}}%
\put(1975,200){\makebox(0,0){0.4}}%
\put(1606,200){\makebox(0,0){0.3}}%
\put(1238,200){\makebox(0,0){0.2}}%
\put(869,200){\makebox(0,0){0.1}}%
\put(500,200){\makebox(0,0){0}}%
\put(450,2060){\makebox(0,0)[r]{0.025}}%
\put(450,1708){\makebox(0,0)[r]{0.02}}%
\put(450,1356){\makebox(0,0)[r]{0.015}}%
\put(450,1004){\makebox(0,0)[r]{0.01}}%
\put(450,652){\makebox(0,0)[r]{0.005}}%
\put(450,300){\makebox(0,0)[r]{0}}%
\end{picture}%
\endgroup
 

%% file: crossspd5.tex
\begingroup%
  \makeatletter%
  \newcommand{\GNUPLOTspecial}{%
    \@sanitize\catcode`\%=14\relax\special}%
  \setlength{\unitlength}{0.1bp}%
{\GNUPLOTspecial{!
/gnudict 256 dict def
gnudict begin
/Color false def
/Solid false def
/gnulinewidth 5.000 def
/userlinewidth gnulinewidth def
/vshift -33 def
/dl {10 mul} def
/hpt_ 31.5 def
/vpt_ 31.5 def
/hpt hpt_ def
/vpt vpt_ def
/M {moveto} bind def
/L {lineto} bind def
/R {rmoveto} bind def
/V {rlineto} bind def
/vpt2 vpt 2 mul def
/hpt2 hpt 2 mul def
/Lshow { currentpoint stroke M
  0 vshift R show } def
/Rshow { currentpoint stroke M
  dup stringwidth pop neg vshift R show } def
/Cshow { currentpoint stroke M
  dup stringwidth pop -2 div vshift R show } def
/UP { dup vpt_ mul /vpt exch def hpt_ mul /hpt exch def
  /hpt2 hpt 2 mul def /vpt2 vpt 2 mul def } def
/DL { Color {setrgbcolor Solid {pop []} if 0 setdash }
 {pop pop pop Solid {pop []} if 0 setdash} ifelse } def
/BL { stroke gnulinewidth 2 mul setlinewidth } def
/AL { stroke gnulinewidth 2 div setlinewidth } def
/UL { gnulinewidth mul /userlinewidth exch def } def
/PL { stroke userlinewidth setlinewidth } def
/LTb { BL [] 0 0 0 DL } def
/LTa { AL [1 dl 2 dl] 0 setdash 0 0 0 setrgbcolor } def
/LT0 { PL [] 1 0 0 DL } def
/LT1 { PL [4 dl 2 dl] 0 1 0 DL } def
/LT2 { PL [2 dl 3 dl] 0 0 1 DL } def
/LT3 { PL [1 dl 1.5 dl] 1 0 1 DL } def
/LT4 { PL [5 dl 2 dl 1 dl 2 dl] 0 1 1 DL } def
/LT5 { PL [4 dl 3 dl 1 dl 3 dl] 1 1 0 DL } def
/LT6 { PL [2 dl 2 dl 2 dl 4 dl] 0 0 0 DL } def
/LT7 { PL [2 dl 2 dl 2 dl 2 dl 2 dl 4 dl] 1 0.3 0 DL } def
/LT8 { PL [2 dl 2 dl 2 dl 2 dl 2 dl 2 dl 2 dl 4 dl] 0.5 0.5 0.5 DL } def
/Pnt { stroke [] 0 setdash
   gsave 1 setlinecap M 0 0 V stroke grestore } def
/Dia { stroke [] 0 setdash 2 copy vpt add M
  hpt neg vpt neg V hpt vpt neg V
  hpt vpt V hpt neg vpt V closepath stroke
  Pnt } def
/Pls { stroke [] 0 setdash vpt sub M 0 vpt2 V
  currentpoint stroke M
  hpt neg vpt neg R hpt2 0 V stroke
  } def
/Box { stroke [] 0 setdash 2 copy exch hpt sub exch vpt add M
  0 vpt2 neg V hpt2 0 V 0 vpt2 V
  hpt2 neg 0 V closepath stroke
  Pnt } def
/Crs { stroke [] 0 setdash exch hpt sub exch vpt add M
  hpt2 vpt2 neg V currentpoint stroke M
  hpt2 neg 0 R hpt2 vpt2 V stroke } def
/TriU { stroke [] 0 setdash 2 copy vpt 1.12 mul add M
  hpt neg vpt -1.62 mul V
  hpt 2 mul 0 V
  hpt neg vpt 1.62 mul V closepath stroke
  Pnt  } def
/Star { 2 copy Pls Crs } def
/BoxF { stroke [] 0 setdash exch hpt sub exch vpt add M
  0 vpt2 neg V  hpt2 0 V  0 vpt2 V
  hpt2 neg 0 V  closepath fill } def
/TriUF { stroke [] 0 setdash vpt 1.12 mul add M
  hpt neg vpt -1.62 mul V
  hpt 2 mul 0 V
  hpt neg vpt 1.62 mul V closepath fill } def
/TriD { stroke [] 0 setdash 2 copy vpt 1.12 mul sub M
  hpt neg vpt 1.62 mul V
  hpt 2 mul 0 V
  hpt neg vpt -1.62 mul V closepath stroke
  Pnt  } def
/TriDF { stroke [] 0 setdash vpt 1.12 mul sub M
  hpt neg vpt 1.62 mul V
  hpt 2 mul 0 V
  hpt neg vpt -1.62 mul V closepath fill} def
/DiaF { stroke [] 0 setdash vpt add M
  hpt neg vpt neg V hpt vpt neg V
  hpt vpt V hpt neg vpt V closepath fill } def
/Pent { stroke [] 0 setdash 2 copy gsave
  translate 0 hpt M 4 {72 rotate 0 hpt L} repeat
  closepath stroke grestore Pnt } def
/PentF { stroke [] 0 setdash gsave
  translate 0 hpt M 4 {72 rotate 0 hpt L} repeat
  closepath fill grestore } def
/Circle { stroke [] 0 setdash 2 copy
  hpt 0 360 arc stroke Pnt } def
/CircleF { stroke [] 0 setdash hpt 0 360 arc fill } def
/C0 { BL [] 0 setdash 2 copy moveto vpt 90 450  arc } bind def
/C1 { BL [] 0 setdash 2 copy        moveto
       2 copy  vpt 0 90 arc closepath fill
               vpt 0 360 arc closepath } bind def
/C2 { BL [] 0 setdash 2 copy moveto
       2 copy  vpt 90 180 arc closepath fill
               vpt 0 360 arc closepath } bind def
/C3 { BL [] 0 setdash 2 copy moveto
       2 copy  vpt 0 180 arc closepath fill
               vpt 0 360 arc closepath } bind def
/C4 { BL [] 0 setdash 2 copy moveto
       2 copy  vpt 180 270 arc closepath fill
               vpt 0 360 arc closepath } bind def
/C5 { BL [] 0 setdash 2 copy moveto
       2 copy  vpt 0 90 arc
       2 copy moveto
       2 copy  vpt 180 270 arc closepath fill
               vpt 0 360 arc } bind def
/C6 { BL [] 0 setdash 2 copy moveto
      2 copy  vpt 90 270 arc closepath fill
              vpt 0 360 arc closepath } bind def
/C7 { BL [] 0 setdash 2 copy moveto
      2 copy  vpt 0 270 arc closepath fill
              vpt 0 360 arc closepath } bind def
/C8 { BL [] 0 setdash 2 copy moveto
      2 copy vpt 270 360 arc closepath fill
              vpt 0 360 arc closepath } bind def
/C9 { BL [] 0 setdash 2 copy moveto
      2 copy  vpt 270 450 arc closepath fill
              vpt 0 360 arc closepath } bind def
/C10 { BL [] 0 setdash 2 copy 2 copy moveto vpt 270 360 arc closepath fill
       2 copy moveto
       2 copy vpt 90 180 arc closepath fill
               vpt 0 360 arc closepath } bind def
/C11 { BL [] 0 setdash 2 copy moveto
       2 copy  vpt 0 180 arc closepath fill
       2 copy moveto
       2 copy  vpt 270 360 arc closepath fill
               vpt 0 360 arc closepath } bind def
/C12 { BL [] 0 setdash 2 copy moveto
       2 copy  vpt 180 360 arc closepath fill
               vpt 0 360 arc closepath } bind def
/C13 { BL [] 0 setdash  2 copy moveto
       2 copy  vpt 0 90 arc closepath fill
       2 copy moveto
       2 copy  vpt 180 360 arc closepath fill
               vpt 0 360 arc closepath } bind def
/C14 { BL [] 0 setdash 2 copy moveto
       2 copy  vpt 90 360 arc closepath fill
               vpt 0 360 arc } bind def
/C15 { BL [] 0 setdash 2 copy vpt 0 360 arc closepath fill
               vpt 0 360 arc closepath } bind def
/Rec   { newpath 4 2 roll moveto 1 index 0 rlineto 0 exch rlineto
       neg 0 rlineto closepath } bind def
/Square { dup Rec } bind def
/Bsquare { vpt sub exch vpt sub exch vpt2 Square } bind def
/S0 { BL [] 0 setdash 2 copy moveto 0 vpt rlineto BL Bsquare } bind def
/S1 { BL [] 0 setdash 2 copy vpt Square fill Bsquare } bind def
/S2 { BL [] 0 setdash 2 copy exch vpt sub exch vpt Square fill Bsquare } bind def
/S3 { BL [] 0 setdash 2 copy exch vpt sub exch vpt2 vpt Rec fill Bsquare } bind def
/S4 { BL [] 0 setdash 2 copy exch vpt sub exch vpt sub vpt Square fill Bsquare } bind def
/S5 { BL [] 0 setdash 2 copy 2 copy vpt Square fill
       exch vpt sub exch vpt sub vpt Square fill Bsquare } bind def
/S6 { BL [] 0 setdash 2 copy exch vpt sub exch vpt sub vpt vpt2 Rec fill Bsquare } bind def
/S7 { BL [] 0 setdash 2 copy exch vpt sub exch vpt sub vpt vpt2 Rec fill
       2 copy vpt Square fill
       Bsquare } bind def
/S8 { BL [] 0 setdash 2 copy vpt sub vpt Square fill Bsquare } bind def
/S9 { BL [] 0 setdash 2 copy vpt sub vpt vpt2 Rec fill Bsquare } bind def
/S10 { BL [] 0 setdash 2 copy vpt sub vpt Square fill 2 copy exch vpt sub exch vpt Square fill
       Bsquare } bind def
/S11 { BL [] 0 setdash 2 copy vpt sub vpt Square fill 2 copy exch vpt sub exch vpt2 vpt Rec fill
       Bsquare } bind def
/S12 { BL [] 0 setdash 2 copy exch vpt sub exch vpt sub vpt2 vpt Rec fill Bsquare } bind def
/S13 { BL [] 0 setdash 2 copy exch vpt sub exch vpt sub vpt2 vpt Rec fill
       2 copy vpt Square fill Bsquare } bind def
/S14 { BL [] 0 setdash 2 copy exch vpt sub exch vpt sub vpt2 vpt Rec fill
       2 copy exch vpt sub exch vpt Square fill Bsquare } bind def
/S15 { BL [] 0 setdash 2 copy Bsquare fill Bsquare } bind def
/D0 { gsave translate 45 rotate 0 0 S0 stroke grestore } bind def
/D1 { gsave translate 45 rotate 0 0 S1 stroke grestore } bind def
/D2 { gsave translate 45 rotate 0 0 S2 stroke grestore } bind def
/D3 { gsave translate 45 rotate 0 0 S3 stroke grestore } bind def
/D4 { gsave translate 45 rotate 0 0 S4 stroke grestore } bind def
/D5 { gsave translate 45 rotate 0 0 S5 stroke grestore } bind def
/D6 { gsave translate 45 rotate 0 0 S6 stroke grestore } bind def
/D7 { gsave translate 45 rotate 0 0 S7 stroke grestore } bind def
/D8 { gsave translate 45 rotate 0 0 S8 stroke grestore } bind def
/D9 { gsave translate 45 rotate 0 0 S9 stroke grestore } bind def
/D10 { gsave translate 45 rotate 0 0 S10 stroke grestore } bind def
/D11 { gsave translate 45 rotate 0 0 S11 stroke grestore } bind def
/D12 { gsave translate 45 rotate 0 0 S12 stroke grestore } bind def
/D13 { gsave translate 45 rotate 0 0 S13 stroke grestore } bind def
/D14 { gsave translate 45 rotate 0 0 S14 stroke grestore } bind def
/D15 { gsave translate 45 rotate 0 0 S15 stroke grestore } bind def
/DiaE { stroke [] 0 setdash vpt add M
  hpt neg vpt neg V hpt vpt neg V
  hpt vpt V hpt neg vpt V closepath stroke } def
/BoxE { stroke [] 0 setdash exch hpt sub exch vpt add M
  0 vpt2 neg V hpt2 0 V 0 vpt2 V
  hpt2 neg 0 V closepath stroke } def
/TriUE { stroke [] 0 setdash vpt 1.12 mul add M
  hpt neg vpt -1.62 mul V
  hpt 2 mul 0 V
  hpt neg vpt 1.62 mul V closepath stroke } def
/TriDE { stroke [] 0 setdash vpt 1.12 mul sub M
  hpt neg vpt 1.62 mul V
  hpt 2 mul 0 V
  hpt neg vpt -1.62 mul V closepath stroke } def
/PentE { stroke [] 0 setdash gsave
  translate 0 hpt M 4 {72 rotate 0 hpt L} repeat
  closepath stroke grestore } def
/CircE { stroke [] 0 setdash 
  hpt 0 360 arc stroke } def
/Opaque { gsave closepath 1 setgray fill grestore 0 setgray closepath } def
/DiaW { stroke [] 0 setdash vpt add M
  hpt neg vpt neg V hpt vpt neg V
  hpt vpt V hpt neg vpt V Opaque stroke } def
/BoxW { stroke [] 0 setdash exch hpt sub exch vpt add M
  0 vpt2 neg V hpt2 0 V 0 vpt2 V
  hpt2 neg 0 V Opaque stroke } def
/TriUW { stroke [] 0 setdash vpt 1.12 mul add M
  hpt neg vpt -1.62 mul V
  hpt 2 mul 0 V
  hpt neg vpt 1.62 mul V Opaque stroke } def
/TriDW { stroke [] 0 setdash vpt 1.12 mul sub M
  hpt neg vpt 1.62 mul V
  hpt 2 mul 0 V
  hpt neg vpt -1.62 mul V Opaque stroke } def
/PentW { stroke [] 0 setdash gsave
  translate 0 hpt M 4 {72 rotate 0 hpt L} repeat
  Opaque stroke grestore } def
/CircW { stroke [] 0 setdash 
  hpt 0 360 arc Opaque stroke } def
/BoxFill { gsave Rec 1 setgray fill grestore } def
end
}}%
\begin{picture}(3600,2160)(0,0)%
{\GNUPLOTspecial{"
gnudict begin
gsave
0 0 translate
0.100 0.100 scale
0 setgray
newpath
1.000 UL
LTb
400 300 M
63 0 V
2987 0 R
-63 0 V
400 652 M
63 0 V
2987 0 R
-63 0 V
400 1004 M
63 0 V
2987 0 R
-63 0 V
400 1356 M
63 0 V
2987 0 R
-63 0 V
400 1708 M
63 0 V
2987 0 R
-63 0 V
400 2060 M
63 0 V
2987 0 R
-63 0 V
400 300 M
0 63 V
0 1697 R
0 -63 V
908 300 M
0 63 V
0 1697 R
0 -63 V
1417 300 M
0 63 V
0 1697 R
0 -63 V
1925 300 M
0 63 V
0 1697 R
0 -63 V
2433 300 M
0 63 V
0 1697 R
0 -63 V
2942 300 M
0 63 V
0 1697 R
0 -63 V
3450 300 M
0 63 V
0 1697 R
0 -63 V
1.000 UL
LTb
400 300 M
3050 0 V
0 1760 V
-3050 0 V
400 300 L
1.000 UL
LT0
1163 1849 M
263 0 V
502 316 M
101 38 V
102 44 V
102 48 V
101 50 V
102 53 V
102 56 V
101 57 V
102 59 V
102 60 V
101 62 V
102 64 V
102 67 V
101 65 V
102 66 V
102 66 V
101 66 V
102 66 V
102 66 V
101 66 V
102 66 V
102 64 V
101 66 V
102 62 V
102 65 V
101 66 V
102 60 V
102 65 V
101 64 V
75 47 V
1.000 UL
LT1
1163 1749 M
263 0 V
502 317 M
101 42 V
102 49 V
102 54 V
101 57 V
102 60 V
102 63 V
101 66 V
102 67 V
102 69 V
101 71 V
102 73 V
102 77 V
101 80 V
102 78 V
102 78 V
101 78 V
102 77 V
102 76 V
101 76 V
102 74 V
102 75 V
101 76 V
102 73 V
102 74 V
101 68 V
16 12 V
1.000 UL
LT2
1163 1649 M
263 0 V
502 342 M
101 53 V
102 63 V
102 67 V
101 72 V
102 74 V
102 78 V
101 80 V
102 82 V
102 85 V
101 86 V
102 88 V
102 94 V
101 96 V
102 94 V
102 94 V
101 93 V
102 93 V
102 91 V
101 91 V
102 88 V
64 56 V
1.000 UP
1.000 UL
LT3
1163 1549 M
263 0 V
-263 31 R
0 -62 V
263 62 R
0 -62 V
715 484 M
0 -89 V
-31 89 R
62 0 V
684 395 M
62 0 V
870 545 M
0 -71 V
-31 71 R
62 0 V
839 474 M
62 0 V
69 169 R
0 -99 V
-31 99 R
62 0 V
939 544 M
62 0 V
50 165 R
0 -121 V
-31 121 R
62 0 V
1020 588 M
62 0 V
55 141 R
0 -128 V
-31 128 R
62 0 V
1106 601 M
62 0 V
61 216 R
0 -156 V
-31 156 R
62 0 V
1198 661 M
62 0 V
75 197 R
0 -164 V
-31 164 R
62 0 V
1304 694 M
62 0 V
104 146 R
0 -171 V
-31 171 R
62 0 V
1439 669 M
62 0 V
255 387 R
0 -258 V
-31 258 R
62 0 V
1725 798 M
62 0 V
111 175 R
0 -211 V
-31 211 R
62 0 V
1867 762 M
62 0 V
222 398 R
0 -256 V
-31 256 R
62 0 V
2120 904 M
62 0 V
222 349 R
0 -306 V
-31 306 R
62 0 V
2373 947 M
62 0 V
226 336 R
0 -333 V
-31 333 R
62 0 V
2630 950 M
62 0 V
223 360 R
0 -346 V
-31 346 R
62 0 V
2884 964 M
62 0 V
223 667 R
0 -453 V
-31 453 R
62 0 V
-62 -453 R
62 0 V
715 439 Pls
870 510 Pls
970 594 Pls
1051 648 Pls
1137 665 Pls
1229 739 Pls
1335 776 Pls
1470 755 Pls
1756 927 Pls
1898 867 Pls
2151 1032 Pls
2404 1100 Pls
2661 1117 Pls
2915 1137 Pls
3169 1405 Pls
1294 1549 Pls
1.000 UP
1.000 UL
LT4
1163 1449 M
263 0 V
-263 31 R
0 -62 V
263 62 R
0 -62 V
807 621 M
0 -74 V
-31 74 R
62 0 V
776 547 M
62 0 V
172 182 R
0 -98 V
-31 98 R
62 0 V
979 631 M
62 0 V
172 197 R
0 -125 V
-31 125 R
62 0 V
1182 703 M
62 0 V
173 163 R
0 -133 V
-31 133 R
62 0 V
1386 733 M
62 0 V
172 223 R
0 -157 V
-31 157 R
62 0 V
1589 799 M
62 0 V
172 301 R
0 -190 V
-31 190 R
62 0 V
1792 910 M
62 0 V
173 216 R
0 -237 V
-31 237 R
62 0 V
1996 889 M
62 0 V
807 588 Crs
1010 678 Crs
1213 762 Crs
1417 800 Crs
1620 882 Crs
1823 1002 Crs
2027 1003 Crs
1294 1449 Crs
1.000 UP
1.000 UL
LT5
1163 1349 M
263 0 V
-263 31 R
0 -62 V
263 62 R
0 -62 V
542 395 M
0 -34 V
-31 34 R
62 0 V
511 361 M
62 0 V
542 378 Star
1294 1349 Star
1.000 UP
1.000 UL
LT6
1163 1249 M
263 0 V
-263 31 R
0 -62 V
263 62 R
0 -62 V
520 399 M
0 -42 V
-31 42 R
62 0 V
489 357 M
62 0 V
7 72 R
0 -68 V
-31 68 R
62 0 V
527 361 M
62 0 V
14 102 R
0 -76 V
-31 76 R
62 0 V
572 387 M
62 0 V
15 114 R
0 -91 V
-31 91 R
62 0 V
618 410 M
62 0 V
520 378 Box
558 395 Box
603 425 Box
649 456 Box
1294 1249 Box
stroke
grestore
end
showpage
}}%
\put(1113,1249){\makebox(0,0)[r]{E401}}%
\put(1113,1349){\makebox(0,0)[r]{E516}}%
\put(1113,1449){\makebox(0,0)[r]{ZEUS prel.}}%
\put(1113,1549){\makebox(0,0)[r]{H1}}%
\put(1113,1649){\makebox(0,0)[r]{SPD, $\beta$}}%
\put(1113,1749){\makebox(0,0)[r]{SPD, $\beta =0$}}%
\put(1113,1849){\makebox(0,0)[r]{PDF, $\beta =0$}}%
\put(1925,50){\makebox(0,0){\Large $W$ ~(GeV)}}%
\put(100,1180){%
\special{ps: gsave currentpoint currentpoint translate
270 rotate neg exch neg exch translate}%
\makebox(0,0)[b]{\shortstack{\Large$ \sigma (\gamma P \rightarrow J/\psi P) $(nb)}}%
\special{ps: currentpoint grestore moveto}%
}%
\put(3450,200){\makebox(0,0){300}}%
\put(2942,200){\makebox(0,0){250}}%
\put(2433,200){\makebox(0,0){200}}%
\put(1925,200){\makebox(0,0){150}}%
\put(1417,200){\makebox(0,0){100}}%
\put(908,200){\makebox(0,0){50}}%
\put(400,200){\makebox(0,0){0}}%
\put(350,2060){\makebox(0,0)[r]{250}}%
\put(350,1708){\makebox(0,0)[r]{200}}%
\put(350,1356){\makebox(0,0)[r]{150}}%
\put(350,1004){\makebox(0,0)[r]{100}}%
\put(350,652){\makebox(0,0)[r]{50}}%
\put(350,300){\makebox(0,0)[r]{0}}%
\end{picture}%
\endgroup
 

%% file: betanew.tex
\begingroup%
  \makeatletter%
  \newcommand{\GNUPLOTspecial}{%
    \@sanitize\catcode`\%=14\relax\special}%
  \setlength{\unitlength}{0.1bp}%
{\GNUPLOTspecial{!
/gnudict 256 dict def
gnudict begin
/Color false def
/Solid false def
/gnulinewidth 5.000 def
/userlinewidth gnulinewidth def
/vshift -33 def
/dl {10 mul} def
/hpt_ 31.5 def
/vpt_ 31.5 def
/hpt hpt_ def
/vpt vpt_ def
/M {moveto} bind def
/L {lineto} bind def
/R {rmoveto} bind def
/V {rlineto} bind def
/vpt2 vpt 2 mul def
/hpt2 hpt 2 mul def
/Lshow { currentpoint stroke M
  0 vshift R show } def
/Rshow { currentpoint stroke M
  dup stringwidth pop neg vshift R show } def
/Cshow { currentpoint stroke M
  dup stringwidth pop -2 div vshift R show } def
/UP { dup vpt_ mul /vpt exch def hpt_ mul /hpt exch def
  /hpt2 hpt 2 mul def /vpt2 vpt 2 mul def } def
/DL { Color {setrgbcolor Solid {pop []} if 0 setdash }
 {pop pop pop Solid {pop []} if 0 setdash} ifelse } def
/BL { stroke gnulinewidth 2 mul setlinewidth } def
/AL { stroke gnulinewidth 2 div setlinewidth } def
/UL { gnulinewidth mul /userlinewidth exch def } def
/PL { stroke userlinewidth setlinewidth } def
/LTb { BL [] 0 0 0 DL } def
/LTa { AL [1 dl 2 dl] 0 setdash 0 0 0 setrgbcolor } def
/LT0 { PL [] 1 0 0 DL } def
/LT1 { PL [4 dl 2 dl] 0 1 0 DL } def
/LT2 { PL [2 dl 3 dl] 0 0 1 DL } def
/LT3 { PL [1 dl 1.5 dl] 1 0 1 DL } def
/LT4 { PL [5 dl 2 dl 1 dl 2 dl] 0 1 1 DL } def
/LT5 { PL [4 dl 3 dl 1 dl 3 dl] 1 1 0 DL } def
/LT6 { PL [2 dl 2 dl 2 dl 4 dl] 0 0 0 DL } def
/LT7 { PL [2 dl 2 dl 2 dl 2 dl 2 dl 4 dl] 1 0.3 0 DL } def
/LT8 { PL [2 dl 2 dl 2 dl 2 dl 2 dl 2 dl 2 dl 4 dl] 0.5 0.5 0.5 DL } def
/Pnt { stroke [] 0 setdash
   gsave 1 setlinecap M 0 0 V stroke grestore } def
/Dia { stroke [] 0 setdash 2 copy vpt add M
  hpt neg vpt neg V hpt vpt neg V
  hpt vpt V hpt neg vpt V closepath stroke
  Pnt } def
/Pls { stroke [] 0 setdash vpt sub M 0 vpt2 V
  currentpoint stroke M
  hpt neg vpt neg R hpt2 0 V stroke
  } def
/Box { stroke [] 0 setdash 2 copy exch hpt sub exch vpt add M
  0 vpt2 neg V hpt2 0 V 0 vpt2 V
  hpt2 neg 0 V closepath stroke
  Pnt } def
/Crs { stroke [] 0 setdash exch hpt sub exch vpt add M
  hpt2 vpt2 neg V currentpoint stroke M
  hpt2 neg 0 R hpt2 vpt2 V stroke } def
/TriU { stroke [] 0 setdash 2 copy vpt 1.12 mul add M
  hpt neg vpt -1.62 mul V
  hpt 2 mul 0 V
  hpt neg vpt 1.62 mul V closepath stroke
  Pnt  } def
/Star { 2 copy Pls Crs } def
/BoxF { stroke [] 0 setdash exch hpt sub exch vpt add M
  0 vpt2 neg V  hpt2 0 V  0 vpt2 V
  hpt2 neg 0 V  closepath fill } def
/TriUF { stroke [] 0 setdash vpt 1.12 mul add M
  hpt neg vpt -1.62 mul V
  hpt 2 mul 0 V
  hpt neg vpt 1.62 mul V closepath fill } def
/TriD { stroke [] 0 setdash 2 copy vpt 1.12 mul sub M
  hpt neg vpt 1.62 mul V
  hpt 2 mul 0 V
  hpt neg vpt -1.62 mul V closepath stroke
  Pnt  } def
/TriDF { stroke [] 0 setdash vpt 1.12 mul sub M
  hpt neg vpt 1.62 mul V
  hpt 2 mul 0 V
  hpt neg vpt -1.62 mul V closepath fill} def
/DiaF { stroke [] 0 setdash vpt add M
  hpt neg vpt neg V hpt vpt neg V
  hpt vpt V hpt neg vpt V closepath fill } def
/Pent { stroke [] 0 setdash 2 copy gsave
  translate 0 hpt M 4 {72 rotate 0 hpt L} repeat
  closepath stroke grestore Pnt } def
/PentF { stroke [] 0 setdash gsave
  translate 0 hpt M 4 {72 rotate 0 hpt L} repeat
  closepath fill grestore } def
/Circle { stroke [] 0 setdash 2 copy
  hpt 0 360 arc stroke Pnt } def
/CircleF { stroke [] 0 setdash hpt 0 360 arc fill } def
/C0 { BL [] 0 setdash 2 copy moveto vpt 90 450  arc } bind def
/C1 { BL [] 0 setdash 2 copy        moveto
       2 copy  vpt 0 90 arc closepath fill
               vpt 0 360 arc closepath } bind def
/C2 { BL [] 0 setdash 2 copy moveto
       2 copy  vpt 90 180 arc closepath fill
               vpt 0 360 arc closepath } bind def
/C3 { BL [] 0 setdash 2 copy moveto
       2 copy  vpt 0 180 arc closepath fill
               vpt 0 360 arc closepath } bind def
/C4 { BL [] 0 setdash 2 copy moveto
       2 copy  vpt 180 270 arc closepath fill
               vpt 0 360 arc closepath } bind def
/C5 { BL [] 0 setdash 2 copy moveto
       2 copy  vpt 0 90 arc
       2 copy moveto
       2 copy  vpt 180 270 arc closepath fill
               vpt 0 360 arc } bind def
/C6 { BL [] 0 setdash 2 copy moveto
      2 copy  vpt 90 270 arc closepath fill
              vpt 0 360 arc closepath } bind def
/C7 { BL [] 0 setdash 2 copy moveto
      2 copy  vpt 0 270 arc closepath fill
              vpt 0 360 arc closepath } bind def
/C8 { BL [] 0 setdash 2 copy moveto
      2 copy vpt 270 360 arc closepath fill
              vpt 0 360 arc closepath } bind def
/C9 { BL [] 0 setdash 2 copy moveto
      2 copy  vpt 270 450 arc closepath fill
              vpt 0 360 arc closepath } bind def
/C10 { BL [] 0 setdash 2 copy 2 copy moveto vpt 270 360 arc closepath fill
       2 copy moveto
       2 copy vpt 90 180 arc closepath fill
               vpt 0 360 arc closepath } bind def
/C11 { BL [] 0 setdash 2 copy moveto
       2 copy  vpt 0 180 arc closepath fill
       2 copy moveto
       2 copy  vpt 270 360 arc closepath fill
               vpt 0 360 arc closepath } bind def
/C12 { BL [] 0 setdash 2 copy moveto
       2 copy  vpt 180 360 arc closepath fill
               vpt 0 360 arc closepath } bind def
/C13 { BL [] 0 setdash  2 copy moveto
       2 copy  vpt 0 90 arc closepath fill
       2 copy moveto
       2 copy  vpt 180 360 arc closepath fill
               vpt 0 360 arc closepath } bind def
/C14 { BL [] 0 setdash 2 copy moveto
       2 copy  vpt 90 360 arc closepath fill
               vpt 0 360 arc } bind def
/C15 { BL [] 0 setdash 2 copy vpt 0 360 arc closepath fill
               vpt 0 360 arc closepath } bind def
/Rec   { newpath 4 2 roll moveto 1 index 0 rlineto 0 exch rlineto
       neg 0 rlineto closepath } bind def
/Square { dup Rec } bind def
/Bsquare { vpt sub exch vpt sub exch vpt2 Square } bind def
/S0 { BL [] 0 setdash 2 copy moveto 0 vpt rlineto BL Bsquare } bind def
/S1 { BL [] 0 setdash 2 copy vpt Square fill Bsquare } bind def
/S2 { BL [] 0 setdash 2 copy exch vpt sub exch vpt Square fill Bsquare } bind def
/S3 { BL [] 0 setdash 2 copy exch vpt sub exch vpt2 vpt Rec fill Bsquare } bind def
/S4 { BL [] 0 setdash 2 copy exch vpt sub exch vpt sub vpt Square fill Bsquare } bind def
/S5 { BL [] 0 setdash 2 copy 2 copy vpt Square fill
       exch vpt sub exch vpt sub vpt Square fill Bsquare } bind def
/S6 { BL [] 0 setdash 2 copy exch vpt sub exch vpt sub vpt vpt2 Rec fill Bsquare } bind def
/S7 { BL [] 0 setdash 2 copy exch vpt sub exch vpt sub vpt vpt2 Rec fill
       2 copy vpt Square fill
       Bsquare } bind def
/S8 { BL [] 0 setdash 2 copy vpt sub vpt Square fill Bsquare } bind def
/S9 { BL [] 0 setdash 2 copy vpt sub vpt vpt2 Rec fill Bsquare } bind def
/S10 { BL [] 0 setdash 2 copy vpt sub vpt Square fill 2 copy exch vpt sub exch vpt Square fill
       Bsquare } bind def
/S11 { BL [] 0 setdash 2 copy vpt sub vpt Square fill 2 copy exch vpt sub exch vpt2 vpt Rec fill
       Bsquare } bind def
/S12 { BL [] 0 setdash 2 copy exch vpt sub exch vpt sub vpt2 vpt Rec fill Bsquare } bind def
/S13 { BL [] 0 setdash 2 copy exch vpt sub exch vpt sub vpt2 vpt Rec fill
       2 copy vpt Square fill Bsquare } bind def
/S14 { BL [] 0 setdash 2 copy exch vpt sub exch vpt sub vpt2 vpt Rec fill
       2 copy exch vpt sub exch vpt Square fill Bsquare } bind def
/S15 { BL [] 0 setdash 2 copy Bsquare fill Bsquare } bind def
/D0 { gsave translate 45 rotate 0 0 S0 stroke grestore } bind def
/D1 { gsave translate 45 rotate 0 0 S1 stroke grestore } bind def
/D2 { gsave translate 45 rotate 0 0 S2 stroke grestore } bind def
/D3 { gsave translate 45 rotate 0 0 S3 stroke grestore } bind def
/D4 { gsave translate 45 rotate 0 0 S4 stroke grestore } bind def
/D5 { gsave translate 45 rotate 0 0 S5 stroke grestore } bind def
/D6 { gsave translate 45 rotate 0 0 S6 stroke grestore } bind def
/D7 { gsave translate 45 rotate 0 0 S7 stroke grestore } bind def
/D8 { gsave translate 45 rotate 0 0 S8 stroke grestore } bind def
/D9 { gsave translate 45 rotate 0 0 S9 stroke grestore } bind def
/D10 { gsave translate 45 rotate 0 0 S10 stroke grestore } bind def
/D11 { gsave translate 45 rotate 0 0 S11 stroke grestore } bind def
/D12 { gsave translate 45 rotate 0 0 S12 stroke grestore } bind def
/D13 { gsave translate 45 rotate 0 0 S13 stroke grestore } bind def
/D14 { gsave translate 45 rotate 0 0 S14 stroke grestore } bind def
/D15 { gsave translate 45 rotate 0 0 S15 stroke grestore } bind def
/DiaE { stroke [] 0 setdash vpt add M
  hpt neg vpt neg V hpt vpt neg V
  hpt vpt V hpt neg vpt V closepath stroke } def
/BoxE { stroke [] 0 setdash exch hpt sub exch vpt add M
  0 vpt2 neg V hpt2 0 V 0 vpt2 V
  hpt2 neg 0 V closepath stroke } def
/TriUE { stroke [] 0 setdash vpt 1.12 mul add M
  hpt neg vpt -1.62 mul V
  hpt 2 mul 0 V
  hpt neg vpt 1.62 mul V closepath stroke } def
/TriDE { stroke [] 0 setdash vpt 1.12 mul sub M
  hpt neg vpt 1.62 mul V
  hpt 2 mul 0 V
  hpt neg vpt -1.62 mul V closepath stroke } def
/PentE { stroke [] 0 setdash gsave
  translate 0 hpt M 4 {72 rotate 0 hpt L} repeat
  closepath stroke grestore } def
/CircE { stroke [] 0 setdash 
  hpt 0 360 arc stroke } def
/Opaque { gsave closepath 1 setgray fill grestore 0 setgray closepath } def
/DiaW { stroke [] 0 setdash vpt add M
  hpt neg vpt neg V hpt vpt neg V
  hpt vpt V hpt neg vpt V Opaque stroke } def
/BoxW { stroke [] 0 setdash exch hpt sub exch vpt add M
  0 vpt2 neg V hpt2 0 V 0 vpt2 V
  hpt2 neg 0 V Opaque stroke } def
/TriUW { stroke [] 0 setdash vpt 1.12 mul add M
  hpt neg vpt -1.62 mul V
  hpt 2 mul 0 V
  hpt neg vpt 1.62 mul V Opaque stroke } def
/TriDW { stroke [] 0 setdash vpt 1.12 mul sub M
  hpt neg vpt 1.62 mul V
  hpt 2 mul 0 V
  hpt neg vpt -1.62 mul V Opaque stroke } def
/PentW { stroke [] 0 setdash gsave
  translate 0 hpt M 4 {72 rotate 0 hpt L} repeat
  Opaque stroke grestore } def
/CircW { stroke [] 0 setdash 
  hpt 0 360 arc Opaque stroke } def
/BoxFill { gsave Rec 1 setgray fill grestore } def
end
}}%
\begin{picture}(3600,2160)(0,0)%
{\GNUPLOTspecial{"
gnudict begin
gsave
0 0 translate
0.100 0.100 scale
0 setgray
newpath
1.000 UL
LTb
400 300 M
63 0 V
2987 0 R
-63 0 V
400 507 M
63 0 V
2987 0 R
-63 0 V
400 714 M
63 0 V
2987 0 R
-63 0 V
400 921 M
63 0 V
2987 0 R
-63 0 V
400 1128 M
63 0 V
2987 0 R
-63 0 V
400 1335 M
63 0 V
2987 0 R
-63 0 V
400 1542 M
63 0 V
2987 0 R
-63 0 V
400 1749 M
63 0 V
2987 0 R
-63 0 V
400 1956 M
63 0 V
2987 0 R
-63 0 V
684 300 M
0 63 V
0 1697 R
0 -63 V
1038 300 M
0 63 V
0 1697 R
0 -63 V
1393 300 M
0 63 V
0 1697 R
0 -63 V
1748 300 M
0 63 V
0 1697 R
0 -63 V
2102 300 M
0 63 V
0 1697 R
0 -63 V
2457 300 M
0 63 V
0 1697 R
0 -63 V
2812 300 M
0 63 V
0 1697 R
0 -63 V
3166 300 M
0 63 V
0 1697 R
0 -63 V
1.000 UL
LTb
400 300 M
3050 0 V
0 1760 V
-3050 0 V
400 300 L
1.000 UL
LT0
2918 1749 M
263 0 V
400 1920 M
35 -216 V
36 -106 V
35 -64 V
36 -45 V
35 -33 V
36 -26 V
35 -20 V
36 -18 V
35 -14 V
36 -12 V
35 -11 V
36 -10 V
35 -8 V
36 -8 V
35 -7 V
35 -6 V
36 -6 V
35 -6 V
36 -4 V
35 -5 V
36 -4 V
35 -4 V
36 -4 V
35 -3 V
36 -4 V
35 -3 V
36 -3 V
35 -3 V
71 -5 V
71 -5 V
71 -4 V
71 -4 V
71 -4 V
71 -3 V
71 -3 V
70 -3 V
71 -3 V
71 -2 V
71 -3 V
71 -2 V
71 -2 V
71 -3 V
71 -2 V
71 -1 V
71 -2 V
71 -2 V
71 -2 V
71 -1 V
71 -2 V
70 -1 V
71 -2 V
71 -1 V
71 -1 V
71 -1 V
71 -2 V
71 -1 V
71 -1 V
1.000 UL
LT1
2918 1649 M
263 0 V
400 1567 M
35 -315 V
36 -138 V
35 -79 V
36 -52 V
35 -38 V
36 -28 V
35 -22 V
36 -19 V
35 -15 V
36 -13 V
35 -11 V
36 -10 V
35 -8 V
36 -8 V
35 -7 V
35 -6 V
36 -6 V
35 -5 V
36 -4 V
35 -5 V
36 -4 V
35 -4 V
36 -3 V
35 -4 V
36 -3 V
35 -3 V
36 -2 V
35 -3 V
71 -5 V
71 -4 V
71 -4 V
71 -4 V
71 -3 V
71 -3 V
71 -3 V
70 -3 V
71 -2 V
71 -3 V
71 -2 V
71 -2 V
71 -2 V
71 -2 V
71 -2 V
71 -1 V
71 -2 V
71 -1 V
71 -2 V
71 -1 V
71 -2 V
70 -1 V
71 -1 V
71 -1 V
71 -1 V
71 -2 V
71 -1 V
71 -1 V
71 -1 V
1.000 UL
LT2
2918 1549 M
263 0 V
400 1085 M
35 35 V
36 23 V
35 16 V
36 12 V
35 10 V
36 8 V
35 7 V
36 6 V
35 5 V
36 4 V
35 4 V
36 3 V
35 4 V
36 3 V
35 2 V
35 3 V
36 2 V
35 2 V
36 2 V
35 2 V
36 1 V
35 2 V
36 2 V
35 1 V
36 1 V
35 2 V
36 1 V
35 1 V
71 2 V
71 2 V
71 2 V
71 1 V
71 2 V
71 1 V
71 2 V
70 1 V
71 1 V
71 1 V
71 1 V
71 1 V
71 1 V
71 1 V
71 1 V
71 1 V
71 0 V
71 1 V
71 1 V
71 1 V
71 0 V
70 1 V
71 0 V
71 1 V
71 0 V
71 1 V
71 0 V
71 1 V
71 0 V
1.000 UL
LT3
2918 1449 M
263 0 V
400 598 M
35 27 V
36 18 V
35 13 V
36 11 V
35 8 V
36 7 V
35 6 V
36 5 V
35 4 V
36 4 V
35 3 V
36 3 V
35 3 V
36 3 V
35 2 V
35 2 V
36 2 V
35 2 V
36 2 V
35 2 V
36 1 V
35 2 V
36 1 V
35 1 V
36 2 V
35 1 V
36 1 V
35 1 V
71 2 V
71 2 V
71 1 V
71 2 V
71 1 V
71 2 V
71 1 V
70 1 V
71 1 V
71 1 V
71 1 V
71 1 V
71 1 V
71 1 V
71 0 V
71 1 V
71 1 V
71 1 V
71 0 V
71 1 V
71 0 V
70 1 V
71 1 V
71 0 V
71 1 V
71 0 V
71 1 V
71 0 V
71 0 V
stroke
grestore
end
showpage
}}%
\put(2868,1449){\makebox(0,0)[r]{$\lambda =~4,~\beta^2$}}%
\put(2868,1549){\makebox(0,0)[r]{$\lambda = ~4,~\beta$}}%
\put(2868,1649){\makebox(0,0)[r]{$\lambda = 10,~\beta^2$}}%
\put(2868,1749){\makebox(0,0)[r]{$\lambda = 10,~\beta$}}%
\put(584,1849){\makebox(0,0)[l]{${\cal R} e {\cal A}$ from two power fit to ${\cal I} m {\cal A}$}}%
\put(1925,50){\makebox(0,0){\Large $W$~(GeV)}}%
\put(100,1180){%
\special{ps: gsave currentpoint currentpoint translate
270 rotate neg exch neg exch translate}%
\makebox(0,0)[b]{\shortstack{\Large$ \beta = {\cal R} e {\cal A}/ {\cal I} m {\cal A}, ~\beta^2$}}%
\special{ps: currentpoint grestore moveto}%
}%
\put(3166,200){\makebox(0,0){800}}%
\put(2812,200){\makebox(0,0){700}}%
\put(2457,200){\makebox(0,0){600}}%
\put(2102,200){\makebox(0,0){500}}%
\put(1748,200){\makebox(0,0){400}}%
\put(1393,200){\makebox(0,0){300}}%
\put(1038,200){\makebox(0,0){200}}%
\put(684,200){\makebox(0,0){100}}%
\put(350,1956){\makebox(0,0)[r]{0.8}}%
\put(350,1749){\makebox(0,0)[r]{0.7}}%
\put(350,1542){\makebox(0,0)[r]{0.6}}%
\put(350,1335){\makebox(0,0)[r]{0.5}}%
\put(350,1128){\makebox(0,0)[r]{0.4}}%
\put(350,921){\makebox(0,0)[r]{0.3}}%
\put(350,714){\makebox(0,0)[r]{0.2}}%
\put(350,507){\makebox(0,0)[r]{0.1}}%
\put(350,300){\makebox(0,0)[r]{0}}%
\end{picture}%
\endgroup
 

%% file: crossspd4.tex
\begingroup%
  \makeatletter%
  \newcommand{\GNUPLOTspecial}{%
    \@sanitize\catcode`\%=14\relax\special}%
  \setlength{\unitlength}{0.1bp}%
{\GNUPLOTspecial{!
/gnudict 256 dict def
gnudict begin
/Color false def
/Solid false def
/gnulinewidth 5.000 def
/userlinewidth gnulinewidth def
/vshift -33 def
/dl {10 mul} def
/hpt_ 31.5 def
/vpt_ 31.5 def
/hpt hpt_ def
/vpt vpt_ def
/M {moveto} bind def
/L {lineto} bind def
/R {rmoveto} bind def
/V {rlineto} bind def
/vpt2 vpt 2 mul def
/hpt2 hpt 2 mul def
/Lshow { currentpoint stroke M
  0 vshift R show } def
/Rshow { currentpoint stroke M
  dup stringwidth pop neg vshift R show } def
/Cshow { currentpoint stroke M
  dup stringwidth pop -2 div vshift R show } def
/UP { dup vpt_ mul /vpt exch def hpt_ mul /hpt exch def
  /hpt2 hpt 2 mul def /vpt2 vpt 2 mul def } def
/DL { Color {setrgbcolor Solid {pop []} if 0 setdash }
 {pop pop pop Solid {pop []} if 0 setdash} ifelse } def
/BL { stroke gnulinewidth 2 mul setlinewidth } def
/AL { stroke gnulinewidth 2 div setlinewidth } def
/UL { gnulinewidth mul /userlinewidth exch def } def
/PL { stroke userlinewidth setlinewidth } def
/LTb { BL [] 0 0 0 DL } def
/LTa { AL [1 dl 2 dl] 0 setdash 0 0 0 setrgbcolor } def
/LT0 { PL [] 1 0 0 DL } def
/LT1 { PL [4 dl 2 dl] 0 1 0 DL } def
/LT2 { PL [2 dl 3 dl] 0 0 1 DL } def
/LT3 { PL [1 dl 1.5 dl] 1 0 1 DL } def
/LT4 { PL [5 dl 2 dl 1 dl 2 dl] 0 1 1 DL } def
/LT5 { PL [4 dl 3 dl 1 dl 3 dl] 1 1 0 DL } def
/LT6 { PL [2 dl 2 dl 2 dl 4 dl] 0 0 0 DL } def
/LT7 { PL [2 dl 2 dl 2 dl 2 dl 2 dl 4 dl] 1 0.3 0 DL } def
/LT8 { PL [2 dl 2 dl 2 dl 2 dl 2 dl 2 dl 2 dl 4 dl] 0.5 0.5 0.5 DL } def
/Pnt { stroke [] 0 setdash
   gsave 1 setlinecap M 0 0 V stroke grestore } def
/Dia { stroke [] 0 setdash 2 copy vpt add M
  hpt neg vpt neg V hpt vpt neg V
  hpt vpt V hpt neg vpt V closepath stroke
  Pnt } def
/Pls { stroke [] 0 setdash vpt sub M 0 vpt2 V
  currentpoint stroke M
  hpt neg vpt neg R hpt2 0 V stroke
  } def
/Box { stroke [] 0 setdash 2 copy exch hpt sub exch vpt add M
  0 vpt2 neg V hpt2 0 V 0 vpt2 V
  hpt2 neg 0 V closepath stroke
  Pnt } def
/Crs { stroke [] 0 setdash exch hpt sub exch vpt add M
  hpt2 vpt2 neg V currentpoint stroke M
  hpt2 neg 0 R hpt2 vpt2 V stroke } def
/TriU { stroke [] 0 setdash 2 copy vpt 1.12 mul add M
  hpt neg vpt -1.62 mul V
  hpt 2 mul 0 V
  hpt neg vpt 1.62 mul V closepath stroke
  Pnt  } def
/Star { 2 copy Pls Crs } def
/BoxF { stroke [] 0 setdash exch hpt sub exch vpt add M
  0 vpt2 neg V  hpt2 0 V  0 vpt2 V
  hpt2 neg 0 V  closepath fill } def
/TriUF { stroke [] 0 setdash vpt 1.12 mul add M
  hpt neg vpt -1.62 mul V
  hpt 2 mul 0 V
  hpt neg vpt 1.62 mul V closepath fill } def
/TriD { stroke [] 0 setdash 2 copy vpt 1.12 mul sub M
  hpt neg vpt 1.62 mul V
  hpt 2 mul 0 V
  hpt neg vpt -1.62 mul V closepath stroke
  Pnt  } def
/TriDF { stroke [] 0 setdash vpt 1.12 mul sub M
  hpt neg vpt 1.62 mul V
  hpt 2 mul 0 V
  hpt neg vpt -1.62 mul V closepath fill} def
/DiaF { stroke [] 0 setdash vpt add M
  hpt neg vpt neg V hpt vpt neg V
  hpt vpt V hpt neg vpt V closepath fill } def
/Pent { stroke [] 0 setdash 2 copy gsave
  translate 0 hpt M 4 {72 rotate 0 hpt L} repeat
  closepath stroke grestore Pnt } def
/PentF { stroke [] 0 setdash gsave
  translate 0 hpt M 4 {72 rotate 0 hpt L} repeat
  closepath fill grestore } def
/Circle { stroke [] 0 setdash 2 copy
  hpt 0 360 arc stroke Pnt } def
/CircleF { stroke [] 0 setdash hpt 0 360 arc fill } def
/C0 { BL [] 0 setdash 2 copy moveto vpt 90 450  arc } bind def
/C1 { BL [] 0 setdash 2 copy        moveto
       2 copy  vpt 0 90 arc closepath fill
               vpt 0 360 arc closepath } bind def
/C2 { BL [] 0 setdash 2 copy moveto
       2 copy  vpt 90 180 arc closepath fill
               vpt 0 360 arc closepath } bind def
/C3 { BL [] 0 setdash 2 copy moveto
       2 copy  vpt 0 180 arc closepath fill
               vpt 0 360 arc closepath } bind def
/C4 { BL [] 0 setdash 2 copy moveto
       2 copy  vpt 180 270 arc closepath fill
               vpt 0 360 arc closepath } bind def
/C5 { BL [] 0 setdash 2 copy moveto
       2 copy  vpt 0 90 arc
       2 copy moveto
       2 copy  vpt 180 270 arc closepath fill
               vpt 0 360 arc } bind def
/C6 { BL [] 0 setdash 2 copy moveto
      2 copy  vpt 90 270 arc closepath fill
              vpt 0 360 arc closepath } bind def
/C7 { BL [] 0 setdash 2 copy moveto
      2 copy  vpt 0 270 arc closepath fill
              vpt 0 360 arc closepath } bind def
/C8 { BL [] 0 setdash 2 copy moveto
      2 copy vpt 270 360 arc closepath fill
              vpt 0 360 arc closepath } bind def
/C9 { BL [] 0 setdash 2 copy moveto
      2 copy  vpt 270 450 arc closepath fill
              vpt 0 360 arc closepath } bind def
/C10 { BL [] 0 setdash 2 copy 2 copy moveto vpt 270 360 arc closepath fill
       2 copy moveto
       2 copy vpt 90 180 arc closepath fill
               vpt 0 360 arc closepath } bind def
/C11 { BL [] 0 setdash 2 copy moveto
       2 copy  vpt 0 180 arc closepath fill
       2 copy moveto
       2 copy  vpt 270 360 arc closepath fill
               vpt 0 360 arc closepath } bind def
/C12 { BL [] 0 setdash 2 copy moveto
       2 copy  vpt 180 360 arc closepath fill
               vpt 0 360 arc closepath } bind def
/C13 { BL [] 0 setdash  2 copy moveto
       2 copy  vpt 0 90 arc closepath fill
       2 copy moveto
       2 copy  vpt 180 360 arc closepath fill
               vpt 0 360 arc closepath } bind def
/C14 { BL [] 0 setdash 2 copy moveto
       2 copy  vpt 90 360 arc closepath fill
               vpt 0 360 arc } bind def
/C15 { BL [] 0 setdash 2 copy vpt 0 360 arc closepath fill
               vpt 0 360 arc closepath } bind def
/Rec   { newpath 4 2 roll moveto 1 index 0 rlineto 0 exch rlineto
       neg 0 rlineto closepath } bind def
/Square { dup Rec } bind def
/Bsquare { vpt sub exch vpt sub exch vpt2 Square } bind def
/S0 { BL [] 0 setdash 2 copy moveto 0 vpt rlineto BL Bsquare } bind def
/S1 { BL [] 0 setdash 2 copy vpt Square fill Bsquare } bind def
/S2 { BL [] 0 setdash 2 copy exch vpt sub exch vpt Square fill Bsquare } bind def
/S3 { BL [] 0 setdash 2 copy exch vpt sub exch vpt2 vpt Rec fill Bsquare } bind def
/S4 { BL [] 0 setdash 2 copy exch vpt sub exch vpt sub vpt Square fill Bsquare } bind def
/S5 { BL [] 0 setdash 2 copy 2 copy vpt Square fill
       exch vpt sub exch vpt sub vpt Square fill Bsquare } bind def
/S6 { BL [] 0 setdash 2 copy exch vpt sub exch vpt sub vpt vpt2 Rec fill Bsquare } bind def
/S7 { BL [] 0 setdash 2 copy exch vpt sub exch vpt sub vpt vpt2 Rec fill
       2 copy vpt Square fill
       Bsquare } bind def
/S8 { BL [] 0 setdash 2 copy vpt sub vpt Square fill Bsquare } bind def
/S9 { BL [] 0 setdash 2 copy vpt sub vpt vpt2 Rec fill Bsquare } bind def
/S10 { BL [] 0 setdash 2 copy vpt sub vpt Square fill 2 copy exch vpt sub exch vpt Square fill
       Bsquare } bind def
/S11 { BL [] 0 setdash 2 copy vpt sub vpt Square fill 2 copy exch vpt sub exch vpt2 vpt Rec fill
       Bsquare } bind def
/S12 { BL [] 0 setdash 2 copy exch vpt sub exch vpt sub vpt2 vpt Rec fill Bsquare } bind def
/S13 { BL [] 0 setdash 2 copy exch vpt sub exch vpt sub vpt2 vpt Rec fill
       2 copy vpt Square fill Bsquare } bind def
/S14 { BL [] 0 setdash 2 copy exch vpt sub exch vpt sub vpt2 vpt Rec fill
       2 copy exch vpt sub exch vpt Square fill Bsquare } bind def
/S15 { BL [] 0 setdash 2 copy Bsquare fill Bsquare } bind def
/D0 { gsave translate 45 rotate 0 0 S0 stroke grestore } bind def
/D1 { gsave translate 45 rotate 0 0 S1 stroke grestore } bind def
/D2 { gsave translate 45 rotate 0 0 S2 stroke grestore } bind def
/D3 { gsave translate 45 rotate 0 0 S3 stroke grestore } bind def
/D4 { gsave translate 45 rotate 0 0 S4 stroke grestore } bind def
/D5 { gsave translate 45 rotate 0 0 S5 stroke grestore } bind def
/D6 { gsave translate 45 rotate 0 0 S6 stroke grestore } bind def
/D7 { gsave translate 45 rotate 0 0 S7 stroke grestore } bind def
/D8 { gsave translate 45 rotate 0 0 S8 stroke grestore } bind def
/D9 { gsave translate 45 rotate 0 0 S9 stroke grestore } bind def
/D10 { gsave translate 45 rotate 0 0 S10 stroke grestore } bind def
/D11 { gsave translate 45 rotate 0 0 S11 stroke grestore } bind def
/D12 { gsave translate 45 rotate 0 0 S12 stroke grestore } bind def
/D13 { gsave translate 45 rotate 0 0 S13 stroke grestore } bind def
/D14 { gsave translate 45 rotate 0 0 S14 stroke grestore } bind def
/D15 { gsave translate 45 rotate 0 0 S15 stroke grestore } bind def
/DiaE { stroke [] 0 setdash vpt add M
  hpt neg vpt neg V hpt vpt neg V
  hpt vpt V hpt neg vpt V closepath stroke } def
/BoxE { stroke [] 0 setdash exch hpt sub exch vpt add M
  0 vpt2 neg V hpt2 0 V 0 vpt2 V
  hpt2 neg 0 V closepath stroke } def
/TriUE { stroke [] 0 setdash vpt 1.12 mul add M
  hpt neg vpt -1.62 mul V
  hpt 2 mul 0 V
  hpt neg vpt 1.62 mul V closepath stroke } def
/TriDE { stroke [] 0 setdash vpt 1.12 mul sub M
  hpt neg vpt 1.62 mul V
  hpt 2 mul 0 V
  hpt neg vpt -1.62 mul V closepath stroke } def
/PentE { stroke [] 0 setdash gsave
  translate 0 hpt M 4 {72 rotate 0 hpt L} repeat
  closepath stroke grestore } def
/CircE { stroke [] 0 setdash 
  hpt 0 360 arc stroke } def
/Opaque { gsave closepath 1 setgray fill grestore 0 setgray closepath } def
/DiaW { stroke [] 0 setdash vpt add M
  hpt neg vpt neg V hpt vpt neg V
  hpt vpt V hpt neg vpt V Opaque stroke } def
/BoxW { stroke [] 0 setdash exch hpt sub exch vpt add M
  0 vpt2 neg V hpt2 0 V 0 vpt2 V
  hpt2 neg 0 V Opaque stroke } def
/TriUW { stroke [] 0 setdash vpt 1.12 mul add M
  hpt neg vpt -1.62 mul V
  hpt 2 mul 0 V
  hpt neg vpt 1.62 mul V Opaque stroke } def
/TriDW { stroke [] 0 setdash vpt 1.12 mul sub M
  hpt neg vpt 1.62 mul V
  hpt 2 mul 0 V
  hpt neg vpt -1.62 mul V Opaque stroke } def
/PentW { stroke [] 0 setdash gsave
  translate 0 hpt M 4 {72 rotate 0 hpt L} repeat
  Opaque stroke grestore } def
/CircW { stroke [] 0 setdash 
  hpt 0 360 arc Opaque stroke } def
/BoxFill { gsave Rec 1 setgray fill grestore } def
end
}}%
\begin{picture}(3600,2160)(0,0)%
{\GNUPLOTspecial{"
gnudict begin
gsave
0 0 translate
0.100 0.100 scale
0 setgray
newpath
1.000 UL
LTb
400 300 M
63 0 V
2987 0 R
-63 0 V
400 520 M
63 0 V
2987 0 R
-63 0 V
400 740 M
63 0 V
2987 0 R
-63 0 V
400 960 M
63 0 V
2987 0 R
-63 0 V
400 1180 M
63 0 V
2987 0 R
-63 0 V
400 1400 M
63 0 V
2987 0 R
-63 0 V
400 1620 M
63 0 V
2987 0 R
-63 0 V
400 1840 M
63 0 V
2987 0 R
-63 0 V
400 2060 M
63 0 V
2987 0 R
-63 0 V
400 300 M
0 63 V
0 1697 R
0 -63 V
908 300 M
0 63 V
0 1697 R
0 -63 V
1417 300 M
0 63 V
0 1697 R
0 -63 V
1925 300 M
0 63 V
0 1697 R
0 -63 V
2433 300 M
0 63 V
0 1697 R
0 -63 V
2942 300 M
0 63 V
0 1697 R
0 -63 V
3450 300 M
0 63 V
0 1697 R
0 -63 V
1.000 UL
LTb
400 300 M
3050 0 V
0 1760 V
-3050 0 V
400 300 L
1.000 UL
LT0
1620 1752 M
263 0 V
502 326 M
101 33 V
102 40 V
102 42 V
101 44 V
102 47 V
102 49 V
101 50 V
102 51 V
102 53 V
101 54 V
102 55 V
102 58 V
101 60 V
102 59 V
102 59 V
101 58 V
102 58 V
102 57 V
101 57 V
102 55 V
102 56 V
101 57 V
102 54 V
102 54 V
101 51 V
102 56 V
102 54 V
101 53 V
102 53 V
1.000 UL
LT1
1620 1652 M
263 0 V
502 346 M
101 41 V
102 38 V
102 36 V
101 36 V
102 35 V
102 35 V
101 35 V
102 35 V
102 35 V
101 35 V
102 35 V
102 36 V
101 35 V
102 36 V
102 35 V
101 36 V
102 36 V
102 35 V
101 36 V
102 36 V
102 37 V
101 36 V
102 36 V
102 36 V
101 37 V
102 37 V
102 36 V
101 37 V
102 37 V
1.000 UL
LT2
1620 1552 M
263 0 V
603 349 M
204 66 V
203 38 V
203 29 V
204 27 V
203 25 V
203 24 V
204 24 V
203 23 V
203 22 V
204 23 V
203 22 V
203 22 V
204 21 V
203 21 V
1.000 UP
1.000 UL
LT3
1620 1452 M
263 0 V
-263 31 R
0 -62 V
263 62 R
0 -62 V
715 415 M
0 -56 V
-31 56 R
62 0 V
684 359 M
62 0 V
124 94 R
0 -44 V
-31 44 R
62 0 V
839 409 M
62 0 V
69 106 R
0 -63 V
-31 63 R
62 0 V
939 452 M
62 0 V
50 104 R
0 -76 V
-31 76 R
62 0 V
-62 -76 R
62 0 V
55 88 R
0 -80 V
-31 80 R
62 0 V
-62 -80 R
62 0 V
61 135 R
0 -97 V
-31 97 R
62 0 V
-62 -97 R
62 0 V
75 123 R
0 -103 V
-31 103 R
62 0 V
1304 546 M
62 0 V
104 92 R
0 -107 V
-31 107 R
62 0 V
1439 531 M
62 0 V
255 241 R
0 -161 V
-31 161 R
62 0 V
1725 611 M
62 0 V
111 109 R
0 -131 V
-31 131 R
62 0 V
1867 589 M
62 0 V
222 249 R
0 -160 V
-31 160 R
62 0 V
2120 678 M
62 0 V
222 218 R
0 -192 V
-31 192 R
62 0 V
2373 704 M
62 0 V
226 211 R
0 -209 V
-31 209 R
62 0 V
2630 706 M
62 0 V
223 225 R
0 -216 V
-31 216 R
62 0 V
2884 715 M
62 0 V
223 417 R
0 -283 V
-31 283 R
62 0 V
3138 849 M
62 0 V
715 387 Pls
870 431 Pls
970 483 Pls
1051 518 Pls
1137 528 Pls
1229 575 Pls
1335 597 Pls
1470 584 Pls
1756 692 Pls
1898 655 Pls
2151 758 Pls
2404 800 Pls
2661 810 Pls
2915 823 Pls
3169 990 Pls
1751 1452 Pls
1.000 UP
1.000 UL
LT4
1620 1352 M
263 0 V
-263 31 R
0 -62 V
263 62 R
0 -62 V
807 501 M
0 -47 V
-31 47 R
62 0 V
776 454 M
62 0 V
172 114 R
0 -61 V
-31 61 R
62 0 V
979 507 M
62 0 V
172 123 R
0 -78 V
-31 78 R
62 0 V
-62 -78 R
62 0 V
173 102 R
0 -84 V
-31 84 R
62 0 V
-62 -84 R
62 0 V
172 140 R
0 -98 V
-31 98 R
62 0 V
-62 -98 R
62 0 V
172 188 R
0 -119 V
-31 119 R
62 0 V
1792 681 M
62 0 V
173 135 R
0 -148 V
-31 148 R
62 0 V
1996 668 M
62 0 V
807 480 Crs
1010 536 Crs
1213 589 Crs
1417 612 Crs
1620 664 Crs
1823 739 Crs
2027 740 Crs
1751 1352 Crs
1.000 UP
1.000 UL
LT5
1620 1252 M
263 0 V
-263 31 R
0 -62 V
263 62 R
0 -62 V
542 359 M
0 -21 V
-31 21 R
62 0 V
511 338 M
62 0 V
542 349 Star
1751 1252 Star
1.000 UP
1.000 UL
LT6
1620 1152 M
263 0 V
-263 31 R
0 -62 V
263 62 R
0 -62 V
520 362 M
0 -26 V
-31 26 R
62 0 V
489 336 M
62 0 V
7 45 R
0 -43 V
-31 43 R
62 0 V
527 338 M
62 0 V
14 64 R
0 -48 V
-31 48 R
62 0 V
572 354 M
62 0 V
15 72 R
0 -57 V
-31 57 R
62 0 V
618 369 M
62 0 V
520 349 Box
558 359 Box
603 378 Box
649 397 Box
1751 1152 Box
stroke
grestore
end
showpage
}}%
\put(1570,1152){\makebox(0,0)[r]{E401}}%
\put(1570,1252){\makebox(0,0)[r]{E516}}%
\put(1570,1352){\makebox(0,0)[r]{ZEUS 96+97 prel.}}%
\put(1570,1452){\makebox(0,0)[r]{H1}}%
\put(1570,1552){\makebox(0,0)[r]{MRSTLO, $\lambda =4$}}%
\put(1570,1652){\makebox(0,0)[r]{CTEQ4L,$\lambda =4$}}%
\put(1570,1752){\makebox(0,0)[r]{CTEQ4L, $\lambda = 10$}}%
\put(1925,50){\makebox(0,0){\Large $W$ ~(GeV)}}%
\put(100,1180){%
\special{ps: gsave currentpoint currentpoint translate
270 rotate neg exch neg exch translate}%
\makebox(0,0)[b]{\shortstack{\Large$ \sigma (\gamma P \rightarrow J/\psi P) $(nb)}}%
\special{ps: currentpoint grestore moveto}%
}%
\put(3450,200){\makebox(0,0){300}}%
\put(2942,200){\makebox(0,0){250}}%
\put(2433,200){\makebox(0,0){200}}%
\put(1925,200){\makebox(0,0){150}}%
\put(1417,200){\makebox(0,0){100}}%
\put(908,200){\makebox(0,0){50}}%
\put(400,200){\makebox(0,0){0}}%
\put(350,2060){\makebox(0,0)[r]{400}}%
\put(350,1840){\makebox(0,0)[r]{350}}%
\put(350,1620){\makebox(0,0)[r]{300}}%
\put(350,1400){\makebox(0,0)[r]{250}}%
\put(350,1180){\makebox(0,0)[r]{200}}%
\put(350,960){\makebox(0,0)[r]{150}}%
\put(350,740){\makebox(0,0)[r]{100}}%
\put(350,520){\makebox(0,0)[r]{50}}%
\put(350,300){\makebox(0,0)[r]{0}}%
\end{picture}%
\endgroup
 

%% file: sighat.tex
\begingroup%
  \makeatletter%
  \newcommand{\GNUPLOTspecial}{%
    \@sanitize\catcode`\%=14\relax\special}%
  \setlength{\unitlength}{0.1bp}%
{\GNUPLOTspecial{!
/gnudict 256 dict def
gnudict begin
/Color false def
/Solid false def
/gnulinewidth 5.000 def
/userlinewidth gnulinewidth def
/vshift -33 def
/dl {10 mul} def
/hpt_ 31.5 def
/vpt_ 31.5 def
/hpt hpt_ def
/vpt vpt_ def
/M {moveto} bind def
/L {lineto} bind def
/R {rmoveto} bind def
/V {rlineto} bind def
/vpt2 vpt 2 mul def
/hpt2 hpt 2 mul def
/Lshow { currentpoint stroke M
  0 vshift R show } def
/Rshow { currentpoint stroke M
  dup stringwidth pop neg vshift R show } def
/Cshow { currentpoint stroke M
  dup stringwidth pop -2 div vshift R show } def
/UP { dup vpt_ mul /vpt exch def hpt_ mul /hpt exch def
  /hpt2 hpt 2 mul def /vpt2 vpt 2 mul def } def
/DL { Color {setrgbcolor Solid {pop []} if 0 setdash }
 {pop pop pop Solid {pop []} if 0 setdash} ifelse } def
/BL { stroke gnulinewidth 2 mul setlinewidth } def
/AL { stroke gnulinewidth 2 div setlinewidth } def
/UL { gnulinewidth mul /userlinewidth exch def } def
/PL { stroke userlinewidth setlinewidth } def
/LTb { BL [] 0 0 0 DL } def
/LTa { AL [1 dl 2 dl] 0 setdash 0 0 0 setrgbcolor } def
/LT0 { PL [] 1 0 0 DL } def
/LT1 { PL [4 dl 2 dl] 0 1 0 DL } def
/LT2 { PL [2 dl 3 dl] 0 0 1 DL } def
/LT3 { PL [1 dl 1.5 dl] 1 0 1 DL } def
/LT4 { PL [5 dl 2 dl 1 dl 2 dl] 0 1 1 DL } def
/LT5 { PL [4 dl 3 dl 1 dl 3 dl] 1 1 0 DL } def
/LT6 { PL [2 dl 2 dl 2 dl 4 dl] 0 0 0 DL } def
/LT7 { PL [2 dl 2 dl 2 dl 2 dl 2 dl 4 dl] 1 0.3 0 DL } def
/LT8 { PL [2 dl 2 dl 2 dl 2 dl 2 dl 2 dl 2 dl 4 dl] 0.5 0.5 0.5 DL } def
/Pnt { stroke [] 0 setdash
   gsave 1 setlinecap M 0 0 V stroke grestore } def
/Dia { stroke [] 0 setdash 2 copy vpt add M
  hpt neg vpt neg V hpt vpt neg V
  hpt vpt V hpt neg vpt V closepath stroke
  Pnt } def
/Pls { stroke [] 0 setdash vpt sub M 0 vpt2 V
  currentpoint stroke M
  hpt neg vpt neg R hpt2 0 V stroke
  } def
/Box { stroke [] 0 setdash 2 copy exch hpt sub exch vpt add M
  0 vpt2 neg V hpt2 0 V 0 vpt2 V
  hpt2 neg 0 V closepath stroke
  Pnt } def
/Crs { stroke [] 0 setdash exch hpt sub exch vpt add M
  hpt2 vpt2 neg V currentpoint stroke M
  hpt2 neg 0 R hpt2 vpt2 V stroke } def
/TriU { stroke [] 0 setdash 2 copy vpt 1.12 mul add M
  hpt neg vpt -1.62 mul V
  hpt 2 mul 0 V
  hpt neg vpt 1.62 mul V closepath stroke
  Pnt  } def
/Star { 2 copy Pls Crs } def
/BoxF { stroke [] 0 setdash exch hpt sub exch vpt add M
  0 vpt2 neg V  hpt2 0 V  0 vpt2 V
  hpt2 neg 0 V  closepath fill } def
/TriUF { stroke [] 0 setdash vpt 1.12 mul add M
  hpt neg vpt -1.62 mul V
  hpt 2 mul 0 V
  hpt neg vpt 1.62 mul V closepath fill } def
/TriD { stroke [] 0 setdash 2 copy vpt 1.12 mul sub M
  hpt neg vpt 1.62 mul V
  hpt 2 mul 0 V
  hpt neg vpt -1.62 mul V closepath stroke
  Pnt  } def
/TriDF { stroke [] 0 setdash vpt 1.12 mul sub M
  hpt neg vpt 1.62 mul V
  hpt 2 mul 0 V
  hpt neg vpt -1.62 mul V closepath fill} def
/DiaF { stroke [] 0 setdash vpt add M
  hpt neg vpt neg V hpt vpt neg V
  hpt vpt V hpt neg vpt V closepath fill } def
/Pent { stroke [] 0 setdash 2 copy gsave
  translate 0 hpt M 4 {72 rotate 0 hpt L} repeat
  closepath stroke grestore Pnt } def
/PentF { stroke [] 0 setdash gsave
  translate 0 hpt M 4 {72 rotate 0 hpt L} repeat
  closepath fill grestore } def
/Circle { stroke [] 0 setdash 2 copy
  hpt 0 360 arc stroke Pnt } def
/CircleF { stroke [] 0 setdash hpt 0 360 arc fill } def
/C0 { BL [] 0 setdash 2 copy moveto vpt 90 450  arc } bind def
/C1 { BL [] 0 setdash 2 copy        moveto
       2 copy  vpt 0 90 arc closepath fill
               vpt 0 360 arc closepath } bind def
/C2 { BL [] 0 setdash 2 copy moveto
       2 copy  vpt 90 180 arc closepath fill
               vpt 0 360 arc closepath } bind def
/C3 { BL [] 0 setdash 2 copy moveto
       2 copy  vpt 0 180 arc closepath fill
               vpt 0 360 arc closepath } bind def
/C4 { BL [] 0 setdash 2 copy moveto
       2 copy  vpt 180 270 arc closepath fill
               vpt 0 360 arc closepath } bind def
/C5 { BL [] 0 setdash 2 copy moveto
       2 copy  vpt 0 90 arc
       2 copy moveto
       2 copy  vpt 180 270 arc closepath fill
               vpt 0 360 arc } bind def
/C6 { BL [] 0 setdash 2 copy moveto
      2 copy  vpt 90 270 arc closepath fill
              vpt 0 360 arc closepath } bind def
/C7 { BL [] 0 setdash 2 copy moveto
      2 copy  vpt 0 270 arc closepath fill
              vpt 0 360 arc closepath } bind def
/C8 { BL [] 0 setdash 2 copy moveto
      2 copy vpt 270 360 arc closepath fill
              vpt 0 360 arc closepath } bind def
/C9 { BL [] 0 setdash 2 copy moveto
      2 copy  vpt 270 450 arc closepath fill
              vpt 0 360 arc closepath } bind def
/C10 { BL [] 0 setdash 2 copy 2 copy moveto vpt 270 360 arc closepath fill
       2 copy moveto
       2 copy vpt 90 180 arc closepath fill
               vpt 0 360 arc closepath } bind def
/C11 { BL [] 0 setdash 2 copy moveto
       2 copy  vpt 0 180 arc closepath fill
       2 copy moveto
       2 copy  vpt 270 360 arc closepath fill
               vpt 0 360 arc closepath } bind def
/C12 { BL [] 0 setdash 2 copy moveto
       2 copy  vpt 180 360 arc closepath fill
               vpt 0 360 arc closepath } bind def
/C13 { BL [] 0 setdash  2 copy moveto
       2 copy  vpt 0 90 arc closepath fill
       2 copy moveto
       2 copy  vpt 180 360 arc closepath fill
               vpt 0 360 arc closepath } bind def
/C14 { BL [] 0 setdash 2 copy moveto
       2 copy  vpt 90 360 arc closepath fill
               vpt 0 360 arc } bind def
/C15 { BL [] 0 setdash 2 copy vpt 0 360 arc closepath fill
               vpt 0 360 arc closepath } bind def
/Rec   { newpath 4 2 roll moveto 1 index 0 rlineto 0 exch rlineto
       neg 0 rlineto closepath } bind def
/Square { dup Rec } bind def
/Bsquare { vpt sub exch vpt sub exch vpt2 Square } bind def
/S0 { BL [] 0 setdash 2 copy moveto 0 vpt rlineto BL Bsquare } bind def
/S1 { BL [] 0 setdash 2 copy vpt Square fill Bsquare } bind def
/S2 { BL [] 0 setdash 2 copy exch vpt sub exch vpt Square fill Bsquare } bind def
/S3 { BL [] 0 setdash 2 copy exch vpt sub exch vpt2 vpt Rec fill Bsquare } bind def
/S4 { BL [] 0 setdash 2 copy exch vpt sub exch vpt sub vpt Square fill Bsquare } bind def
/S5 { BL [] 0 setdash 2 copy 2 copy vpt Square fill
       exch vpt sub exch vpt sub vpt Square fill Bsquare } bind def
/S6 { BL [] 0 setdash 2 copy exch vpt sub exch vpt sub vpt vpt2 Rec fill Bsquare } bind def
/S7 { BL [] 0 setdash 2 copy exch vpt sub exch vpt sub vpt vpt2 Rec fill
       2 copy vpt Square fill
       Bsquare } bind def
/S8 { BL [] 0 setdash 2 copy vpt sub vpt Square fill Bsquare } bind def
/S9 { BL [] 0 setdash 2 copy vpt sub vpt vpt2 Rec fill Bsquare } bind def
/S10 { BL [] 0 setdash 2 copy vpt sub vpt Square fill 2 copy exch vpt sub exch vpt Square fill
       Bsquare } bind def
/S11 { BL [] 0 setdash 2 copy vpt sub vpt Square fill 2 copy exch vpt sub exch vpt2 vpt Rec fill
       Bsquare } bind def
/S12 { BL [] 0 setdash 2 copy exch vpt sub exch vpt sub vpt2 vpt Rec fill Bsquare } bind def
/S13 { BL [] 0 setdash 2 copy exch vpt sub exch vpt sub vpt2 vpt Rec fill
       2 copy vpt Square fill Bsquare } bind def
/S14 { BL [] 0 setdash 2 copy exch vpt sub exch vpt sub vpt2 vpt Rec fill
       2 copy exch vpt sub exch vpt Square fill Bsquare } bind def
/S15 { BL [] 0 setdash 2 copy Bsquare fill Bsquare } bind def
/D0 { gsave translate 45 rotate 0 0 S0 stroke grestore } bind def
/D1 { gsave translate 45 rotate 0 0 S1 stroke grestore } bind def
/D2 { gsave translate 45 rotate 0 0 S2 stroke grestore } bind def
/D3 { gsave translate 45 rotate 0 0 S3 stroke grestore } bind def
/D4 { gsave translate 45 rotate 0 0 S4 stroke grestore } bind def
/D5 { gsave translate 45 rotate 0 0 S5 stroke grestore } bind def
/D6 { gsave translate 45 rotate 0 0 S6 stroke grestore } bind def
/D7 { gsave translate 45 rotate 0 0 S7 stroke grestore } bind def
/D8 { gsave translate 45 rotate 0 0 S8 stroke grestore } bind def
/D9 { gsave translate 45 rotate 0 0 S9 stroke grestore } bind def
/D10 { gsave translate 45 rotate 0 0 S10 stroke grestore } bind def
/D11 { gsave translate 45 rotate 0 0 S11 stroke grestore } bind def
/D12 { gsave translate 45 rotate 0 0 S12 stroke grestore } bind def
/D13 { gsave translate 45 rotate 0 0 S13 stroke grestore } bind def
/D14 { gsave translate 45 rotate 0 0 S14 stroke grestore } bind def
/D15 { gsave translate 45 rotate 0 0 S15 stroke grestore } bind def
/DiaE { stroke [] 0 setdash vpt add M
  hpt neg vpt neg V hpt vpt neg V
  hpt vpt V hpt neg vpt V closepath stroke } def
/BoxE { stroke [] 0 setdash exch hpt sub exch vpt add M
  0 vpt2 neg V hpt2 0 V 0 vpt2 V
  hpt2 neg 0 V closepath stroke } def
/TriUE { stroke [] 0 setdash vpt 1.12 mul add M
  hpt neg vpt -1.62 mul V
  hpt 2 mul 0 V
  hpt neg vpt 1.62 mul V closepath stroke } def
/TriDE { stroke [] 0 setdash vpt 1.12 mul sub M
  hpt neg vpt 1.62 mul V
  hpt 2 mul 0 V
  hpt neg vpt -1.62 mul V closepath stroke } def
/PentE { stroke [] 0 setdash gsave
  translate 0 hpt M 4 {72 rotate 0 hpt L} repeat
  closepath stroke grestore } def
/CircE { stroke [] 0 setdash 
  hpt 0 360 arc stroke } def
/Opaque { gsave closepath 1 setgray fill grestore 0 setgray closepath } def
/DiaW { stroke [] 0 setdash vpt add M
  hpt neg vpt neg V hpt vpt neg V
  hpt vpt V hpt neg vpt V Opaque stroke } def
/BoxW { stroke [] 0 setdash exch hpt sub exch vpt add M
  0 vpt2 neg V hpt2 0 V 0 vpt2 V
  hpt2 neg 0 V Opaque stroke } def
/TriUW { stroke [] 0 setdash vpt 1.12 mul add M
  hpt neg vpt -1.62 mul V
  hpt 2 mul 0 V
  hpt neg vpt 1.62 mul V Opaque stroke } def
/TriDW { stroke [] 0 setdash vpt 1.12 mul sub M
  hpt neg vpt 1.62 mul V
  hpt 2 mul 0 V
  hpt neg vpt -1.62 mul V Opaque stroke } def
/PentW { stroke [] 0 setdash gsave
  translate 0 hpt M 4 {72 rotate 0 hpt L} repeat
  Opaque stroke grestore } def
/CircW { stroke [] 0 setdash 
  hpt 0 360 arc Opaque stroke } def
/BoxFill { gsave Rec 1 setgray fill grestore } def
end
}}%
\begin{picture}(3600,2160)(0,0)%
{\GNUPLOTspecial{"
gnudict begin
gsave
0 0 translate
0.100 0.100 scale
0 setgray
newpath
1.000 UL
LTb
350 300 M
63 0 V
3037 0 R
-63 0 V
350 496 M
63 0 V
3037 0 R
-63 0 V
350 691 M
63 0 V
3037 0 R
-63 0 V
350 887 M
63 0 V
3037 0 R
-63 0 V
350 1082 M
63 0 V
3037 0 R
-63 0 V
350 1278 M
63 0 V
3037 0 R
-63 0 V
350 1473 M
63 0 V
3037 0 R
-63 0 V
350 1669 M
63 0 V
3037 0 R
-63 0 V
350 1864 M
63 0 V
3037 0 R
-63 0 V
350 2060 M
63 0 V
3037 0 R
-63 0 V
350 300 M
0 63 V
0 1697 R
0 -63 V
660 300 M
0 63 V
0 1697 R
0 -63 V
970 300 M
0 63 V
0 1697 R
0 -63 V
1280 300 M
0 63 V
0 1697 R
0 -63 V
1590 300 M
0 63 V
0 1697 R
0 -63 V
1900 300 M
0 63 V
0 1697 R
0 -63 V
2210 300 M
0 63 V
0 1697 R
0 -63 V
2520 300 M
0 63 V
0 1697 R
0 -63 V
2830 300 M
0 63 V
0 1697 R
0 -63 V
3140 300 M
0 63 V
0 1697 R
0 -63 V
3450 300 M
0 63 V
0 1697 R
0 -63 V
1.000 UL
LTb
350 300 M
3100 0 V
0 1760 V
-3100 0 V
350 300 L
1.000 UL
LT0
970 1669 M
263 0 V
353 300 M
10 0 V
11 0 V
10 0 V
10 0 V
11 0 V
10 1 V
10 0 V
11 1 V
10 1 V
10 1 V
11 1 V
10 2 V
10 1 V
11 2 V
10 2 V
10 3 V
11 2 V
10 3 V
10 3 V
11 4 V
10 3 V
10 4 V
11 4 V
10 5 V
10 4 V
10 5 V
11 5 V
10 5 V
10 6 V
11 5 V
10 6 V
10 6 V
11 6 V
10 6 V
10 7 V
11 6 V
10 7 V
10 7 V
11 6 V
10 7 V
10 8 V
11 7 V
10 7 V
10 8 V
11 7 V
10 7 V
10 7 V
11 8 V
10 8 V
10 8 V
11 7 V
10 8 V
10 7 V
11 7 V
10 8 V
10 7 V
11 7 V
10 8 V
10 7 V
10 8 V
11 7 V
10 8 V
10 7 V
11 7 V
10 7 V
10 7 V
11 6 V
10 7 V
10 6 V
11 6 V
10 6 V
10 6 V
11 6 V
10 7 V
10 8 V
11 8 V
10 8 V
10 8 V
11 8 V
10 8 V
10 8 V
11 8 V
10 8 V
10 8 V
11 8 V
10 8 V
10 8 V
11 8 V
10 8 V
10 8 V
10 8 V
11 8 V
10 8 V
10 8 V
11 8 V
10 8 V
10 8 V
11 8 V
10 8 V
10 8 V
11 8 V
10 8 V
10 8 V
11 8 V
10 8 V
10 8 V
11 7 V
10 8 V
10 8 V
11 8 V
10 8 V
10 8 V
11 8 V
10 8 V
10 8 V
11 8 V
10 8 V
10 8 V
11 8 V
10 8 V
10 8 V
11 8 V
10 8 V
10 8 V
10 8 V
11 8 V
10 8 V
10 8 V
11 8 V
10 8 V
10 8 V
11 8 V
10 8 V
10 8 V
11 8 V
10 8 V
10 8 V
11 8 V
10 8 V
10 8 V
11 8 V
10 8 V
10 8 V
11 8 V
10 8 V
10 8 V
11 8 V
10 8 V
10 8 V
11 8 V
10 8 V
10 8 V
11 8 V
10 8 V
10 8 V
10 8 V
11 8 V
10 8 V
10 8 V
11 8 V
10 8 V
10 8 V
11 8 V
10 8 V
10 8 V
11 8 V
10 8 V
10 8 V
11 8 V
10 8 V
10 8 V
11 8 V
10 8 V
10 8 V
11 7 V
10 8 V
10 8 V
11 8 V
10 8 V
10 8 V
11 8 V
10 8 V
10 8 V
11 8 V
10 8 V
10 8 V
11 8 V
10 8 V
10 8 V
10 8 V
11 8 V
10 8 V
10 8 V
11 8 V
10 8 V
10 4 V
11 5 V
10 4 V
10 5 V
11 4 V
10 5 V
10 4 V
11 4 V
10 5 V
10 4 V
11 4 V
10 4 V
10 4 V
11 4 V
10 4 V
10 4 V
11 4 V
10 4 V
10 4 V
11 4 V
10 4 V
10 4 V
11 4 V
10 3 V
10 4 V
10 4 V
11 3 V
10 4 V
10 4 V
11 3 V
10 4 V
10 3 V
11 4 V
10 3 V
10 4 V
11 3 V
10 3 V
10 4 V
11 3 V
10 3 V
10 4 V
11 3 V
10 3 V
10 3 V
11 3 V
10 3 V
10 3 V
11 4 V
10 3 V
10 3 V
11 3 V
10 3 V
10 3 V
11 2 V
10 3 V
10 3 V
10 3 V
11 3 V
10 3 V
10 2 V
11 3 V
10 3 V
10 3 V
11 2 V
10 3 V
10 3 V
11 2 V
10 3 V
10 2 V
11 3 V
10 3 V
10 2 V
11 3 V
10 2 V
10 2 V
11 3 V
10 2 V
10 3 V
11 2 V
10 2 V
10 3 V
11 2 V
10 2 V
10 3 V
11 2 V
10 2 V
10 2 V
11 3 V
10 2 V
10 2 V
10 2 V
11 2 V
10 3 V
10 2 V
11 2 V
10 2 V
10 2 V
11 2 V
10 2 V
10 2 V
11 2 V
10 2 V
10 2 V
11 2 V
10 2 V
1.000 UL
LT1
970 1569 M
263 0 V
353 300 M
10 0 V
11 0 V
10 0 V
10 0 V
11 0 V
10 0 V
10 1 V
11 0 V
10 1 V
10 0 V
11 1 V
10 1 V
10 1 V
11 1 V
10 2 V
10 1 V
11 2 V
10 3 V
10 2 V
11 2 V
10 3 V
10 3 V
11 4 V
10 3 V
10 4 V
10 4 V
11 4 V
10 4 V
10 5 V
11 6 V
10 5 V
10 5 V
11 6 V
10 6 V
10 6 V
11 6 V
10 7 V
10 7 V
11 7 V
10 7 V
10 8 V
11 8 V
10 8 V
10 8 V
11 9 V
10 8 V
10 8 V
11 10 V
10 9 V
10 10 V
11 9 V
10 9 V
10 10 V
11 9 V
10 11 V
10 10 V
11 11 V
10 11 V
10 10 V
10 11 V
11 10 V
10 10 V
10 11 V
11 12 V
10 12 V
10 11 V
11 12 V
10 12 V
10 11 V
11 11 V
10 12 V
10 11 V
11 11 V
10 11 V
10 13 V
11 12 V
10 13 V
10 12 V
11 12 V
10 12 V
10 12 V
11 12 V
10 12 V
10 12 V
11 11 V
10 12 V
10 11 V
11 11 V
10 11 V
10 12 V
10 12 V
11 11 V
10 7 V
10 6 V
11 7 V
10 6 V
10 7 V
11 6 V
10 7 V
10 6 V
11 7 V
10 6 V
10 7 V
11 6 V
10 6 V
10 7 V
11 6 V
10 7 V
10 6 V
11 7 V
10 6 V
10 7 V
11 6 V
10 7 V
10 6 V
11 7 V
10 6 V
10 7 V
11 6 V
10 7 V
10 6 V
11 7 V
10 6 V
10 7 V
10 6 V
11 7 V
10 6 V
10 7 V
11 6 V
10 7 V
10 6 V
11 7 V
10 6 V
10 7 V
11 6 V
10 6 V
10 7 V
11 6 V
10 7 V
10 6 V
11 7 V
10 6 V
10 7 V
11 6 V
10 7 V
10 6 V
11 7 V
10 6 V
10 7 V
11 6 V
10 7 V
10 6 V
11 7 V
10 6 V
10 7 V
10 6 V
11 7 V
10 6 V
10 7 V
11 6 V
10 7 V
10 6 V
11 7 V
10 6 V
10 7 V
11 6 V
10 6 V
10 7 V
11 6 V
10 7 V
10 6 V
11 7 V
10 6 V
10 7 V
11 6 V
10 7 V
10 6 V
11 7 V
10 6 V
10 7 V
11 6 V
10 7 V
10 6 V
11 7 V
10 6 V
10 7 V
11 6 V
10 7 V
10 6 V
10 7 V
11 6 V
10 7 V
10 6 V
11 7 V
10 6 V
10 4 V
11 5 V
10 4 V
10 5 V
11 4 V
10 5 V
10 4 V
11 4 V
10 5 V
10 4 V
11 4 V
10 4 V
10 4 V
11 4 V
10 4 V
10 4 V
11 4 V
10 4 V
10 4 V
11 4 V
10 4 V
10 4 V
11 4 V
10 3 V
10 4 V
10 4 V
11 3 V
10 4 V
10 4 V
11 3 V
10 4 V
10 3 V
11 4 V
10 3 V
10 4 V
11 3 V
10 3 V
10 4 V
11 3 V
10 3 V
10 4 V
11 3 V
10 3 V
10 3 V
11 3 V
10 3 V
10 3 V
11 4 V
10 3 V
10 3 V
11 3 V
10 3 V
10 3 V
11 2 V
10 3 V
10 3 V
10 3 V
11 3 V
10 3 V
10 2 V
11 3 V
10 3 V
10 3 V
11 2 V
10 3 V
10 3 V
11 2 V
10 3 V
10 2 V
11 3 V
10 3 V
10 2 V
11 3 V
10 2 V
10 2 V
11 3 V
10 2 V
10 3 V
11 2 V
10 2 V
10 3 V
11 2 V
10 2 V
10 3 V
11 2 V
10 2 V
10 2 V
11 3 V
10 2 V
10 2 V
10 2 V
11 2 V
10 3 V
10 2 V
11 2 V
10 2 V
10 2 V
11 2 V
10 2 V
10 2 V
11 2 V
10 2 V
10 2 V
11 2 V
10 2 V
1.000 UL
LT0
353 300 M
10 0 V
11 0 V
10 0 V
10 0 V
11 0 V
10 0 V
10 0 V
11 0 V
10 1 V
10 0 V
11 0 V
10 1 V
10 0 V
11 1 V
10 1 V
10 0 V
11 1 V
10 1 V
10 1 V
11 2 V
10 1 V
10 1 V
11 2 V
10 2 V
10 2 V
10 2 V
11 2 V
10 2 V
10 2 V
11 3 V
10 2 V
10 3 V
11 2 V
10 3 V
10 3 V
11 3 V
10 3 V
10 3 V
11 3 V
10 4 V
10 3 V
11 4 V
10 3 V
10 3 V
11 4 V
10 4 V
10 4 V
11 4 V
10 3 V
10 4 V
11 4 V
10 3 V
10 5 V
11 4 V
10 4 V
10 4 V
11 4 V
10 4 V
10 4 V
10 4 V
11 4 V
10 4 V
10 3 V
11 4 V
10 4 V
10 4 V
11 4 V
10 4 V
10 4 V
11 4 V
10 4 V
10 4 V
11 3 V
10 5 V
10 8 V
11 7 V
10 8 V
10 8 V
11 7 V
10 8 V
10 8 V
11 7 V
10 8 V
10 8 V
11 8 V
10 7 V
10 8 V
11 8 V
10 7 V
10 8 V
10 8 V
11 7 V
10 8 V
10 8 V
11 8 V
10 7 V
10 8 V
11 8 V
10 7 V
10 8 V
11 8 V
10 7 V
10 8 V
11 8 V
10 7 V
10 8 V
11 8 V
10 8 V
10 7 V
11 8 V
10 8 V
10 7 V
11 8 V
10 8 V
10 7 V
11 8 V
10 8 V
10 7 V
11 8 V
10 8 V
10 8 V
11 7 V
10 8 V
10 8 V
10 7 V
11 8 V
10 8 V
10 7 V
11 8 V
10 8 V
10 8 V
11 7 V
10 8 V
10 8 V
11 7 V
10 8 V
10 8 V
11 7 V
10 8 V
10 8 V
11 7 V
10 8 V
10 8 V
11 8 V
10 7 V
10 8 V
11 8 V
10 7 V
10 8 V
11 8 V
10 7 V
10 8 V
11 8 V
10 8 V
10 7 V
10 8 V
11 8 V
10 7 V
10 8 V
11 8 V
10 7 V
10 8 V
11 8 V
10 7 V
10 8 V
11 8 V
10 8 V
10 7 V
11 8 V
10 8 V
10 7 V
11 8 V
10 8 V
10 7 V
11 8 V
10 8 V
10 7 V
11 8 V
10 8 V
10 8 V
11 7 V
10 8 V
10 8 V
11 7 V
10 8 V
10 8 V
11 7 V
10 8 V
10 8 V
10 8 V
11 7 V
10 8 V
10 8 V
11 7 V
10 8 V
10 3 V
11 4 V
10 4 V
10 4 V
11 3 V
10 4 V
10 4 V
11 3 V
10 4 V
10 3 V
11 4 V
10 3 V
10 4 V
11 3 V
10 3 V
10 4 V
11 3 V
10 3 V
10 3 V
11 4 V
10 3 V
10 3 V
11 3 V
10 3 V
10 3 V
10 4 V
11 3 V
10 3 V
10 3 V
11 3 V
10 3 V
10 2 V
11 3 V
10 3 V
10 3 V
11 3 V
10 3 V
10 2 V
11 3 V
10 3 V
10 3 V
11 2 V
10 3 V
10 3 V
11 2 V
10 3 V
10 2 V
11 3 V
10 2 V
10 3 V
11 2 V
10 3 V
10 2 V
11 3 V
10 2 V
10 3 V
10 2 V
11 2 V
10 3 V
10 2 V
11 2 V
10 2 V
10 3 V
11 2 V
10 2 V
10 2 V
11 3 V
10 2 V
10 2 V
11 2 V
10 2 V
10 2 V
11 2 V
10 2 V
10 2 V
11 2 V
10 2 V
10 2 V
11 2 V
10 2 V
10 2 V
11 2 V
10 2 V
10 2 V
11 2 V
10 2 V
10 2 V
11 2 V
10 1 V
10 2 V
10 2 V
11 2 V
10 2 V
10 1 V
11 2 V
10 2 V
10 1 V
11 2 V
10 2 V
10 2 V
11 1 V
10 2 V
10 2 V
11 1 V
10 2 V
1.000 UL
LT1
353 300 M
10 0 V
11 0 V
10 0 V
10 0 V
11 0 V
10 0 V
10 0 V
11 0 V
10 0 V
10 0 V
11 0 V
10 1 V
10 0 V
11 0 V
10 1 V
10 0 V
11 1 V
10 0 V
10 1 V
11 1 V
10 1 V
10 1 V
11 1 V
10 1 V
10 1 V
10 2 V
11 1 V
10 2 V
10 1 V
11 2 V
10 2 V
10 2 V
11 2 V
10 3 V
10 2 V
11 2 V
10 3 V
10 3 V
11 3 V
10 3 V
10 3 V
11 3 V
10 3 V
10 4 V
11 3 V
10 4 V
10 4 V
11 4 V
10 3 V
10 4 V
11 4 V
10 5 V
10 4 V
11 5 V
10 4 V
10 5 V
11 4 V
10 5 V
10 5 V
10 5 V
11 5 V
10 5 V
10 6 V
11 5 V
10 5 V
10 6 V
11 5 V
10 5 V
10 6 V
11 5 V
10 6 V
10 6 V
11 6 V
10 6 V
10 6 V
11 6 V
10 6 V
10 6 V
11 6 V
10 6 V
10 6 V
11 6 V
10 6 V
10 7 V
11 6 V
10 7 V
10 6 V
11 7 V
10 6 V
10 6 V
10 7 V
11 6 V
10 6 V
10 7 V
11 6 V
10 6 V
10 6 V
11 6 V
10 6 V
10 6 V
11 6 V
10 7 V
10 6 V
11 6 V
10 7 V
10 6 V
11 7 V
10 6 V
10 6 V
11 6 V
10 6 V
10 6 V
11 6 V
10 6 V
10 6 V
11 6 V
10 6 V
10 8 V
11 9 V
10 8 V
10 8 V
11 9 V
10 8 V
10 8 V
10 9 V
11 8 V
10 9 V
10 8 V
11 8 V
10 9 V
10 8 V
11 8 V
10 9 V
10 8 V
11 8 V
10 9 V
10 8 V
11 8 V
10 9 V
10 8 V
11 9 V
10 8 V
10 8 V
11 9 V
10 8 V
10 8 V
11 9 V
10 8 V
10 8 V
11 9 V
10 8 V
10 9 V
11 8 V
10 8 V
10 9 V
10 8 V
11 8 V
10 9 V
10 8 V
11 8 V
10 9 V
10 8 V
11 9 V
10 8 V
10 8 V
11 9 V
10 8 V
10 8 V
11 9 V
10 8 V
10 8 V
11 9 V
10 8 V
10 8 V
11 9 V
10 8 V
10 9 V
11 8 V
10 8 V
10 9 V
11 8 V
10 8 V
10 9 V
11 8 V
10 8 V
10 9 V
11 8 V
10 9 V
10 8 V
10 8 V
11 9 V
10 8 V
10 8 V
11 9 V
10 8 V
10 3 V
11 4 V
10 4 V
10 4 V
11 3 V
10 4 V
10 4 V
11 3 V
10 4 V
10 3 V
11 4 V
10 3 V
10 4 V
11 3 V
10 3 V
10 4 V
11 3 V
10 3 V
10 3 V
11 4 V
10 3 V
10 3 V
11 3 V
10 3 V
10 3 V
10 4 V
11 3 V
10 3 V
10 3 V
11 3 V
10 3 V
10 2 V
11 3 V
10 3 V
10 3 V
11 3 V
10 3 V
10 2 V
11 3 V
10 3 V
10 3 V
11 2 V
10 3 V
10 3 V
11 2 V
10 3 V
10 2 V
11 3 V
10 2 V
10 3 V
11 2 V
10 3 V
10 2 V
11 3 V
10 2 V
10 3 V
10 2 V
11 2 V
10 3 V
10 2 V
11 2 V
10 2 V
10 3 V
11 2 V
10 2 V
10 2 V
11 3 V
10 2 V
10 2 V
11 2 V
10 2 V
10 2 V
11 2 V
10 2 V
10 2 V
11 2 V
10 2 V
10 2 V
11 2 V
10 2 V
10 2 V
11 2 V
10 2 V
10 2 V
11 2 V
10 2 V
10 2 V
11 2 V
10 1 V
10 2 V
10 2 V
11 2 V
10 2 V
10 1 V
11 2 V
10 2 V
10 1 V
11 2 V
10 2 V
10 2 V
11 1 V
10 2 V
10 2 V
11 1 V
10 2 V
1.000 UL
LT0
353 300 M
10 0 V
11 0 V
10 0 V
10 0 V
11 0 V
10 0 V
10 0 V
11 0 V
10 0 V
10 0 V
11 0 V
10 0 V
10 0 V
11 0 V
10 0 V
10 1 V
11 0 V
10 0 V
10 0 V
11 1 V
10 0 V
10 0 V
11 1 V
10 0 V
10 1 V
10 1 V
11 0 V
10 1 V
10 1 V
11 1 V
10 0 V
10 1 V
11 1 V
10 1 V
10 1 V
11 2 V
10 1 V
10 1 V
11 1 V
10 2 V
10 1 V
11 2 V
10 1 V
10 2 V
11 1 V
10 2 V
10 2 V
11 1 V
10 2 V
10 2 V
11 2 V
10 2 V
10 2 V
11 2 V
10 2 V
10 2 V
11 2 V
10 2 V
10 2 V
10 2 V
11 3 V
10 2 V
10 2 V
11 2 V
10 2 V
10 3 V
11 2 V
10 2 V
10 3 V
11 2 V
10 2 V
10 3 V
11 2 V
10 4 V
10 7 V
11 7 V
10 7 V
10 7 V
11 7 V
10 6 V
10 7 V
11 7 V
10 7 V
10 7 V
11 7 V
10 7 V
10 7 V
11 7 V
10 7 V
10 7 V
10 7 V
11 7 V
10 7 V
10 7 V
11 7 V
10 7 V
10 7 V
11 7 V
10 7 V
10 7 V
11 7 V
10 7 V
10 7 V
11 6 V
10 7 V
10 7 V
11 7 V
10 7 V
10 7 V
11 7 V
10 7 V
10 7 V
11 7 V
10 7 V
10 7 V
11 7 V
10 7 V
10 7 V
11 7 V
10 7 V
10 7 V
11 7 V
10 7 V
10 7 V
10 7 V
11 7 V
10 7 V
10 7 V
11 6 V
10 7 V
10 7 V
11 7 V
10 7 V
10 7 V
11 7 V
10 7 V
10 7 V
11 7 V
10 7 V
10 7 V
11 7 V
10 7 V
10 7 V
11 7 V
10 7 V
10 7 V
11 7 V
10 7 V
10 7 V
11 7 V
10 7 V
10 7 V
11 6 V
10 7 V
10 7 V
10 7 V
11 7 V
10 7 V
10 7 V
11 7 V
10 7 V
10 7 V
11 7 V
10 7 V
10 7 V
11 7 V
10 7 V
10 7 V
11 7 V
10 7 V
10 7 V
11 7 V
10 7 V
10 7 V
11 7 V
10 7 V
10 7 V
11 6 V
10 7 V
10 7 V
11 7 V
10 7 V
10 7 V
11 7 V
10 7 V
10 7 V
11 7 V
10 7 V
10 7 V
10 7 V
11 7 V
10 7 V
10 7 V
11 7 V
10 6 V
10 4 V
11 3 V
10 3 V
10 3 V
11 3 V
10 3 V
10 3 V
11 3 V
10 3 V
10 3 V
11 3 V
10 3 V
10 2 V
11 3 V
10 3 V
10 3 V
11 3 V
10 2 V
10 3 V
11 3 V
10 2 V
10 3 V
11 3 V
10 2 V
10 3 V
10 2 V
11 3 V
10 2 V
10 3 V
11 2 V
10 3 V
10 2 V
11 3 V
10 2 V
10 2 V
11 3 V
10 2 V
10 2 V
11 3 V
10 2 V
10 2 V
11 2 V
10 3 V
10 2 V
11 2 V
10 2 V
10 2 V
11 2 V
10 3 V
10 2 V
11 2 V
10 2 V
10 2 V
11 2 V
10 2 V
10 2 V
10 2 V
11 2 V
10 2 V
10 2 V
11 1 V
10 2 V
10 2 V
11 2 V
10 2 V
10 2 V
11 2 V
10 1 V
10 2 V
11 2 V
10 2 V
10 1 V
11 2 V
10 2 V
10 2 V
11 1 V
10 2 V
10 2 V
11 1 V
10 2 V
10 1 V
11 2 V
10 2 V
10 1 V
11 2 V
10 1 V
10 2 V
11 1 V
10 2 V
10 2 V
10 1 V
11 1 V
10 2 V
10 1 V
11 2 V
10 1 V
10 2 V
11 1 V
10 2 V
10 1 V
11 1 V
10 2 V
10 1 V
11 1 V
10 2 V
1.000 UL
LT1
353 300 M
10 0 V
11 0 V
10 0 V
10 0 V
11 0 V
10 0 V
10 0 V
11 0 V
10 0 V
10 0 V
11 0 V
10 0 V
10 0 V
11 0 V
10 0 V
10 0 V
11 0 V
10 0 V
10 0 V
11 0 V
10 1 V
10 0 V
11 0 V
10 0 V
10 0 V
10 1 V
11 0 V
10 0 V
10 1 V
11 0 V
10 1 V
10 0 V
11 1 V
10 0 V
10 1 V
11 1 V
10 1 V
10 0 V
11 1 V
10 1 V
10 1 V
11 1 V
10 1 V
10 1 V
11 2 V
10 1 V
10 1 V
11 1 V
10 2 V
10 1 V
11 2 V
10 1 V
10 2 V
11 2 V
10 1 V
10 2 V
11 2 V
10 2 V
10 2 V
10 2 V
11 2 V
10 2 V
10 2 V
11 2 V
10 2 V
10 3 V
11 2 V
10 3 V
10 2 V
11 3 V
10 2 V
10 3 V
11 2 V
10 3 V
10 2 V
11 3 V
10 3 V
10 3 V
11 3 V
10 3 V
10 3 V
11 3 V
10 3 V
10 3 V
11 3 V
10 3 V
10 3 V
11 3 V
10 4 V
10 3 V
10 4 V
11 3 V
10 3 V
10 4 V
11 3 V
10 4 V
10 3 V
11 4 V
10 3 V
10 4 V
11 3 V
10 3 V
10 4 V
11 3 V
10 4 V
10 4 V
11 4 V
10 3 V
10 4 V
11 4 V
10 4 V
10 3 V
11 4 V
10 4 V
10 4 V
11 4 V
10 4 V
10 10 V
11 9 V
10 9 V
10 9 V
11 9 V
10 9 V
10 9 V
10 9 V
11 10 V
10 9 V
10 9 V
11 9 V
10 9 V
10 9 V
11 9 V
10 9 V
10 10 V
11 9 V
10 9 V
10 9 V
11 9 V
10 9 V
10 9 V
11 9 V
10 10 V
10 9 V
11 9 V
10 9 V
10 9 V
11 9 V
10 9 V
10 9 V
11 10 V
10 9 V
10 9 V
11 9 V
10 9 V
10 9 V
10 9 V
11 9 V
10 9 V
10 10 V
11 9 V
10 9 V
10 9 V
11 9 V
10 9 V
10 9 V
11 9 V
10 10 V
10 9 V
11 9 V
10 9 V
10 9 V
11 9 V
10 9 V
10 9 V
11 10 V
10 9 V
10 9 V
11 9 V
10 9 V
10 9 V
11 9 V
10 9 V
10 10 V
11 9 V
10 9 V
10 9 V
11 9 V
10 9 V
10 9 V
10 9 V
11 10 V
10 9 V
10 9 V
11 9 V
10 8 V
10 4 V
11 3 V
10 3 V
10 3 V
11 3 V
10 3 V
10 3 V
11 3 V
10 3 V
10 3 V
11 3 V
10 3 V
10 2 V
11 3 V
10 3 V
10 3 V
11 3 V
10 2 V
10 3 V
11 3 V
10 2 V
10 3 V
11 3 V
10 2 V
10 3 V
10 2 V
11 3 V
10 2 V
10 3 V
11 2 V
10 3 V
10 2 V
11 3 V
10 2 V
10 2 V
11 3 V
10 2 V
10 2 V
11 3 V
10 2 V
10 2 V
11 2 V
10 3 V
10 2 V
11 2 V
10 2 V
10 2 V
11 2 V
10 3 V
10 2 V
11 2 V
10 2 V
10 2 V
11 2 V
10 2 V
10 2 V
10 2 V
11 2 V
10 2 V
10 2 V
11 1 V
10 2 V
10 2 V
11 2 V
10 2 V
10 2 V
11 2 V
10 1 V
10 2 V
11 2 V
10 2 V
10 1 V
11 2 V
10 2 V
10 2 V
11 1 V
10 2 V
10 2 V
11 1 V
10 2 V
10 1 V
11 2 V
10 2 V
10 1 V
11 2 V
10 1 V
10 2 V
11 1 V
10 2 V
10 2 V
10 1 V
11 1 V
10 2 V
10 1 V
11 2 V
10 1 V
10 2 V
11 1 V
10 2 V
10 1 V
11 1 V
10 2 V
10 1 V
11 1 V
10 2 V
stroke
grestore
end
showpage
}}%
\put(920,1569){\makebox(0,0)[r]{$\lambda = 10$}}%
\put(920,1669){\makebox(0,0)[r]{$\lambda =4$}}%
\put(2396,1630){\makebox(0,0)[l]{x = 0.0001}}%
\put(2396,1405){\makebox(0,0)[l]{x = 0.001}}%
\put(2396,1200){\makebox(0,0)[l]{x = 0.01}}%
\put(1900,50){\makebox(0,0){\Large $b$ (fm) }}%
\put(100,1180){%
\special{ps: gsave currentpoint currentpoint translate
270 rotate neg exch neg exch translate}%
\makebox(0,0)[b]{\shortstack{\Large ${\hat \sigma} (b,x,Q^2)$~(mb) }}%
\special{ps: currentpoint grestore moveto}%
}%
\put(3450,200){\makebox(0,0){1}}%
\put(3140,200){\makebox(0,0){0.9}}%
\put(2830,200){\makebox(0,0){0.8}}%
\put(2520,200){\makebox(0,0){0.7}}%
\put(2210,200){\makebox(0,0){0.6}}%
\put(1900,200){\makebox(0,0){0.5}}%
\put(1590,200){\makebox(0,0){0.4}}%
\put(1280,200){\makebox(0,0){0.3}}%
\put(970,200){\makebox(0,0){0.2}}%
\put(660,200){\makebox(0,0){0.1}}%
\put(350,200){\makebox(0,0){0}}%
\put(300,2060){\makebox(0,0)[r]{45}}%
\put(300,1864){\makebox(0,0)[r]{40}}%
\put(300,1669){\makebox(0,0)[r]{35}}%
\put(300,1473){\makebox(0,0)[r]{30}}%
\put(300,1278){\makebox(0,0)[r]{25}}%
\put(300,1082){\makebox(0,0)[r]{20}}%
\put(300,887){\makebox(0,0)[r]{15}}%
\put(300,691){\makebox(0,0)[r]{10}}%
\put(300,496){\makebox(0,0)[r]{5}}%
\put(300,300){\makebox(0,0)[r]{0}}%
\end{picture}%
\endgroup
 

%% file: highw4.tex
\begingroup%
  \makeatletter%
  \newcommand{\GNUPLOTspecial}{%
    \@sanitize\catcode`\%=14\relax\special}%
  \setlength{\unitlength}{0.1bp}%
{\GNUPLOTspecial{!
/gnudict 256 dict def
gnudict begin
/Color false def
/Solid false def
/gnulinewidth 5.000 def
/userlinewidth gnulinewidth def
/vshift -33 def
/dl {10 mul} def
/hpt_ 31.5 def
/vpt_ 31.5 def
/hpt hpt_ def
/vpt vpt_ def
/M {moveto} bind def
/L {lineto} bind def
/R {rmoveto} bind def
/V {rlineto} bind def
/vpt2 vpt 2 mul def
/hpt2 hpt 2 mul def
/Lshow { currentpoint stroke M
  0 vshift R show } def
/Rshow { currentpoint stroke M
  dup stringwidth pop neg vshift R show } def
/Cshow { currentpoint stroke M
  dup stringwidth pop -2 div vshift R show } def
/UP { dup vpt_ mul /vpt exch def hpt_ mul /hpt exch def
  /hpt2 hpt 2 mul def /vpt2 vpt 2 mul def } def
/DL { Color {setrgbcolor Solid {pop []} if 0 setdash }
 {pop pop pop Solid {pop []} if 0 setdash} ifelse } def
/BL { stroke gnulinewidth 2 mul setlinewidth } def
/AL { stroke gnulinewidth 2 div setlinewidth } def
/UL { gnulinewidth mul /userlinewidth exch def } def
/PL { stroke userlinewidth setlinewidth } def
/LTb { BL [] 0 0 0 DL } def
/LTa { AL [1 dl 2 dl] 0 setdash 0 0 0 setrgbcolor } def
/LT0 { PL [] 1 0 0 DL } def
/LT1 { PL [4 dl 2 dl] 0 1 0 DL } def
/LT2 { PL [2 dl 3 dl] 0 0 1 DL } def
/LT3 { PL [1 dl 1.5 dl] 1 0 1 DL } def
/LT4 { PL [5 dl 2 dl 1 dl 2 dl] 0 1 1 DL } def
/LT5 { PL [4 dl 3 dl 1 dl 3 dl] 1 1 0 DL } def
/LT6 { PL [2 dl 2 dl 2 dl 4 dl] 0 0 0 DL } def
/LT7 { PL [2 dl 2 dl 2 dl 2 dl 2 dl 4 dl] 1 0.3 0 DL } def
/LT8 { PL [2 dl 2 dl 2 dl 2 dl 2 dl 2 dl 2 dl 4 dl] 0.5 0.5 0.5 DL } def
/Pnt { stroke [] 0 setdash
   gsave 1 setlinecap M 0 0 V stroke grestore } def
/Dia { stroke [] 0 setdash 2 copy vpt add M
  hpt neg vpt neg V hpt vpt neg V
  hpt vpt V hpt neg vpt V closepath stroke
  Pnt } def
/Pls { stroke [] 0 setdash vpt sub M 0 vpt2 V
  currentpoint stroke M
  hpt neg vpt neg R hpt2 0 V stroke
  } def
/Box { stroke [] 0 setdash 2 copy exch hpt sub exch vpt add M
  0 vpt2 neg V hpt2 0 V 0 vpt2 V
  hpt2 neg 0 V closepath stroke
  Pnt } def
/Crs { stroke [] 0 setdash exch hpt sub exch vpt add M
  hpt2 vpt2 neg V currentpoint stroke M
  hpt2 neg 0 R hpt2 vpt2 V stroke } def
/TriU { stroke [] 0 setdash 2 copy vpt 1.12 mul add M
  hpt neg vpt -1.62 mul V
  hpt 2 mul 0 V
  hpt neg vpt 1.62 mul V closepath stroke
  Pnt  } def
/Star { 2 copy Pls Crs } def
/BoxF { stroke [] 0 setdash exch hpt sub exch vpt add M
  0 vpt2 neg V  hpt2 0 V  0 vpt2 V
  hpt2 neg 0 V  closepath fill } def
/TriUF { stroke [] 0 setdash vpt 1.12 mul add M
  hpt neg vpt -1.62 mul V
  hpt 2 mul 0 V
  hpt neg vpt 1.62 mul V closepath fill } def
/TriD { stroke [] 0 setdash 2 copy vpt 1.12 mul sub M
  hpt neg vpt 1.62 mul V
  hpt 2 mul 0 V
  hpt neg vpt -1.62 mul V closepath stroke
  Pnt  } def
/TriDF { stroke [] 0 setdash vpt 1.12 mul sub M
  hpt neg vpt 1.62 mul V
  hpt 2 mul 0 V
  hpt neg vpt -1.62 mul V closepath fill} def
/DiaF { stroke [] 0 setdash vpt add M
  hpt neg vpt neg V hpt vpt neg V
  hpt vpt V hpt neg vpt V closepath fill } def
/Pent { stroke [] 0 setdash 2 copy gsave
  translate 0 hpt M 4 {72 rotate 0 hpt L} repeat
  closepath stroke grestore Pnt } def
/PentF { stroke [] 0 setdash gsave
  translate 0 hpt M 4 {72 rotate 0 hpt L} repeat
  closepath fill grestore } def
/Circle { stroke [] 0 setdash 2 copy
  hpt 0 360 arc stroke Pnt } def
/CircleF { stroke [] 0 setdash hpt 0 360 arc fill } def
/C0 { BL [] 0 setdash 2 copy moveto vpt 90 450  arc } bind def
/C1 { BL [] 0 setdash 2 copy        moveto
       2 copy  vpt 0 90 arc closepath fill
               vpt 0 360 arc closepath } bind def
/C2 { BL [] 0 setdash 2 copy moveto
       2 copy  vpt 90 180 arc closepath fill
               vpt 0 360 arc closepath } bind def
/C3 { BL [] 0 setdash 2 copy moveto
       2 copy  vpt 0 180 arc closepath fill
               vpt 0 360 arc closepath } bind def
/C4 { BL [] 0 setdash 2 copy moveto
       2 copy  vpt 180 270 arc closepath fill
               vpt 0 360 arc closepath } bind def
/C5 { BL [] 0 setdash 2 copy moveto
       2 copy  vpt 0 90 arc
       2 copy moveto
       2 copy  vpt 180 270 arc closepath fill
               vpt 0 360 arc } bind def
/C6 { BL [] 0 setdash 2 copy moveto
      2 copy  vpt 90 270 arc closepath fill
              vpt 0 360 arc closepath } bind def
/C7 { BL [] 0 setdash 2 copy moveto
      2 copy  vpt 0 270 arc closepath fill
              vpt 0 360 arc closepath } bind def
/C8 { BL [] 0 setdash 2 copy moveto
      2 copy vpt 270 360 arc closepath fill
              vpt 0 360 arc closepath } bind def
/C9 { BL [] 0 setdash 2 copy moveto
      2 copy  vpt 270 450 arc closepath fill
              vpt 0 360 arc closepath } bind def
/C10 { BL [] 0 setdash 2 copy 2 copy moveto vpt 270 360 arc closepath fill
       2 copy moveto
       2 copy vpt 90 180 arc closepath fill
               vpt 0 360 arc closepath } bind def
/C11 { BL [] 0 setdash 2 copy moveto
       2 copy  vpt 0 180 arc closepath fill
       2 copy moveto
       2 copy  vpt 270 360 arc closepath fill
               vpt 0 360 arc closepath } bind def
/C12 { BL [] 0 setdash 2 copy moveto
       2 copy  vpt 180 360 arc closepath fill
               vpt 0 360 arc closepath } bind def
/C13 { BL [] 0 setdash  2 copy moveto
       2 copy  vpt 0 90 arc closepath fill
       2 copy moveto
       2 copy  vpt 180 360 arc closepath fill
               vpt 0 360 arc closepath } bind def
/C14 { BL [] 0 setdash 2 copy moveto
       2 copy  vpt 90 360 arc closepath fill
               vpt 0 360 arc } bind def
/C15 { BL [] 0 setdash 2 copy vpt 0 360 arc closepath fill
               vpt 0 360 arc closepath } bind def
/Rec   { newpath 4 2 roll moveto 1 index 0 rlineto 0 exch rlineto
       neg 0 rlineto closepath } bind def
/Square { dup Rec } bind def
/Bsquare { vpt sub exch vpt sub exch vpt2 Square } bind def
/S0 { BL [] 0 setdash 2 copy moveto 0 vpt rlineto BL Bsquare } bind def
/S1 { BL [] 0 setdash 2 copy vpt Square fill Bsquare } bind def
/S2 { BL [] 0 setdash 2 copy exch vpt sub exch vpt Square fill Bsquare } bind def
/S3 { BL [] 0 setdash 2 copy exch vpt sub exch vpt2 vpt Rec fill Bsquare } bind def
/S4 { BL [] 0 setdash 2 copy exch vpt sub exch vpt sub vpt Square fill Bsquare } bind def
/S5 { BL [] 0 setdash 2 copy 2 copy vpt Square fill
       exch vpt sub exch vpt sub vpt Square fill Bsquare } bind def
/S6 { BL [] 0 setdash 2 copy exch vpt sub exch vpt sub vpt vpt2 Rec fill Bsquare } bind def
/S7 { BL [] 0 setdash 2 copy exch vpt sub exch vpt sub vpt vpt2 Rec fill
       2 copy vpt Square fill
       Bsquare } bind def
/S8 { BL [] 0 setdash 2 copy vpt sub vpt Square fill Bsquare } bind def
/S9 { BL [] 0 setdash 2 copy vpt sub vpt vpt2 Rec fill Bsquare } bind def
/S10 { BL [] 0 setdash 2 copy vpt sub vpt Square fill 2 copy exch vpt sub exch vpt Square fill
       Bsquare } bind def
/S11 { BL [] 0 setdash 2 copy vpt sub vpt Square fill 2 copy exch vpt sub exch vpt2 vpt Rec fill
       Bsquare } bind def
/S12 { BL [] 0 setdash 2 copy exch vpt sub exch vpt sub vpt2 vpt Rec fill Bsquare } bind def
/S13 { BL [] 0 setdash 2 copy exch vpt sub exch vpt sub vpt2 vpt Rec fill
       2 copy vpt Square fill Bsquare } bind def
/S14 { BL [] 0 setdash 2 copy exch vpt sub exch vpt sub vpt2 vpt Rec fill
       2 copy exch vpt sub exch vpt Square fill Bsquare } bind def
/S15 { BL [] 0 setdash 2 copy Bsquare fill Bsquare } bind def
/D0 { gsave translate 45 rotate 0 0 S0 stroke grestore } bind def
/D1 { gsave translate 45 rotate 0 0 S1 stroke grestore } bind def
/D2 { gsave translate 45 rotate 0 0 S2 stroke grestore } bind def
/D3 { gsave translate 45 rotate 0 0 S3 stroke grestore } bind def
/D4 { gsave translate 45 rotate 0 0 S4 stroke grestore } bind def
/D5 { gsave translate 45 rotate 0 0 S5 stroke grestore } bind def
/D6 { gsave translate 45 rotate 0 0 S6 stroke grestore } bind def
/D7 { gsave translate 45 rotate 0 0 S7 stroke grestore } bind def
/D8 { gsave translate 45 rotate 0 0 S8 stroke grestore } bind def
/D9 { gsave translate 45 rotate 0 0 S9 stroke grestore } bind def
/D10 { gsave translate 45 rotate 0 0 S10 stroke grestore } bind def
/D11 { gsave translate 45 rotate 0 0 S11 stroke grestore } bind def
/D12 { gsave translate 45 rotate 0 0 S12 stroke grestore } bind def
/D13 { gsave translate 45 rotate 0 0 S13 stroke grestore } bind def
/D14 { gsave translate 45 rotate 0 0 S14 stroke grestore } bind def
/D15 { gsave translate 45 rotate 0 0 S15 stroke grestore } bind def
/DiaE { stroke [] 0 setdash vpt add M
  hpt neg vpt neg V hpt vpt neg V
  hpt vpt V hpt neg vpt V closepath stroke } def
/BoxE { stroke [] 0 setdash exch hpt sub exch vpt add M
  0 vpt2 neg V hpt2 0 V 0 vpt2 V
  hpt2 neg 0 V closepath stroke } def
/TriUE { stroke [] 0 setdash vpt 1.12 mul add M
  hpt neg vpt -1.62 mul V
  hpt 2 mul 0 V
  hpt neg vpt 1.62 mul V closepath stroke } def
/TriDE { stroke [] 0 setdash vpt 1.12 mul sub M
  hpt neg vpt 1.62 mul V
  hpt 2 mul 0 V
  hpt neg vpt -1.62 mul V closepath stroke } def
/PentE { stroke [] 0 setdash gsave
  translate 0 hpt M 4 {72 rotate 0 hpt L} repeat
  closepath stroke grestore } def
/CircE { stroke [] 0 setdash 
  hpt 0 360 arc stroke } def
/Opaque { gsave closepath 1 setgray fill grestore 0 setgray closepath } def
/DiaW { stroke [] 0 setdash vpt add M
  hpt neg vpt neg V hpt vpt neg V
  hpt vpt V hpt neg vpt V Opaque stroke } def
/BoxW { stroke [] 0 setdash exch hpt sub exch vpt add M
  0 vpt2 neg V hpt2 0 V 0 vpt2 V
  hpt2 neg 0 V Opaque stroke } def
/TriUW { stroke [] 0 setdash vpt 1.12 mul add M
  hpt neg vpt -1.62 mul V
  hpt 2 mul 0 V
  hpt neg vpt 1.62 mul V Opaque stroke } def
/TriDW { stroke [] 0 setdash vpt 1.12 mul sub M
  hpt neg vpt 1.62 mul V
  hpt 2 mul 0 V
  hpt neg vpt -1.62 mul V Opaque stroke } def
/PentW { stroke [] 0 setdash gsave
  translate 0 hpt M 4 {72 rotate 0 hpt L} repeat
  Opaque stroke grestore } def
/CircW { stroke [] 0 setdash 
  hpt 0 360 arc Opaque stroke } def
/BoxFill { gsave Rec 1 setgray fill grestore } def
end
}}%
\begin{picture}(3600,2160)(0,0)%
{\GNUPLOTspecial{"
gnudict begin
gsave
0 0 translate
0.100 0.100 scale
0 setgray
newpath
1.000 UL
LTb
400 300 M
63 0 V
2987 0 R
-63 0 V
400 491 M
63 0 V
2987 0 R
-63 0 V
400 683 M
63 0 V
2987 0 R
-63 0 V
400 874 M
63 0 V
2987 0 R
-63 0 V
400 1065 M
63 0 V
2987 0 R
-63 0 V
400 1257 M
63 0 V
2987 0 R
-63 0 V
400 1448 M
63 0 V
2987 0 R
-63 0 V
400 1639 M
63 0 V
2987 0 R
-63 0 V
400 1830 M
63 0 V
2987 0 R
-63 0 V
400 2022 M
63 0 V
2987 0 R
-63 0 V
400 300 M
0 63 V
0 1697 R
0 -63 V
926 300 M
0 63 V
0 1697 R
0 -63 V
1452 300 M
0 63 V
0 1697 R
0 -63 V
1978 300 M
0 63 V
0 1697 R
0 -63 V
2503 300 M
0 63 V
0 1697 R
0 -63 V
3029 300 M
0 63 V
0 1697 R
0 -63 V
1.000 UL
LTb
400 300 M
3050 0 V
0 1760 V
-3050 0 V
400 300 L
1.000 UL
LT0
1452 1830 M
263 0 V
400 975 M
105 44 V
105 44 V
106 43 V
105 42 V
105 42 V
105 41 V
105 40 V
105 40 V
106 39 V
105 38 V
105 38 V
105 37 V
105 37 V
105 35 V
106 36 V
105 36 V
105 35 V
105 35 V
105 34 V
105 34 V
106 34 V
105 33 V
105 33 V
105 33 V
105 32 V
105 32 V
106 32 V
105 31 V
105 31 V
1.000 UL
LT1
1452 1730 M
263 0 V
400 775 M
105 32 V
105 32 V
106 33 V
105 32 V
105 33 V
105 33 V
105 33 V
105 33 V
106 34 V
105 33 V
105 34 V
105 33 V
105 34 V
105 34 V
106 34 V
105 34 V
105 34 V
105 34 V
105 35 V
105 34 V
106 35 V
105 35 V
105 35 V
105 34 V
105 35 V
105 35 V
106 36 V
105 35 V
105 35 V
1.000 UL
LT2
1452 1630 M
263 0 V
400 490 M
105 9 V
105 9 V
106 9 V
105 9 V
105 9 V
105 8 V
105 9 V
105 8 V
106 9 V
105 8 V
105 9 V
105 8 V
105 8 V
105 8 V
106 9 V
105 8 V
105 8 V
105 8 V
105 8 V
105 8 V
106 8 V
105 7 V
105 8 V
105 8 V
105 8 V
105 7 V
106 8 V
105 8 V
105 7 V
stroke
grestore
end
showpage
}}%
\put(1402,1630){\makebox(0,0)[r]{MRSTL $\lambda =4$}}%
\put(1402,1730){\makebox(0,0)[r]{CTEQ4L $\lambda =4$}}%
\put(1402,1830){\makebox(0,0)[r]{CTEQ4L $\lambda = 10$}}%
\put(1925,50){\makebox(0,0){\Large $W$ ~(GeV)}}%
\put(100,1180){%
\special{ps: gsave currentpoint currentpoint translate
270 rotate neg exch neg exch translate}%
\makebox(0,0)[b]{\shortstack{\Large$ \sigma (\gamma P \rightarrow J/\psi P) $(nb)}}%
\special{ps: currentpoint grestore moveto}%
}%
\put(3029,200){\makebox(0,0){800}}%
\put(2503,200){\makebox(0,0){700}}%
\put(1978,200){\makebox(0,0){600}}%
\put(1452,200){\makebox(0,0){500}}%
\put(926,200){\makebox(0,0){400}}%
\put(400,200){\makebox(0,0){300}}%
\put(350,2022){\makebox(0,0)[r]{900}}%
\put(350,1830){\makebox(0,0)[r]{800}}%
\put(350,1639){\makebox(0,0)[r]{700}}%
\put(350,1448){\makebox(0,0)[r]{600}}%
\put(350,1257){\makebox(0,0)[r]{500}}%
\put(350,1065){\makebox(0,0)[r]{400}}%
\put(350,874){\makebox(0,0)[r]{300}}%
\put(350,683){\makebox(0,0)[r]{200}}%
\put(350,491){\makebox(0,0)[r]{100}}%
\put(350,300){\makebox(0,0)[r]{0}}%
\end{picture}%
\endgroup
 

%% file: alpri3.tex
\begingroup%
  \makeatletter%
  \newcommand{\GNUPLOTspecial}{%
    \@sanitize\catcode`\%=14\relax\special}%
  \setlength{\unitlength}{0.1bp}%
{\GNUPLOTspecial{!
/gnudict 256 dict def
gnudict begin
/Color false def
/Solid false def
/gnulinewidth 5.000 def
/userlinewidth gnulinewidth def
/vshift -33 def
/dl {10 mul} def
/hpt_ 31.5 def
/vpt_ 31.5 def
/hpt hpt_ def
/vpt vpt_ def
/M {moveto} bind def
/L {lineto} bind def
/R {rmoveto} bind def
/V {rlineto} bind def
/vpt2 vpt 2 mul def
/hpt2 hpt 2 mul def
/Lshow { currentpoint stroke M
  0 vshift R show } def
/Rshow { currentpoint stroke M
  dup stringwidth pop neg vshift R show } def
/Cshow { currentpoint stroke M
  dup stringwidth pop -2 div vshift R show } def
/UP { dup vpt_ mul /vpt exch def hpt_ mul /hpt exch def
  /hpt2 hpt 2 mul def /vpt2 vpt 2 mul def } def
/DL { Color {setrgbcolor Solid {pop []} if 0 setdash }
 {pop pop pop Solid {pop []} if 0 setdash} ifelse } def
/BL { stroke gnulinewidth 2 mul setlinewidth } def
/AL { stroke gnulinewidth 2 div setlinewidth } def
/UL { gnulinewidth mul /userlinewidth exch def } def
/PL { stroke userlinewidth setlinewidth } def
/LTb { BL [] 0 0 0 DL } def
/LTa { AL [1 dl 2 dl] 0 setdash 0 0 0 setrgbcolor } def
/LT0 { PL [] 1 0 0 DL } def
/LT1 { PL [4 dl 2 dl] 0 1 0 DL } def
/LT2 { PL [2 dl 3 dl] 0 0 1 DL } def
/LT3 { PL [1 dl 1.5 dl] 1 0 1 DL } def
/LT4 { PL [5 dl 2 dl 1 dl 2 dl] 0 1 1 DL } def
/LT5 { PL [4 dl 3 dl 1 dl 3 dl] 1 1 0 DL } def
/LT6 { PL [2 dl 2 dl 2 dl 4 dl] 0 0 0 DL } def
/LT7 { PL [2 dl 2 dl 2 dl 2 dl 2 dl 4 dl] 1 0.3 0 DL } def
/LT8 { PL [2 dl 2 dl 2 dl 2 dl 2 dl 2 dl 2 dl 4 dl] 0.5 0.5 0.5 DL } def
/Pnt { stroke [] 0 setdash
   gsave 1 setlinecap M 0 0 V stroke grestore } def
/Dia { stroke [] 0 setdash 2 copy vpt add M
  hpt neg vpt neg V hpt vpt neg V
  hpt vpt V hpt neg vpt V closepath stroke
  Pnt } def
/Pls { stroke [] 0 setdash vpt sub M 0 vpt2 V
  currentpoint stroke M
  hpt neg vpt neg R hpt2 0 V stroke
  } def
/Box { stroke [] 0 setdash 2 copy exch hpt sub exch vpt add M
  0 vpt2 neg V hpt2 0 V 0 vpt2 V
  hpt2 neg 0 V closepath stroke
  Pnt } def
/Crs { stroke [] 0 setdash exch hpt sub exch vpt add M
  hpt2 vpt2 neg V currentpoint stroke M
  hpt2 neg 0 R hpt2 vpt2 V stroke } def
/TriU { stroke [] 0 setdash 2 copy vpt 1.12 mul add M
  hpt neg vpt -1.62 mul V
  hpt 2 mul 0 V
  hpt neg vpt 1.62 mul V closepath stroke
  Pnt  } def
/Star { 2 copy Pls Crs } def
/BoxF { stroke [] 0 setdash exch hpt sub exch vpt add M
  0 vpt2 neg V  hpt2 0 V  0 vpt2 V
  hpt2 neg 0 V  closepath fill } def
/TriUF { stroke [] 0 setdash vpt 1.12 mul add M
  hpt neg vpt -1.62 mul V
  hpt 2 mul 0 V
  hpt neg vpt 1.62 mul V closepath fill } def
/TriD { stroke [] 0 setdash 2 copy vpt 1.12 mul sub M
  hpt neg vpt 1.62 mul V
  hpt 2 mul 0 V
  hpt neg vpt -1.62 mul V closepath stroke
  Pnt  } def
/TriDF { stroke [] 0 setdash vpt 1.12 mul sub M
  hpt neg vpt 1.62 mul V
  hpt 2 mul 0 V
  hpt neg vpt -1.62 mul V closepath fill} def
/DiaF { stroke [] 0 setdash vpt add M
  hpt neg vpt neg V hpt vpt neg V
  hpt vpt V hpt neg vpt V closepath fill } def
/Pent { stroke [] 0 setdash 2 copy gsave
  translate 0 hpt M 4 {72 rotate 0 hpt L} repeat
  closepath stroke grestore Pnt } def
/PentF { stroke [] 0 setdash gsave
  translate 0 hpt M 4 {72 rotate 0 hpt L} repeat
  closepath fill grestore } def
/Circle { stroke [] 0 setdash 2 copy
  hpt 0 360 arc stroke Pnt } def
/CircleF { stroke [] 0 setdash hpt 0 360 arc fill } def
/C0 { BL [] 0 setdash 2 copy moveto vpt 90 450  arc } bind def
/C1 { BL [] 0 setdash 2 copy        moveto
       2 copy  vpt 0 90 arc closepath fill
               vpt 0 360 arc closepath } bind def
/C2 { BL [] 0 setdash 2 copy moveto
       2 copy  vpt 90 180 arc closepath fill
               vpt 0 360 arc closepath } bind def
/C3 { BL [] 0 setdash 2 copy moveto
       2 copy  vpt 0 180 arc closepath fill
               vpt 0 360 arc closepath } bind def
/C4 { BL [] 0 setdash 2 copy moveto
       2 copy  vpt 180 270 arc closepath fill
               vpt 0 360 arc closepath } bind def
/C5 { BL [] 0 setdash 2 copy moveto
       2 copy  vpt 0 90 arc
       2 copy moveto
       2 copy  vpt 180 270 arc closepath fill
               vpt 0 360 arc } bind def
/C6 { BL [] 0 setdash 2 copy moveto
      2 copy  vpt 90 270 arc closepath fill
              vpt 0 360 arc closepath } bind def
/C7 { BL [] 0 setdash 2 copy moveto
      2 copy  vpt 0 270 arc closepath fill
              vpt 0 360 arc closepath } bind def
/C8 { BL [] 0 setdash 2 copy moveto
      2 copy vpt 270 360 arc closepath fill
              vpt 0 360 arc closepath } bind def
/C9 { BL [] 0 setdash 2 copy moveto
      2 copy  vpt 270 450 arc closepath fill
              vpt 0 360 arc closepath } bind def
/C10 { BL [] 0 setdash 2 copy 2 copy moveto vpt 270 360 arc closepath fill
       2 copy moveto
       2 copy vpt 90 180 arc closepath fill
               vpt 0 360 arc closepath } bind def
/C11 { BL [] 0 setdash 2 copy moveto
       2 copy  vpt 0 180 arc closepath fill
       2 copy moveto
       2 copy  vpt 270 360 arc closepath fill
               vpt 0 360 arc closepath } bind def
/C12 { BL [] 0 setdash 2 copy moveto
       2 copy  vpt 180 360 arc closepath fill
               vpt 0 360 arc closepath } bind def
/C13 { BL [] 0 setdash  2 copy moveto
       2 copy  vpt 0 90 arc closepath fill
       2 copy moveto
       2 copy  vpt 180 360 arc closepath fill
               vpt 0 360 arc closepath } bind def
/C14 { BL [] 0 setdash 2 copy moveto
       2 copy  vpt 90 360 arc closepath fill
               vpt 0 360 arc } bind def
/C15 { BL [] 0 setdash 2 copy vpt 0 360 arc closepath fill
               vpt 0 360 arc closepath } bind def
/Rec   { newpath 4 2 roll moveto 1 index 0 rlineto 0 exch rlineto
       neg 0 rlineto closepath } bind def
/Square { dup Rec } bind def
/Bsquare { vpt sub exch vpt sub exch vpt2 Square } bind def
/S0 { BL [] 0 setdash 2 copy moveto 0 vpt rlineto BL Bsquare } bind def
/S1 { BL [] 0 setdash 2 copy vpt Square fill Bsquare } bind def
/S2 { BL [] 0 setdash 2 copy exch vpt sub exch vpt Square fill Bsquare } bind def
/S3 { BL [] 0 setdash 2 copy exch vpt sub exch vpt2 vpt Rec fill Bsquare } bind def
/S4 { BL [] 0 setdash 2 copy exch vpt sub exch vpt sub vpt Square fill Bsquare } bind def
/S5 { BL [] 0 setdash 2 copy 2 copy vpt Square fill
       exch vpt sub exch vpt sub vpt Square fill Bsquare } bind def
/S6 { BL [] 0 setdash 2 copy exch vpt sub exch vpt sub vpt vpt2 Rec fill Bsquare } bind def
/S7 { BL [] 0 setdash 2 copy exch vpt sub exch vpt sub vpt vpt2 Rec fill
       2 copy vpt Square fill
       Bsquare } bind def
/S8 { BL [] 0 setdash 2 copy vpt sub vpt Square fill Bsquare } bind def
/S9 { BL [] 0 setdash 2 copy vpt sub vpt vpt2 Rec fill Bsquare } bind def
/S10 { BL [] 0 setdash 2 copy vpt sub vpt Square fill 2 copy exch vpt sub exch vpt Square fill
       Bsquare } bind def
/S11 { BL [] 0 setdash 2 copy vpt sub vpt Square fill 2 copy exch vpt sub exch vpt2 vpt Rec fill
       Bsquare } bind def
/S12 { BL [] 0 setdash 2 copy exch vpt sub exch vpt sub vpt2 vpt Rec fill Bsquare } bind def
/S13 { BL [] 0 setdash 2 copy exch vpt sub exch vpt sub vpt2 vpt Rec fill
       2 copy vpt Square fill Bsquare } bind def
/S14 { BL [] 0 setdash 2 copy exch vpt sub exch vpt sub vpt2 vpt Rec fill
       2 copy exch vpt sub exch vpt Square fill Bsquare } bind def
/S15 { BL [] 0 setdash 2 copy Bsquare fill Bsquare } bind def
/D0 { gsave translate 45 rotate 0 0 S0 stroke grestore } bind def
/D1 { gsave translate 45 rotate 0 0 S1 stroke grestore } bind def
/D2 { gsave translate 45 rotate 0 0 S2 stroke grestore } bind def
/D3 { gsave translate 45 rotate 0 0 S3 stroke grestore } bind def
/D4 { gsave translate 45 rotate 0 0 S4 stroke grestore } bind def
/D5 { gsave translate 45 rotate 0 0 S5 stroke grestore } bind def
/D6 { gsave translate 45 rotate 0 0 S6 stroke grestore } bind def
/D7 { gsave translate 45 rotate 0 0 S7 stroke grestore } bind def
/D8 { gsave translate 45 rotate 0 0 S8 stroke grestore } bind def
/D9 { gsave translate 45 rotate 0 0 S9 stroke grestore } bind def
/D10 { gsave translate 45 rotate 0 0 S10 stroke grestore } bind def
/D11 { gsave translate 45 rotate 0 0 S11 stroke grestore } bind def
/D12 { gsave translate 45 rotate 0 0 S12 stroke grestore } bind def
/D13 { gsave translate 45 rotate 0 0 S13 stroke grestore } bind def
/D14 { gsave translate 45 rotate 0 0 S14 stroke grestore } bind def
/D15 { gsave translate 45 rotate 0 0 S15 stroke grestore } bind def
/DiaE { stroke [] 0 setdash vpt add M
  hpt neg vpt neg V hpt vpt neg V
  hpt vpt V hpt neg vpt V closepath stroke } def
/BoxE { stroke [] 0 setdash exch hpt sub exch vpt add M
  0 vpt2 neg V hpt2 0 V 0 vpt2 V
  hpt2 neg 0 V closepath stroke } def
/TriUE { stroke [] 0 setdash vpt 1.12 mul add M
  hpt neg vpt -1.62 mul V
  hpt 2 mul 0 V
  hpt neg vpt 1.62 mul V closepath stroke } def
/TriDE { stroke [] 0 setdash vpt 1.12 mul sub M
  hpt neg vpt 1.62 mul V
  hpt 2 mul 0 V
  hpt neg vpt -1.62 mul V closepath stroke } def
/PentE { stroke [] 0 setdash gsave
  translate 0 hpt M 4 {72 rotate 0 hpt L} repeat
  closepath stroke grestore } def
/CircE { stroke [] 0 setdash 
  hpt 0 360 arc stroke } def
/Opaque { gsave closepath 1 setgray fill grestore 0 setgray closepath } def
/DiaW { stroke [] 0 setdash vpt add M
  hpt neg vpt neg V hpt vpt neg V
  hpt vpt V hpt neg vpt V Opaque stroke } def
/BoxW { stroke [] 0 setdash exch hpt sub exch vpt add M
  0 vpt2 neg V hpt2 0 V 0 vpt2 V
  hpt2 neg 0 V Opaque stroke } def
/TriUW { stroke [] 0 setdash vpt 1.12 mul add M
  hpt neg vpt -1.62 mul V
  hpt 2 mul 0 V
  hpt neg vpt 1.62 mul V Opaque stroke } def
/TriDW { stroke [] 0 setdash vpt 1.12 mul sub M
  hpt neg vpt 1.62 mul V
  hpt 2 mul 0 V
  hpt neg vpt -1.62 mul V Opaque stroke } def
/PentW { stroke [] 0 setdash gsave
  translate 0 hpt M 4 {72 rotate 0 hpt L} repeat
  Opaque stroke grestore } def
/CircW { stroke [] 0 setdash 
  hpt 0 360 arc Opaque stroke } def
/BoxFill { gsave Rec 1 setgray fill grestore } def
end
}}%
\begin{picture}(3600,2160)(0,0)%
{\GNUPLOTspecial{"
gnudict begin
gsave
0 0 translate
0.100 0.100 scale
0 setgray
newpath
1.000 UL
LTb
450 300 M
63 0 V
2937 0 R
-63 0 V
450 520 M
63 0 V
2937 0 R
-63 0 V
450 740 M
63 0 V
2937 0 R
-63 0 V
450 960 M
63 0 V
2937 0 R
-63 0 V
450 1180 M
63 0 V
2937 0 R
-63 0 V
450 1400 M
63 0 V
2937 0 R
-63 0 V
450 1620 M
63 0 V
2937 0 R
-63 0 V
450 1840 M
63 0 V
2937 0 R
-63 0 V
450 2060 M
63 0 V
2937 0 R
-63 0 V
450 300 M
0 63 V
0 1697 R
0 -63 V
950 300 M
0 63 V
0 1697 R
0 -63 V
1450 300 M
0 63 V
0 1697 R
0 -63 V
1950 300 M
0 63 V
0 1697 R
0 -63 V
2450 300 M
0 63 V
0 1697 R
0 -63 V
2950 300 M
0 63 V
0 1697 R
0 -63 V
3450 300 M
0 63 V
0 1697 R
0 -63 V
1.000 UL
LTb
450 300 M
3000 0 V
0 1760 V
-3000 0 V
450 300 L
1.000 UL
LT0
3087 1947 M
263 0 V
550 1499 M
650 1183 L
750 1027 L
850 924 L
950 847 L
100 -60 V
100 -50 V
100 -42 V
100 -36 V
100 -32 V
100 -28 V
100 -26 V
100 -23 V
100 -21 V
100 -19 V
100 -18 V
100 -16 V
100 -16 V
100 -14 V
100 -13 V
100 -13 V
100 -12 V
100 -11 V
100 -11 V
100 -10 V
100 -10 V
100 -9 V
100 -9 V
100 -8 V
100 -8 V
1.000 UL
LT1
3087 1847 M
263 0 V
550 1950 M
650 1294 L
750 1049 L
850 911 L
950 819 L
100 -67 V
100 -52 V
100 -42 V
100 -35 V
100 -30 V
100 -25 V
100 -23 V
100 -16 V
100 -14 V
100 -16 V
100 -15 V
100 -14 V
100 -14 V
100 -14 V
100 -13 V
100 -13 V
100 -12 V
100 -11 V
100 -11 V
100 -11 V
100 -11 V
100 -9 V
100 -9 V
100 -9 V
100 -8 V
stroke
grestore
end
showpage
}}%
\put(3037,1847){\makebox(0,0)[r]{$\lambda = 10$}}%
\put(3037,1947){\makebox(0,0)[r]{$\lambda = 4$}}%
\put(1950,50){\makebox(0,0){\Large $W$ (GeV) }}%
\put(100,1180){%
\special{ps: gsave currentpoint currentpoint translate
270 rotate neg exch neg exch translate}%
\makebox(0,0)[b]{\shortstack{\Large $<\alpha^{\prime}>$ (GeV$^{-2}$)}}%
\special{ps: currentpoint grestore moveto}%
}%
\put(3450,200){\makebox(0,0){300}}%
\put(2950,200){\makebox(0,0){250}}%
\put(2450,200){\makebox(0,0){200}}%
\put(1950,200){\makebox(0,0){150}}%
\put(1450,200){\makebox(0,0){100}}%
\put(950,200){\makebox(0,0){50}}%
\put(450,200){\makebox(0,0){0}}%
\put(400,2060){\makebox(0,0)[r]{0.16}}%
\put(400,1840){\makebox(0,0)[r]{0.15}}%
\put(400,1620){\makebox(0,0)[r]{0.14}}%
\put(400,1400){\makebox(0,0)[r]{0.13}}%
\put(400,1180){\makebox(0,0)[r]{0.12}}%
\put(400,960){\makebox(0,0)[r]{0.11}}%
\put(400,740){\makebox(0,0)[r]{0.1}}%
\put(400,520){\makebox(0,0)[r]{0.09}}%
\put(400,300){\makebox(0,0)[r]{0.08}}%
\end{picture}%
\endgroup
 